\documentclass{aa}

\usepackage{txfonts}
\usepackage{graphicx}
\usepackage{soul}

\begin{document}

   \title{Globular cluster candidates in the Galactic bulge:\\
            Gaia and VVV view of the latest discoveries}
   \titlerunning{Gaia  and VVV view of the last bulge Globular clusters candidates}
   \authorrunning{Gran et al.}

   \author{F. Gran\inst{1}\fnmsep\inst{2},
          M. Zoccali\inst{1}\fnmsep\inst{2},
          R. Contreras Ramos\inst{1}\fnmsep\inst{2},
          E. Valenti\inst{3}\fnmsep\inst{4},
          A. Rojas-Arriagada\inst{1}\fnmsep\inst{2},\\
          J. A. Carballo-Bello\inst{1},
          J. Alonso-Garc\'ia\inst{6}\fnmsep\inst{2},
          D. Minniti\inst{2}\fnmsep\inst{5}\fnmsep\inst{7},
          M. Rejkuba\inst{3}\fnmsep\inst{8},
          F. Surot\inst{1}
          }

   \institute{Instituto de Astrof\'isica, Av. Vicuna Mackenna 4860, Santiago, Chile
            \and
            Instituto Milenio de Astrof\'isica, Santiago, Chile
            \and
            European Southern Observatory, Karl Schwarzschild-Strabe 2, D-85748 Garching bei Munchen, Germany
            \and
            Excellence Cluster ORIGINS, Boltzmann-Strasse 2, D-85748 Garching bei M\"unchen, Germany
            \and
            Departamento de Ciencias F\'isicas, Facultad de Ciencias Exactas, Universidad Andr\'es Bello, Av. Fern\'andez Concha 700, Las Condes, Santiago, Chile
            \and
            Centro de Astronom\'ia (CITEVA), Universidad de Antofagasta, Av. Angamos 601, Antofagasta, Chile
            \and
            Vatican Observatory, V00120 Vatican City State, Italy
            \and
            Excellence Cluster Universe, Boltzmannstr. 2, 85748 Garching, Germany
             }

   \date{Started Aug 2, 2018; Received Dec 27, 2018; Accepted: Apr 14, 2019}


  \abstract
{Thanks to the recent wide-area photometric  surveys, the number of star cluster
  candidates have risen exponentially in  the last few years.   Most detections,
  however, are based only on the presence  of an overdensity of stars in a given  
  region,  or  an overdensity  of variable stars,
  regardless  of their  distance. As  candidates, their  detection has  not been
  dynamically confirmed. Therefore, it is  currently unknown how many, and which ones, 
  of the published candidates, are true clusters, and which ones are chance  alignments.}
{We present  a method  to  detect and confirm  star clusters
  based on the  spatial distribution, coherence in motion and  appearance on the
  color-magnitude diagram.  We explain  and  apply it  to one  new star
  cluster, and several candidate star clusters published in the literature.}
{The presented method is based on data from the Second Data Release of Gaia 
  complemented with data from the VISTA Variables in the V\'\i a L\'actea survey for
  the innermost bulge regions. It consists of a nearest neighbors algorithm applied
  simultaneously over spatial coordinates, star color, and proper motions, in order to detect groups
  of stars that are close in the sky, move coherently and define narrow sequences
  in the color-magnitude diagram, such as a young main sequence or a red giant branch.}
{When tested in the bulge area ($-10<\ell\ {\rm (deg)}<+10$; $-10<b\ {\rm (deg)}<+10$)
    the method successfully recovered several known young and old star clusters.
  We report here the detection of one new, likely old star cluster, while  deferring the 
  others to a forthcoming paper. Additionally, the code has been  applied to the position of  
  93 candidate star clusters published in  the literature.  As a result, only two 
  of them are confirmed as coherently moving groups of stars at their nominal positions.}
{}


   \maketitle

\section{Introduction}
\label{sec:Intro}

Star clusters  are invaluable  astrophysics tools  for a  number of  reasons. In
addition to being a laboratory  for stellar  evolution,  including  chemical
evolution and self-enrichment, they are among the few objects for which a rather
precise age can  be measured. If young and massive,  their mass function closely
resembles the initial mass function \citep{Leigh12,Webb15}.  For all the others,
the radial variation of the present-day  mass function allows us to quantify the
internal  dynamical evolution,  while the  global present-day mass  function is
related   to   the    dynamical   interaction   of   the    cluster   with   its
environment. Although we do not know  exactly how and where massive cluster form \citep{Forbes18}, we do
know that stars do  not form in isolation.  Most of them form  in groups, if not
in massive clusters, and  therefore the struggle to get a  census of the cluster 
population of a  galaxy is motivated by  their relevance for the  building up of
the field star population \citep{Kruijssen18}.

In the  last decades, the number  of candidate star  clusters reported  in the
literature has increased significantly, thanks  to wide area photometric surveys
that  allowed to  literally scan  the  sky in  search  of groups \citep[][and references therein]{Koposov07, 
Belokurov10, Munoz12, Ortolani12, Belokurov14, Laevens14, Bechtol15, Kim15, 
Laevens15a, Laevens15b, Koposov17, Luque17, Ryu18}.   Only in  the 
direction of  the Galactic bulge, from  VISTA Variables in the  V\'\i a L\'actea
\citep[VVV,][]{Minniti10}   survey   images,  \citet{Minniti11}   reported   the
discovery  of the  candidate  globular  cluster (GC) VVV-CL001,  \citet{MoniBidin11}
identified  two more  candidate  clusters CL002  and CL003,  \citet{Borissova14}
lists  58  new  infrared  star  cluster candidates,  and  another  one,  already
catalogued as cluster candidate, was further analysed in search for variables by
\citet{Minniti17a}.   Another  84  old   cluster  candidates  were  reported  by
\citet{Minniti17b}, \citet{Minniti17c} and \citet{Minniti17d} based on detection
of  spatial overdensities,  projected overdensities  of RR  Lyrae variables  and
projected overdensities of RR Lyrae and type II Cepheids, respectively.
Finally, another five globular cluster candidates were identified by \citet{Camargo18}
by visual inspection Wide-field Infrared Survey Explorer (WISE) images.

Although new  star clusters  can be initially  identified as  overdensities, the 
only way  to confirm  their cluster nature  is to verify  that their  stars move
coherently in space,  i.e., they are gravitationally bound. This  can be done by
measuring either radial velocities or proper  motions (PMs) of stars in a region
centered  at the  center of  the spatial  overdensity. In  the present  paper, we
describe a method to identify  unusual concentrations of stars simultaneously in
the  plane of  the sky,  in the  vector  point diagram  (VPD) and  in the  
color-magnitude diagram (CMD).

The data  used here come mostly  from the second  data release (DR2) of  the Gaia 
mission  \citep{GaiaMission,Gaia}, including  positions, PMs \citep{Lindegren18} and  magnitudes in three
photometric  bands for  all  of the  stars \citep{Riello18, Evans18}.  
The  work  of  \cite{Pancino17}, \cite{Helmi18}, and \cite{Vasiliev18}
illustrates the potential of Gaia  to characterize the globular cluster known
up to date.  In the region close to the Galactic plane, at latitudes $|b|<3^\circ$
the Gaia  catalogue is highly incomplete  due to the large interstellar extinction
affecting optical  fluxes, and, to a  minor extent, to the  higher stellar surface
density coupled with the limited transmission  bandpass of the satellite.  In this
region, Gaia  detects  almost exclusively the brightest blue disk  stars, while it
is virtually blind  to the bulge red  giants.  On the contrary,  the near-infrared
(near-IR) VVV observations are optimized for  the reddest bulge giants, and the PM
catalogues obtained with the method described in \citet{ContrerasRamos17} are both
deeper and more precise  than the Gaia  ones.  We use the VVV  PMs, in addition to
Gaia  to analyze candidate clusters at latitudes $|b|<3^\circ$.

The   paper  is   organized  as   follows:  Sec.~\ref{sec:method}   describes  the 
automated method  to detect --and  simultaneously confirm-- new  star clusters,
including  the   detection  of  a  new   old  star  cluster  labelled   {\tt  Gran
1}. Sec~\ref{sec:candidates}  presents an  analysis of the  PM of  stars within
1~arcmin  from the  nominal position  of  a sample  of 93  old cluster  candidates
reported  by \citet{Minniti11,  MoniBidin11,  Minniti17b, Minniti17c,  Minniti17d,
Camargo18,  Bica18}.   The present  clustering  method  was then  applied,  with
relaxed parameters, within a region of 2~arcmins across the center of each cluster
candidate,  in order  to double  check against  possible errors  in the  candidate
estimated centers.   Finally, Sec.~\ref{sec:summary}  summarizes our  results. The
VPDs  and CMDs  of all  the  unconfirmed candidate  clusters are  included in  the
Appendix~\ref{sec:Appendix}.

\section{A method to detect coherent groups}
\label{sec:method}

The region selected  to search for star  clusters was the whole  bulge area within
$-10^\circ  \leq (\ell,\  b) \leq  10^\circ$.  This  region was  divided in  small
circles  of 0.8$^\circ$  radius,  centered  on every  integer  degree in  Galactic
latitude and  longitude. An algorithm was  developed in order to  search for stars
that have  an unusually  large number  of neighbors in  a 5  dimension phase-space
including  coordinates,  color,  and  PMs.   For  each  star   with  complete 
information, the  algorithm counts the number  of neighbors within 1  arcmin in
space, one mas/yr in  PM and 1 mag in color using  the K-Dimensional Tree (KDTree)
implementation on {\tt scikit-learn}  \citep{scikit-learn}.  For each given field,
a minimum  threshold of  10 neighbors  per star  was imposed,  in order  to ensure
statistical significance of  the results.  Stars with more than  10 neighbors 
within  the complete  magnitude range  were then  searched for  groups, in  the
phase-space  mentioned  above,  with   the  Density-Based  Spatial  Clustering  of
Applications with Noise (DBSCAN) algorithm  \citep{DBSCAN} also implemented in the
{\it scikit-learn} package. Groups were kept as {\it candidate clusters}, and then
visually inspected, only if they contained at least 20 members.  Note that a group
of neighbor  stars define ``a neighborhood''  that may end up  being significantly
larger than the phase-space radius (1 arcmin,  1 ${\rm mas\ yr^{-1}}$, 1 mag) used
for the initial search around each star.

It should be  noted that several false positives are  detected by the algorithm,
close to the mean  PM of bulge ($-6,-0.2$) and disk  ($-2,-0.5$) field stars, in
the VPD.  After visual inspection,  however, we  keep only clusters  that define
narrow sequences in the CMD, significantly different -by shape and/or tightness-
from the main branches of the bulge+disk CMD.

A very first step for the validation of the algorithm described above is that it
must be able to rediscover all the clusters known to exist in the area explored.
In order to verify this, the catalog of overdensities was cross-matched with the
latest  version  of  the  {\tt  Global   survey  of  Milky  Way  star  clusters}
\citep[MWSC,][]{MWSC} catalog  in order to  identify all the clusters  that were
previously known. Indeed, we detect all  the 45 known bulge globular clusters in
this region and $\sim  17\%$ (22/129) of the open clusters up  to 5~arcmins from
their  nominal center.   Five known  GCs located  in the  region $|b|<2.5^\circ$
(Terzan~4, Terzan~6,  Terzan~10, 2MASS-GC002 and  Djorg~2) were not  detected in
Gaia, but  they were detected  in VVV.   Another 4 that  are listed in  the MWSC
catalogue as  {\it candidate} GCs,  namely ESO~456-09, ESO~373-12,  FSR~0019, and 
FSR~0025, were not detected in any of the two surveys.  Based on our data, we do
not find any evidence allowing us to confirm their cluster nature.

\begin{figure}
  \centering
  \includegraphics[width=9cm,angle=0]{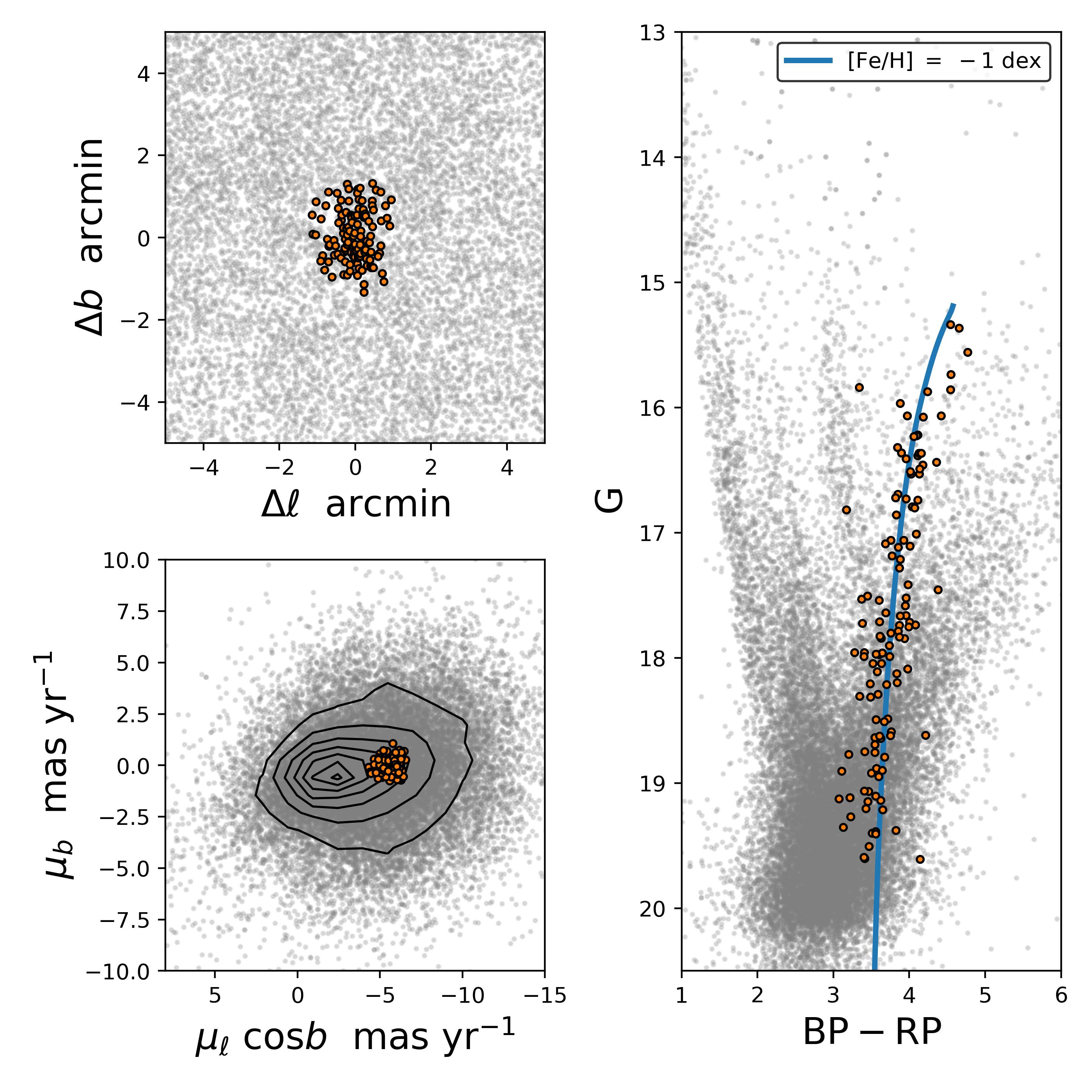}
  \caption{The  globular  cluster  Terzan~1 as detected by  our algorithm.  Marked
    (orange)  points represent  the detected overdensity along with background
    stars (gray points) within 15~arcmins from the detected cluster centroid.
    Gray stars define three populated sequences, from left to right: the disk MS,
    the disk red clump [both spread along the line of sight] and the bulge upper
    red giant branch, with a PGPUC isochrone overplotted.}
  \label{fig:Ter1}
\end{figure}

Figure~\ref{fig:Ter1}   shows  the   detection   of   Terzan~1  \citep[$\ell   =
  -2.44^\circ, \ b  = 0.99^\circ$,][]{Ortolani99}.  The mean PM  of this cluster
is very similar to the mean PM of bulge stars (lower left panel).  Nonetheless,
the algorithm detects a higher concentration  of stars, in the VPD, with respect
to a  smoother background.  Of  course, the detection  is also triggered  by the
clustering of Terzan~1  stars both in the  plane of the sky (upper  left) and in
the  CMD (right).   Note  that, as  explained above,  the  algorithm can  detect
compact sequences of  stars, in addition to roundish groups.  We overplot here a
PGPUC  isochrone \cite{PGPUC}  for  a  ${\rm [Fe/H]} = -1$ dex, 12 Gyr simple 
stellar population, at  the cluster  distance  quoted  in  \citet{MWSC},   
in  order  to  confirm  that  the overdensity is indeed Terzan~1.

\begin{figure}
  \centering
  \includegraphics[width=9cm,angle=0]{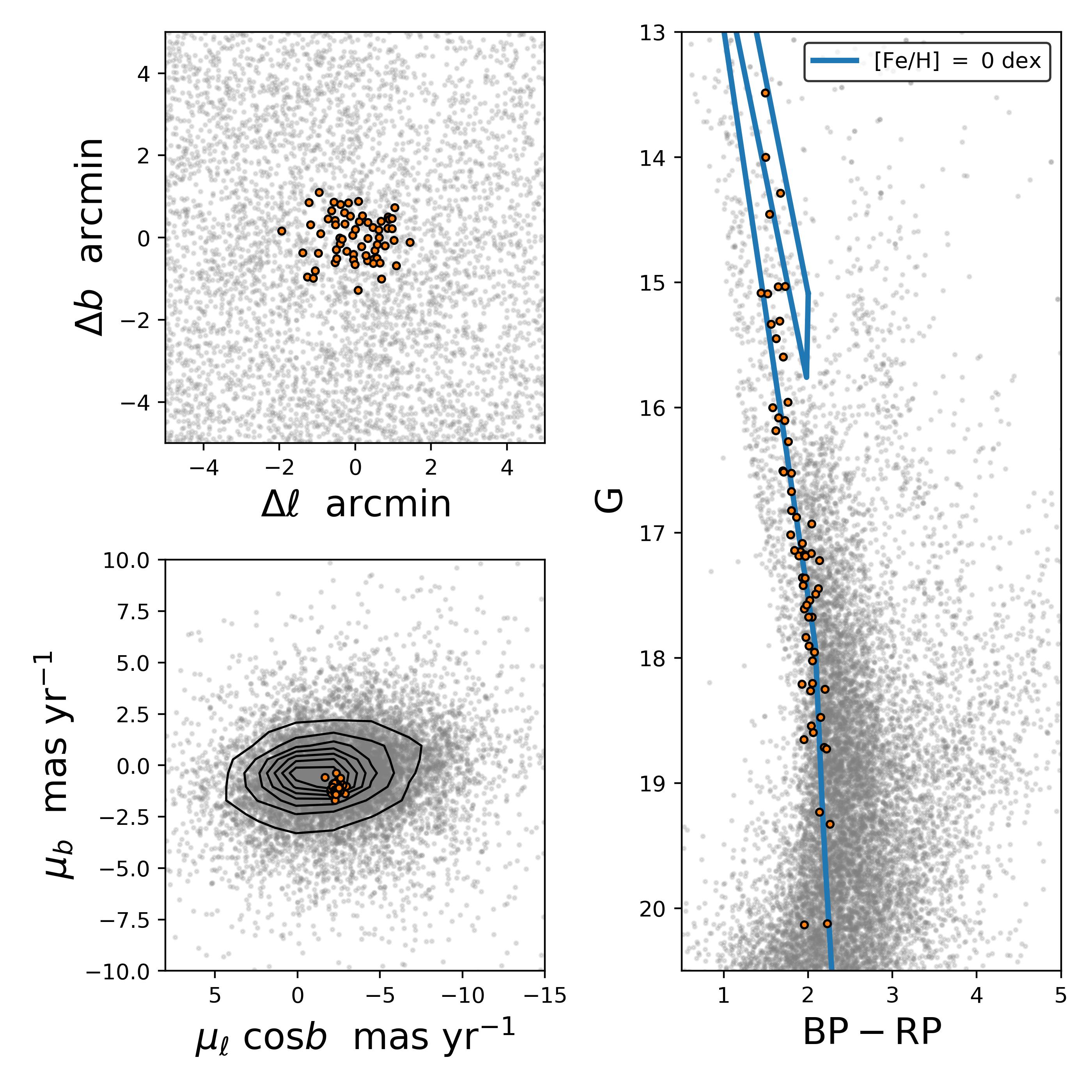}
  \caption{The open  cluster  ESO~589-26 as detected by  our algorithm.  The
    background sequences are the same as in Fig.~\ref{fig:Ter1}, although the bulge
    red giant branch is highly incomplete because this cluster is located very
    close to the Galactic plane ($b=0.33^\circ$), in a region  of the sky with large
    interstellar extinction.}
  \label{fig:ESO589}
\end{figure}

The  algorithm is  also  able to  detect  open star  clusters.   As an  example,
Fig.~\ref{fig:ESO589} shows  the recovery  of ESO-589-26,  a young  star cluster
reported at  RA=18:02:14 DEC=$-$21:54:54  ($l,b$)=(7.9517$^\circ$, 0.3279$^\circ$)
that our code independently detected 7  arcsec away from its nominal position. A
total of 62  cluster members were identified with our  method, defining a narrow
main sequence in the CMD and a very coherent group in the VPD. A  3~Myr  PARSEC
\citep{PARSEC}  isochrone of solar metallicity is included in the CMD in order to
guide the eye. The isochrone was shifted to a distance of $\sim 2.4$ kpc, as
reported in \citet{MWSC}.

Application  of this  neighbor  algorithm  to the  selected  bulge area  yielded
several cluster candidates not listed in the MWSC catalog. We visually inspected
the region within a radius of 5~arcmins from the center of each of them, and end
up with at least 7 new globular  clusters. The  new clusters will  be published  
in dedicated forthcoming  papers once they have been spectroscopically confirmed. We
show here only one  of the new globular clusters, that we name  {\tt Gran 1}, as
an example of  the ability of the  algorithm to find new  clusters.  Further, we
examine the candidate cluster VVV-CL001 \citep{Minniti11}, VVV-CL002, VVV-CL003
and  VVV-CL004  discussed in  \citet{MoniBidin11},  the  84 candidate  clusters
published in  the series of  papers by \cite{Minniti17b},  \cite{Minniti17c} and
\cite{Minniti17d}, plus another 5  candidates presented in \cite{Camargo18}, one
of which is further analysed by \cite{Bica18}.

\subsection{The new globular cluster Gran 1}

\begin{figure}
  \centering
  \includegraphics[width=9cm, angle=0]{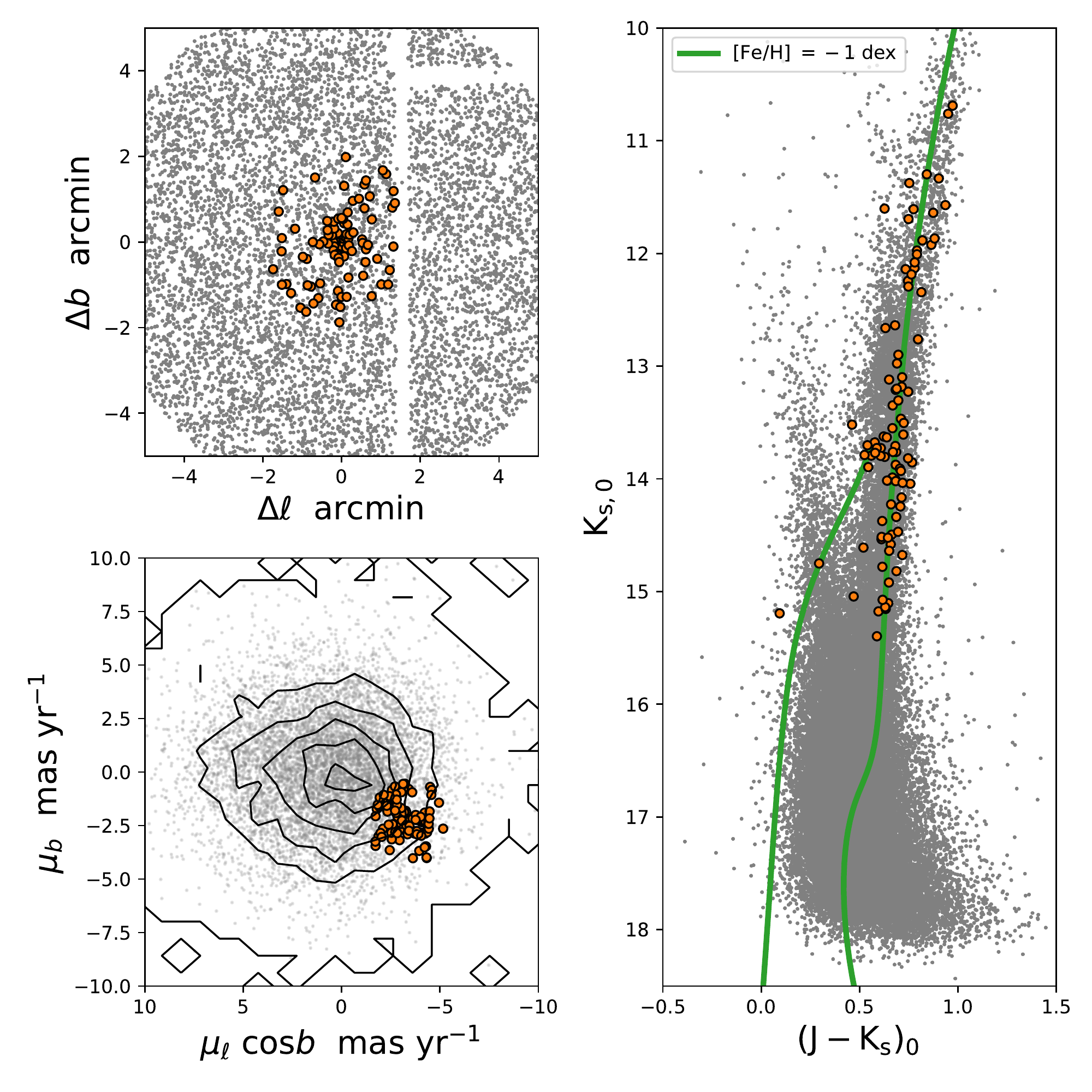}
  \caption{The new GC Gran~1 as detected in the plane of the sky (upper left),
    in the VPD (lower  left) and in the dereddened CMD (right). All data are from VVV, using
    the dereddened PSF photometry by \citet{surot19} and the PMs from \citet{ContrerasRamos17}.
    Small, light grey dots are
    field stars within 5 arcmin from the cluster center given in Table~\ref{table:Params}, 
    while large orange  dots  are bona fide cluster members. Blank stripes in the upper 
    left panel denote the chip separations where stars were rejected because the proper 
    motion values were not well constrained \citep{ContrerasRamos17}.} 
  \label{fig:Gran1}
\end{figure}

A new cluster, not present in the  MWSC catalog, was initially detected as a group
of  24 stars  within  0.731~arcmins  from ($\ell,b$)  =($-1.22^\circ,-3.98^\circ$)
(RA=17:58:36.61  DEC=$-$32:01:10.72).   As  we  believe that  this  is  the  first
detection of a cluster in this position, we  name it {\it Gran 1}.  The plots that
triggered its discovery are shown  in Fig.~\ref{fig:Gran1}. The 24 clustered stars
initially found  by the code  were used  to define a  cluster center, both  in the
plane of the sky and in the VPD. These positions were used to select all the stars
included both within 2 arcmin in the sky  and within 2 mas\ yr$^{-1}$ in the VPD,
which   yielded  95   stars,   shown  as   filled   orange  circles   in 
Fig.~\ref{fig:Gran1}.  The  CMD in  the right panel  demonstrates that  they are 
compatible with the RGB and horizontal branch (HB) of a cluster with [Fe/H]=$-1$ dex, 
located at a distance of $\approx$ 8.8 kpc, or m$-$M+A$_{K_{\rm s}}$= 15 mag, 
i.e., within the Galactic bulge.   We emphasize that the  cluster metal content, 
distance, and age, cannot  be constrained by the present data, as the turnoff 
cannot be identified in  the CMD, future spectroscopic follow-up will derive more 
precise parameters of this cluster.  As a reference, we show a 12 Gyr PGPUC 
isochrone overplotted to the data, using the putative red clump and HB 
stars (J-K$_{\rm s}\sim 0.25$ mag, and K$_{\rm s}\sim 15$ mag.)
as  an anchor  to estimate  a reddening  of  E(J$-$K$_{\rm s}$)  
and, adopting the  extinction law by  \cite{Nishiyama09}, a total extinction 
of A$_{K_{\rm s}}$. The age was assumed to be larger than 10 Gyr, due to the 
presence of the two blue HB stars, whose nature would need confirmation.
If this two stars are proved to be field stars, the age of the candidate cluster
could be lower than the estimate we give in Table~\ref{table:Params}.

\begin{figure}
  \centering
  \includegraphics[width=9cm, angle=0]{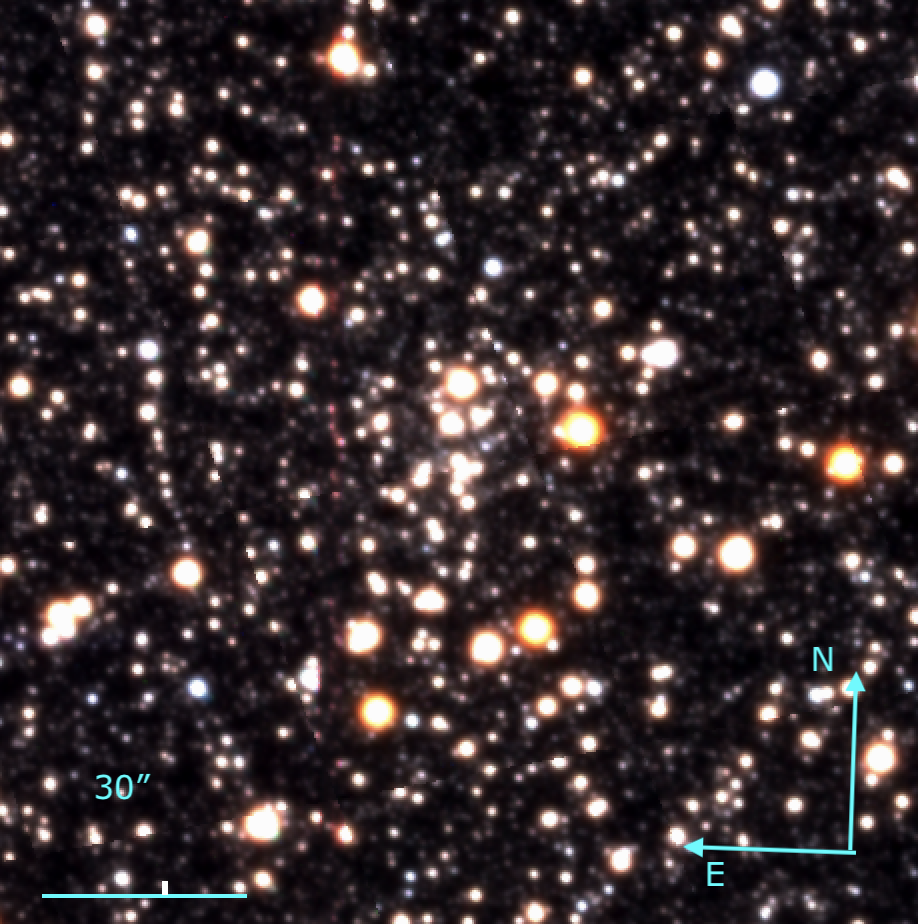}
  \caption{JYZ Color image of the new GC Gran~1 from the VVV Data Release 4.}
  \label{fig:Gran1_rgb}
\end{figure}

It should  be noted that  this cluster  would not have  been discovered as  a high
spatial concentration  only, as it is an overdensity of only 3.4 sigma above the 
mean stellar density of the field in this region.  It is detected here because the selected
cluster member  stars are, simultaneously, clustered  in the plane of  the sky and
they share a coherent motion with a median value of ($\mu_{\ell}\cos{b}, \mu_{b}$)
= ($-11$,+3) ${\rm mas\ yr^{-1}}$, which  is off-centered with respect to the mean
proper motion of field stars.
The integrated magnitude of Gran 1, down to the limit magnitude of the VVV catalogue, 
is $\rm K_{\rm s}=8.07$, while that of Terzan~1, in the same range, is $\rm K_{\rm s}=5.24$.
Gran~1 is similar in luminosity to Whiting~1, AM~4 or Koposov~1 \citep{Harris10}.
A JYZ color image of the newly discovered cluster is shown in Fig.~\ref{fig:Gran1_rgb}.

\begin{table}
  \caption{Basic parameters for the newly discovered GC Gran 1, derived from
    the present analysis. We emphasize that the metallicity, age and distance for
    the cluster are very uncertain, as they were derived using 76 stars by 
    comparing the observed CMD with a PGPUC isochrone.}  
  \label{table:Params}     
  \centering                               
  \begin{tabular}{c|c|c}         
    \hline\hline                       
    Parameter & Value & Unit\\    
    \hline                                  
    $\ell$                     & $-$1.2320        & deg    \\     
    $b$                        & $-$3.9776        & deg    \\
    RA (J2015.5)               & 17:58:36.61      & hh:mm:ss  \\
    Dec (J2015.5)              & $-$32:01:10.72   & dd:mm:ss  \\
    $E(J-K_{\rm s})$           & $\sim 0.45$      & mag    \\
    $A_{K_{\rm s}}$            & $\sim 0.24$      & mag    \\
    $d_\odot$                  & $\sim 8.8$       & kpc    \\
    $\mu_{\ell}\cos{b}$        & $-$10.9426       & ${\rm mas\ yr^{-1}}$ \\
    $\mu_{b}$                  & 3.0252           & ${\rm mas\ yr^{-1}}$ \\
    $\mu_{\alpha}\cos{\delta}$ & $-$8.0583        & ${\rm mas\ yr^{-1}}$ \\
    $\mu_{\delta}$             & $-$8.0833        & ${\rm mas\ yr^{-1}}$ \\
    Age                        & $\sim 8-12$      & Gyr    \\
    ${\rm [Fe/H]}$             & $\sim -1$        & dex    \\
    \hline                                             
  \end{tabular}
\end{table}

\section{Application to candidate globular clusters in the literature}
\label{sec:candidates}

Hereafter we  examine different sets  of candidate clusters recently  published in
the literature.   Specifically, \citet{Minniti11} reported  on the discovery  of a
low mass globular cluster, named  VVV GC001, at coordinates ($l,b$)=(5.25$^\circ$,
0.78$^\circ$), (RA=17:54:42.5, DEC=$-$24:00:53), that is, approximately 10 arcmins
away from  the known  GC UKS~1.   On the  same year,  \citet{MoniBidin11} detected
three new candidate  clusters tentatively named VVV GC002, GC003,  and GC004. Upon
analysis of  their CMD, the  same authors concluded that  they were most  likely a
globular cluster, a stellar association, and an overdensity whose nature could not
be established, respectively.

In 2017, Minniti et  al.  compiled a catalog containing 84  GC candidates based on
overdensities   of   stars   in   the   plane  of   the   sky   \cite[Minni~1   to
  Minni~21;][]{Minniti17b} and on  the coincidence, in the plane of  the sky, of a
couple  or more  RRL  \cite[Minni~22 to  Minni~60;][]{Minniti17c}  and/or Type  II
Cepheids  \cite[Minni~61 to  Minni~84;][]{Minniti17d}  within 2  arcmins.  In  the
latter two  papers, they checked  that at least two  of the variables  within each
group had magnitude consistent with similar distances.

Finally, \citet{Camargo18} published  a list  of 5  clusters visually  selected in
multicolor  images  from  the  WISE  satellite,  and  then  analysed  using  2MASS
photometry and Gaia  DR2 PMs.

The  algorithm developed  here, applied  to the  Gaia PM  catalog did  not blindly
detect any  of the 93  cluster candidates mentioned above  (4 VVV GCs,  84 Minnis,
plus 5  Camargos) in the  Gaia DR2 catalogue. Because  of the much  higher stellar
density  close  to  the  plane,  a   blind  search  across  the  VVV  area  within
$|b|\lesssim3^\circ$  yielded  a  large  number of  overdensities,  that  will  be
examined and validated  in a forthcoming paper. For the  present purpose, in order
to validate  previously published candidate clusters  we only need to  search in a
very restricted  area around their  nominal centers.   Therefore, we have  run the
original code, this time  allowing for a lower minimum number  of member stars (15
instead of 20) only  in a region of 2~arcmins radius  around the candidate centers
for stars brighter than ${\rm G = 19}$ mag. This restricted search was run for all
the 93  clusters, on the  Gaia catalog, and  for the 54  candidates (4 VVV  GC, 46
Minnis, and 4 Camargos) within $|b|\lesssim3^\circ$ on the VVV PM catalog.

\begin{figure}
  \centering
  \includegraphics[scale=0.5]{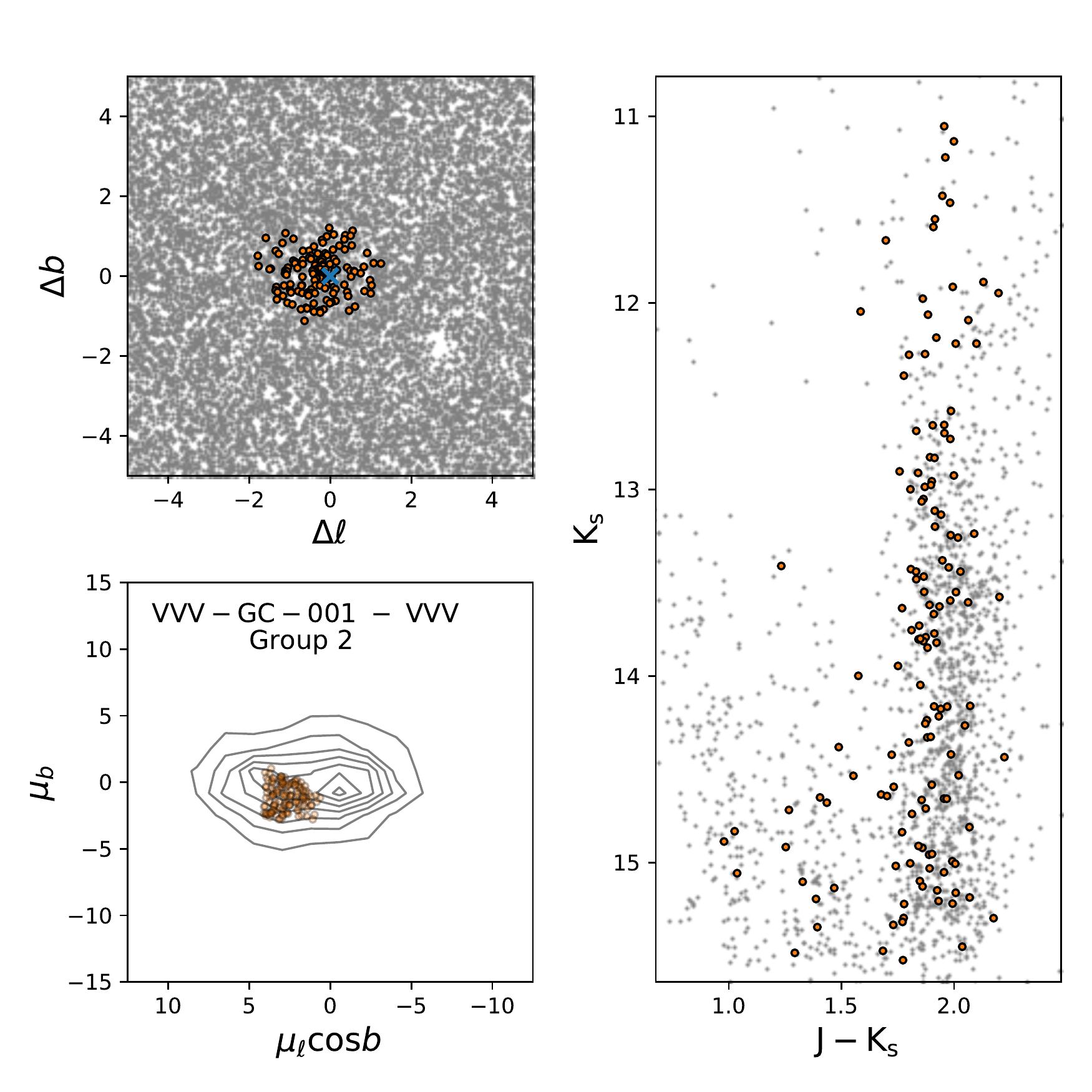}
  \caption{Diagnostic plots for the globular cluster GC001. Stars within
    2~arcmins from the nominal cluster center were used to perform the clustering algorithm.
    The spatial distribution of stars (Group 2 defined by the code) that comprise the 
    GC are showed in orange points (upper left), plotted the VPD (bottom left) 
    and in the CMD. Note that the distribution of cluster stars is shifted with 
    respect to the contours of the total distribution of stars in the VPD.}
  \label{fig:GC001}
\end{figure}

\begin{figure}
  \centering
  \includegraphics[scale=0.5]{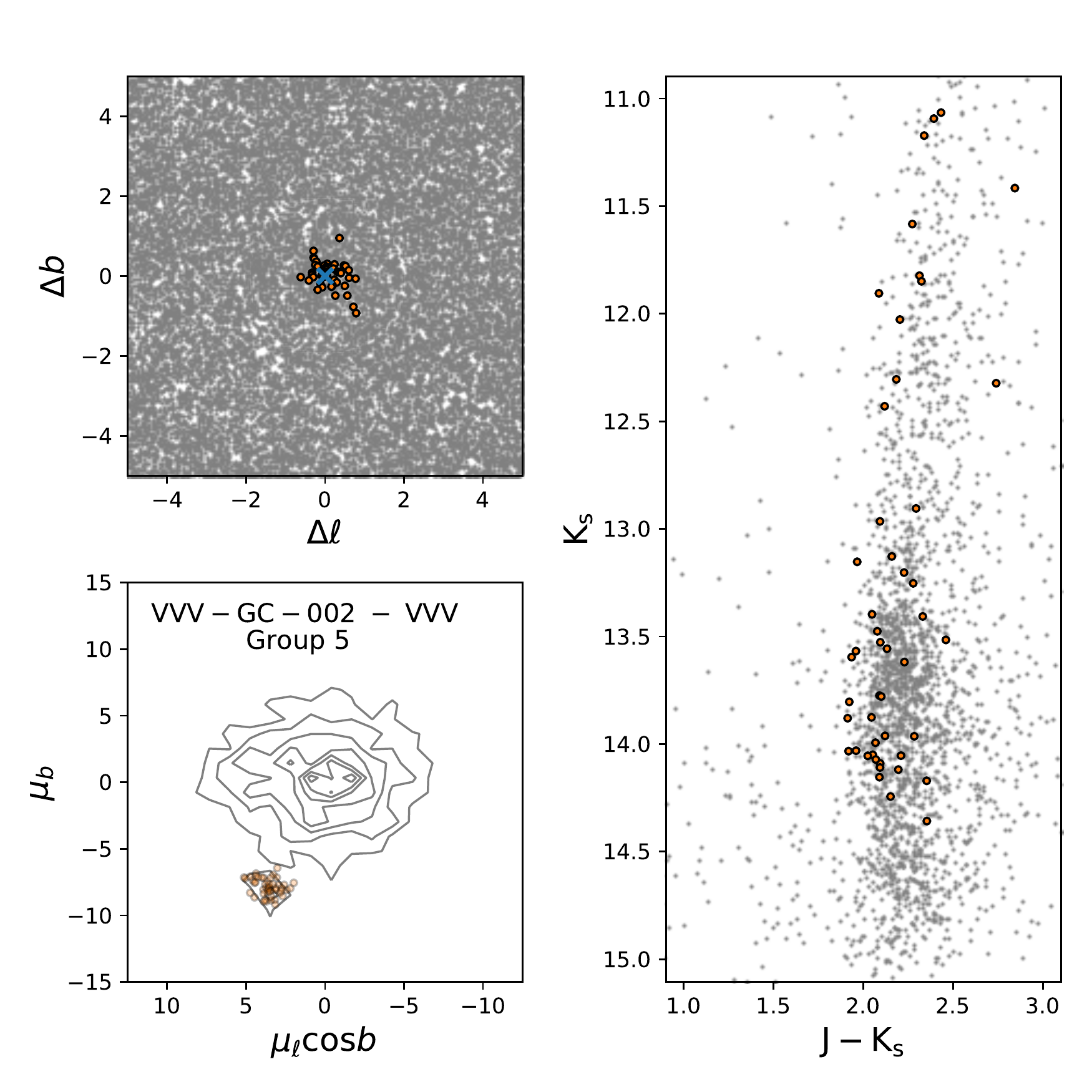}
  \caption{Same as Fig.~\ref{fig:GC001} for the globular cluster VVV GC002.}
  \label{fig:GC002}
\end{figure}

As a result, only the candidate VVV GC001 and GC002 seem to be real clusters. Both
of them  lie relatively close to  the plane, and  nothing is detected in  the Gaia
catalog.  In  VVV data, however,  the algorithm picks  up an overdensity  of stars
whose mean PM is offset with respect to the mean PM of field stars, and who define
a rather narrower RGB sequence, compared with  that of field stars within the same
spatial    region.   This    is    illustrated    in   Fig.~\ref{fig:GC001}    and
Fig.~\ref{fig:GC002}.

For all the others, the algorithm either did not detect any overdensity, or it did
detect a broad peak, but it was centered,  in the VPD, at the same position of the
median PM of field stars and the CMD  did not show anything different from the CMD
of field  stars. The  Appendix~\ref{sec:Appendix} shows  the spatial  selection of
stars within 1  arcmin from the candidate center, together  with their position in
the  VPD, compared  to that  of  field stars  within 10~arcmins  of the  published
center, and the  CMD of both. For  each cluster, we show these  diagrams from Gaia
data on the  left and, if available, from  VVV data on the right.  We also include
the two confirmed candidates VVV GC001 and  VVV GC002 in order to demonstrate that
we  would  have been  able  to  confirm them  even  without  using the  clustering
algorithm. In fact, by  selecting stars within 1 arcmin from  their center, in VVV
data, the field contamination  is larger but a group of stars  with mean PM offset
from that of  field stars is visible in  both of them.  The offset  is smaller for
GC001, where the  CMD is better defined, and  it is larger in GC002  where the CMD
alone would be more ambiguous. On the contrary, by means of these plots, we reject
all  the clusters  were nor  the VPD  nor the  CMD allow  seeing anything  clearly
different from  the dominant  field population. With  this argument,  we concluded
that, by means of the present data, we cannot confirm the cluster nature of any of
the other 91 candidates.

A few  considerations are in  order. First, it is  expected that, by  pure Poisson
statistics applied  to the  VPD, one  would expect  larger fluctuations  where the
density of  stars is  higher. That is,  at the  peak PM of  field stars.  In other
words, we do expect a large number  of false positives whose mean PMs is identical
to that of the dominant population of  field stars, either bulge or disk.  Second,
we used the catalogue of 150  GC PMs provided by \citet{Vasiliev18}, together with
the   mean   bulge   PM    from   \citet{Reid04}\footnote{   The   authors   quote
  ($\mu_{\ell}\cos{b}$,$\mu_b$)=($-6.379,-0.202$)  ${\rm  mas\  yr^{-1}}$  as  the
  absolute PM of Sagittarius  A, which must coincide with the  mean absolute PM of
  bulge  stars.   Converted  to  equatorial  coordinates,  it  gives  ($\mu_\alpha
  \cos{\delta}$,$\mu_\delta$)=($-3.15,-5.55$)  ${\rm   mas\  yr^{-1}}$,   that  we
  subtract from the values in Table~B1 from \citet{Vasiliev18}.}  to estimate what
fraction of  known clusters are expected  to have a  mean PM centered at  the mean
value for bulge stars.   The result is that only 7 clusters have  mean PM within 1
${\rm mas\ yr^{-1}}$ from the mean bulge  PM.  Of those, only 3 are located within
the area explored here, where there are in total 49 clusters. In other words, only
7$\%$ of the clusters  located in the bulge are expected to have  the same mean PM
as bulge stars.

If the candidate clusters were massive enough,  we would detect them even if their
PM   would  be   identical  to   the  peak   of  bulge   stars,  as   we  do   for
Terzan~1. Therefore, we can  safely exclude that any of the  91 candidates that we
do not confirm here  is as massive as Terzan~1. In order for  them to be real, they
must have a very low  mass, and/or low central concentration {\it and}  their PM
must be identical  to that  of bulge stars.  This is expected to be true for at most 6 of them.

\section{Summary}
\label{sec:summary}

We present a clustering algorithm  that looks for overdensities simultaneously
in  a  five dimension  space  including  the two  coordinates  in  the sky,  the
corresponding stellar PMs  in each coordinate, and one color.  The algorithm was
independently applied  to the Gaia   DR2 catalog of a  region of the  sky within
$-10^\circ \leq  (\ell,\ b) \leq 10^\circ$,  and to a VVV  catalog including PSF
photometry  and  PMs for  stars  within  $|b|<3^\circ$,  where Gaia   is  highly
incomplete, for bulge stars, due to large interstellar extinction.

With the adopted parameters, discussed in Sec.~\ref{sec:method} the algorithm is
able to recover all  the clusters known to be present in this  area, and to find
several new young and old star clusters.  We presented here the detection of one
old cluster named {\it Gran~1}, and  deferred the discussion of  the other new
clusters to forthcoming, dedicated paper.

The PMs of Gaia   and VVV were then used to confirm/dismiss  the cluster nature of
93 cluster candidates recently presented in  the literature. Based on the 
requirement that the putative cluster members must move coherently, we could 
confirm 2 of them (VVV GC001 and GC002) and discarded another 91.  

The  present result  emphasizes  that statistical  fluctuations  of the  projected
stellar density, in  the plane of the  sky, or chance alignment of  a few variable
stars can  be relatively frequent, in  a huge survey  as Gaia  and VVV. They can  be easily
mistaken  for a  star cluster,  because: {\it  i)} the  stellar population  of the
Galactic bulge has a (relatively low) spread in metallicity and age that makes its
RGB  only slightly  wider than  that of  a globular  cluster; and  {\it ii)}  when
selecting stars in  a small region of the  sky, that is within a few~arcmins from a
spatial overdensity peak, the CMD sequences appear always narrower than those of a
wider region,  because they are affected  by a lower differential  extinction. For
this reason,  the list of  candidate new star clusters  published in the  last few
years has  increased enormously.  Only stellar motions,  however, can  confirm the
real cluster nature of a detected overdensity and they have proven to dismiss the 
large majority of  candidates. Given the present availability of  Gaia  PMs across
almost the whole sky, claimed detection of new candidate clusters should always be
supported by the  kinematics analysis, proving that the cluster  member stars move
coherently in space.

By means  of the present data, we cannot exclude that  a few of the  candidates are
extremely low  mass clusters  that happen to  have zero PM  with respect  to bulge
stars, are located at a distance very  close to 8 kpc and present relatively broad
CMD sequences,  possibly due to differential  extinction. We can only  state that,
because of these  characteristics, they would be invisible to  both Gaia  and VVV.
It is important to  keep this in mind, however, when  deciding whether to allocate
telescope time to followup studies of such candidates.

\begin{acknowledgements}

This  work is  part  of the Ph.D. thesis  of F.G.,  funded  by grant  CONICYT-PCHA
Doctorado Nacional 2017-21171485. F.G. also acknowledge CONICYT-Pasant\'ia 
Doctoral en el Extranjero 2019-75190166 and ESO SSDF 19/20 (ST) GAR funding. 
We acknowledge support from  the Ministry for the Economy, Development, and 
Tourism's Programa Iniciativa Cient\'\i fica Milenio through grant IC120009, 
awarded to  Millenium Institute of Astrophysics (MAS), the BASAL  CATA Center 
for  Astrophysics and  Associated  Technologies through  grant AFB-170002, and 
from FONDECYT Regular 1150345. JAC-B acknowledges financial support to CAS-CONICYT 17003.\\

Based on observations taken within the ESO VISTA Public Survey VVV, Program ID 179.B-2002.
This work has made use of data from the European Space Agency (ESA) mission
{\it Gaia} (\url{https://www.cosmos.esa.int/gaia}), processed by the {\it Gaia}
Data Processing and Analysis Consortium (DPAC,
\url{https://www.cosmos.esa.int/web/gaia/dpac/consortium}). Funding for the DPAC
has been provided by national institutions, in particular the institutions
participating in the {\it Gaia} Multilateral Agreement.\\

This research made use of: TOPCAT \citep{topcat}, GitHub, IPython \citep{ipython},
numpy \citep{numpy}, matplotlib \citep{matplotlib}, Astropy, a community-developed
core   Python   package   for    Astronomy   \citep{astropy}, scikit-leran
\cite{scikit-learn}, galpy: A Python Library for Galactic Dynamics \cite{galpy}, 
and "Aladin sky atlas" developed at CDS, Strasbourg Observatory, France \citep{Aladin1, Aladin2}.   
This research  has  made use  of  NASA's Astrophysics  Data System.

\end{acknowledgements}

\bibliography{bulge_gc_vr2}

\newpage

\begin{appendix}
\section{Appendix} 
Diagnostic plots of the candidate globular  clusters found in Gaia (left side) and
VVV (right side).   Each panel contains the sky position  (upper left), VPD (lower
left) and CMD (right) for the inner 1 arcmin (highlighted in orange) and the field
stars within  10~arcmins from  the nominal  cluster center  up to  ${\rm G  = 19}$
mag. All the 93 candidate clusters but one (Minni 02) were located in the Gaia DR2
catalog and 53 in our VVV catalog. We recall that PMs are available from VVV 
only  within  latitudes  $|b|<3^\circ$,  plus  a  few  fields  where  they  were 
calculated specifically for other projects.
\label{sec:Appendix}

\begin{table*}
\begin{tabular}{cc}
\includegraphics[width=8cm]{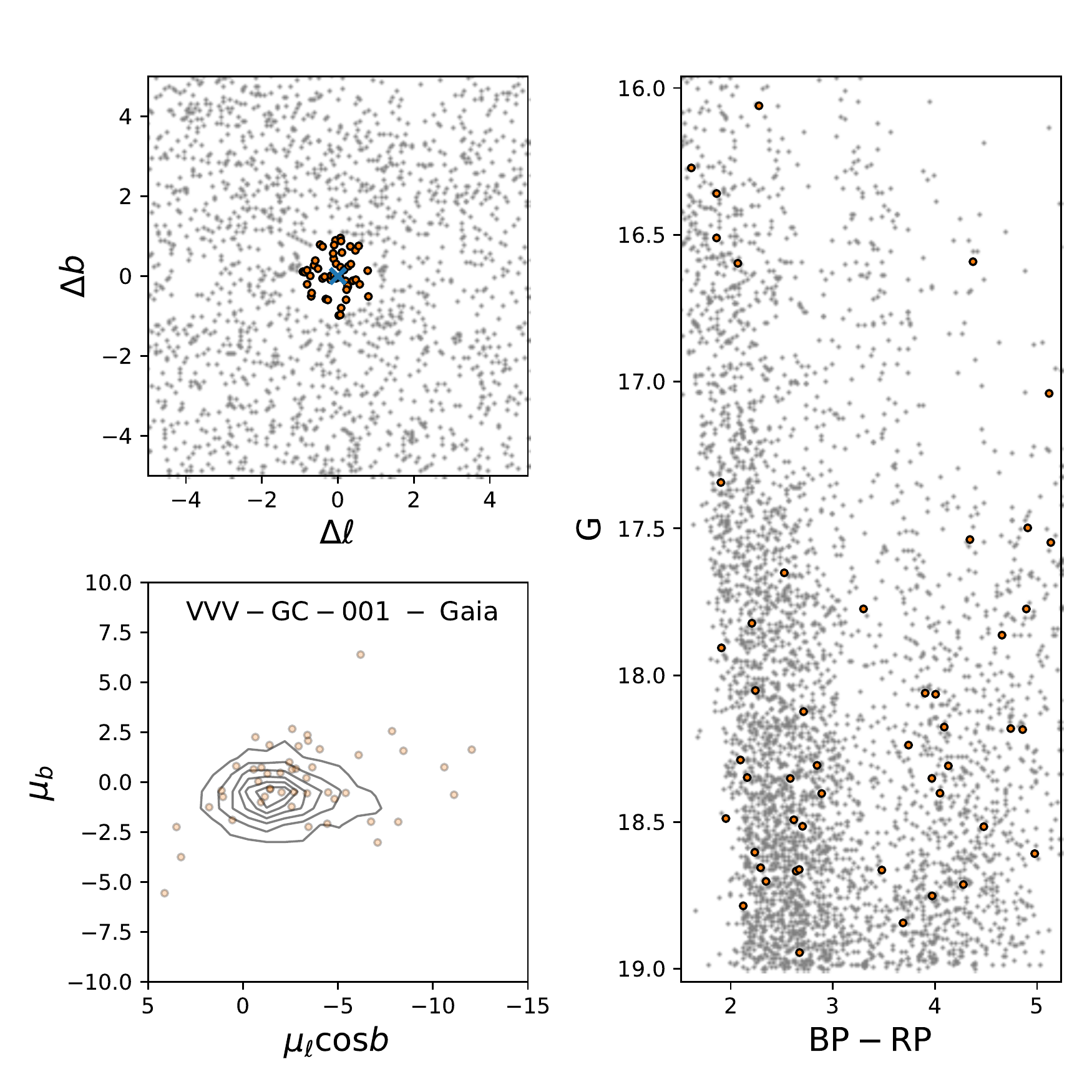} &
\includegraphics[width=8cm]{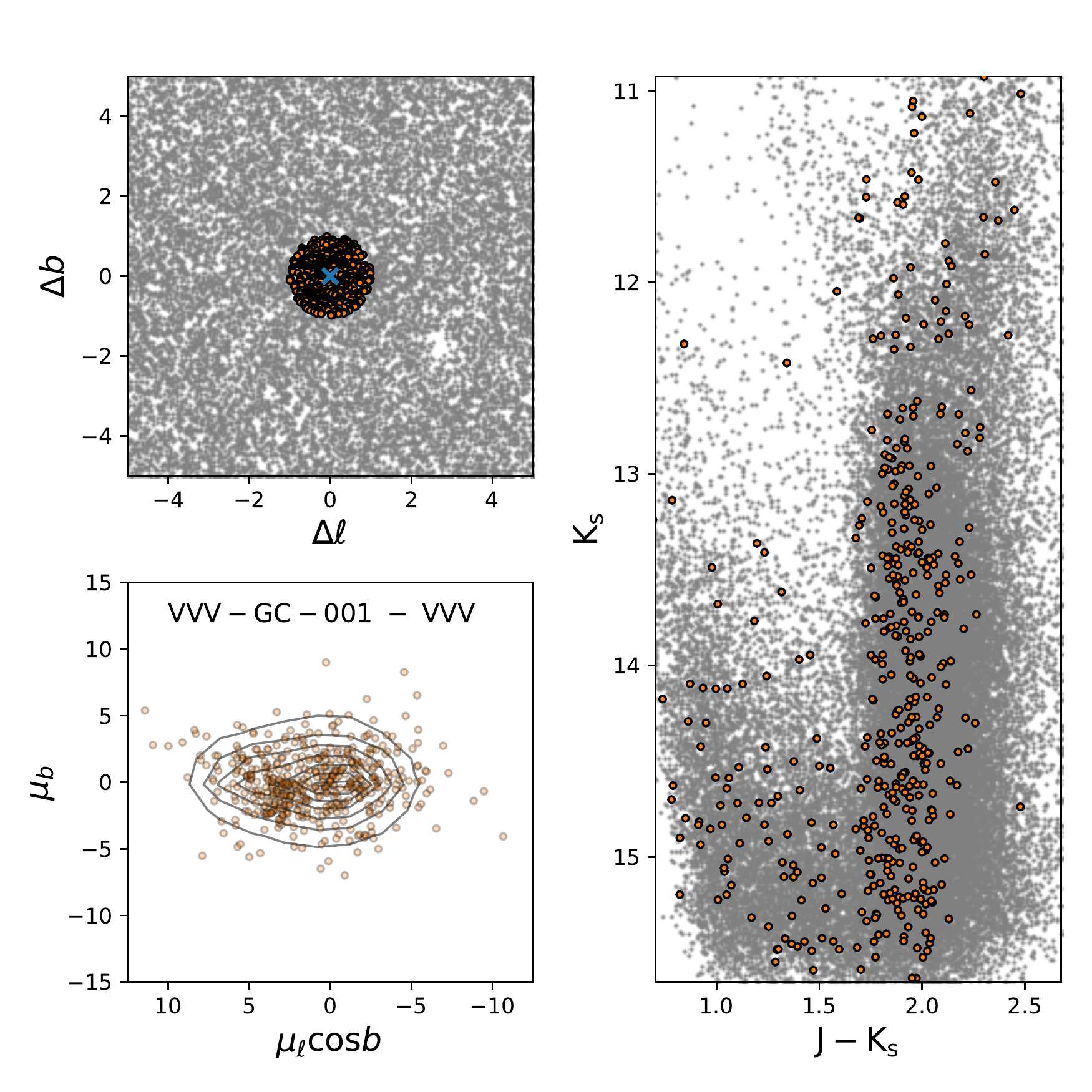} \\
\includegraphics[width=8cm]{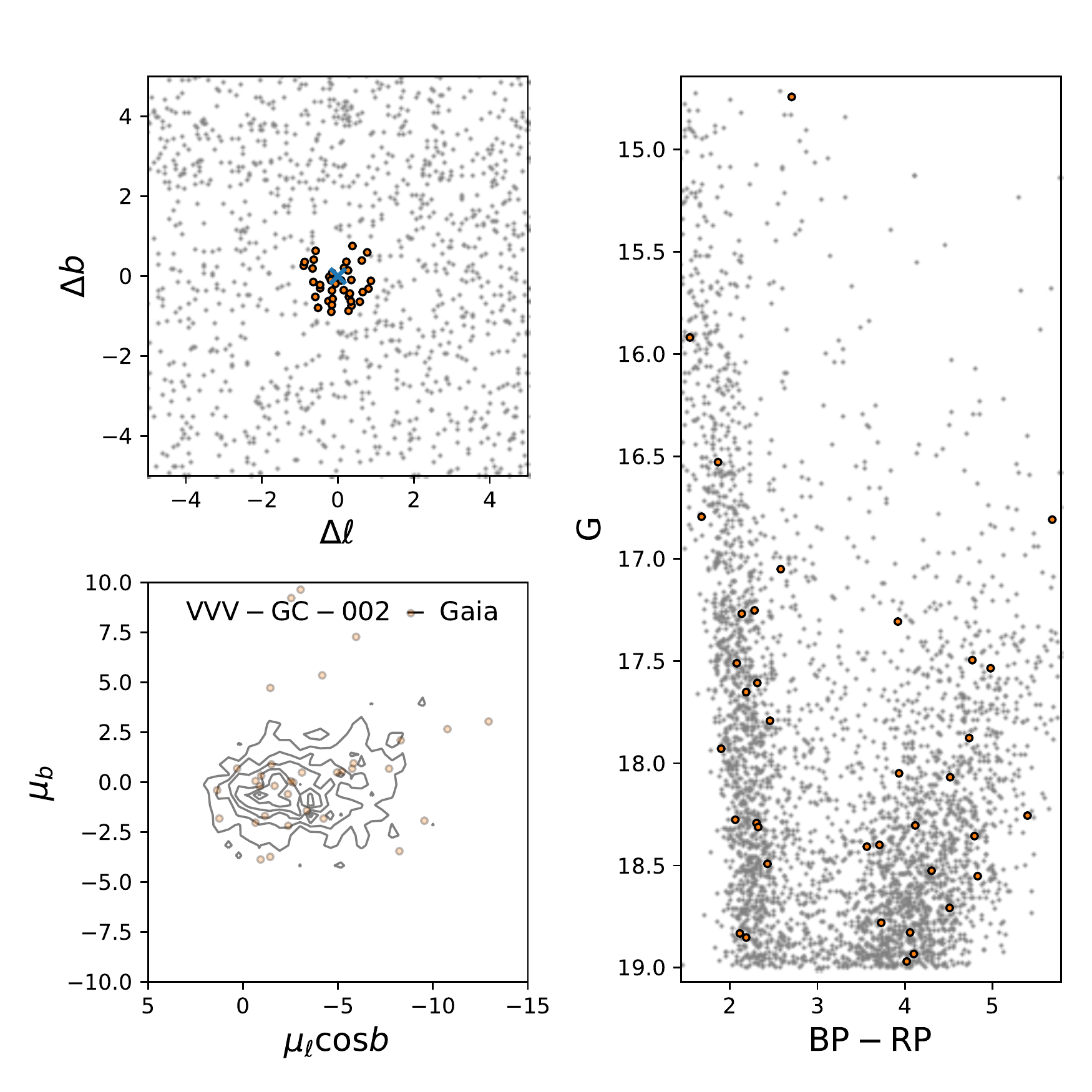} &  
\includegraphics[width=8cm]{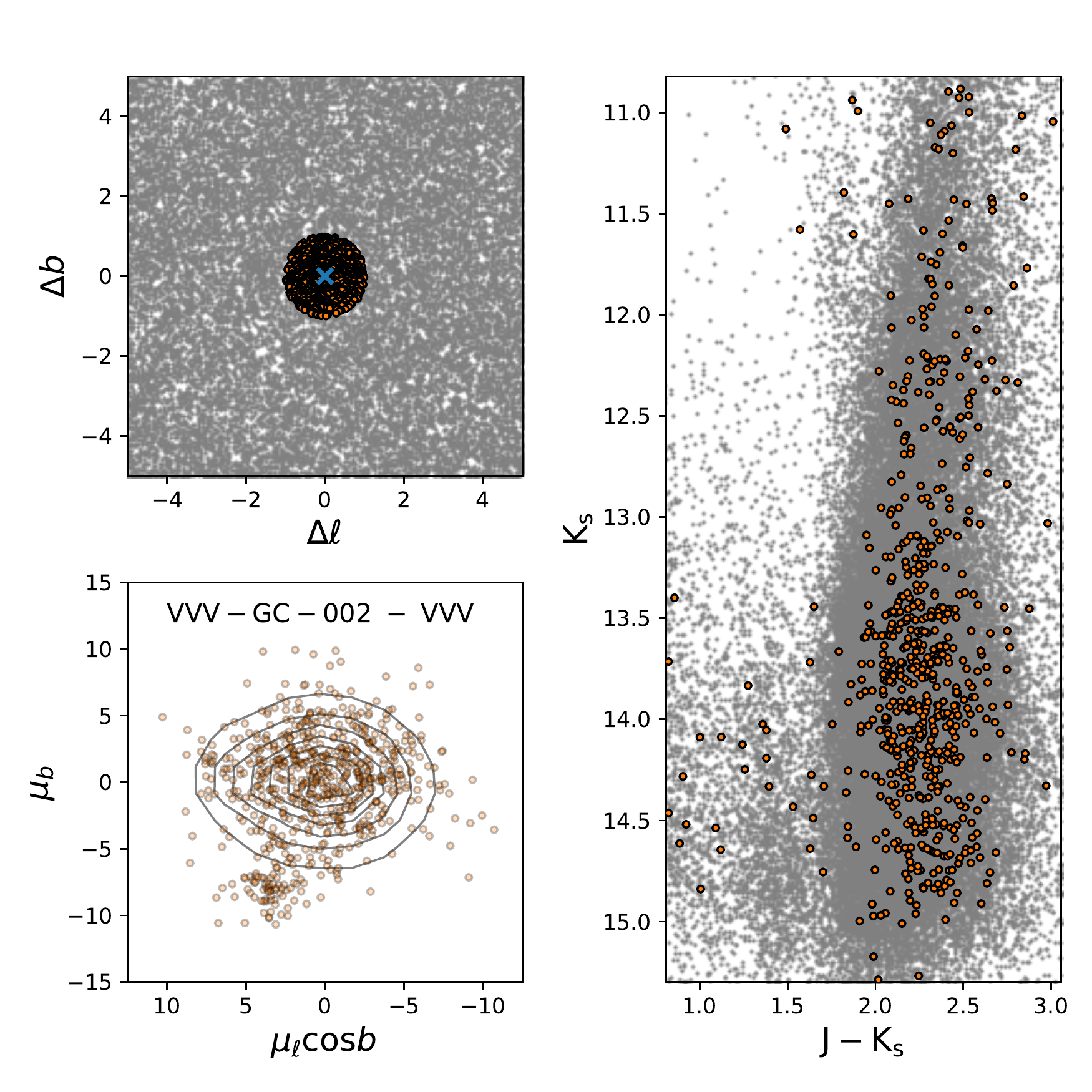}\\
\includegraphics[width=8cm]{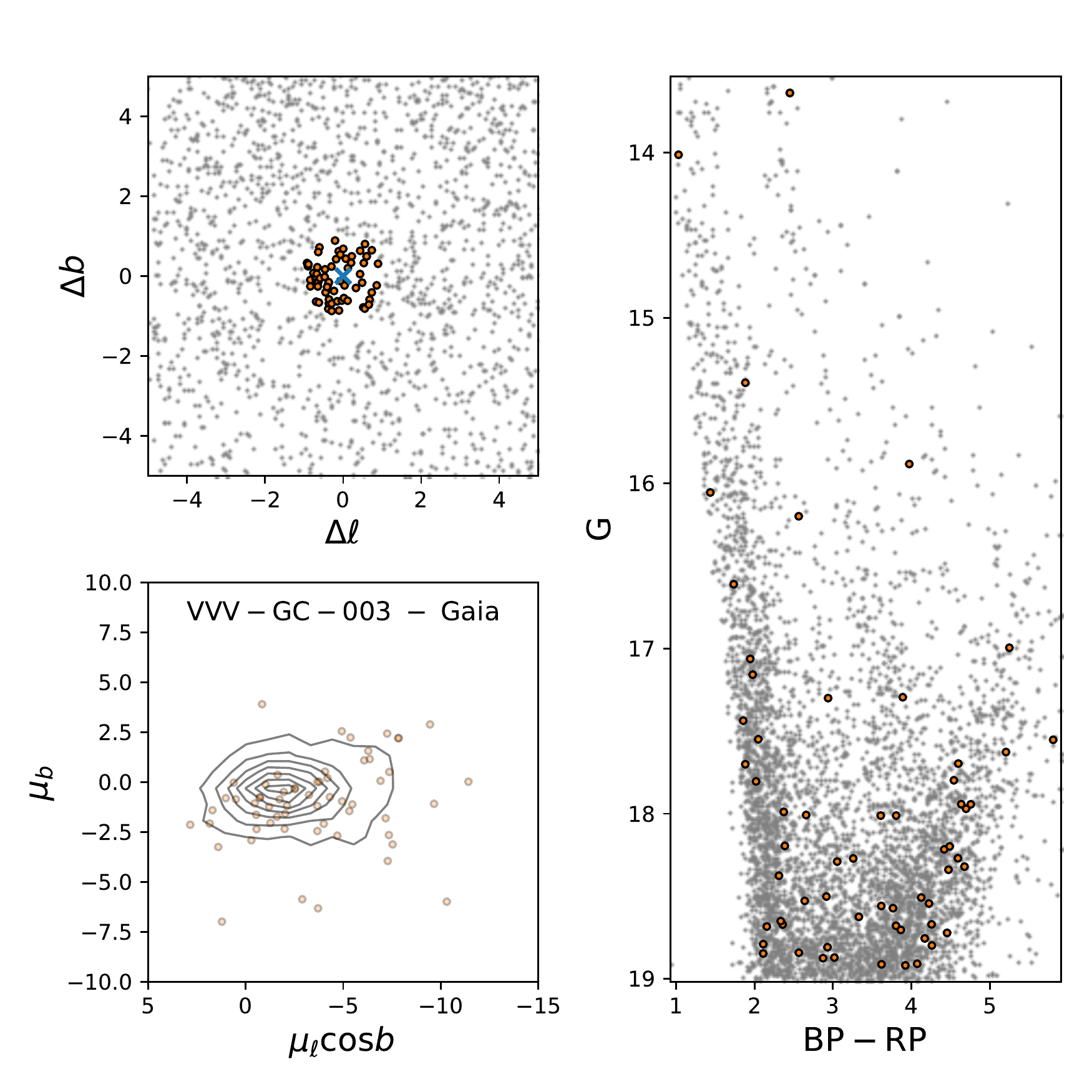} &  
\includegraphics[width=8cm]{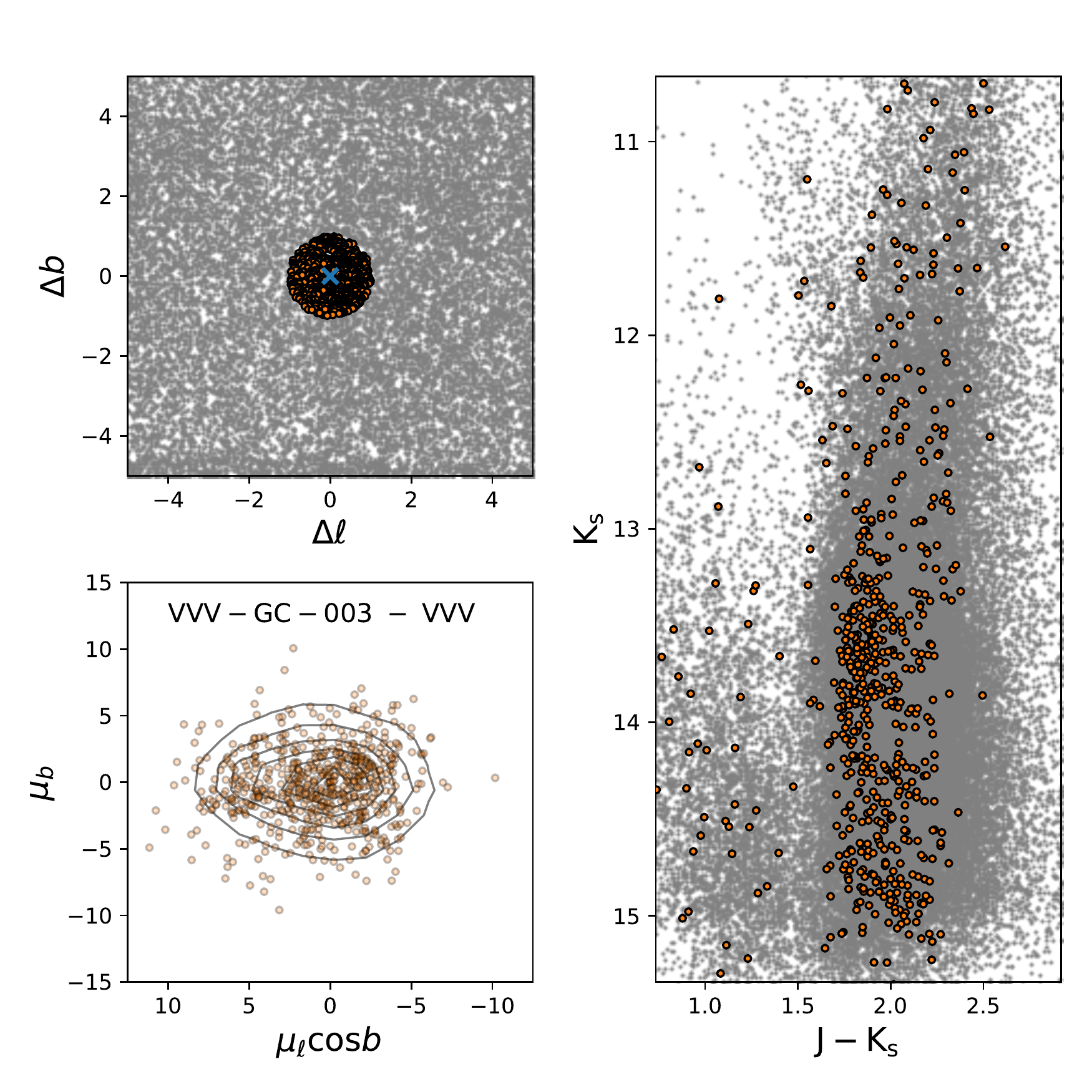} \\
\end{tabular}
\end{table*}
\newpage
\begin{table*}
\begin{tabular}{cc}
\includegraphics[width=8cm]{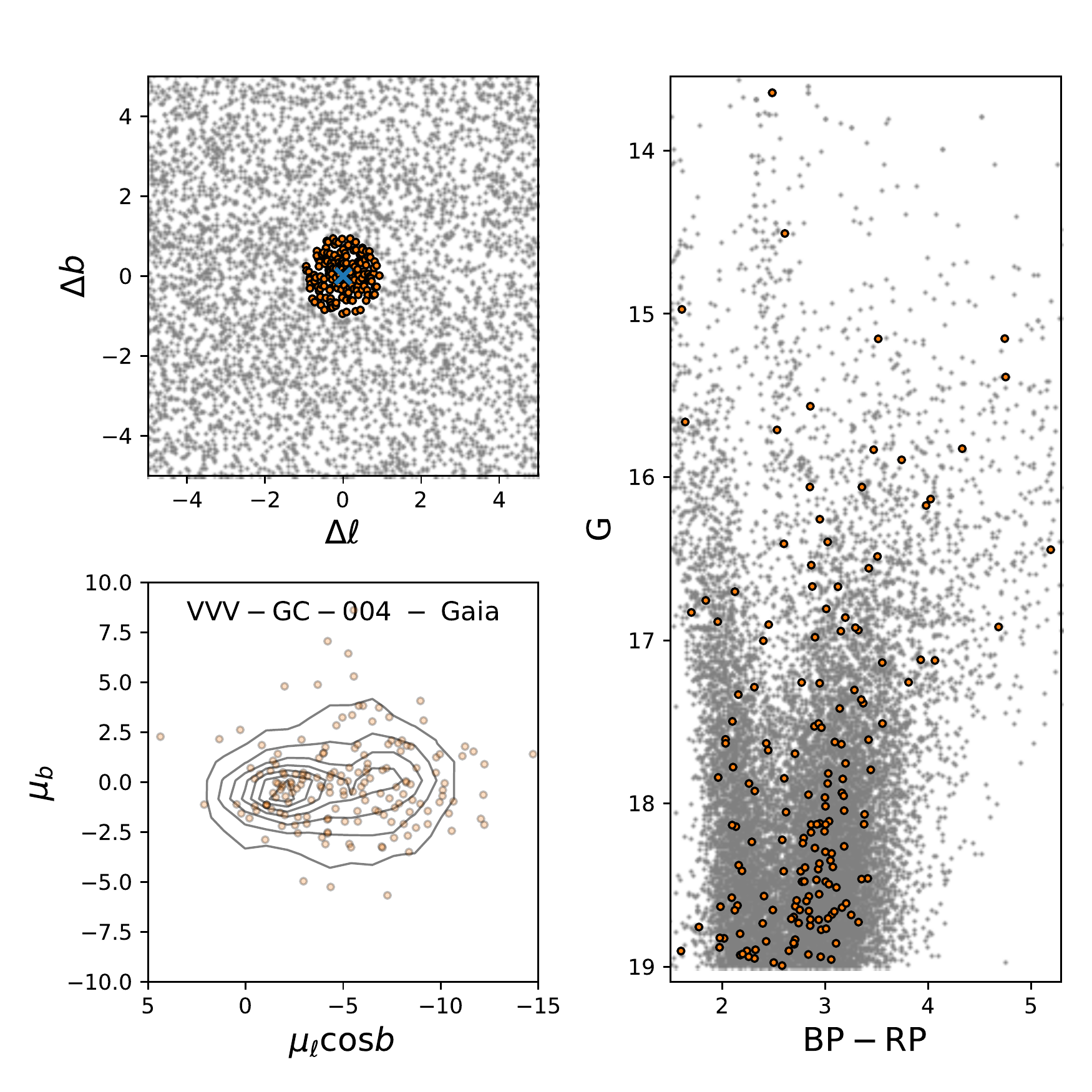} &
\includegraphics[width=8cm]{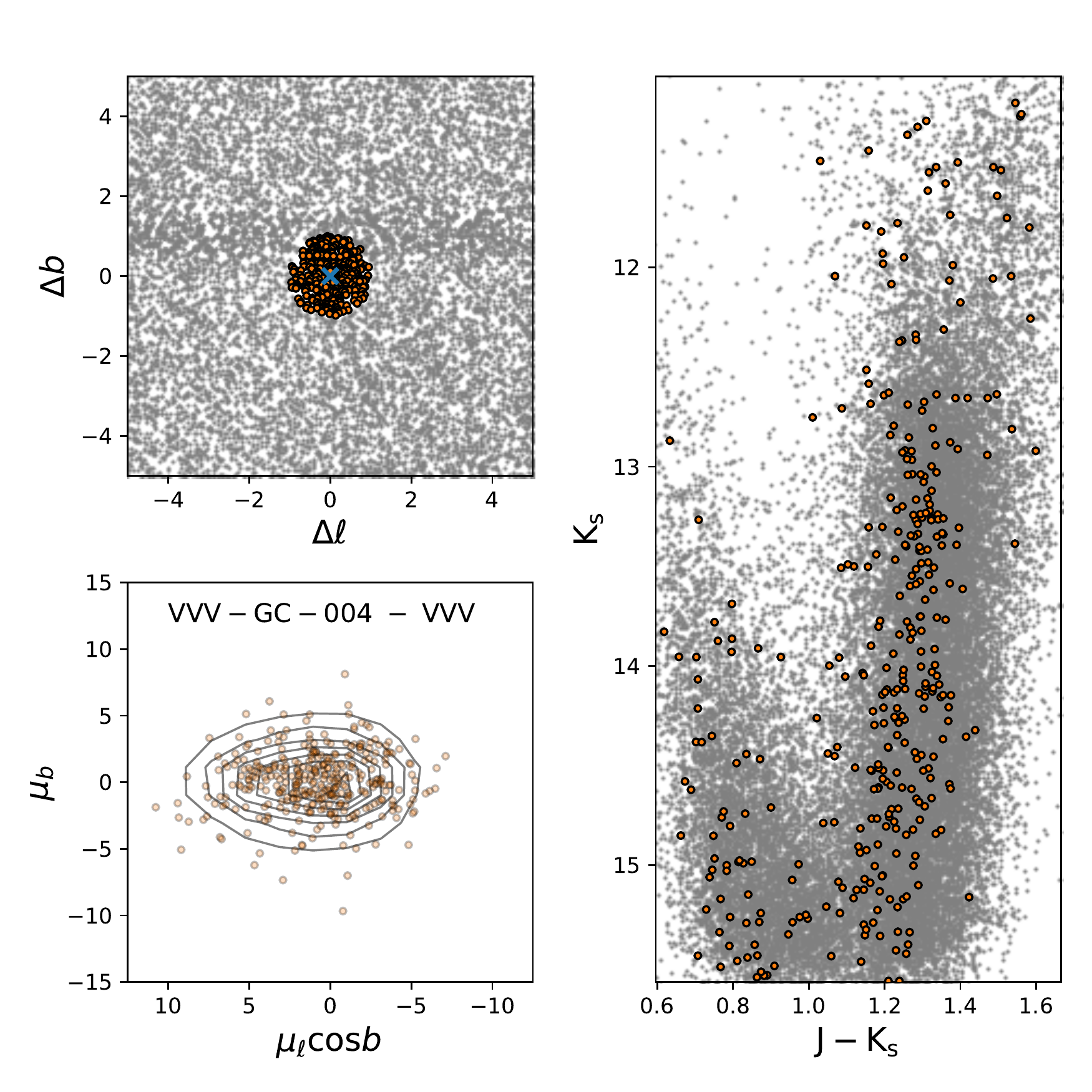} \\
\includegraphics[width=8cm]{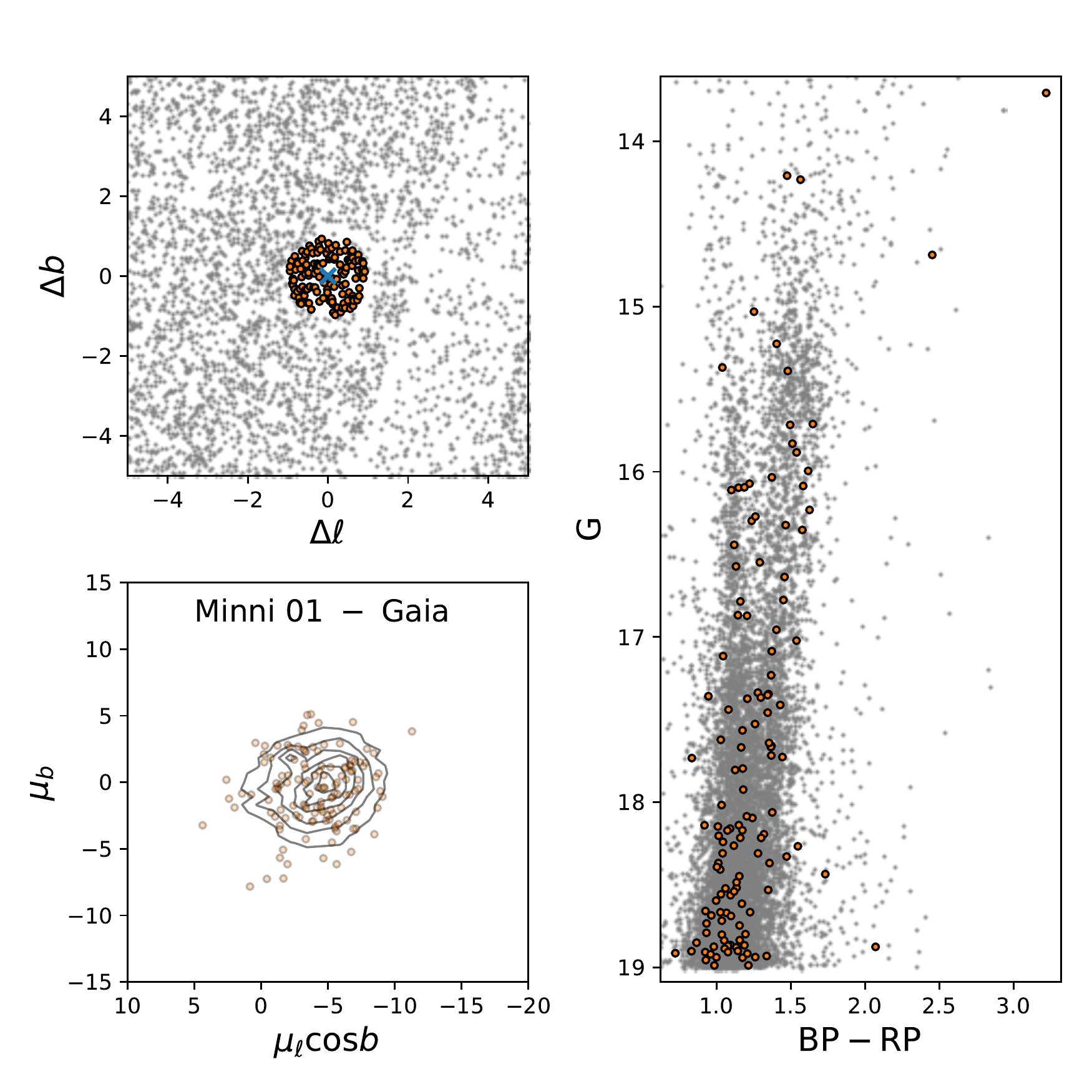} &
\includegraphics[width=8cm]{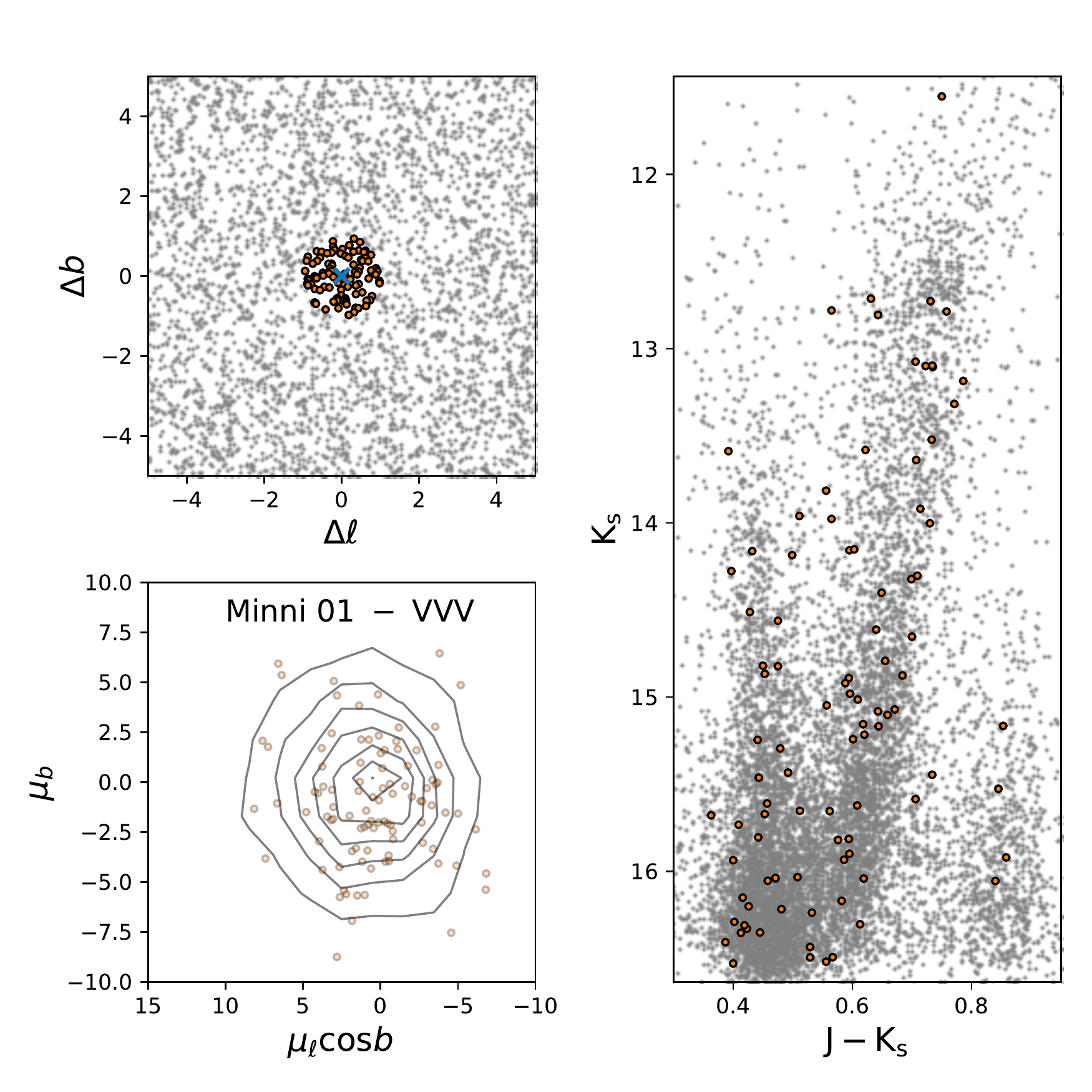} \\
& 
\includegraphics[width=8cm]{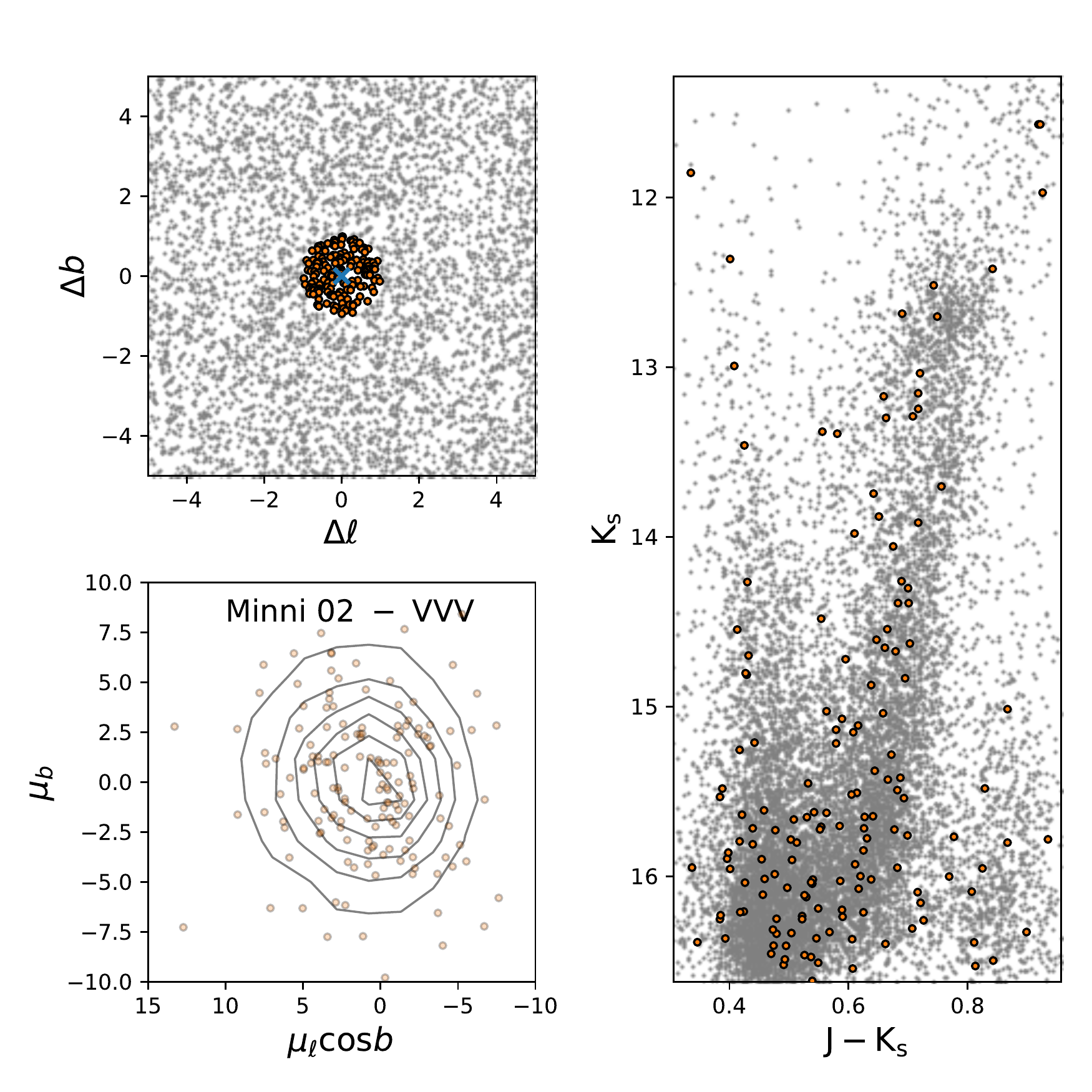} \\ 
\end{tabular}
\end{table*}
\newpage
\begin{table*}
\begin{tabular}{cc}
\includegraphics[width=8cm]{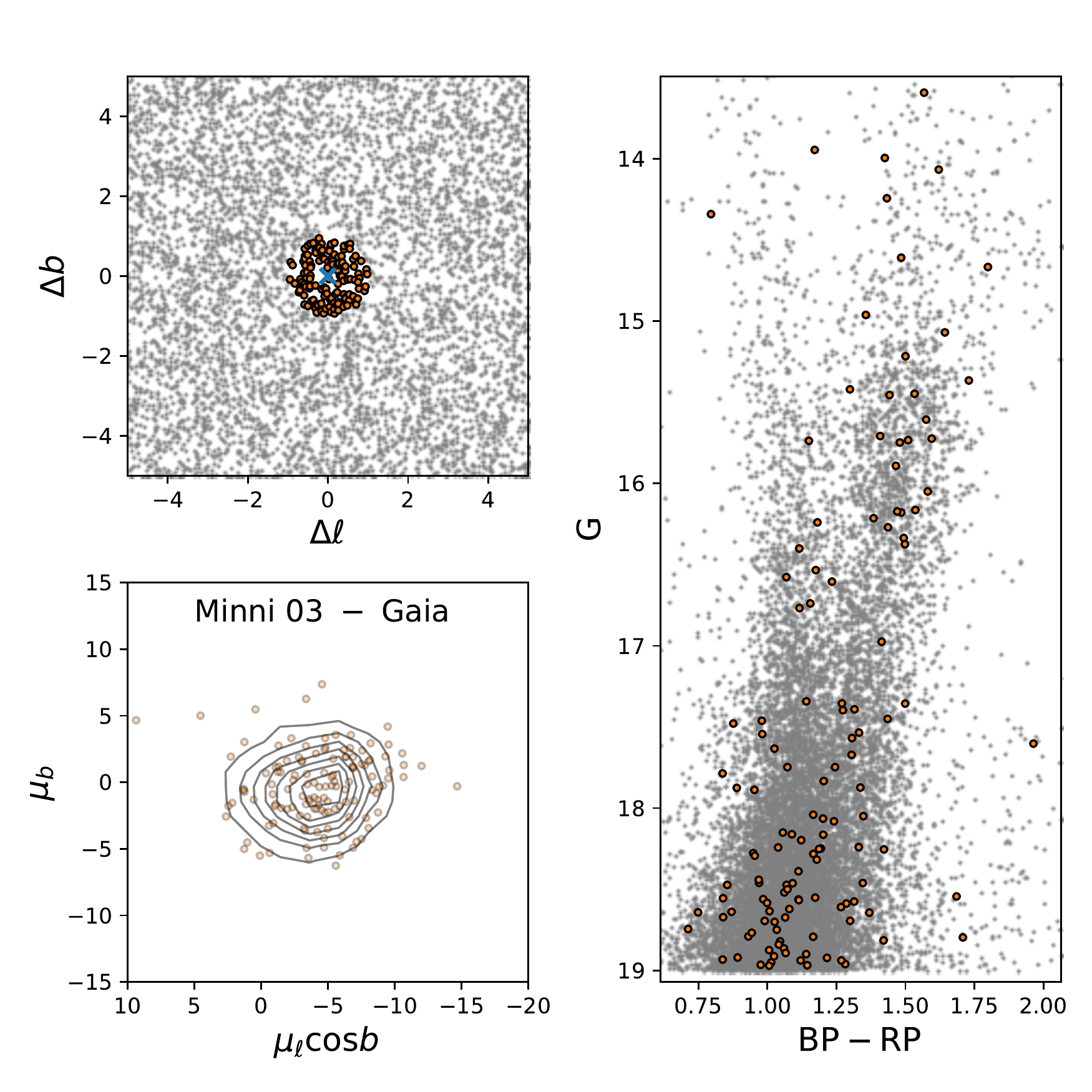} & \\ 
\includegraphics[width=8cm]{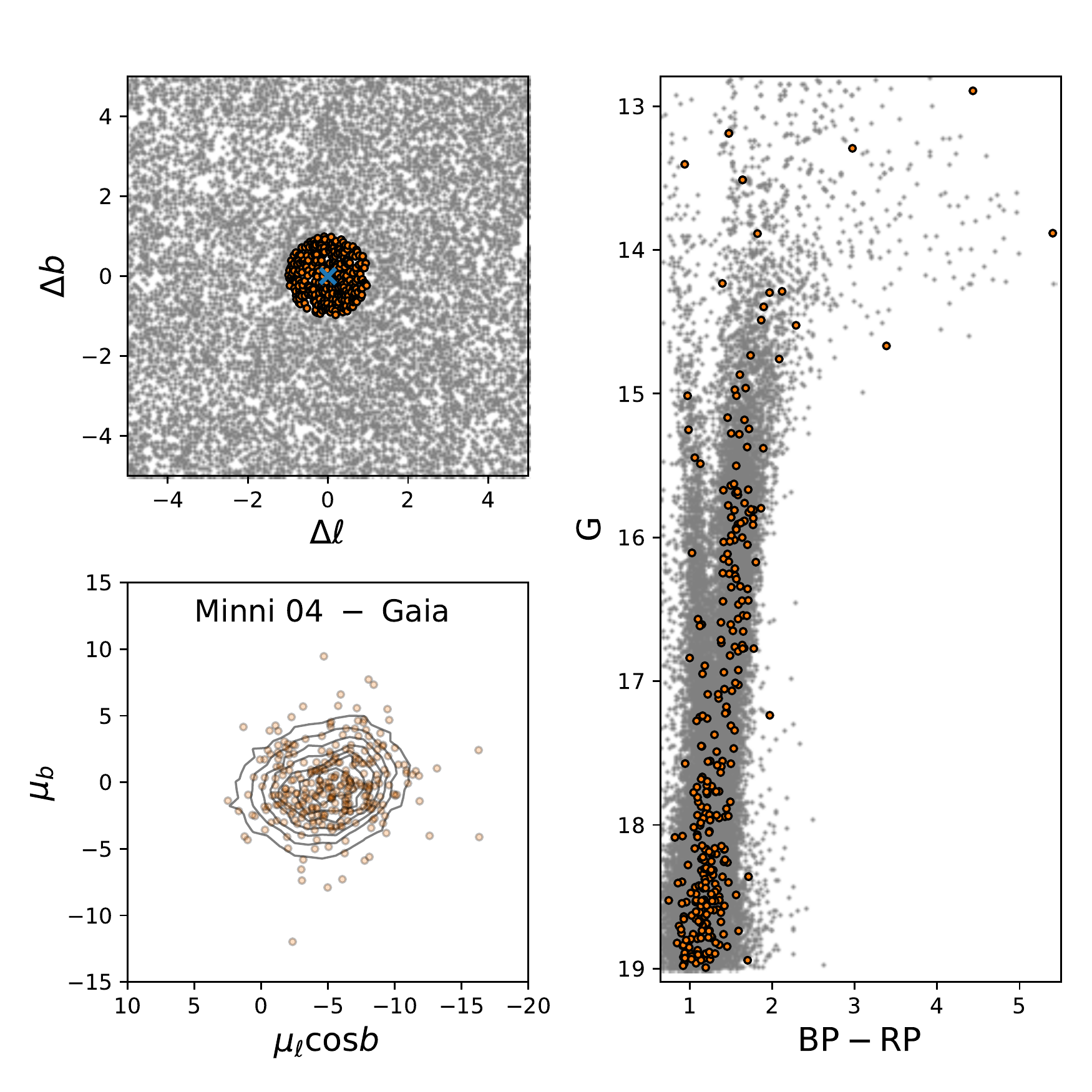} & \\
\includegraphics[width=8cm]{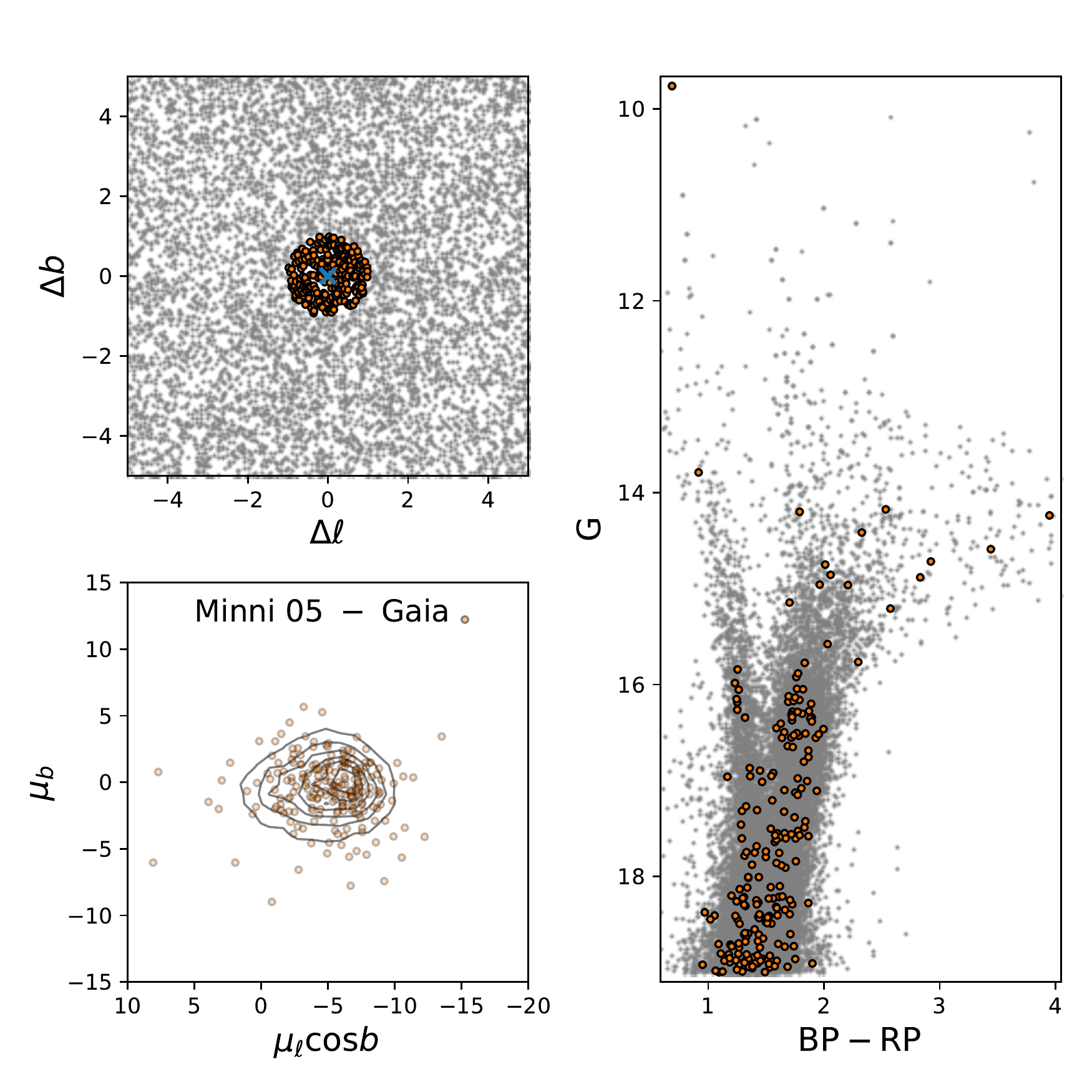} & \\
\end{tabular}
\end{table*}
\newpage
\begin{table*}
\begin{tabular}{cc}
\includegraphics[width=8cm]{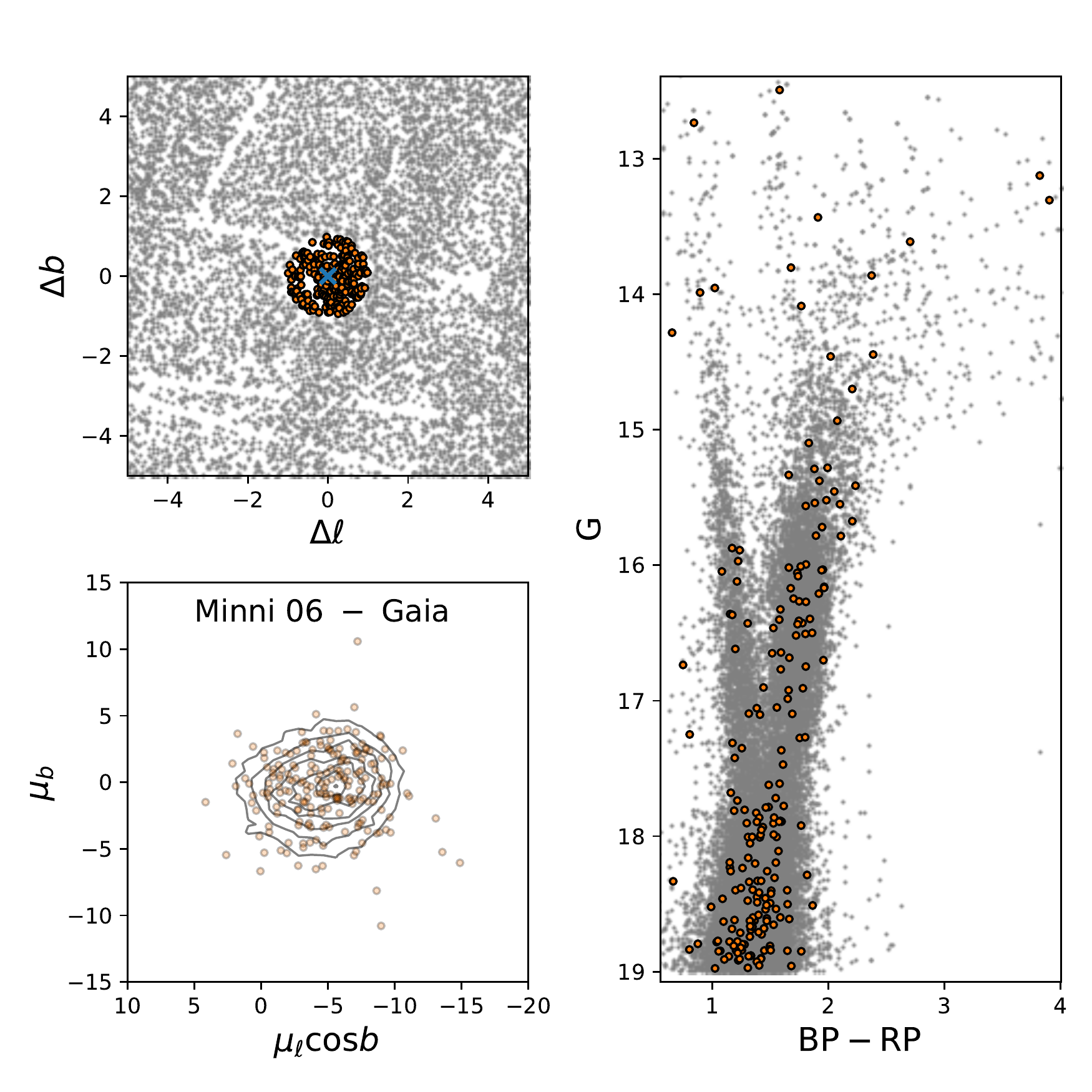} & \\
\includegraphics[width=8cm]{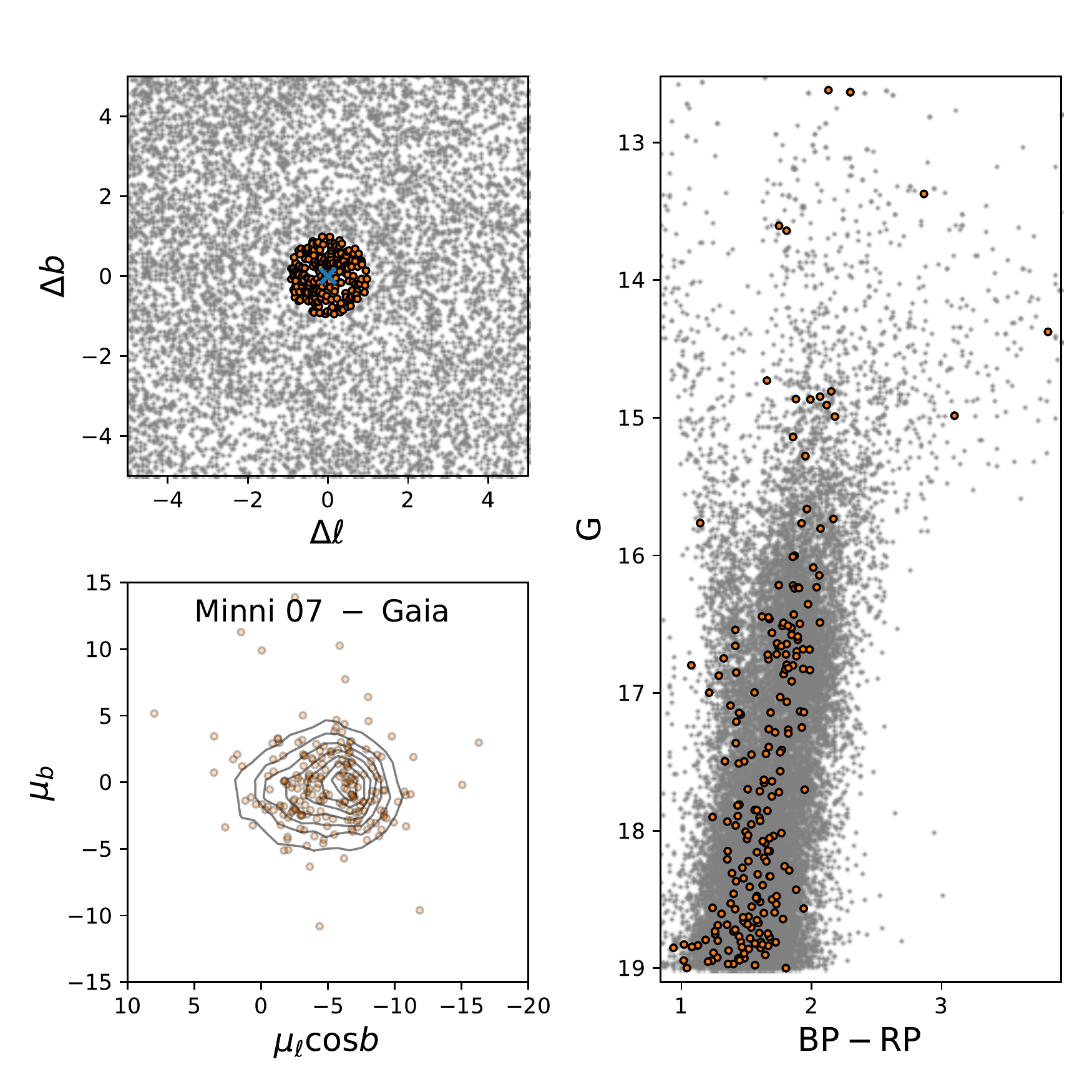} & \\ 
\includegraphics[width=8cm]{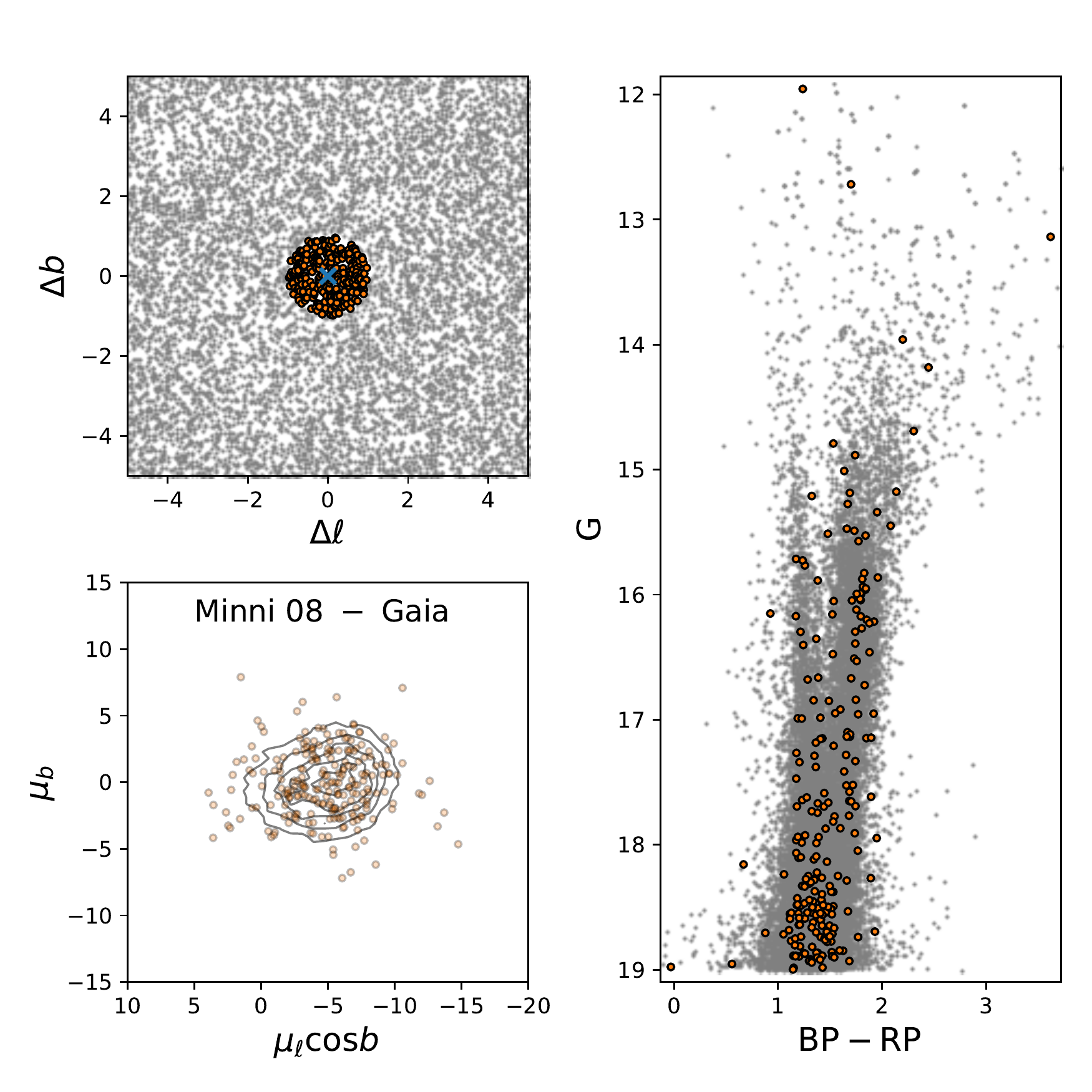} & \\
\end{tabular}
\end{table*}
\newpage
\begin{table*}
\begin{tabular}{cc}
\includegraphics[width=8cm]{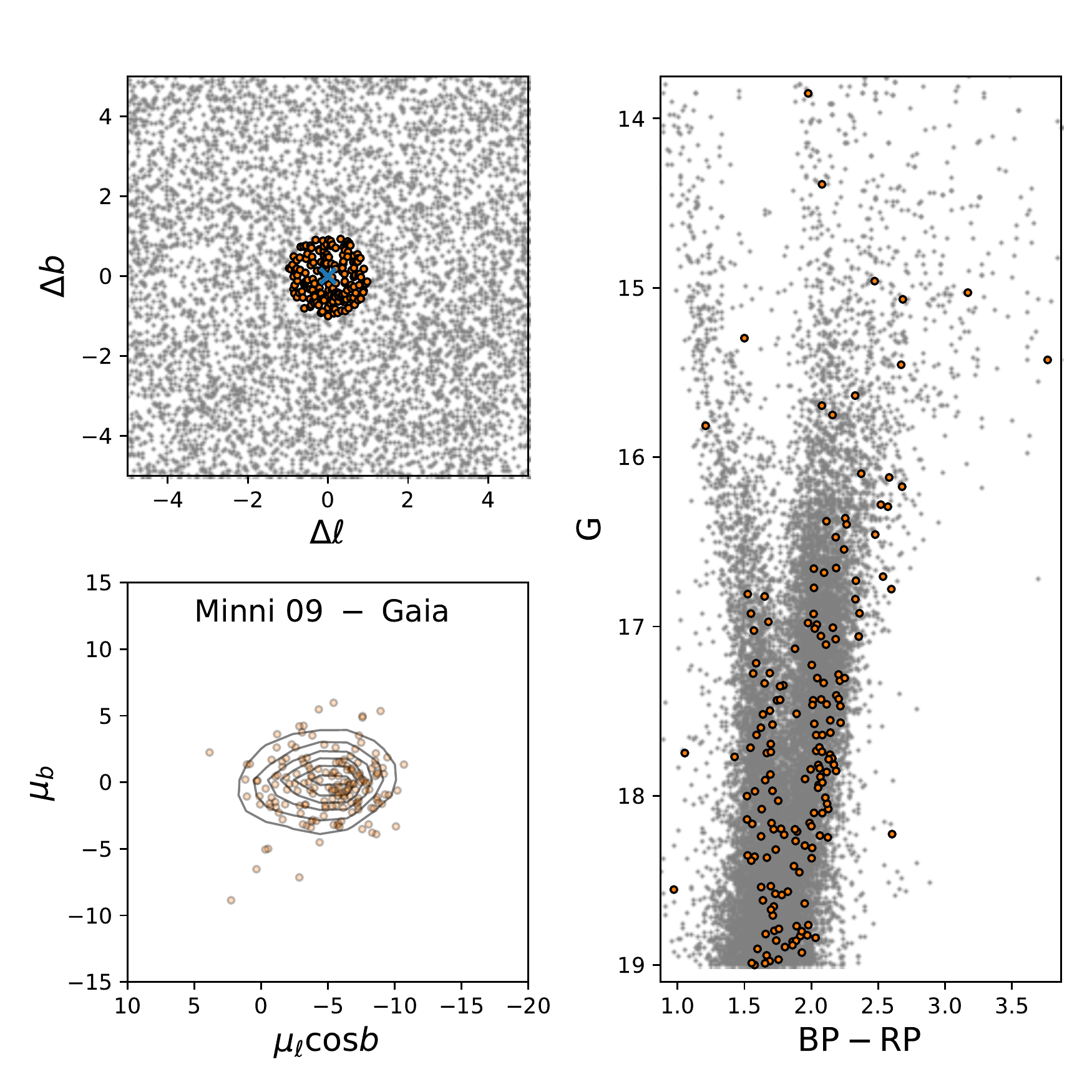} & \\
\includegraphics[width=8cm]{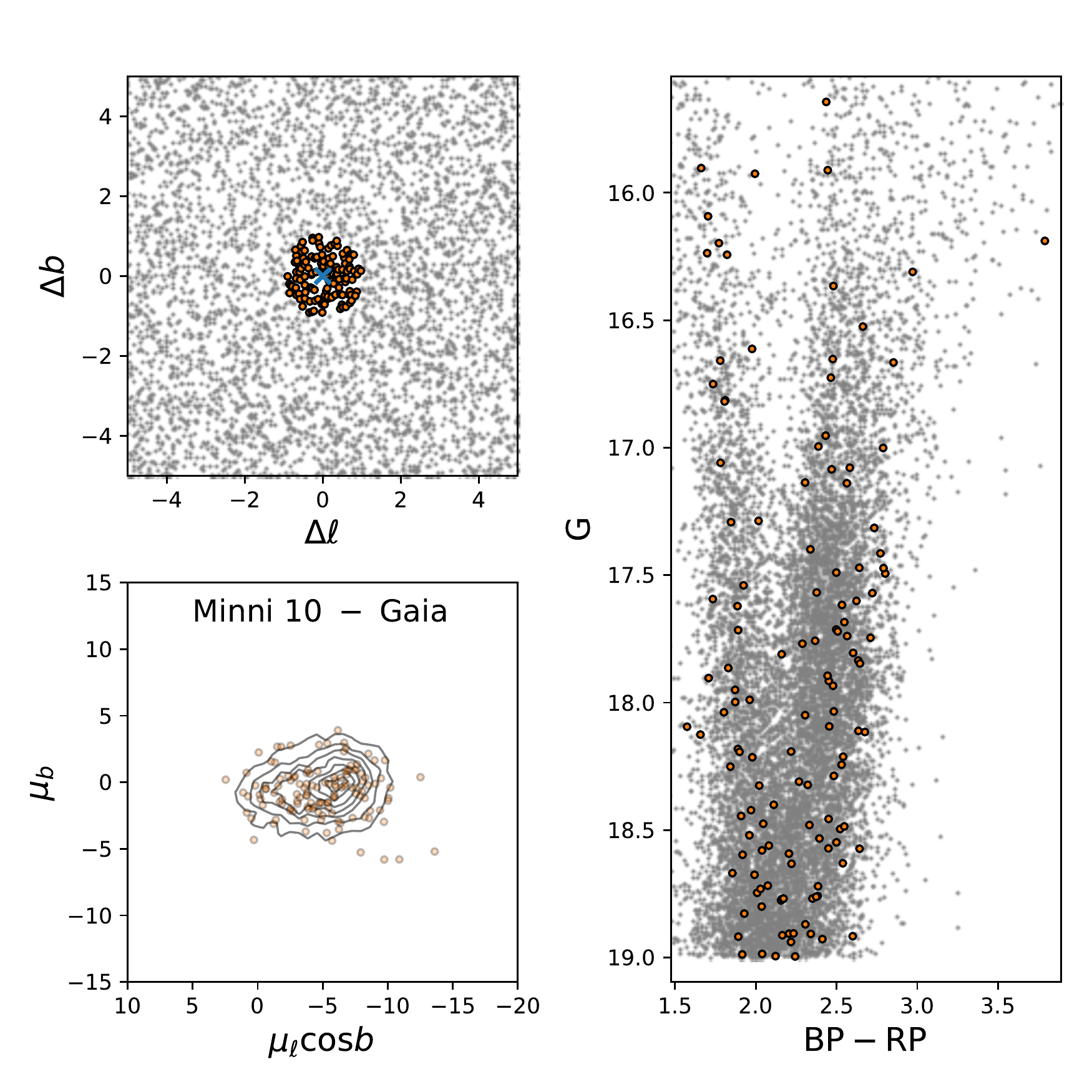} & \\ 
\includegraphics[width=8cm]{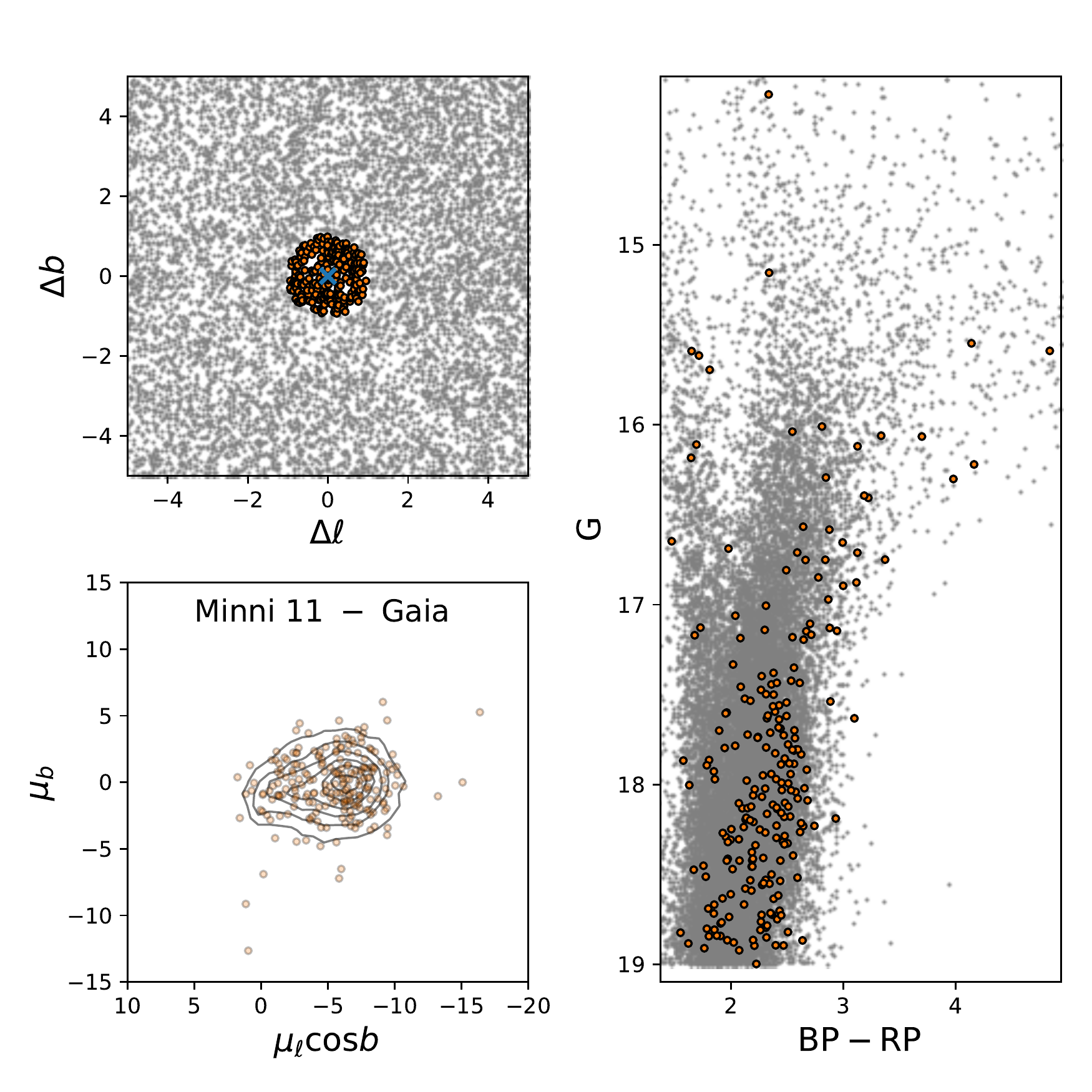} & \\
\end{tabular}
\end{table*}
\newpage
\begin{table*}
\begin{tabular}{cc}
\includegraphics[width=8cm]{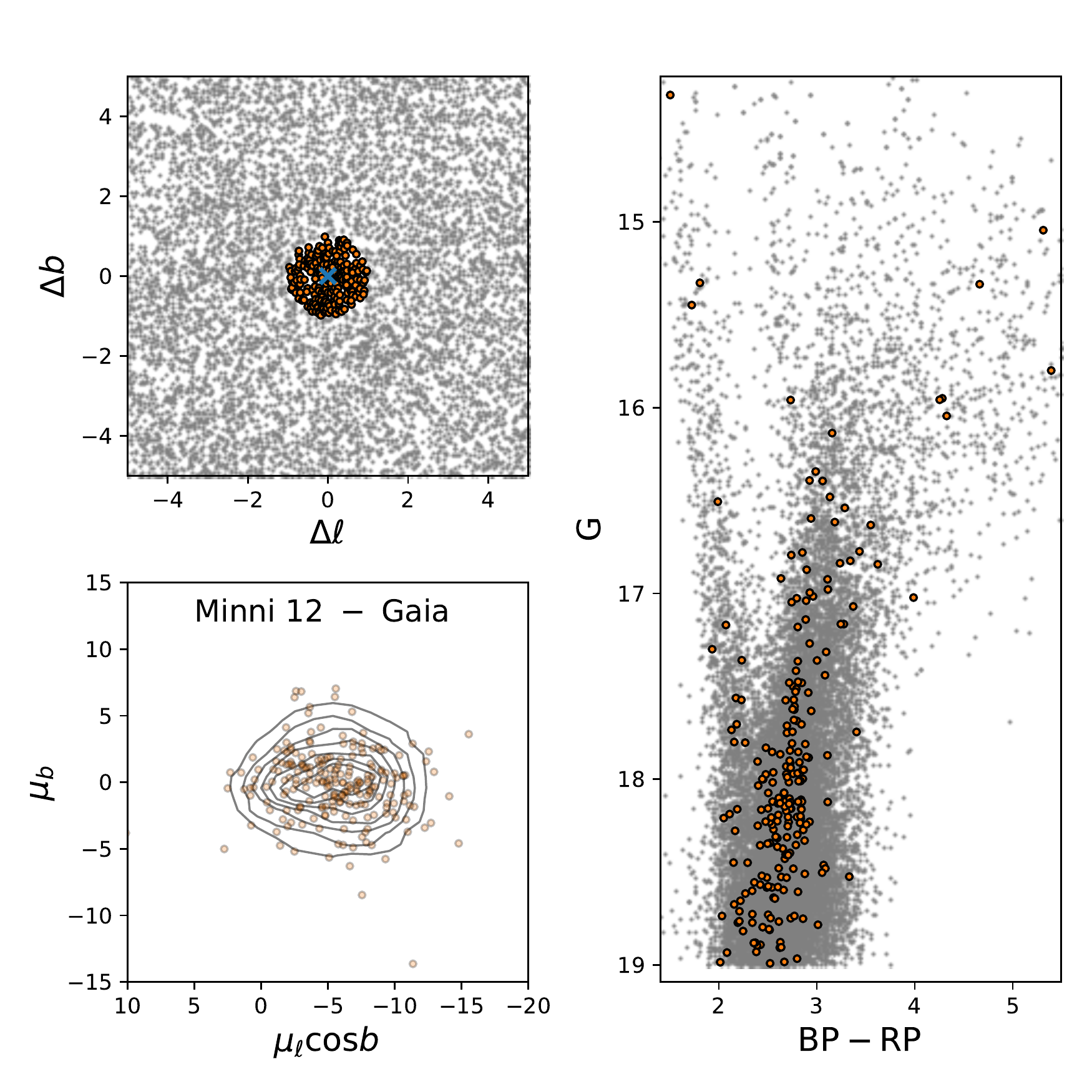} & 
\includegraphics[width=8cm]{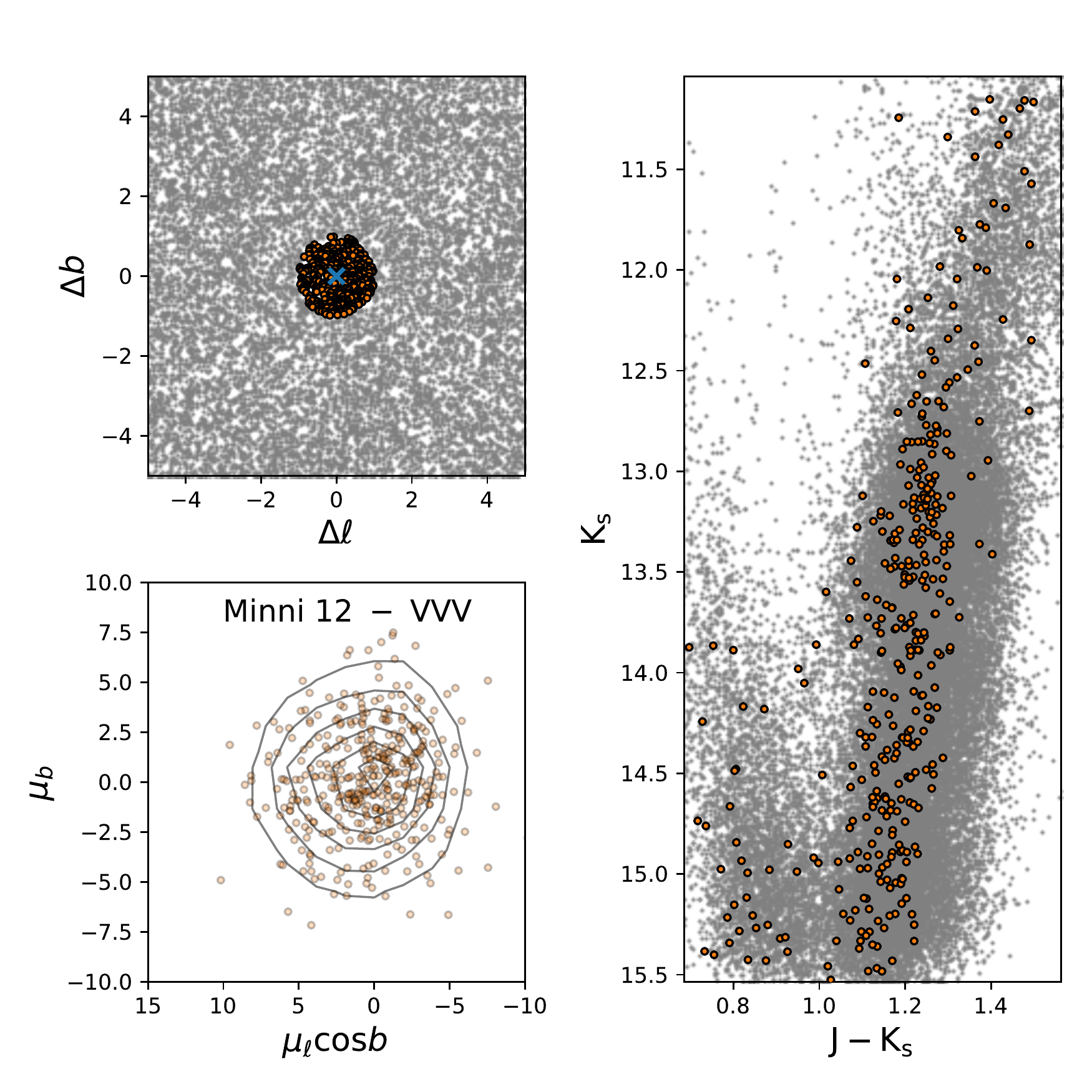} \\  
\includegraphics[width=8cm]{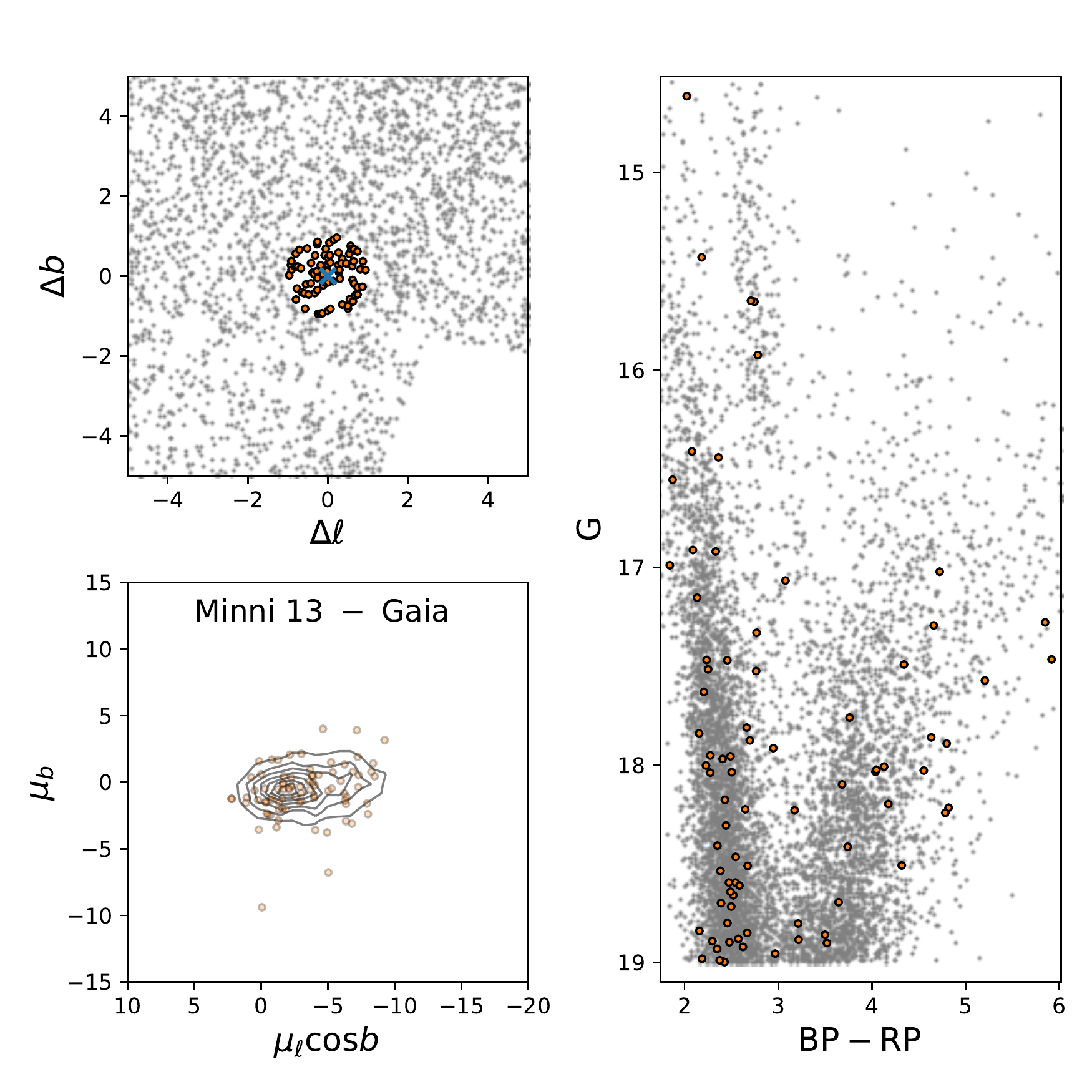} &
\includegraphics[width=8cm]{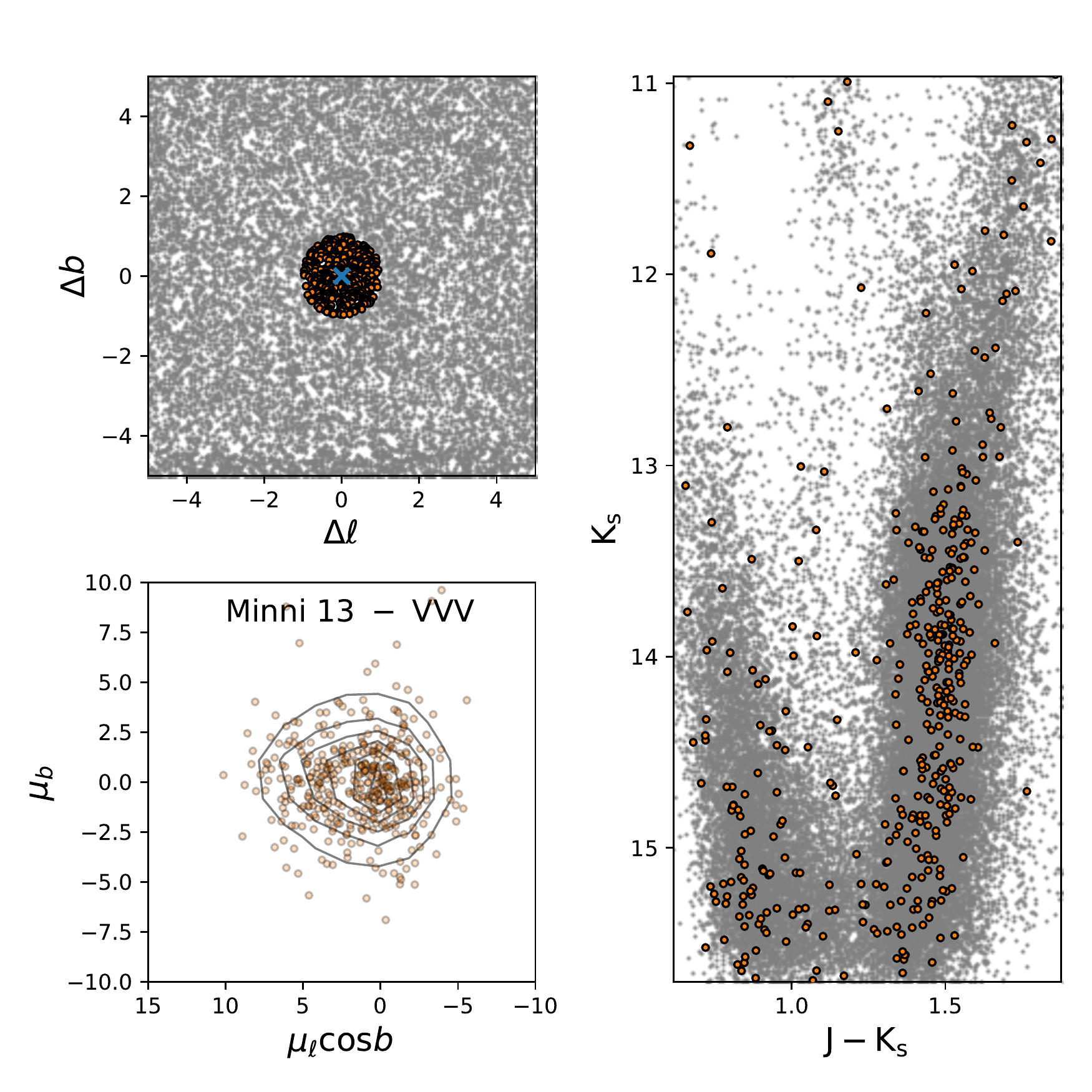} \\
\includegraphics[width=8cm]{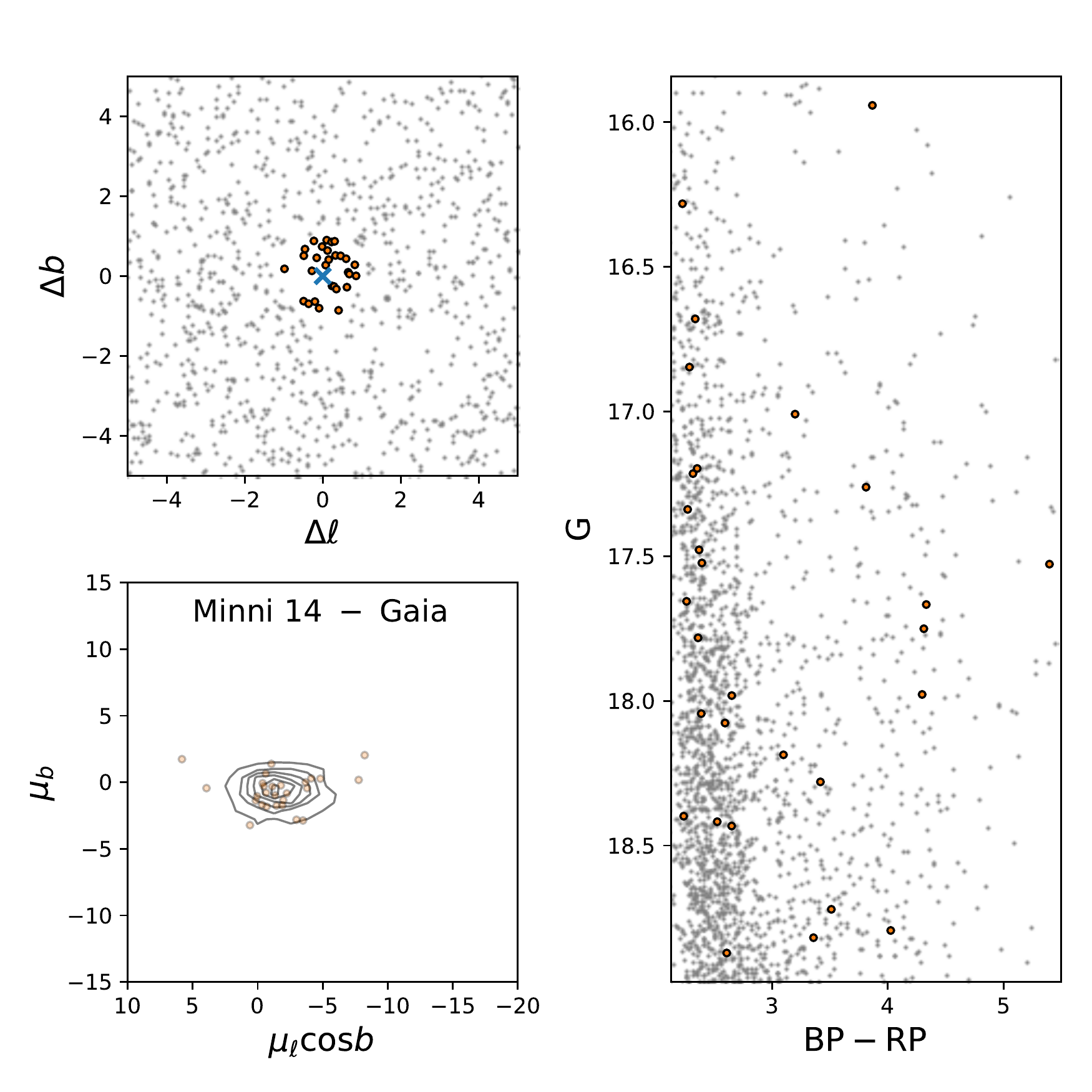} & 
\includegraphics[width=8cm]{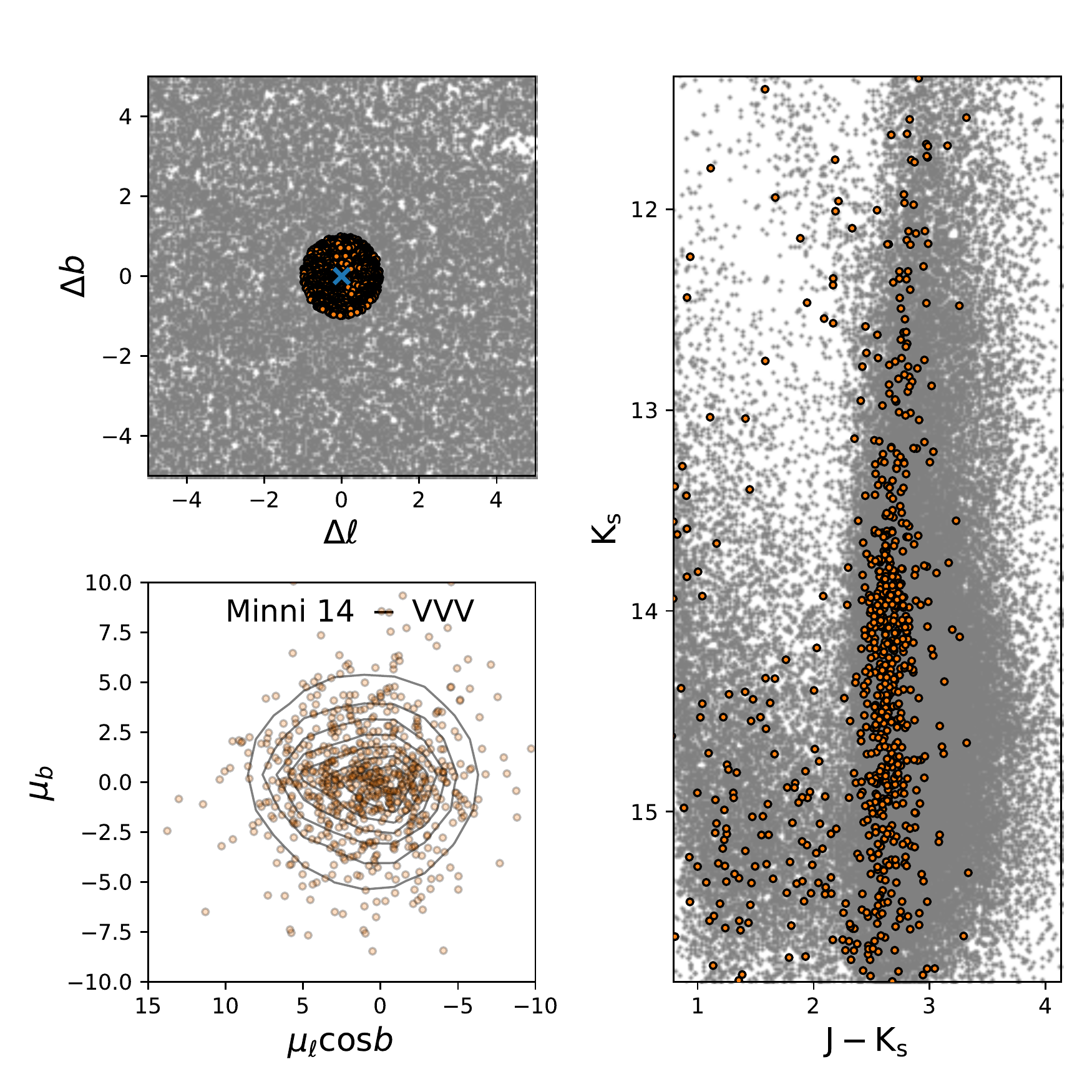} \\
\end{tabular}
\end{table*}
\newpage
\begin{table*}
\begin{tabular}{cc}
\includegraphics[width=8cm]{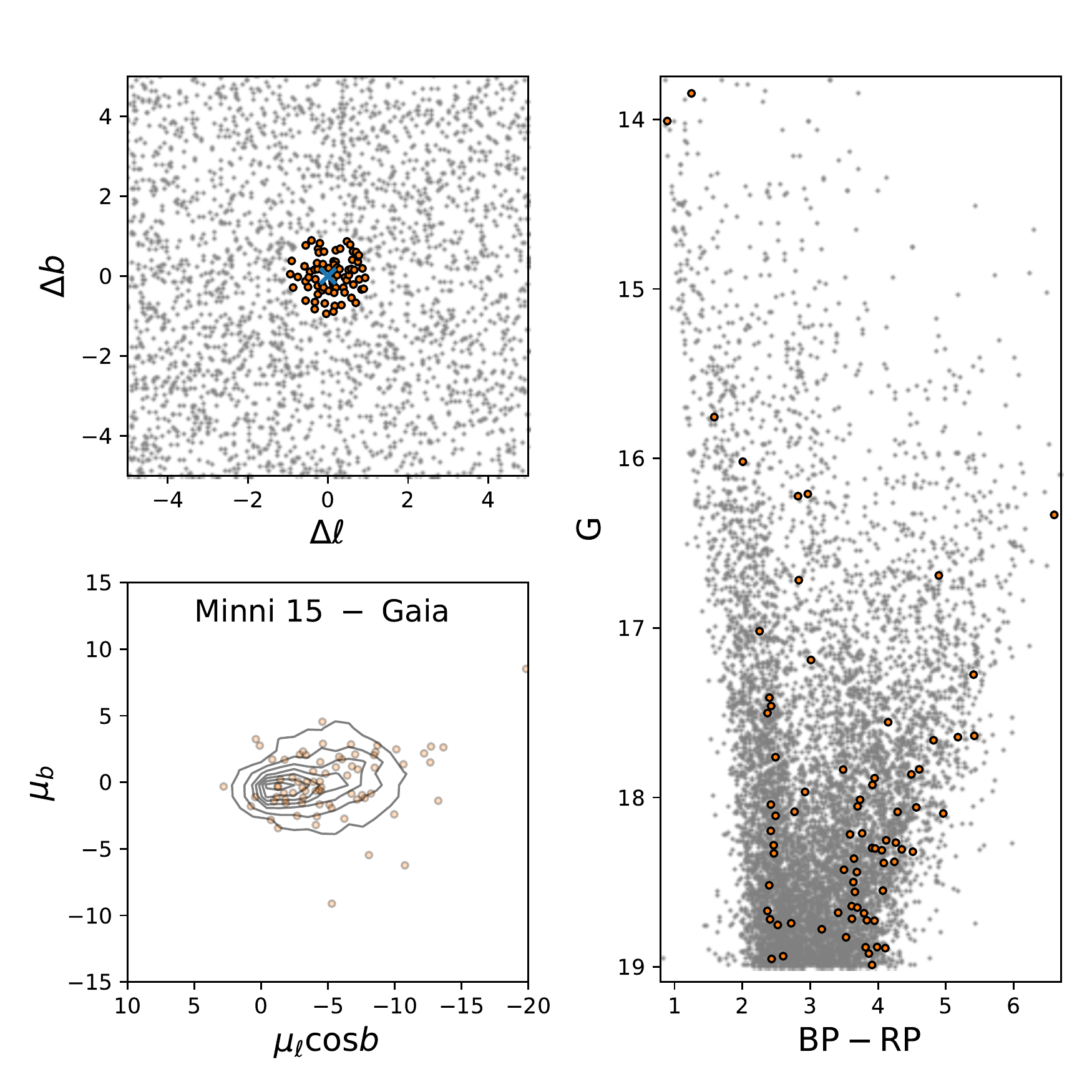} &  
\includegraphics[width=8cm]{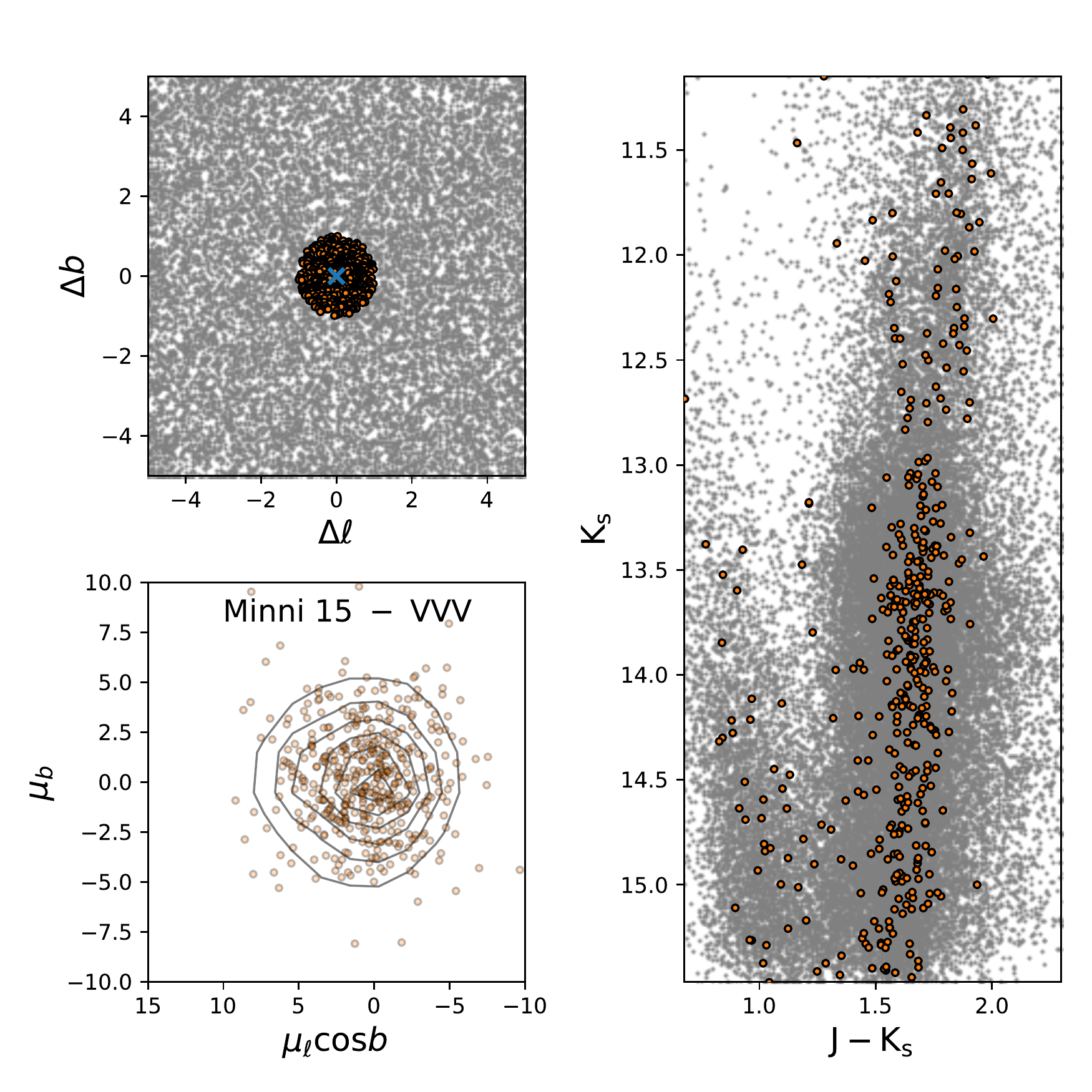} \\  
\includegraphics[width=8cm]{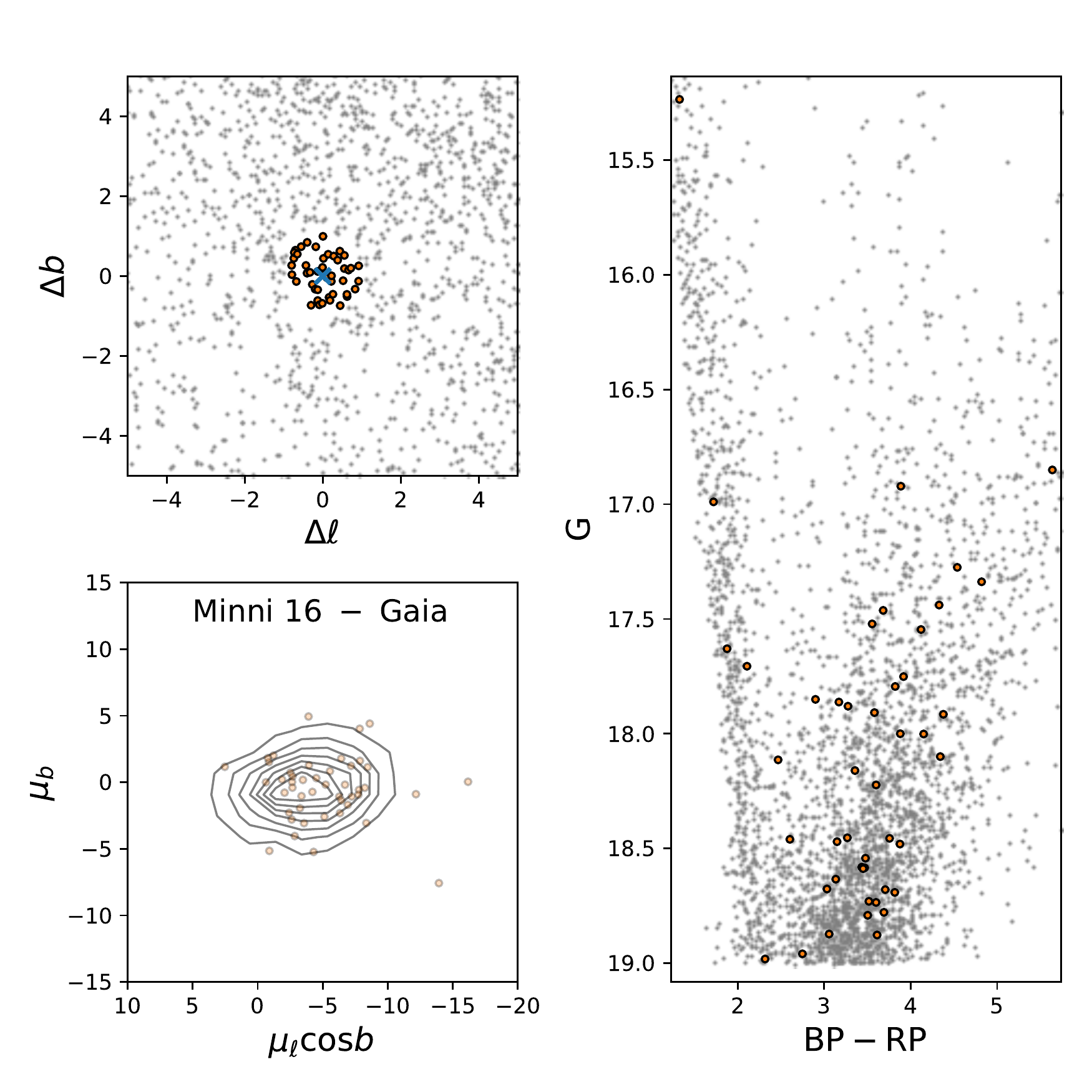} & 
\includegraphics[width=8cm]{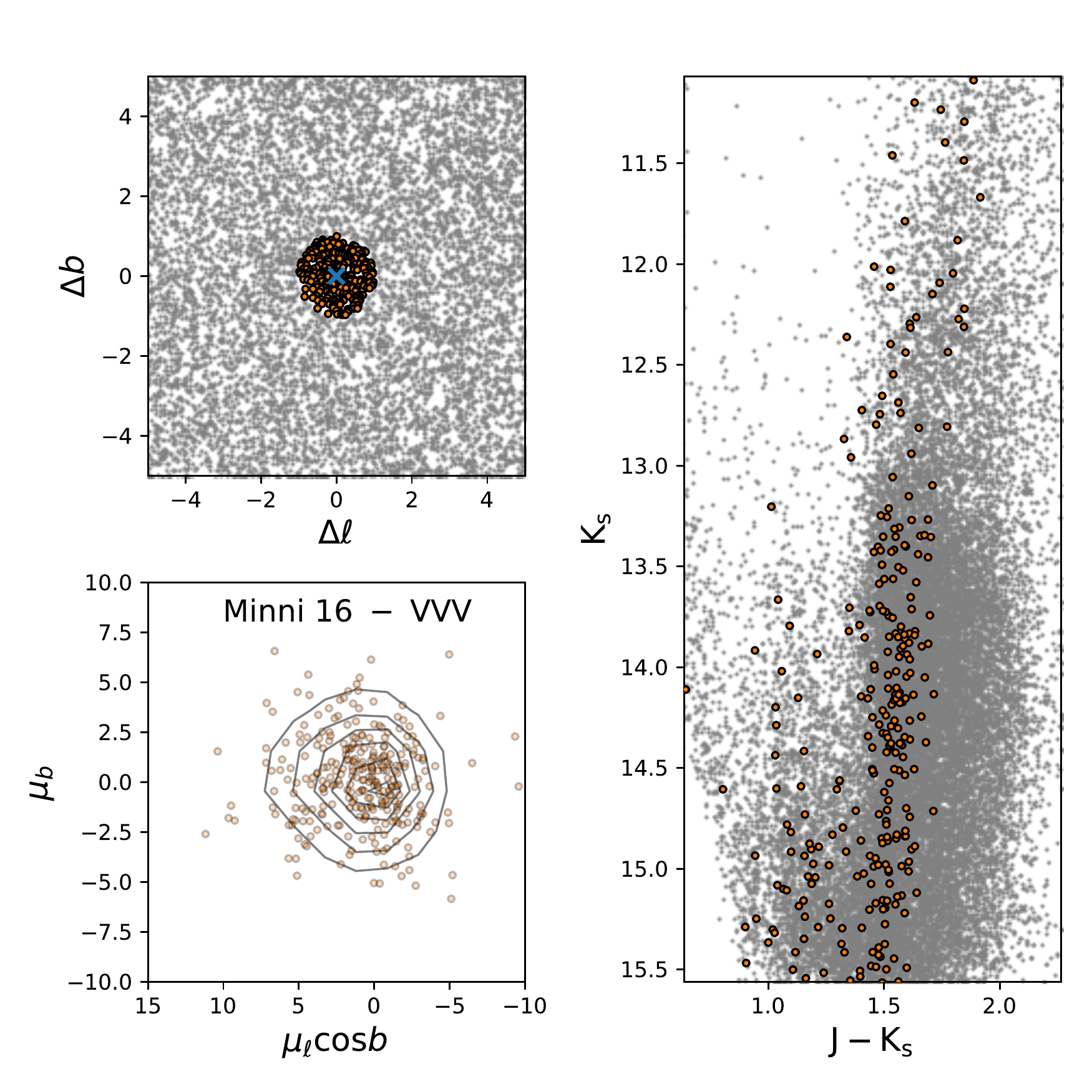} \\  
\includegraphics[width=8cm]{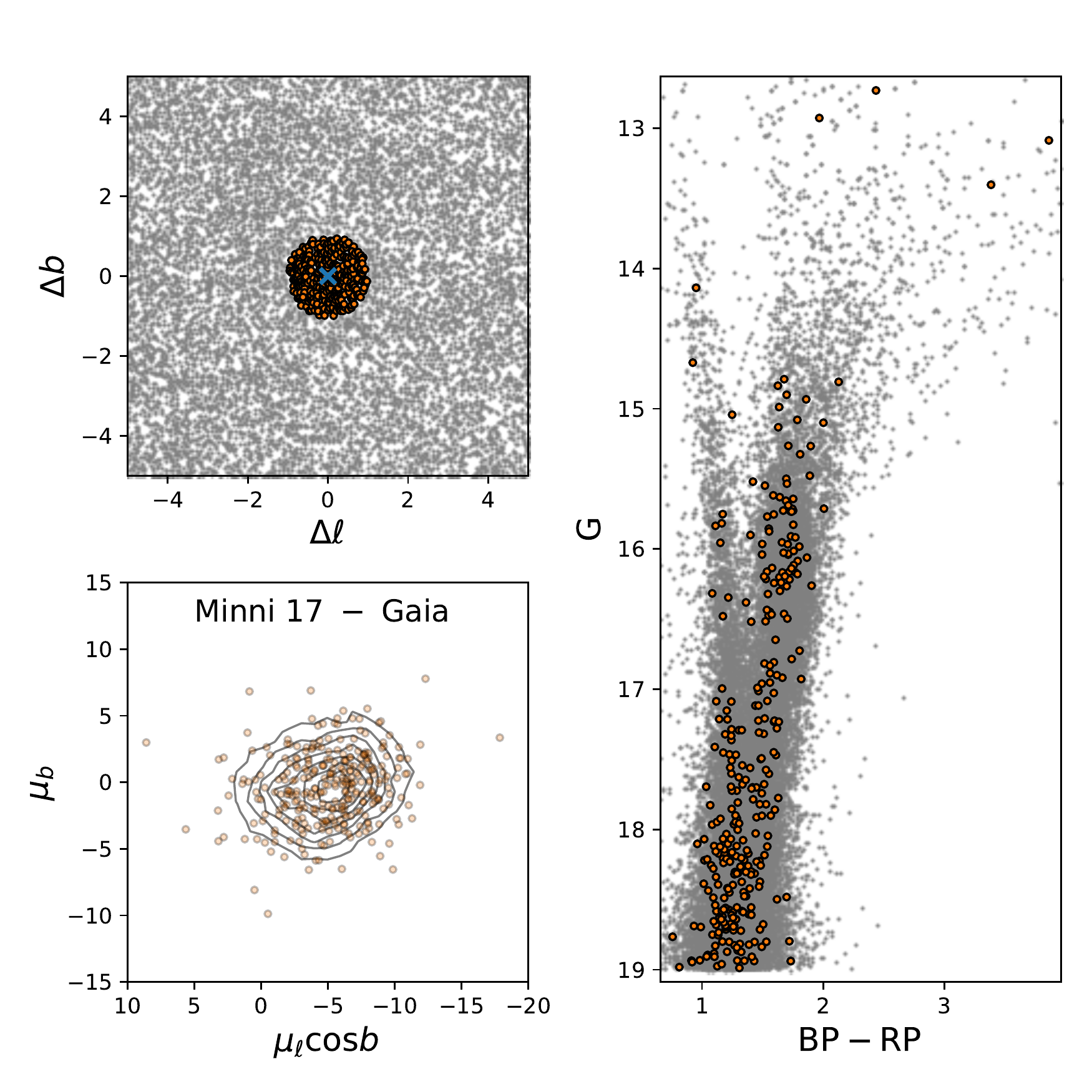} & \\
\end{tabular}
\end{table*}
\newpage
\begin{table*}
\begin{tabular}{cc}
\includegraphics[width=8cm]{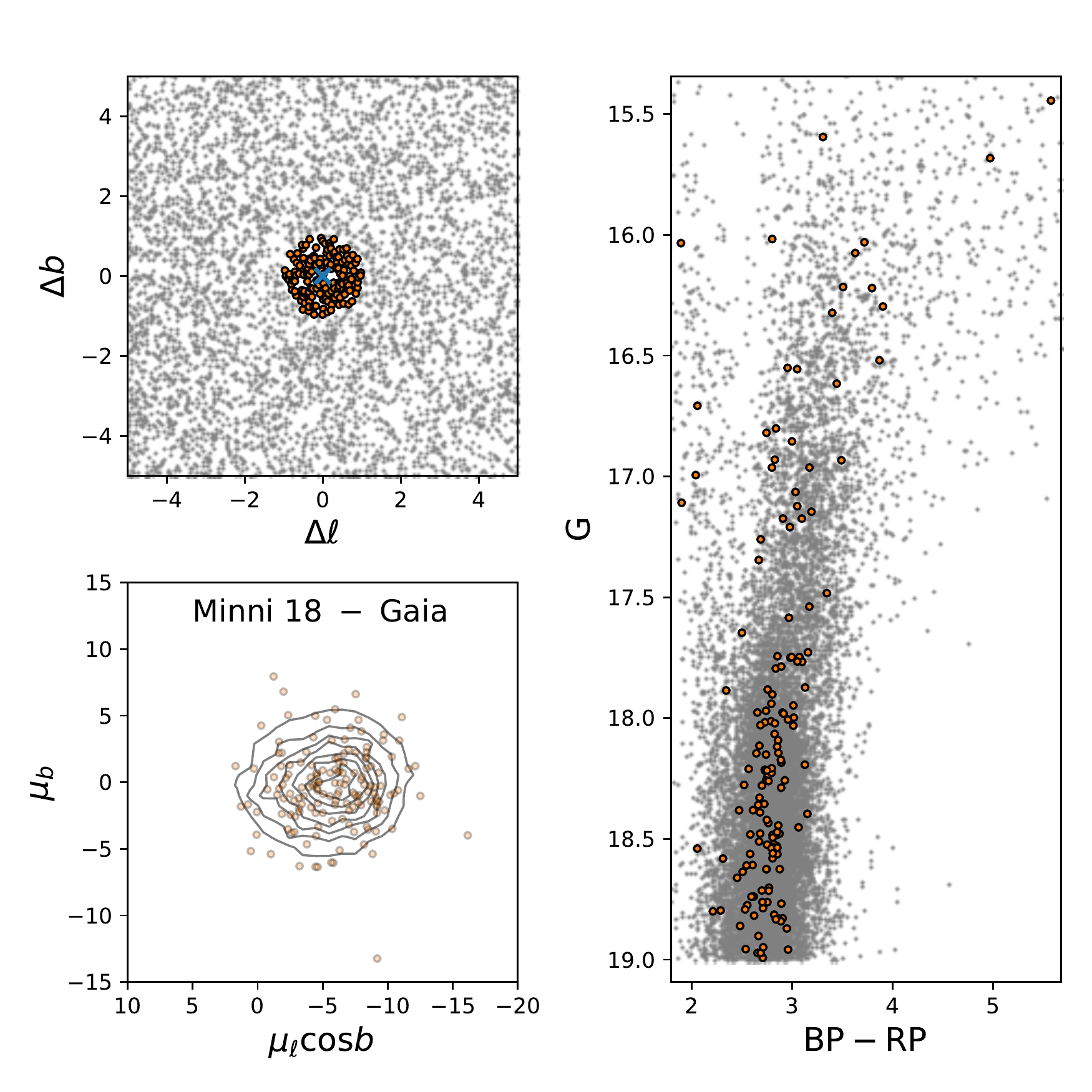} & \\
\includegraphics[width=8cm]{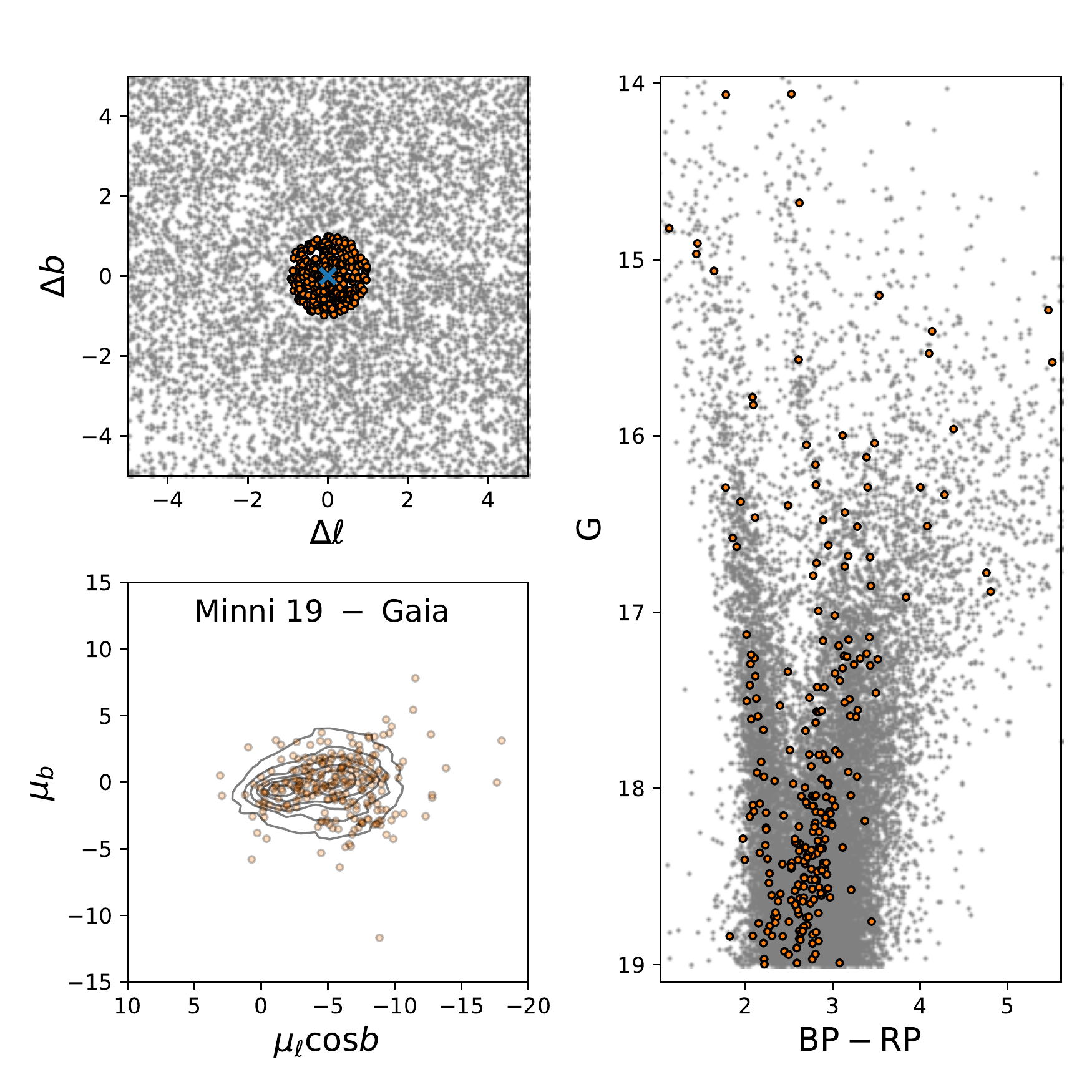} &  
\includegraphics[width=8cm]{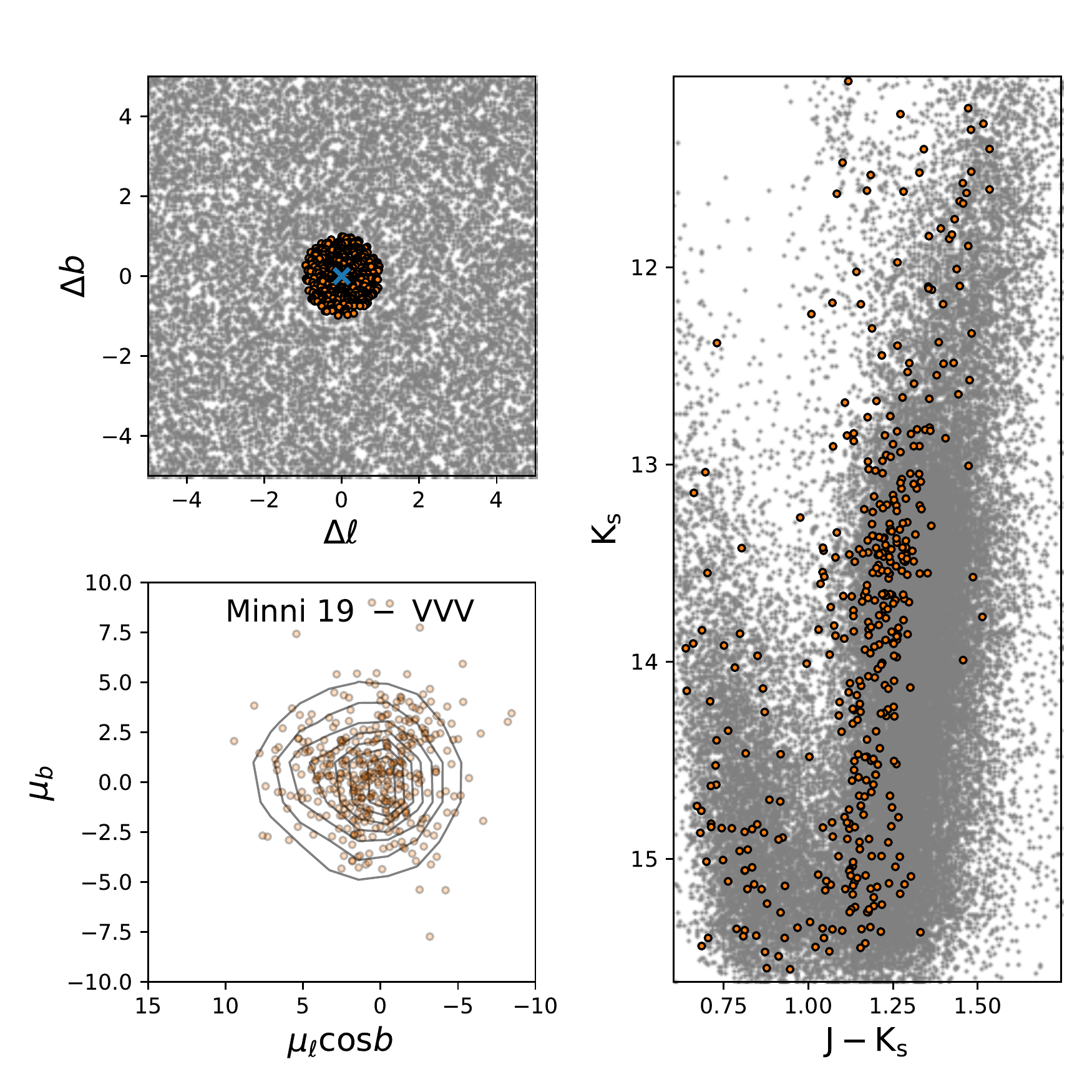} \\ 
\includegraphics[width=8cm]{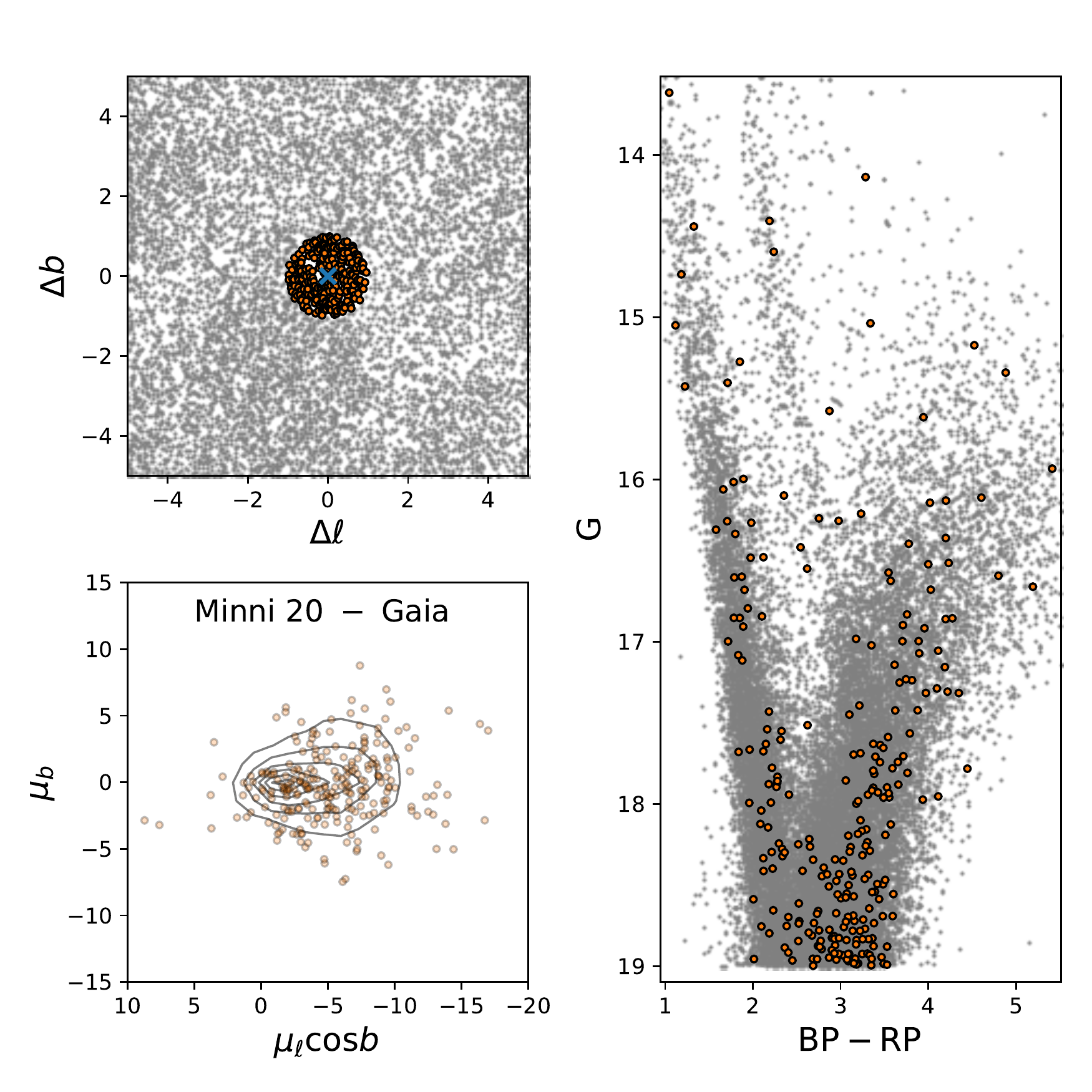} &
\includegraphics[width=8cm]{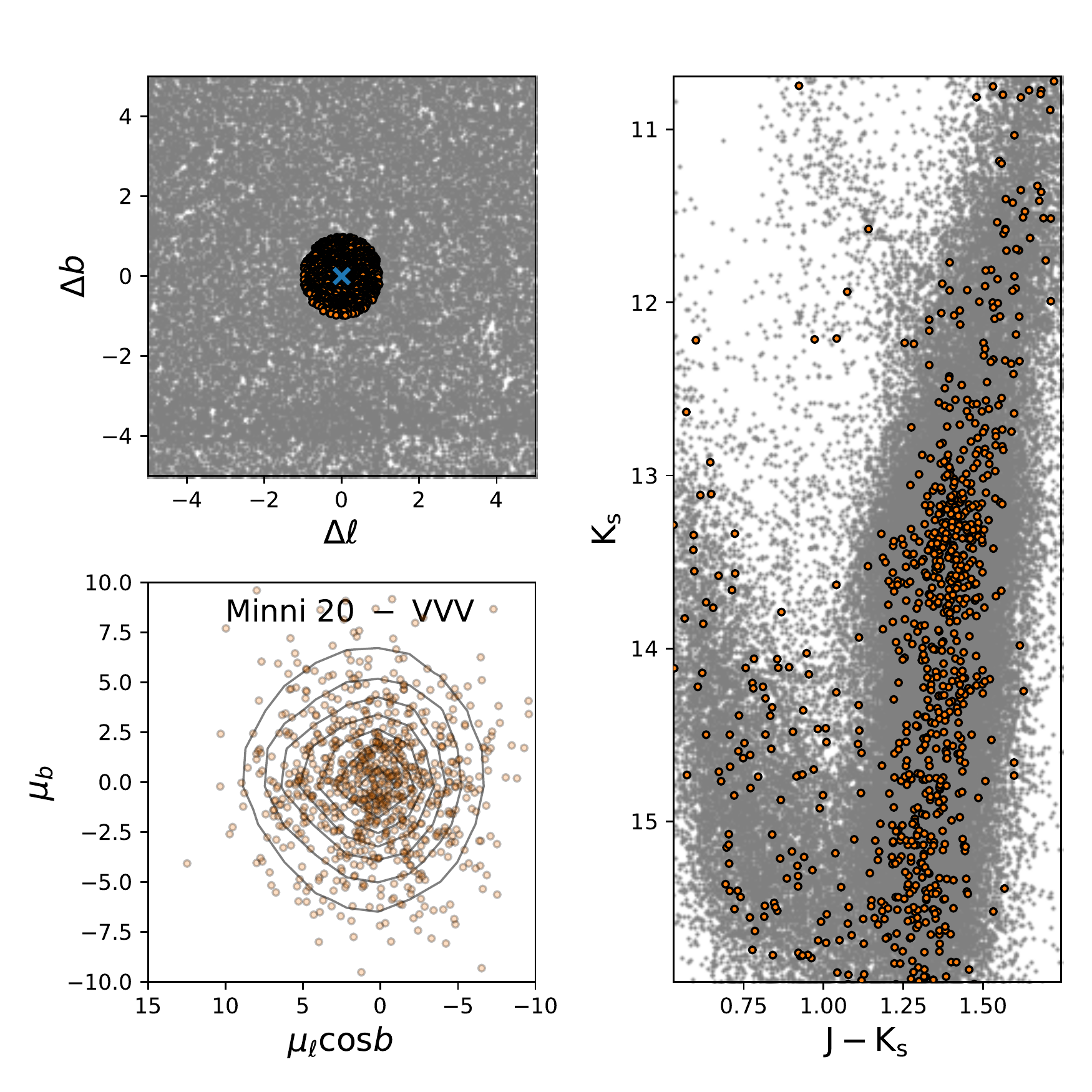} \\
\end{tabular}
\end{table*}
\newpage
\begin{table*}
\begin{tabular}{cc}
\includegraphics[width=8cm]{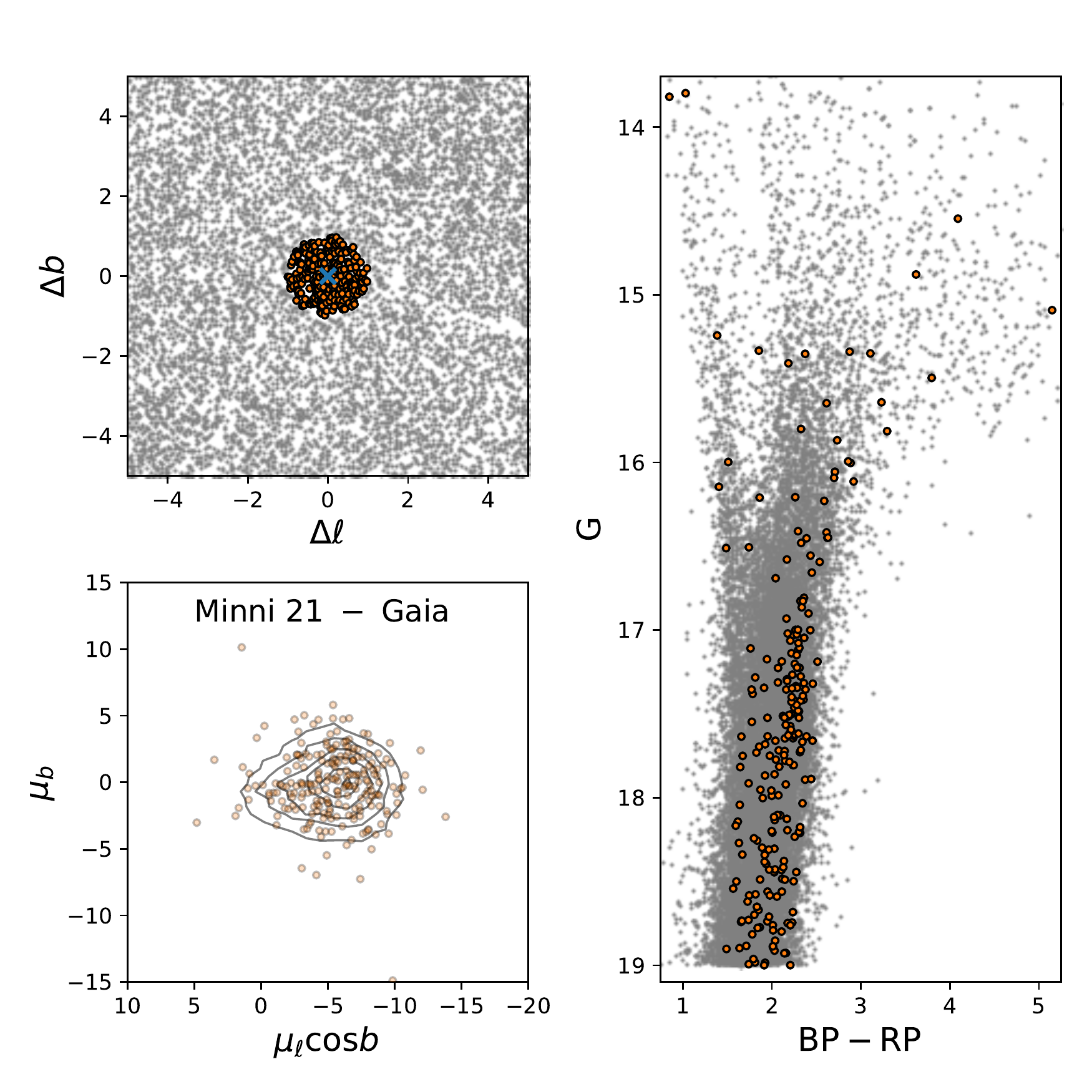} & \\
\includegraphics[width=8cm]{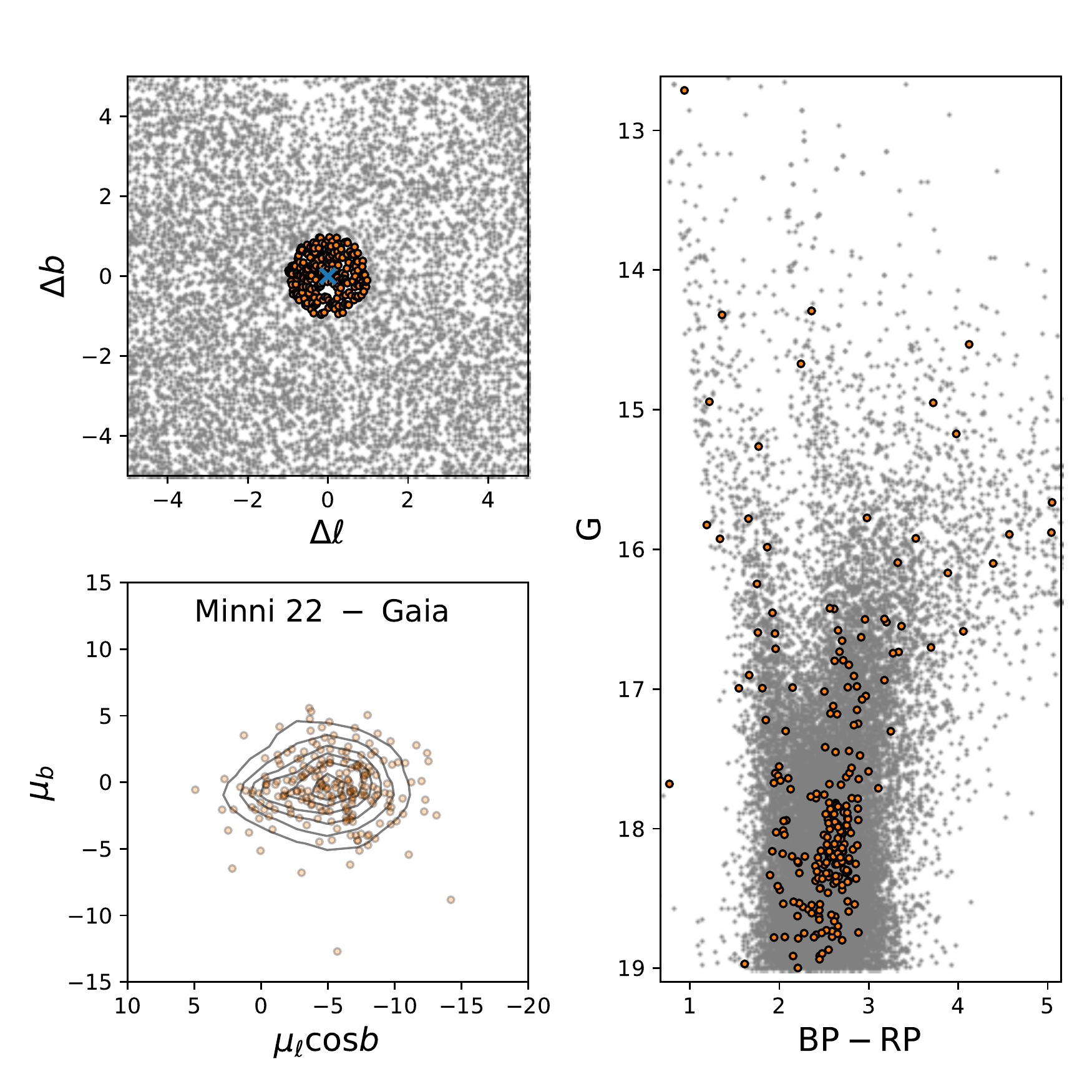} & \\ 
\includegraphics[width=8cm]{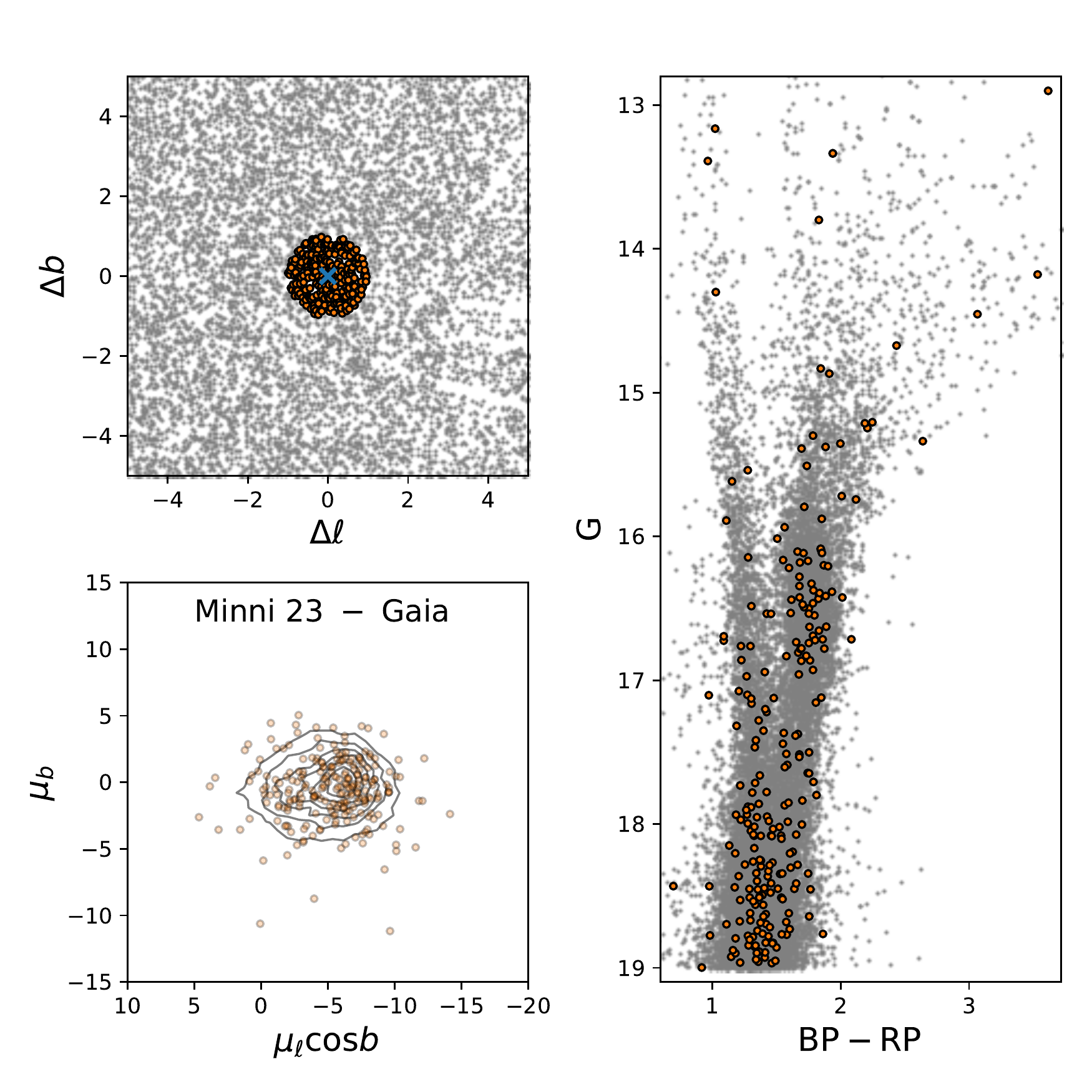} & \\
\end{tabular}
\end{table*}
\newpage
\begin{table*}
\begin{tabular}{cc}
\includegraphics[width=8cm]{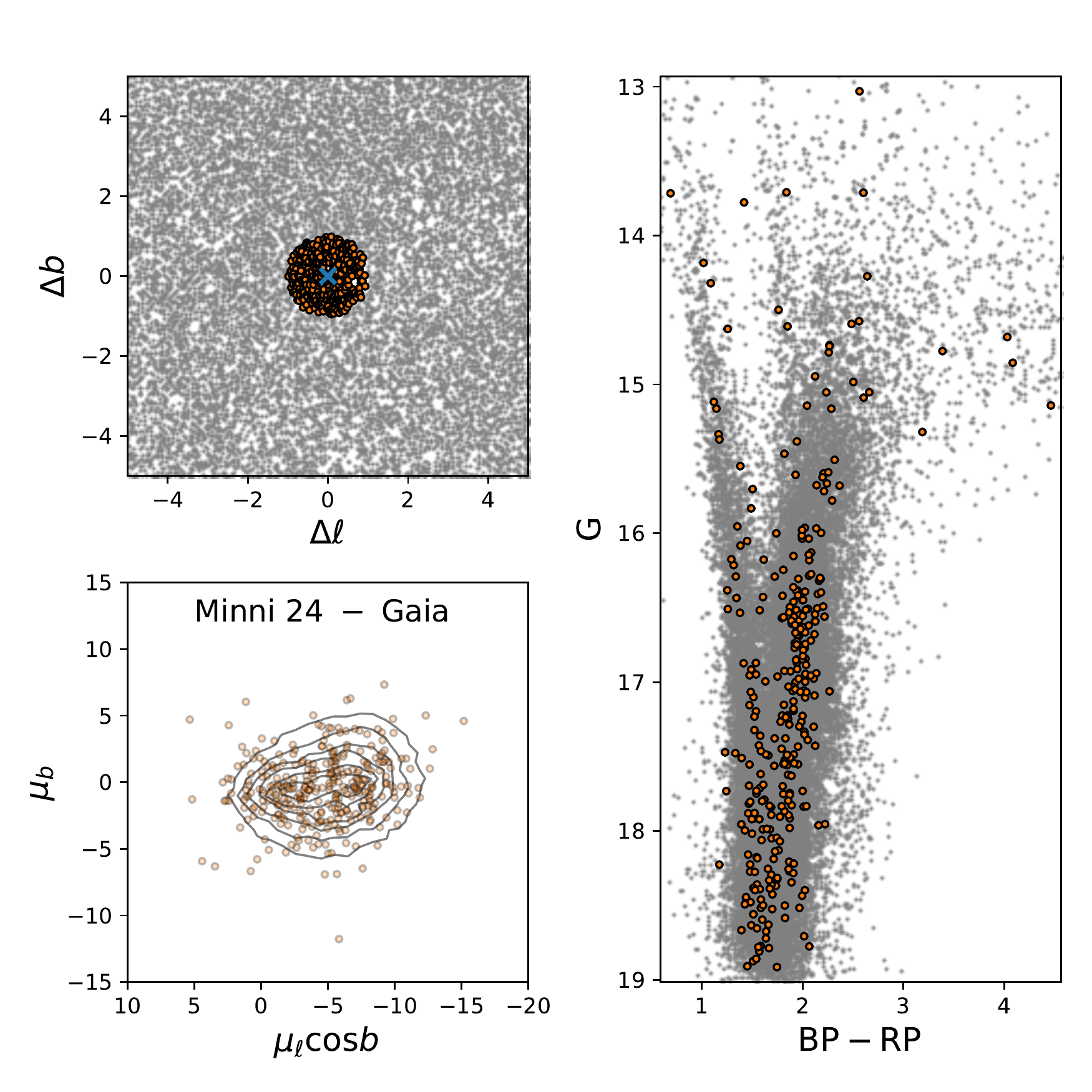} & \\ 
\includegraphics[width=8cm]{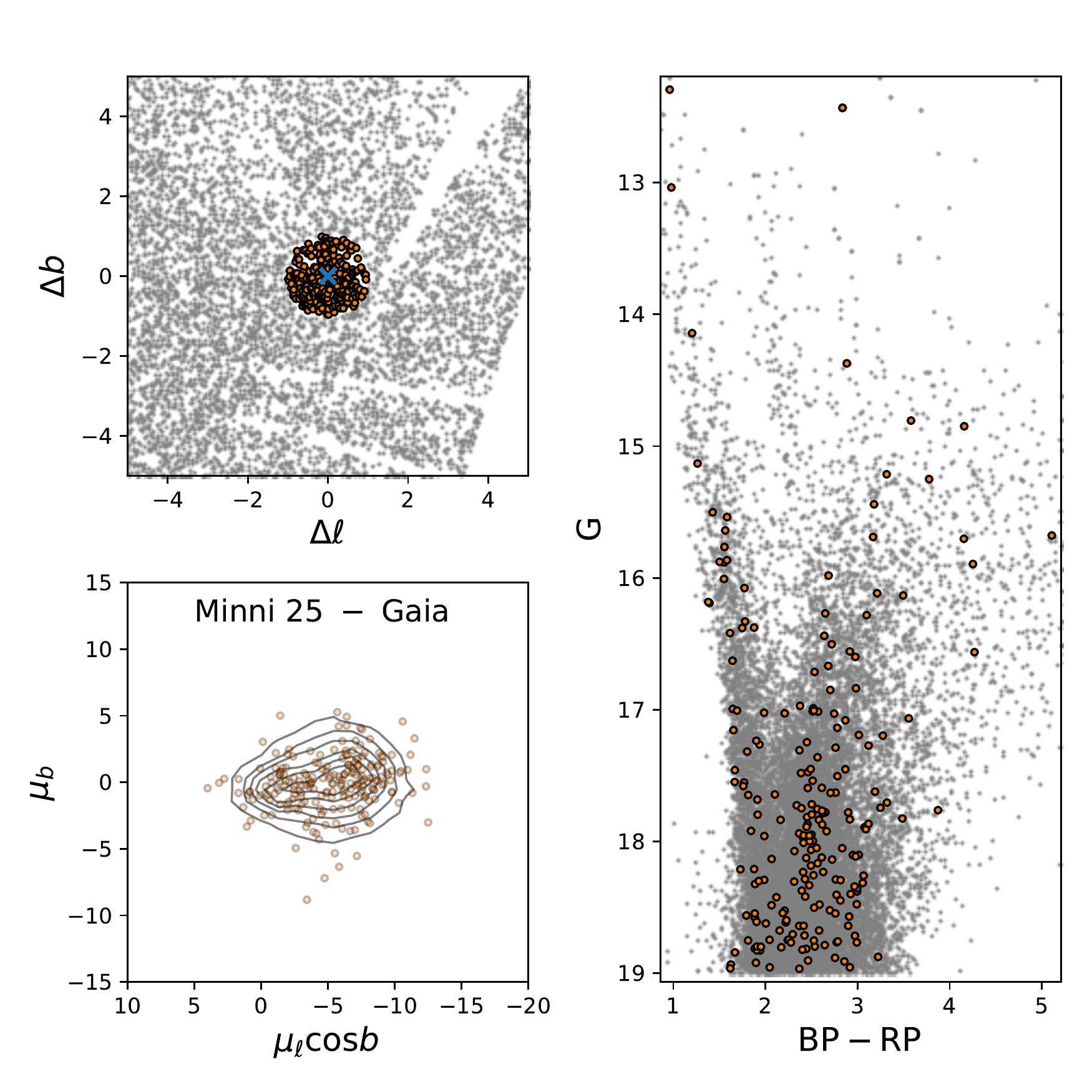} &
\includegraphics[width=8cm]{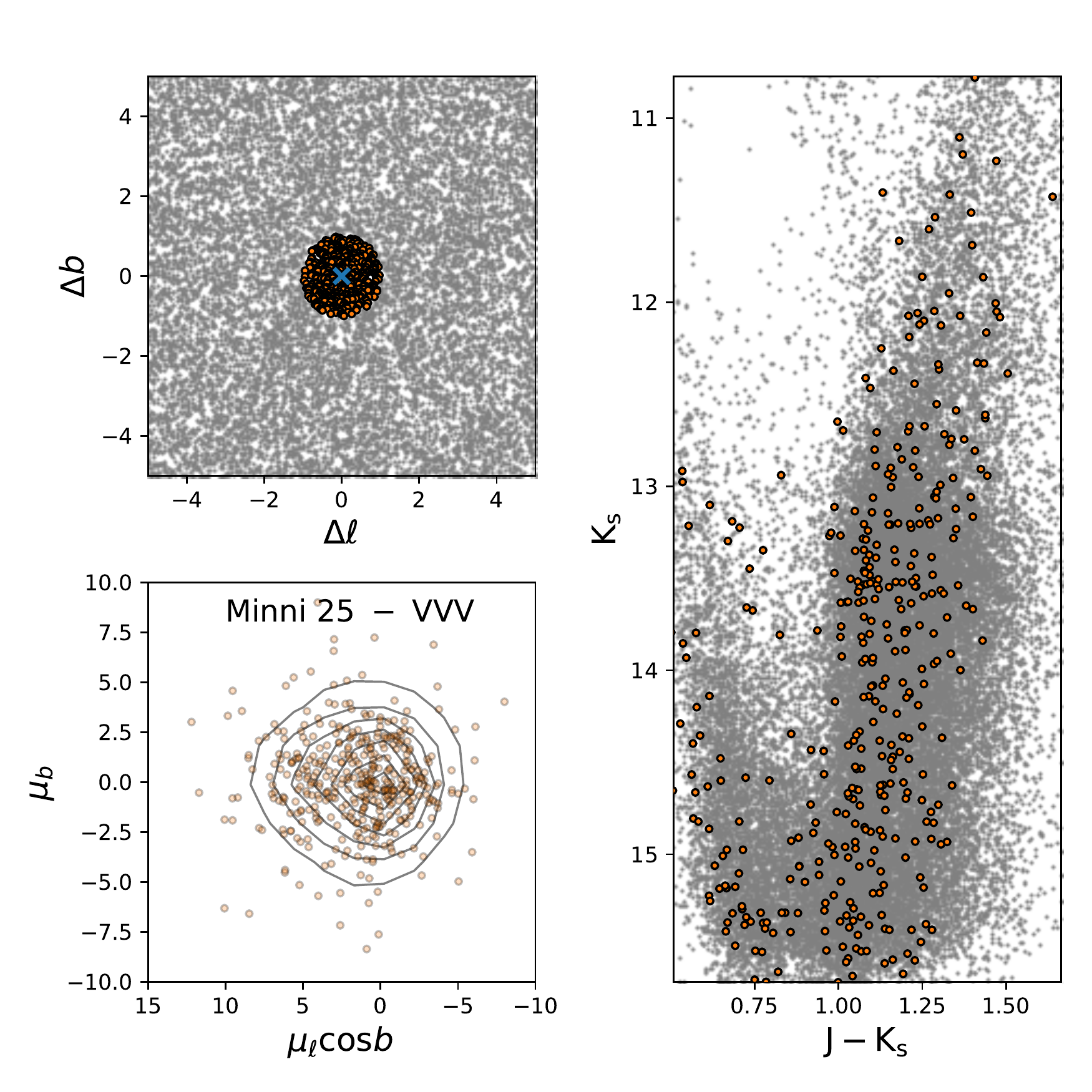} \\  
\includegraphics[width=8cm]{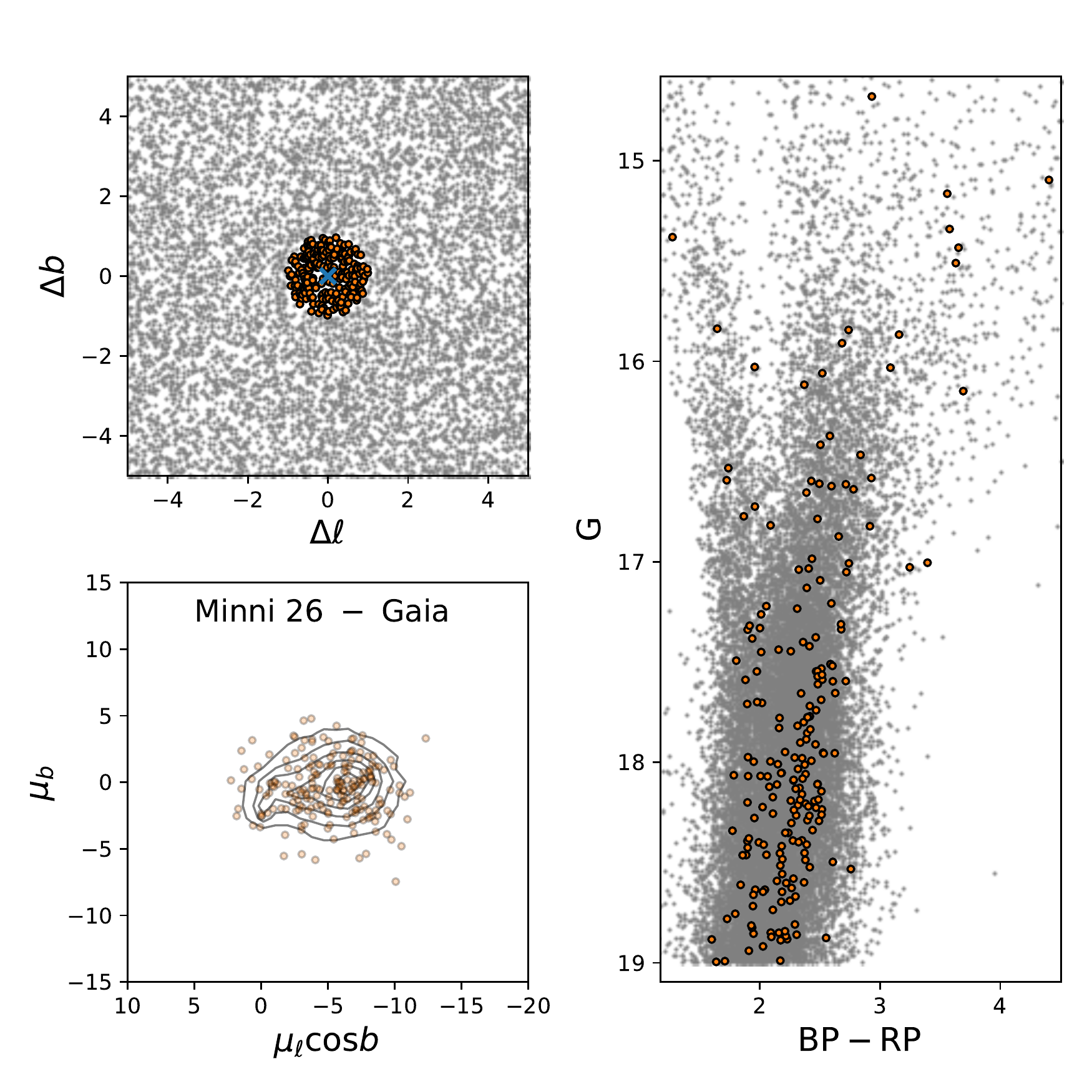} & \\
\end{tabular}
\end{table*}
\newpage
\begin{table*}
\begin{tabular}{cc}
\includegraphics[width=8cm]{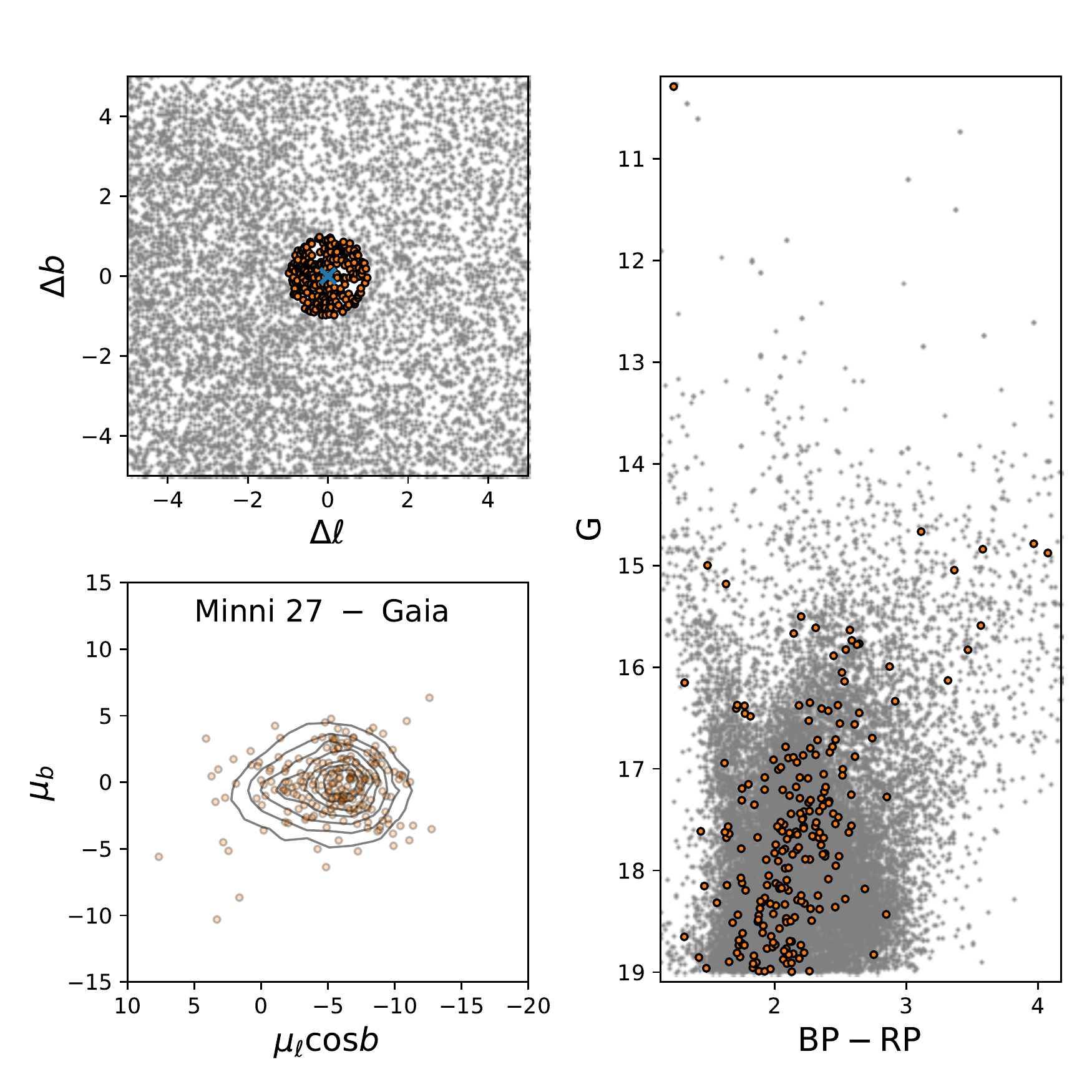} & \\ 
\includegraphics[width=8cm]{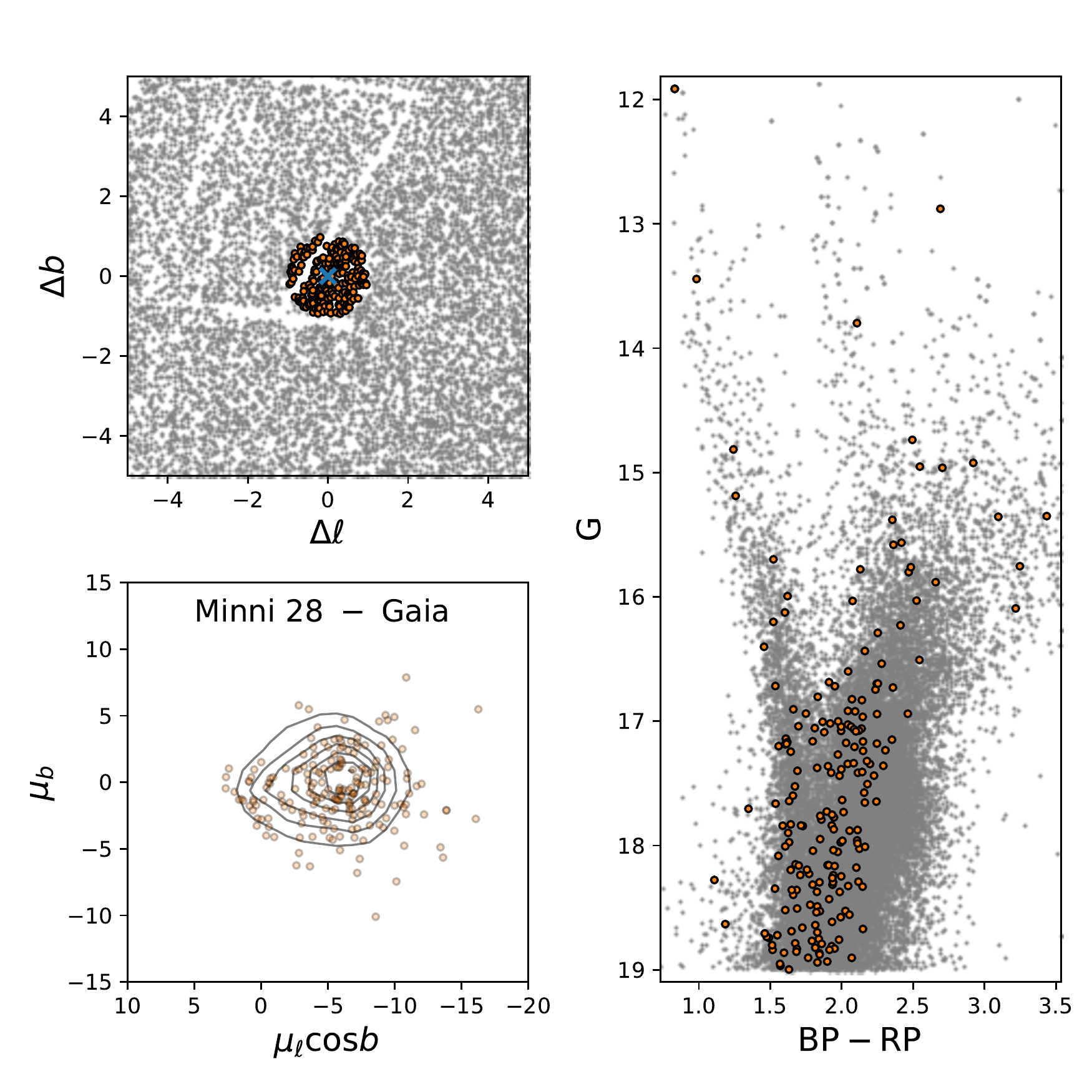} & \\ 
\includegraphics[width=8cm]{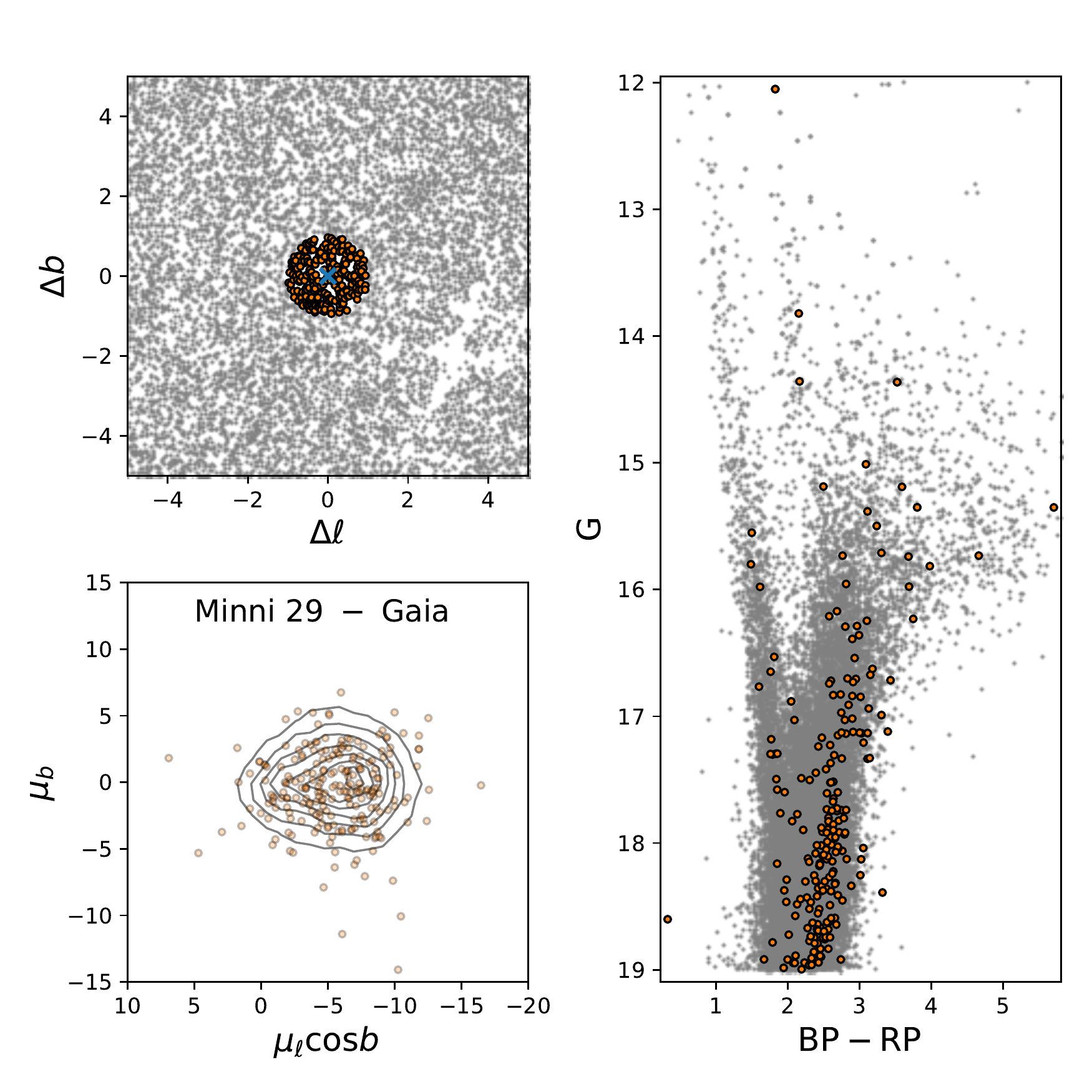} & \\
\end{tabular}
\end{table*}
\newpage
\begin{table*}
\begin{tabular}{cc}
\includegraphics[width=8cm]{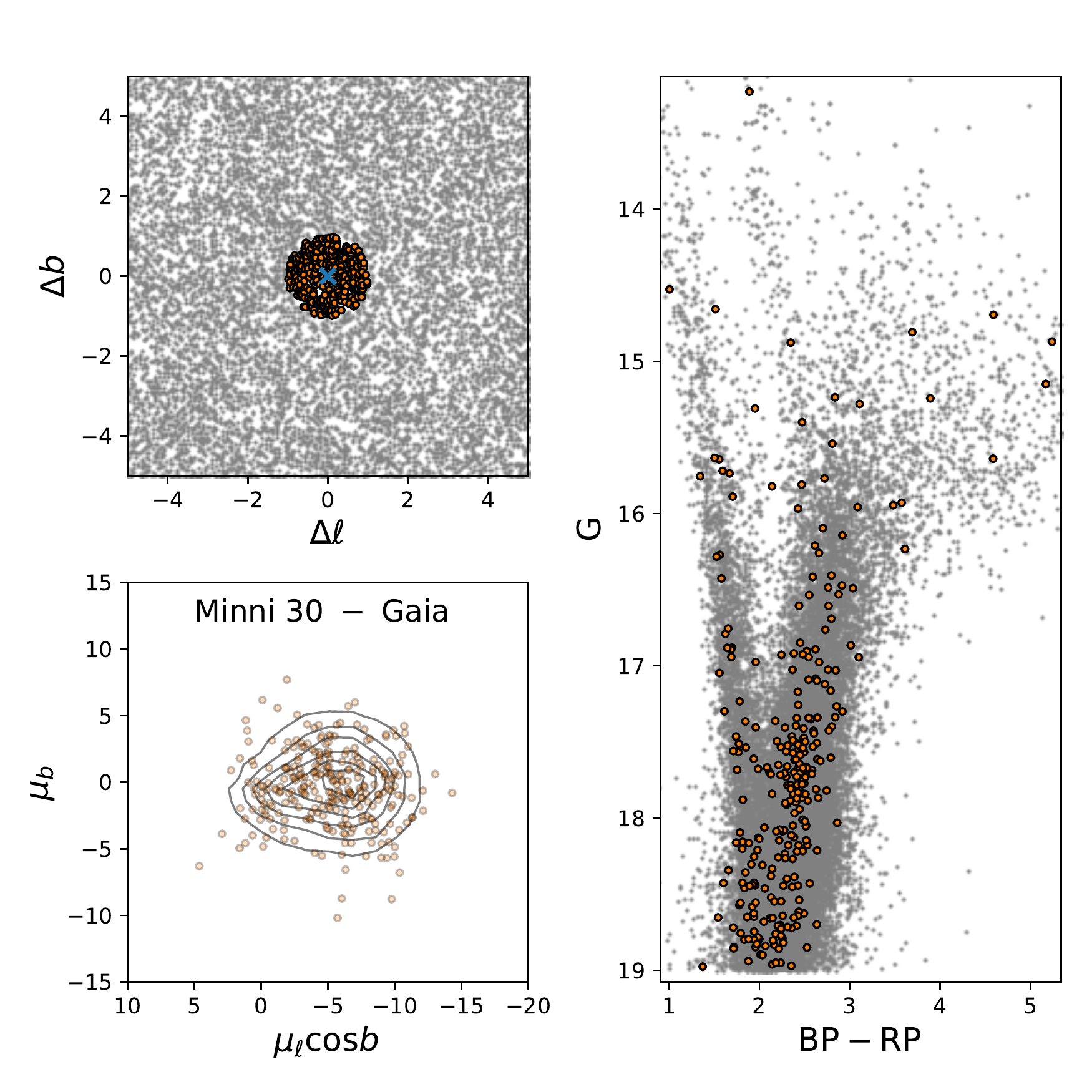} & \\  
\includegraphics[width=8cm]{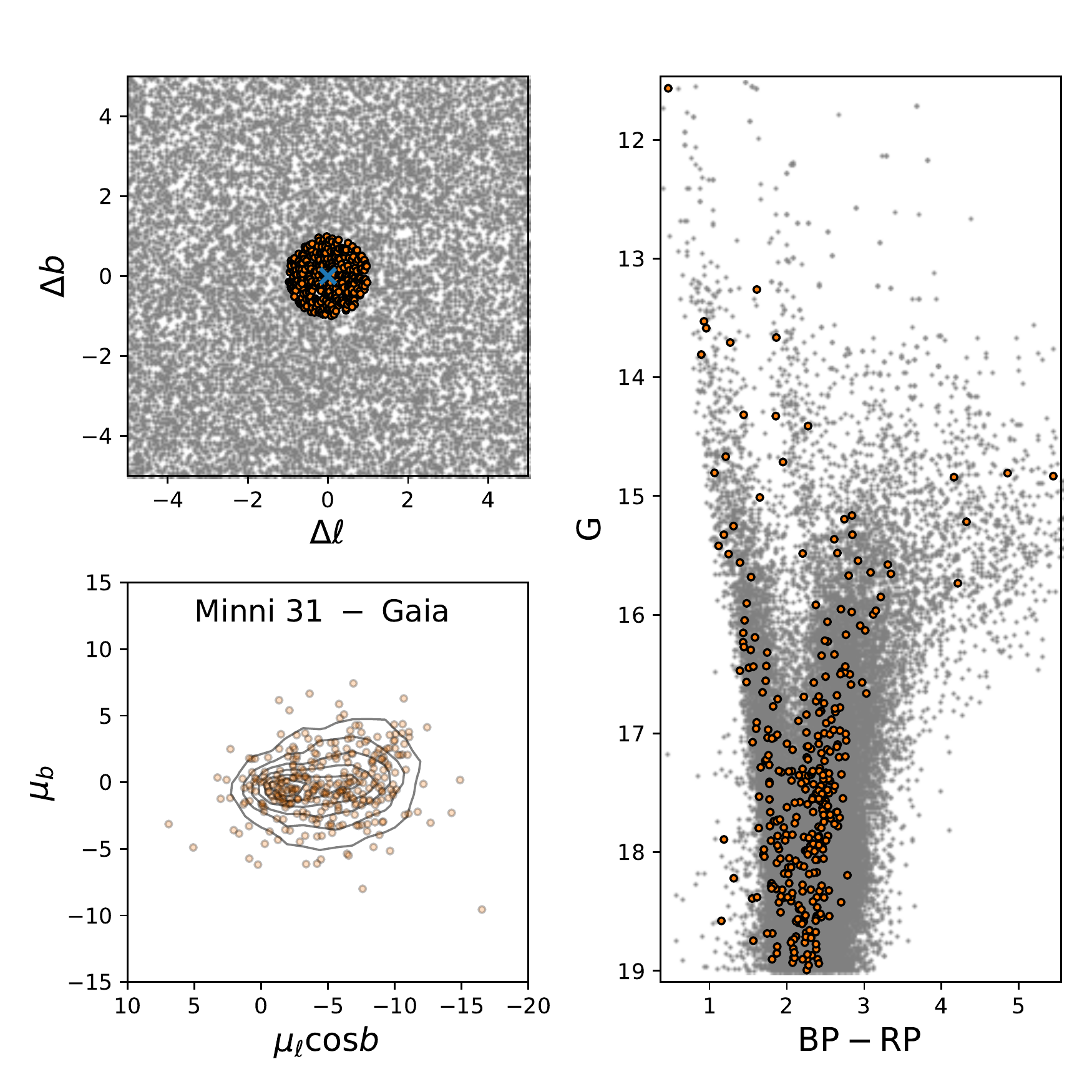} &  
\includegraphics[width=8cm]{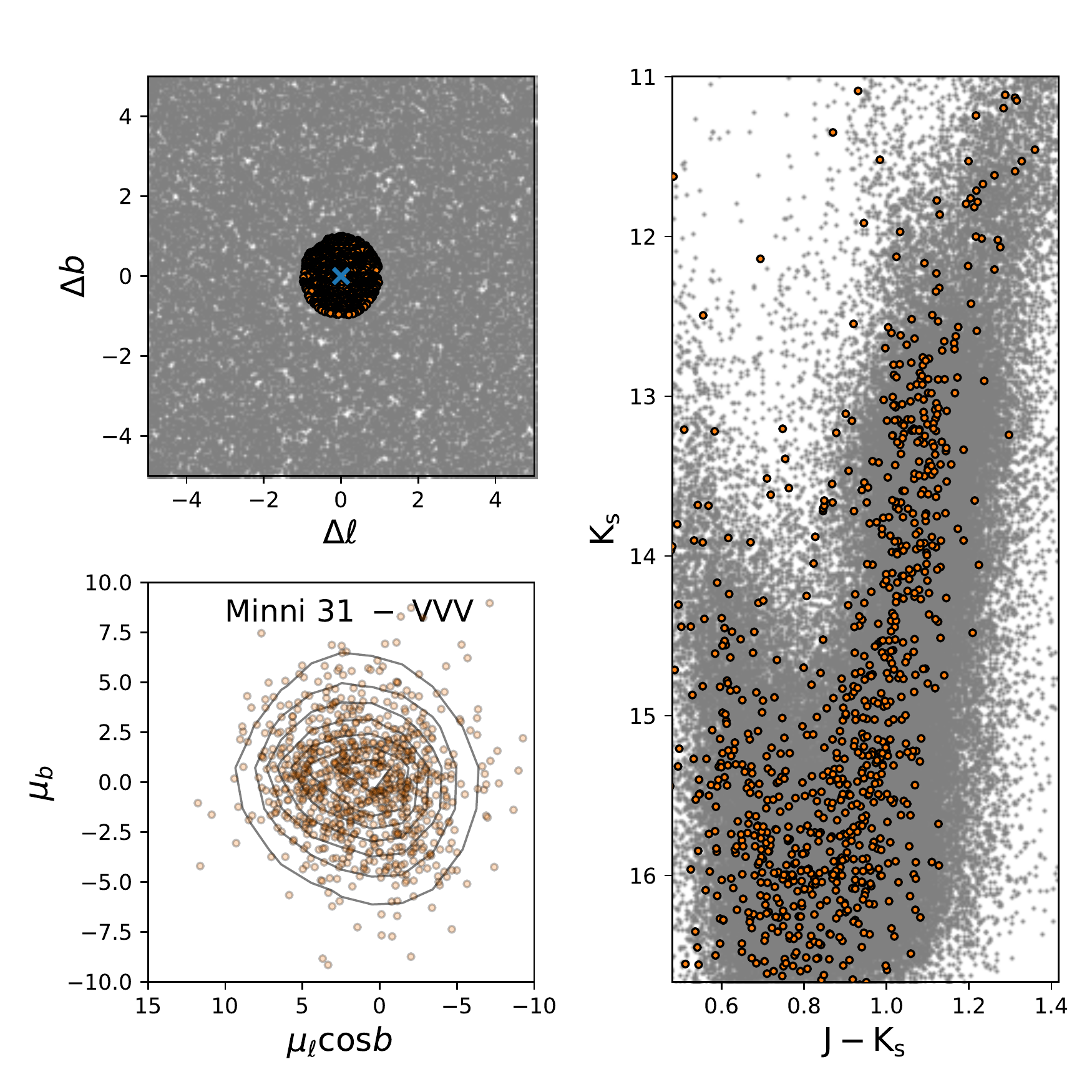} \\
\includegraphics[width=8cm]{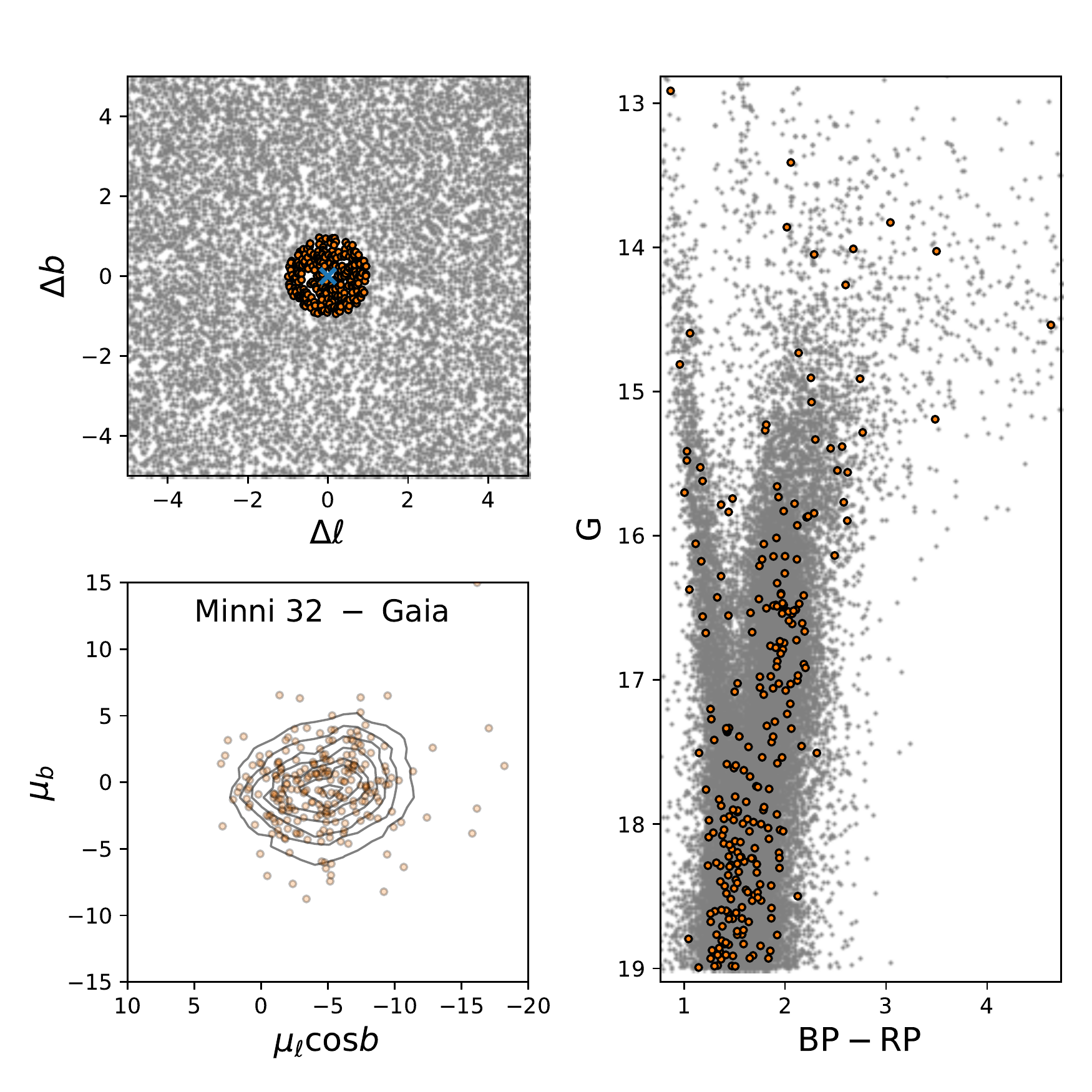} & \\
\end{tabular}
\end{table*}
\newpage
\begin{table*}
\begin{tabular}{cc}
\includegraphics[width=8cm]{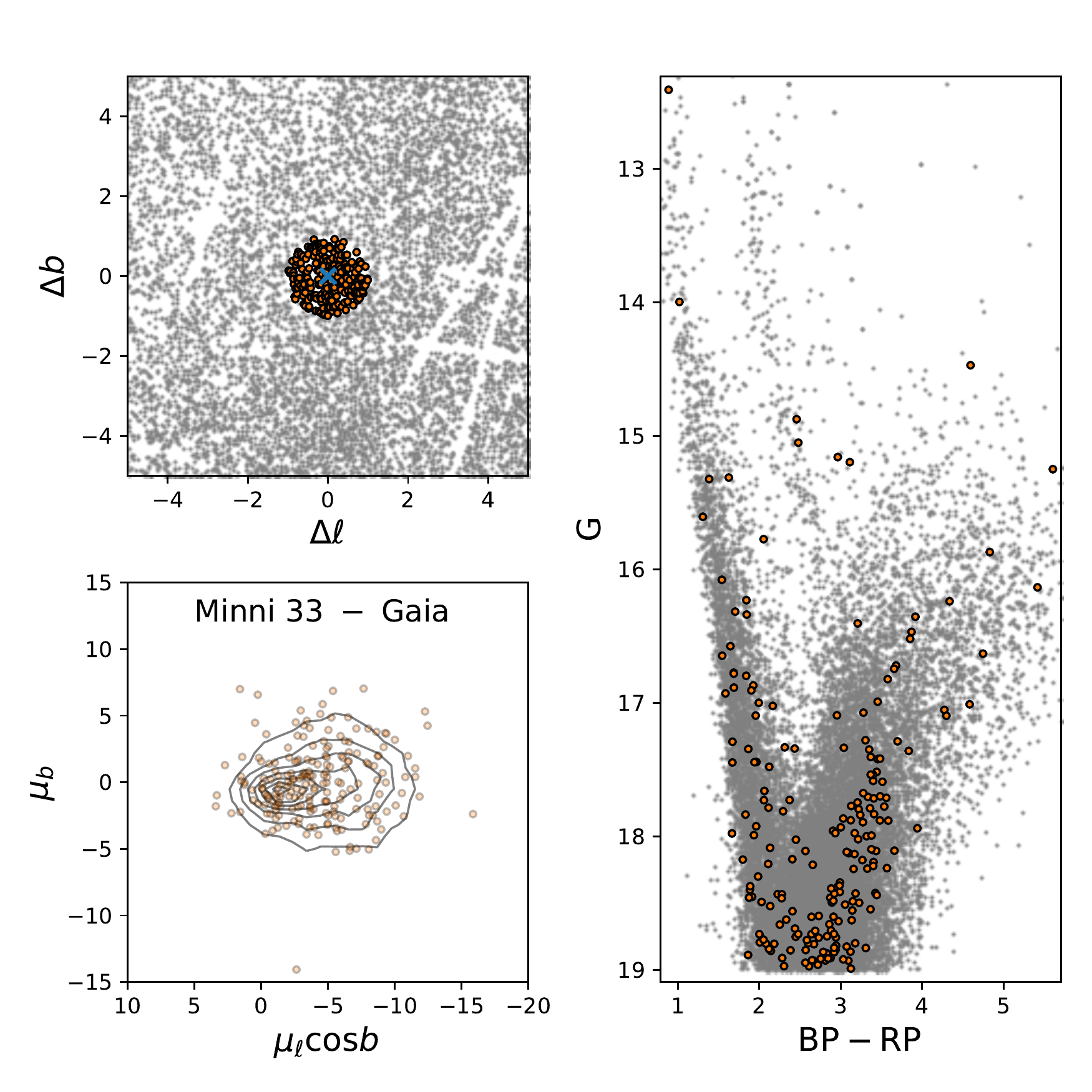} &
\includegraphics[width=8cm]{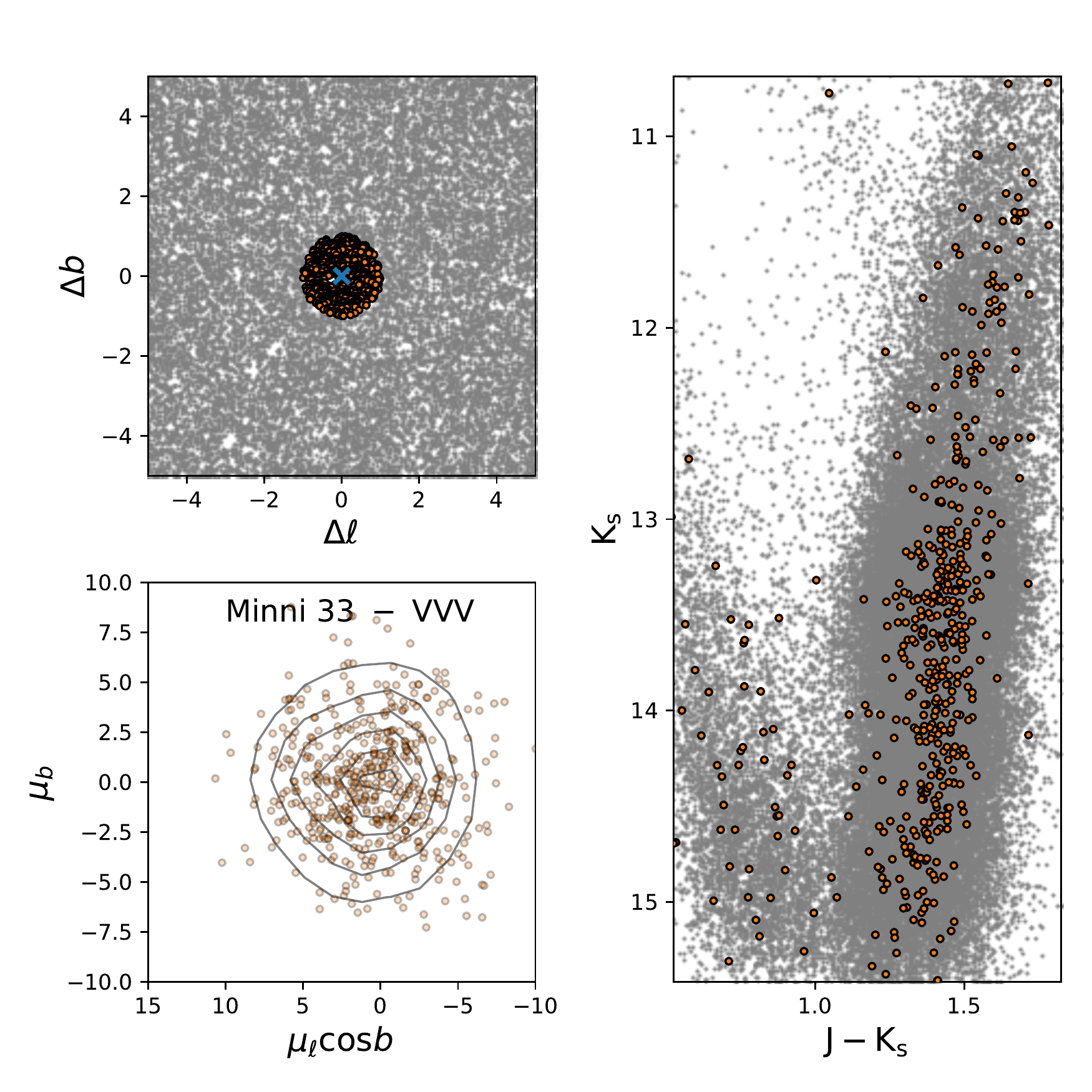} \\  
\includegraphics[width=8cm]{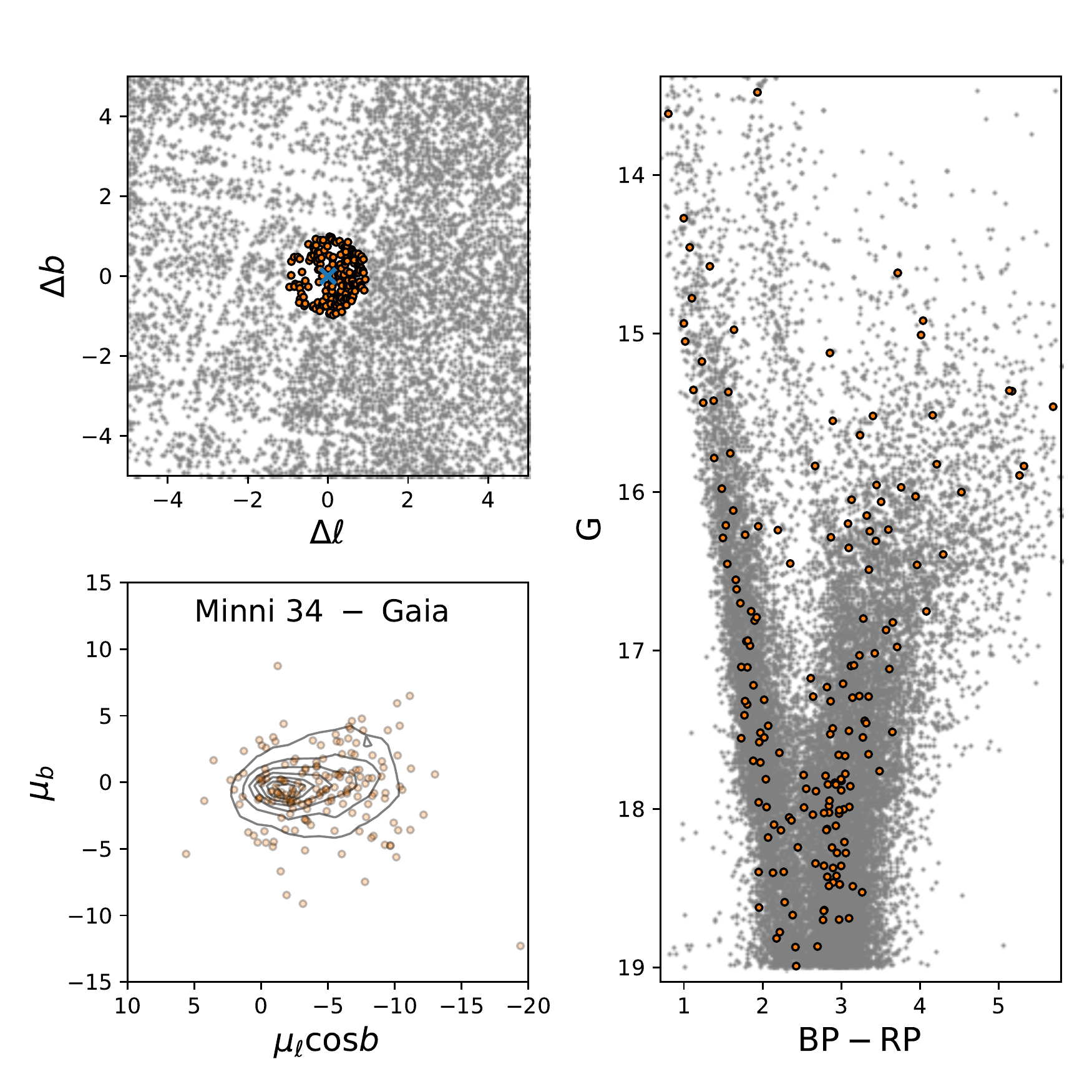} &  
\includegraphics[width=8cm]{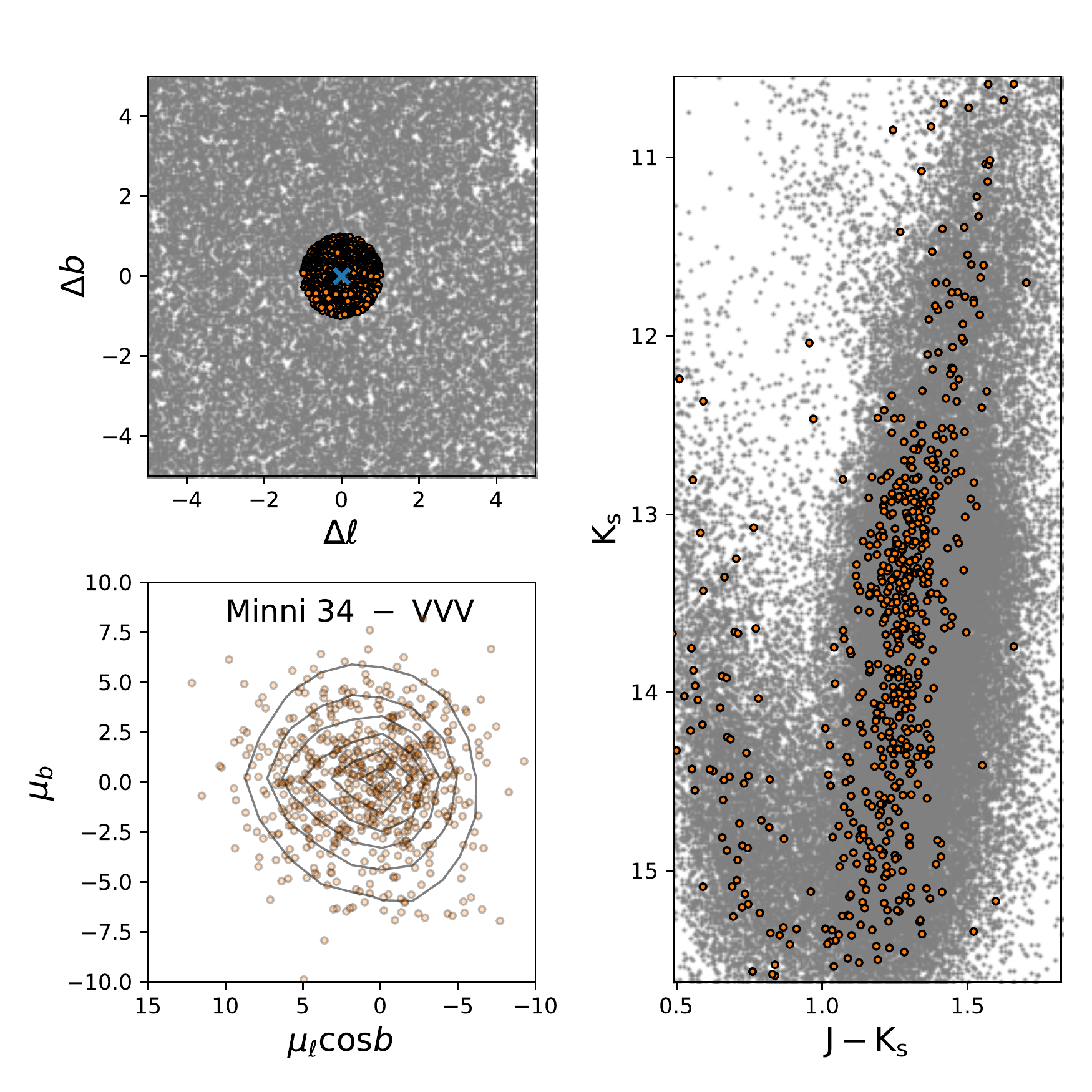} \\  
\includegraphics[width=8cm]{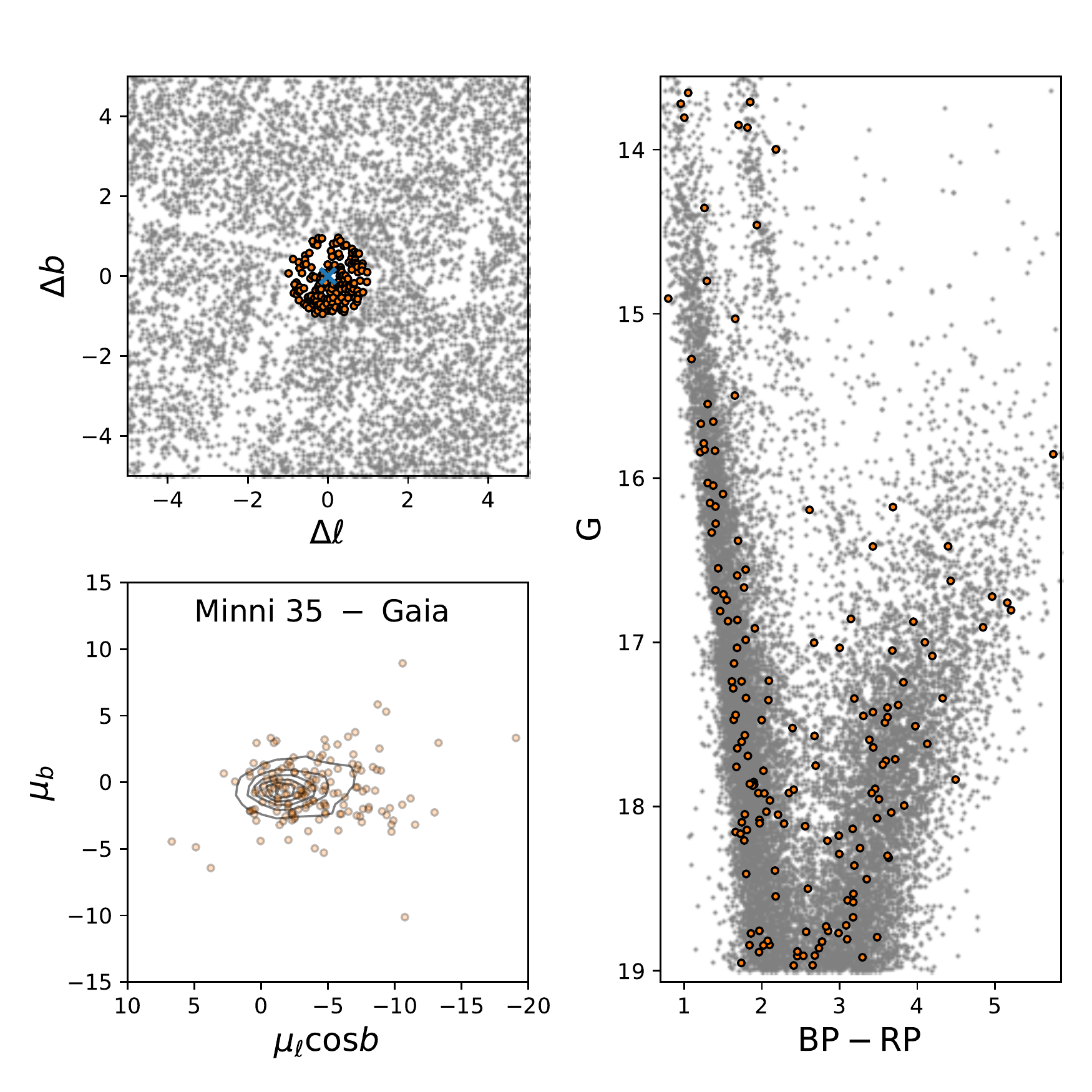} &  
\includegraphics[width=8cm]{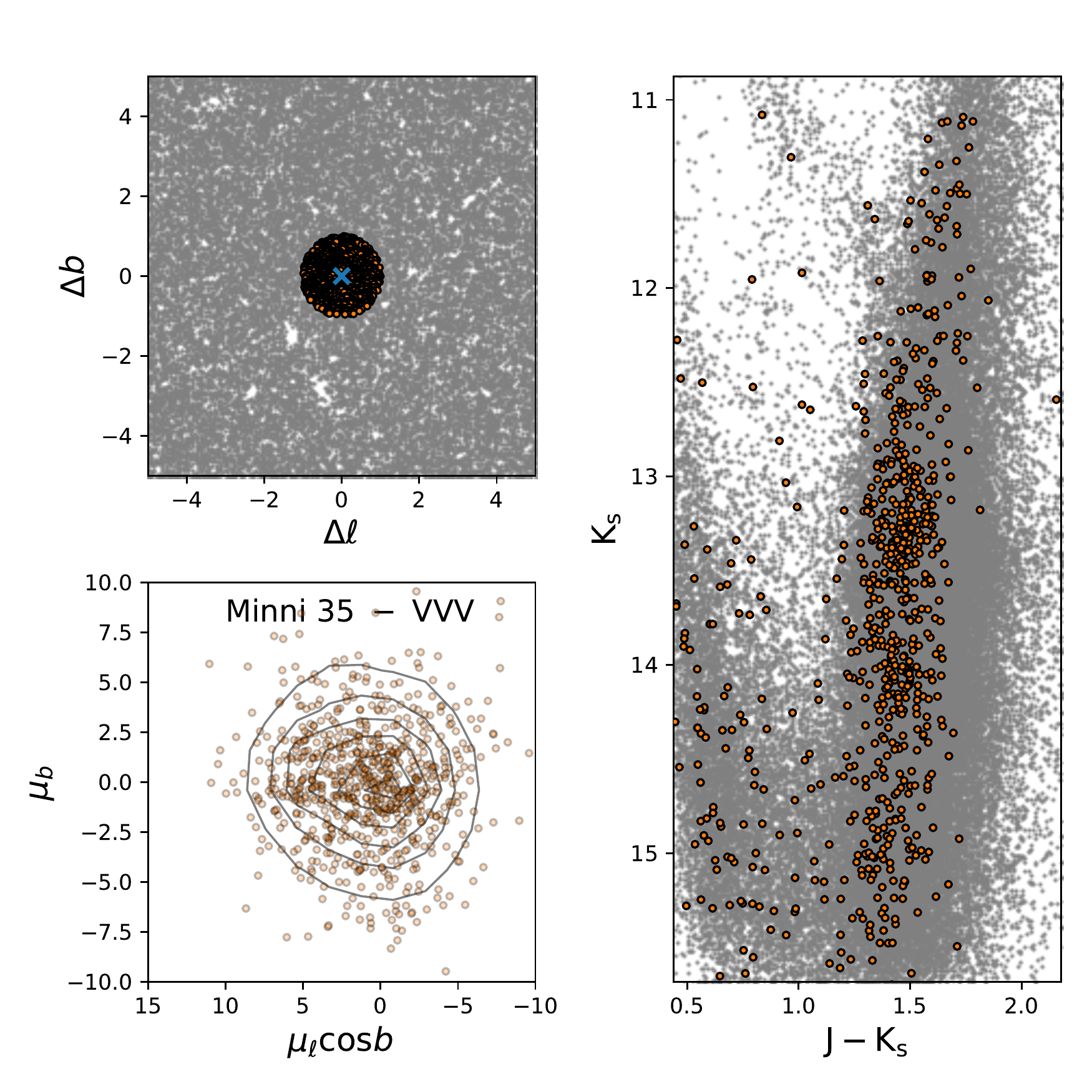} \\
\end{tabular}
\end{table*}
\newpage
\begin{table*}
\begin{tabular}{cc}
\includegraphics[width=8cm]{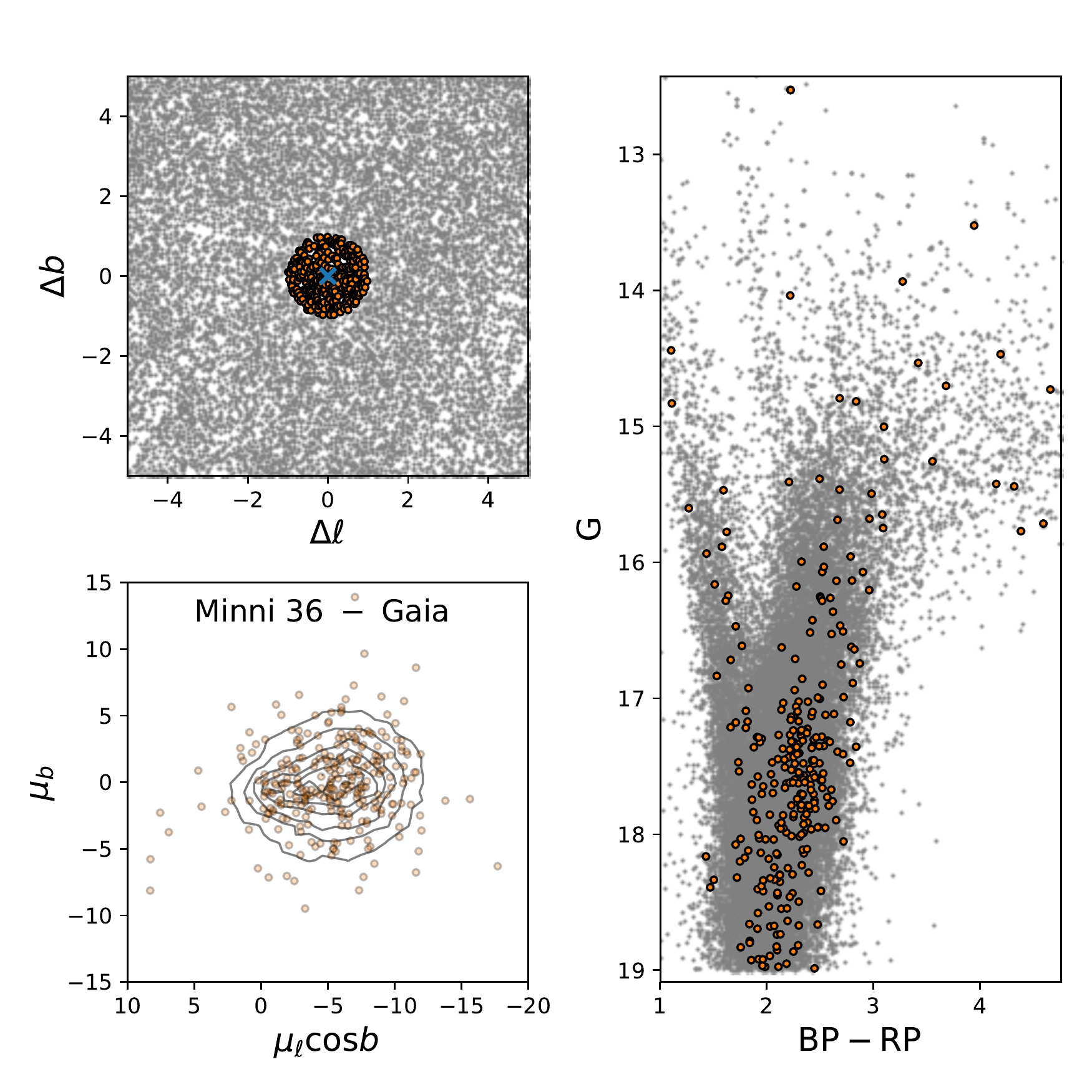} &
\includegraphics[width=8cm]{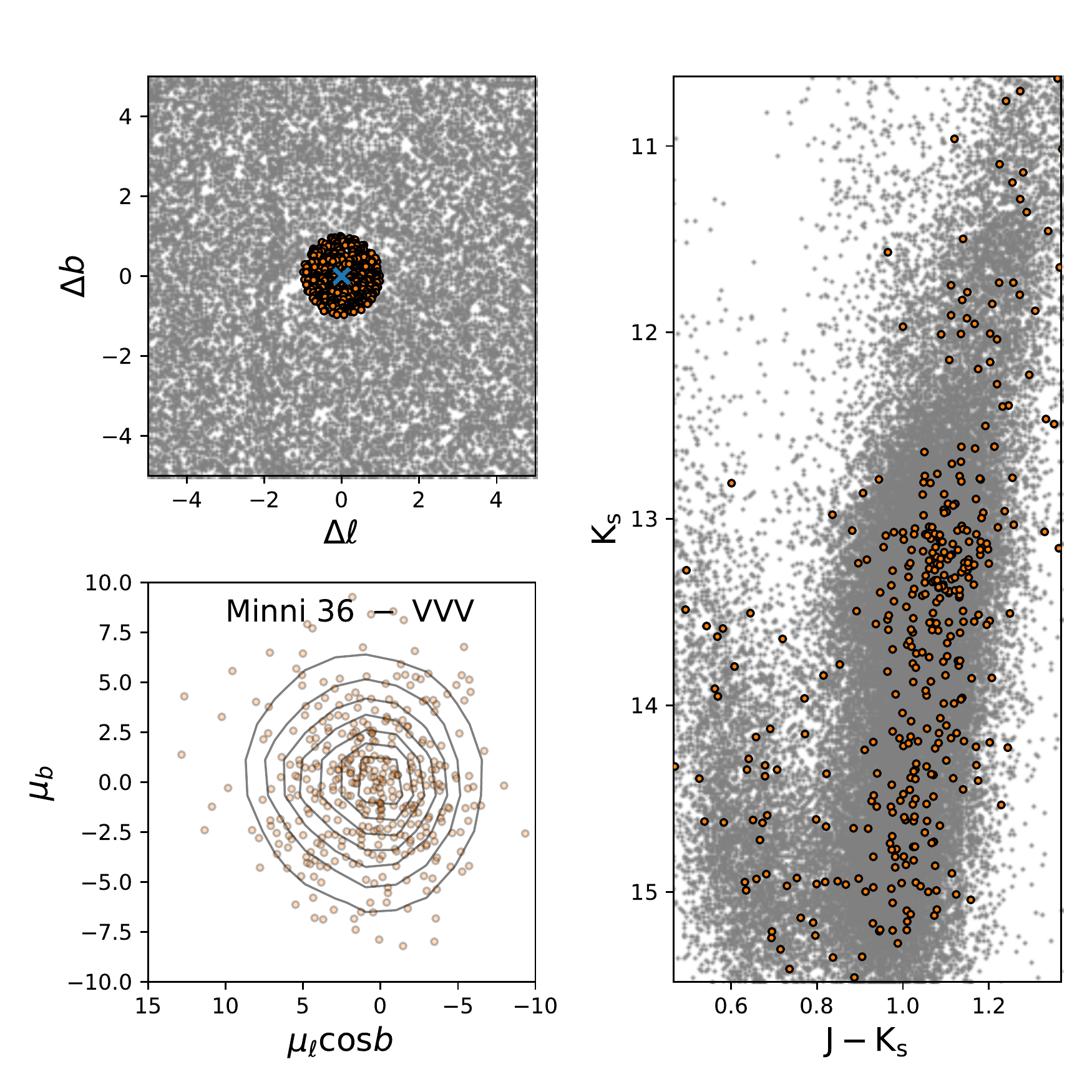} \\  
\includegraphics[width=8cm]{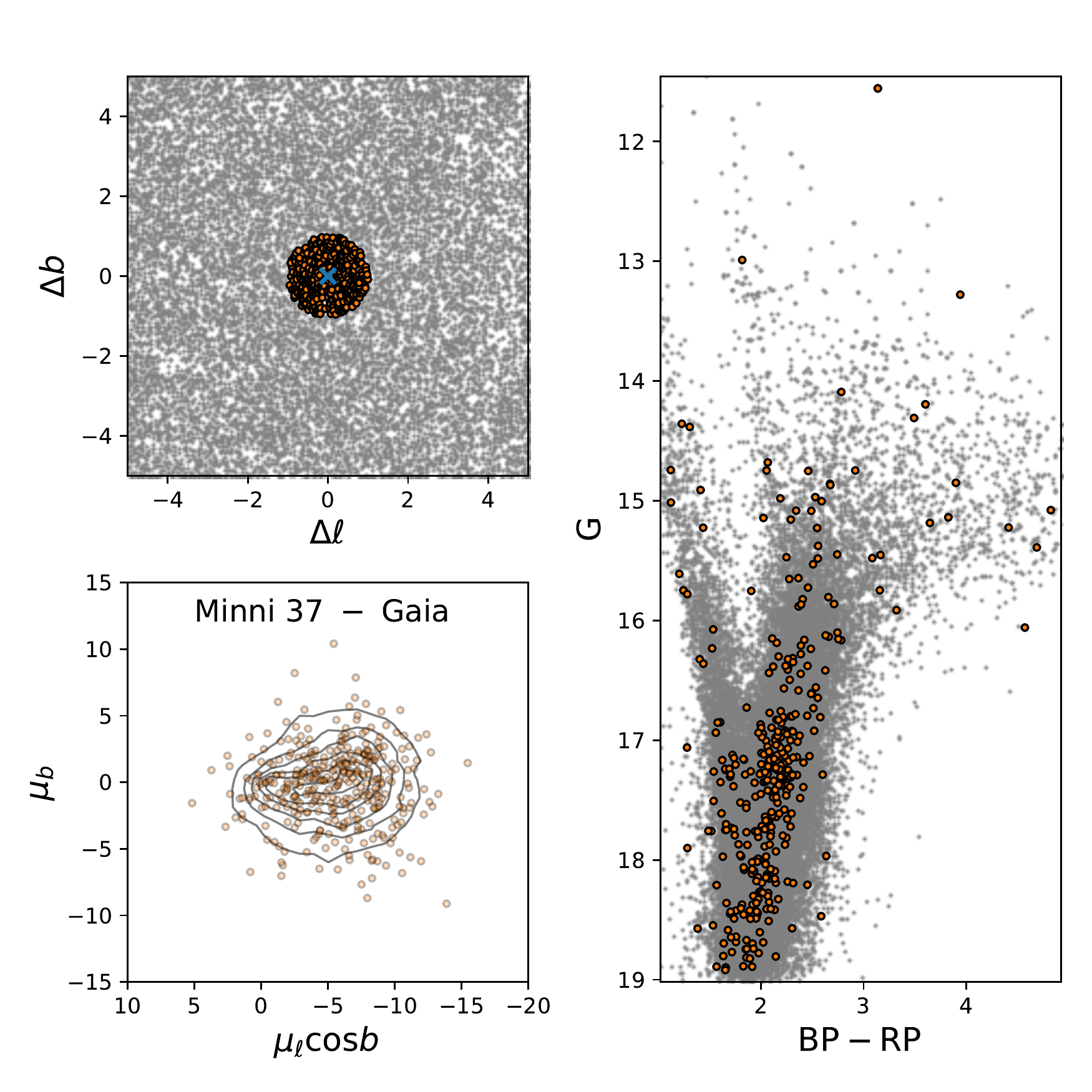} &
\includegraphics[width=8cm]{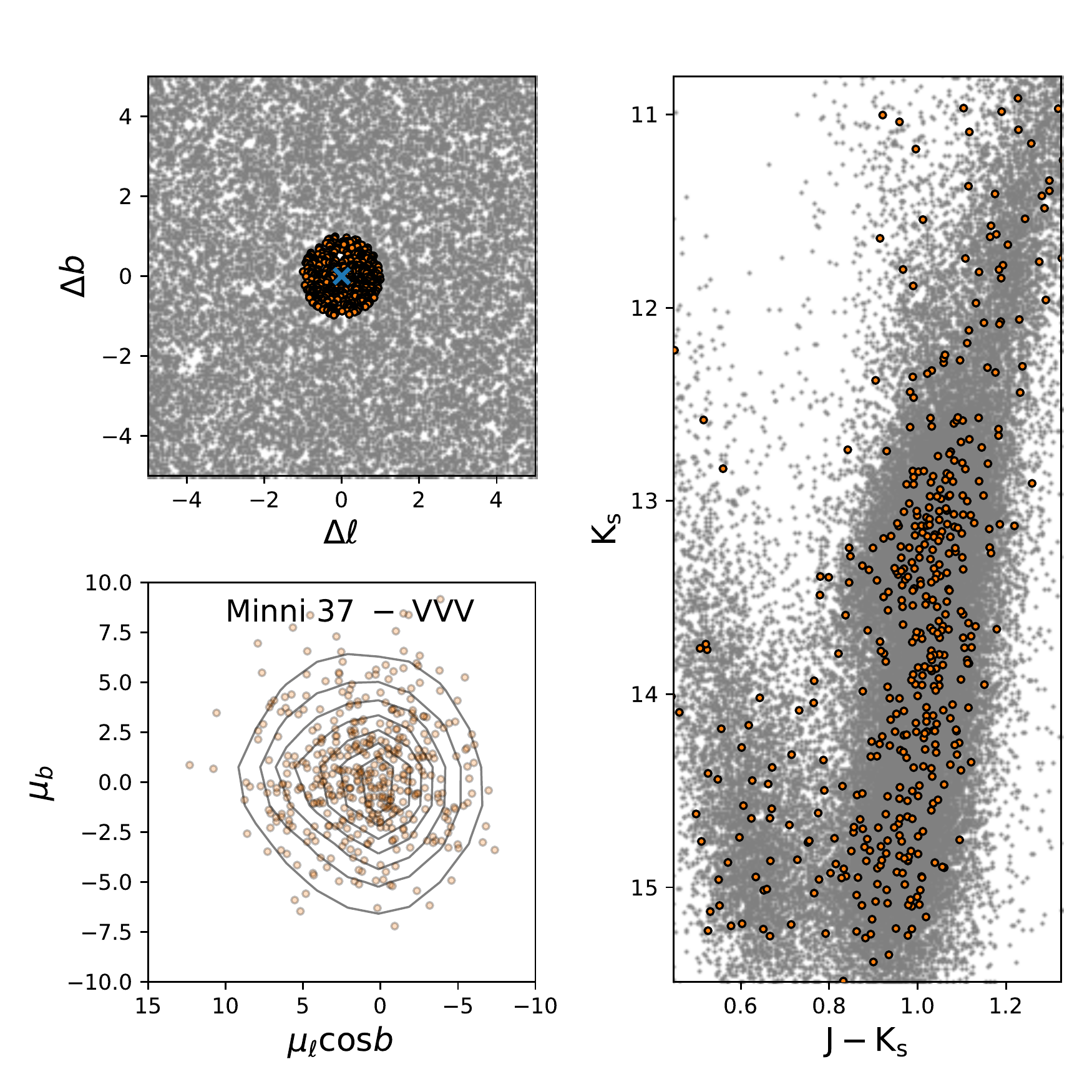} \\ 
\includegraphics[width=8cm]{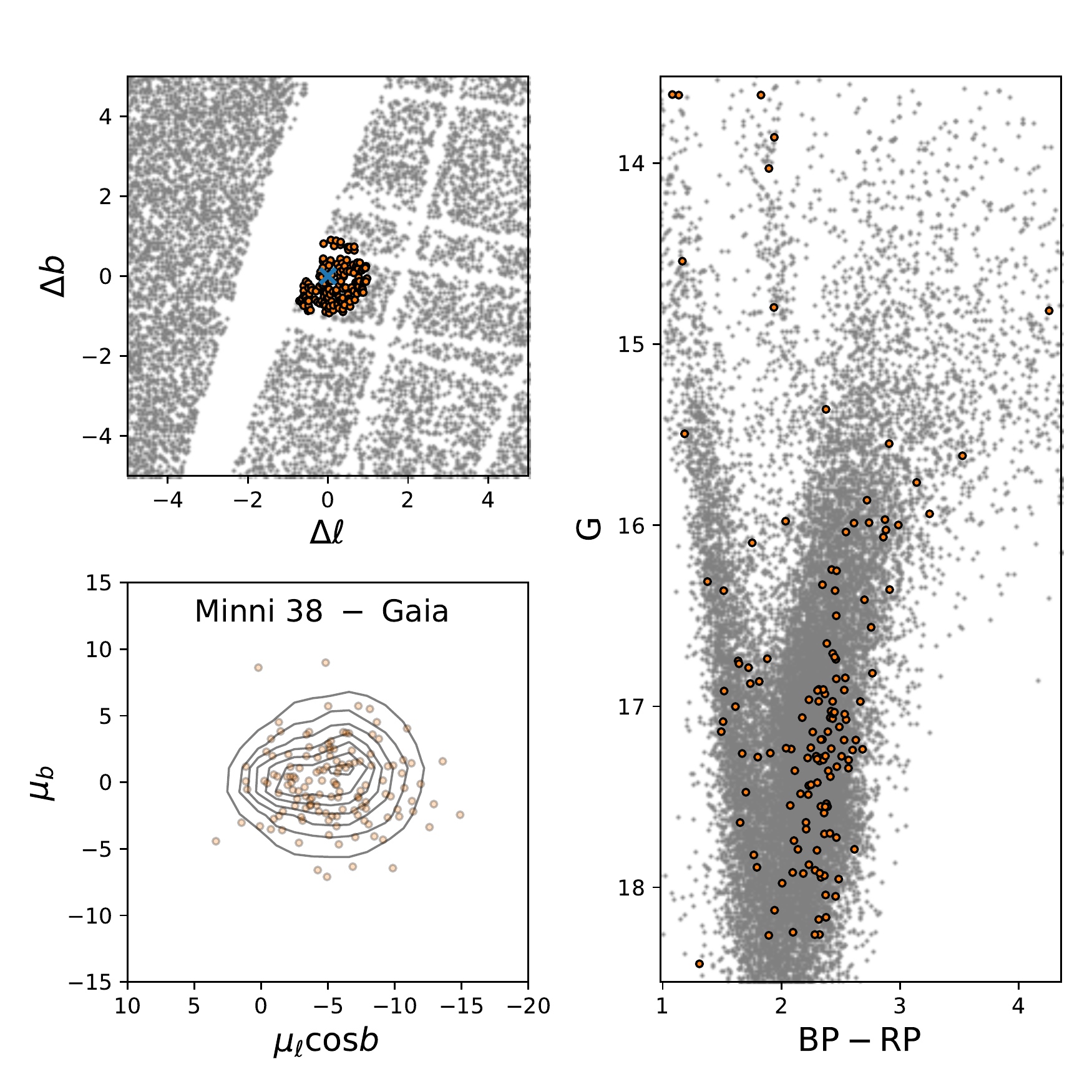} &  
\includegraphics[width=8cm]{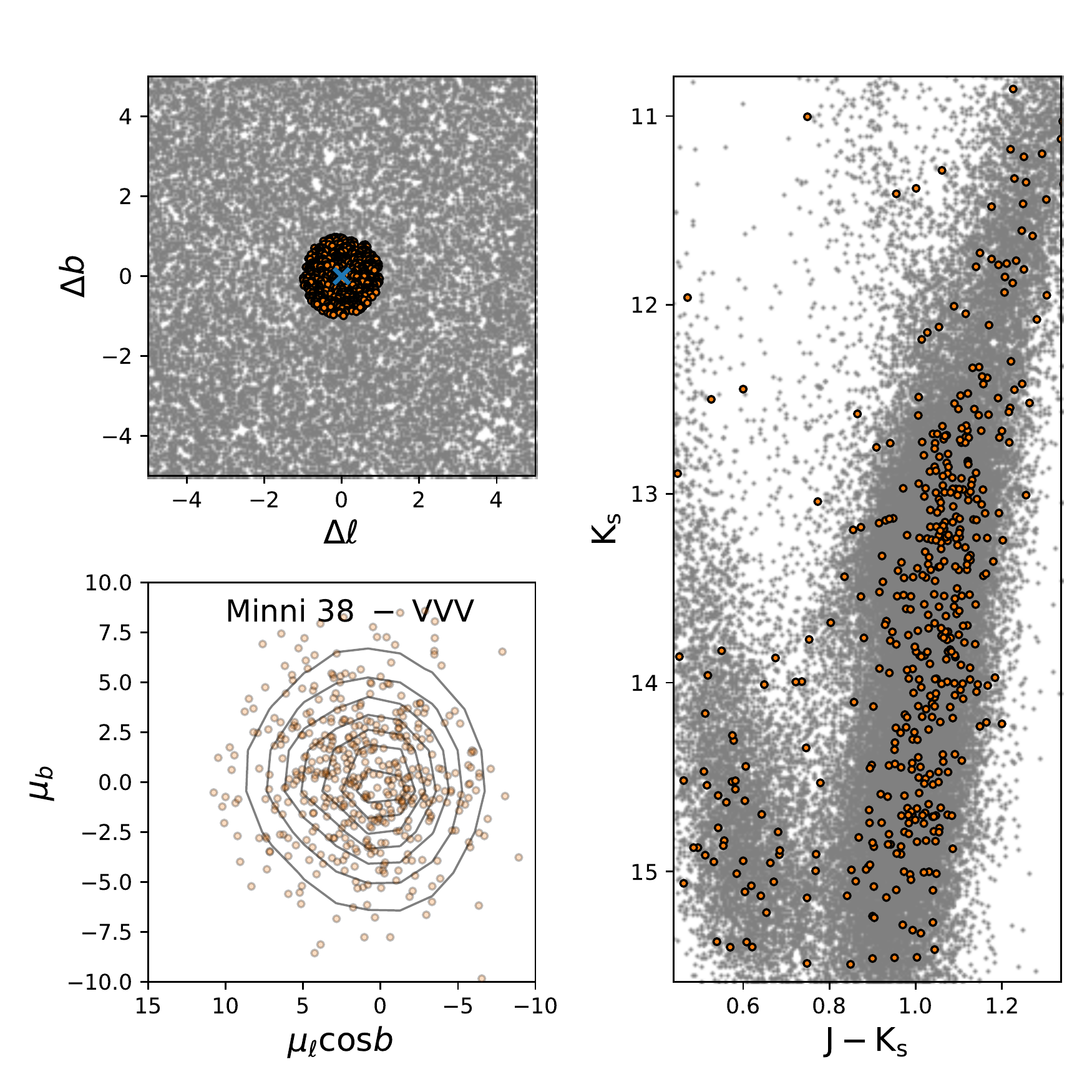} \\
\end{tabular}
\end{table*}
\newpage
\begin{table*}
\begin{tabular}{cccc}
\includegraphics[width=8cm]{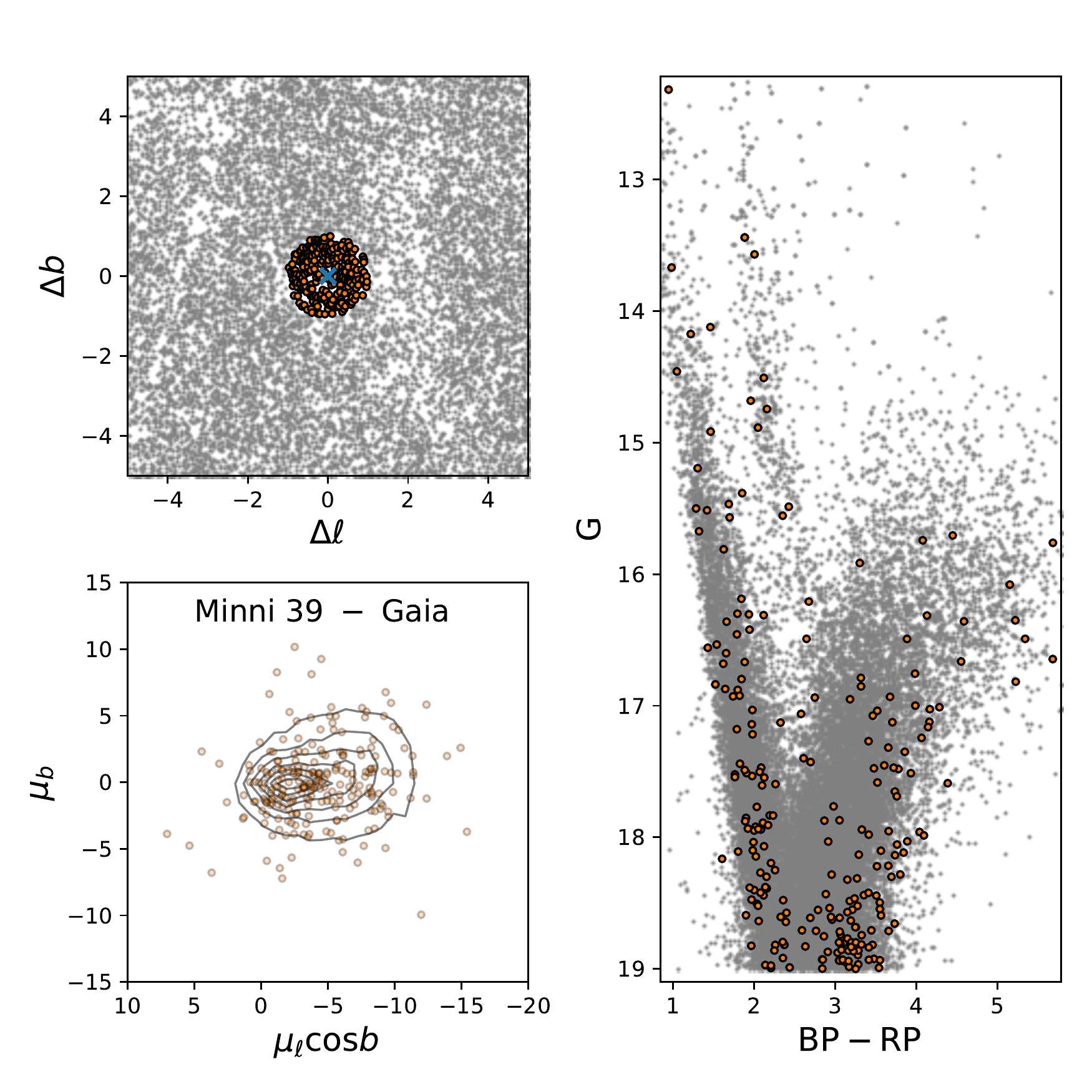} &  
\includegraphics[width=8cm]{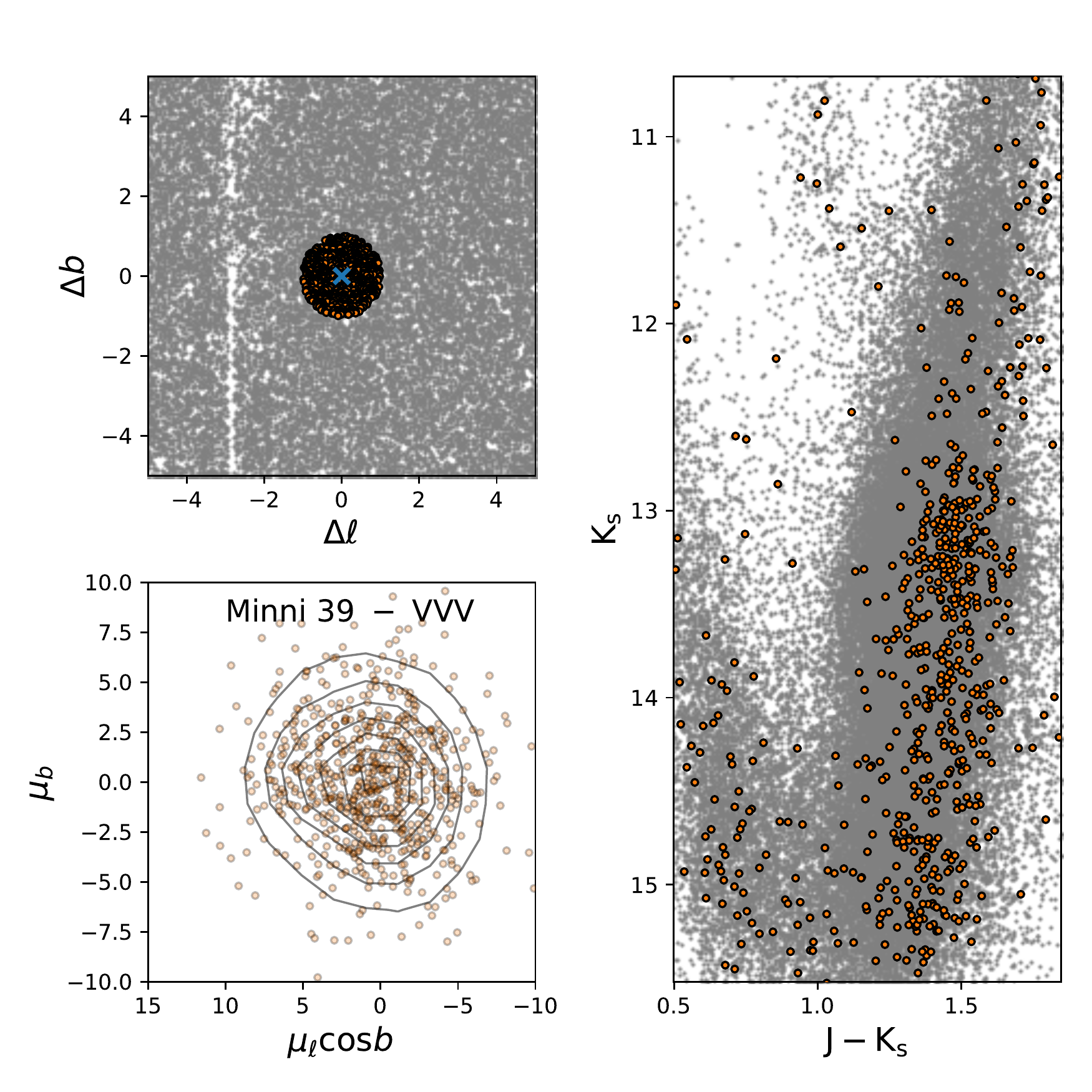} \\  
\includegraphics[width=8cm]{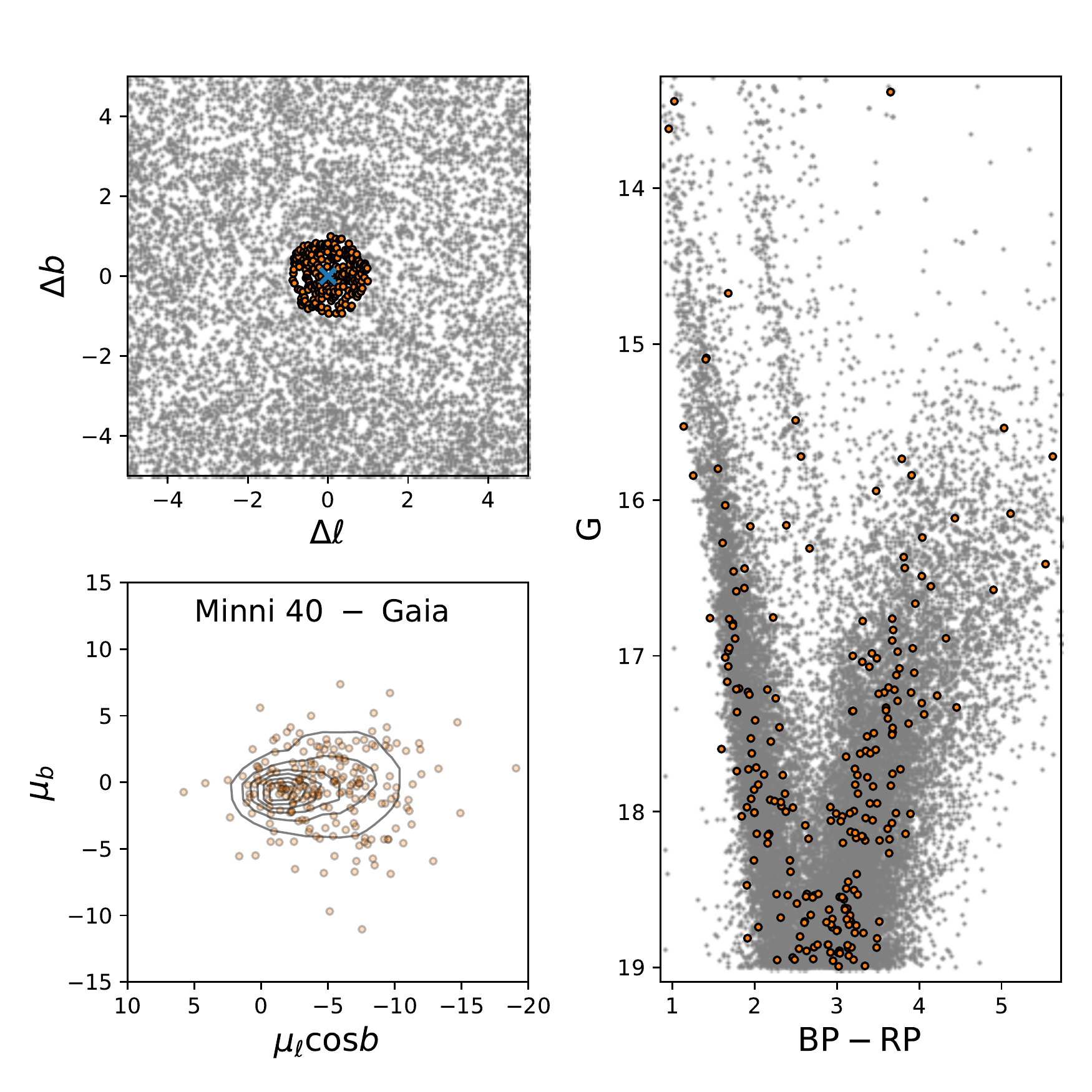} & 
\includegraphics[width=8cm]{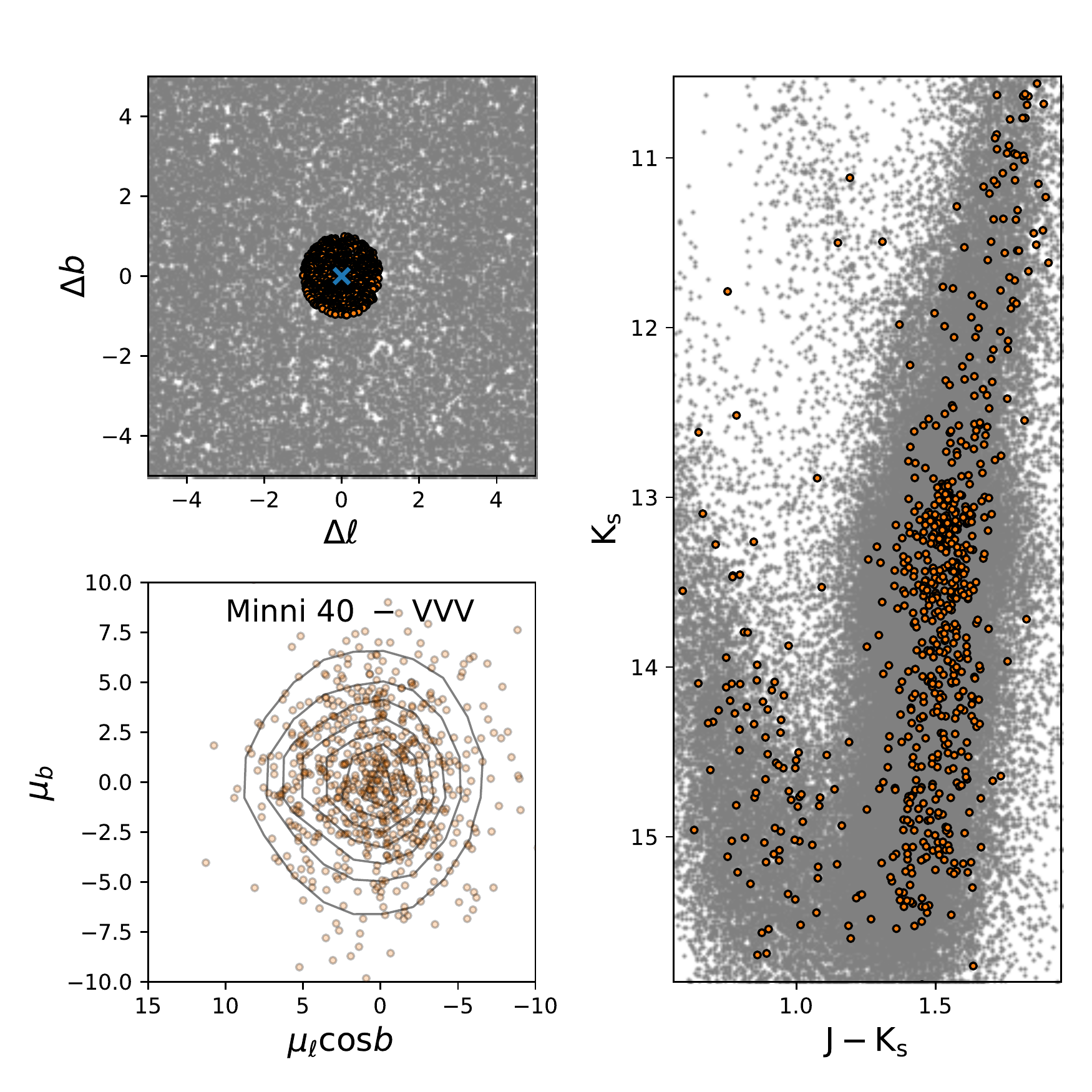} \\  
\end{tabular}
\end{table*}
\newpage
\begin{table*}
\begin{tabular}{cccc}
\includegraphics[width=8cm]{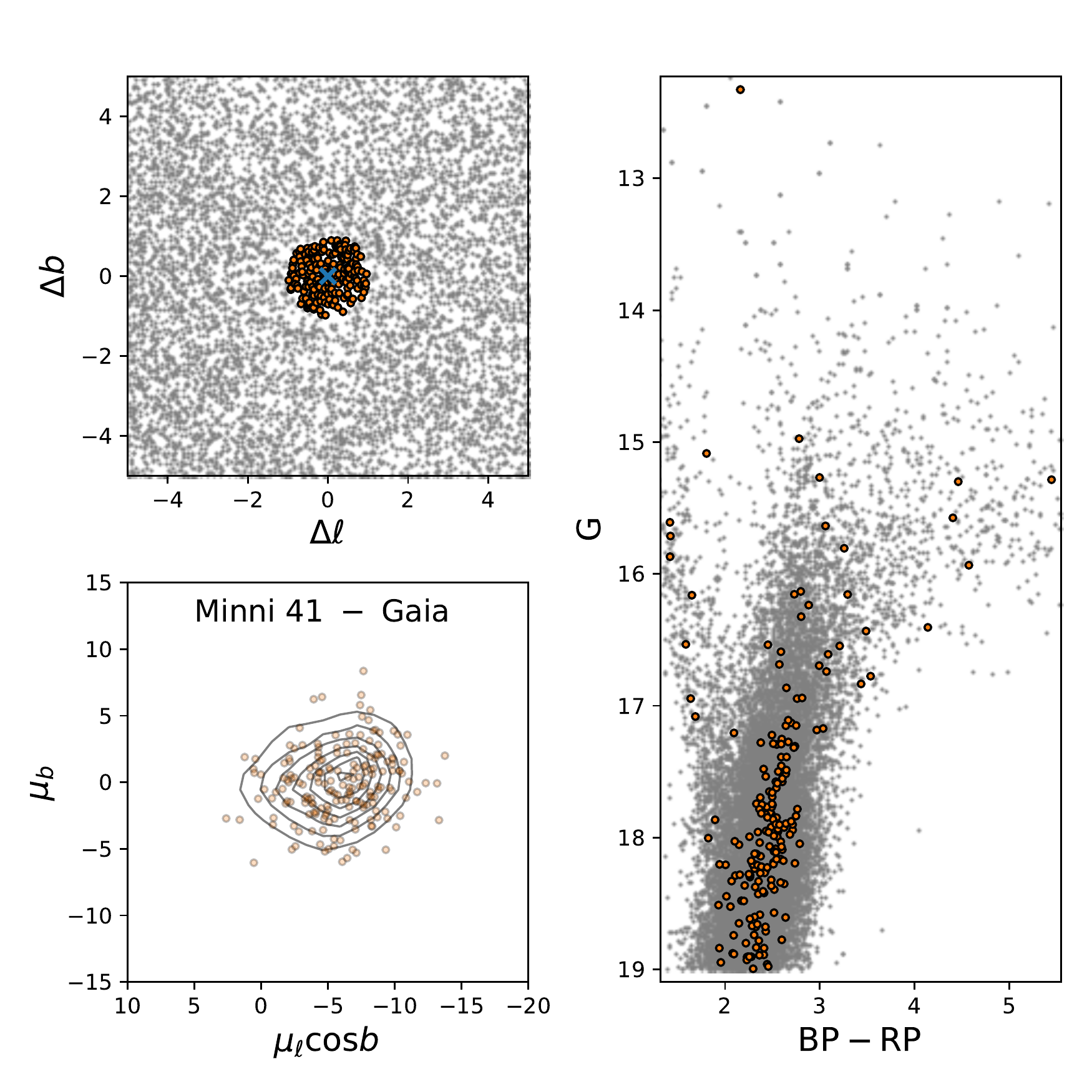} & \\
\includegraphics[width=8cm]{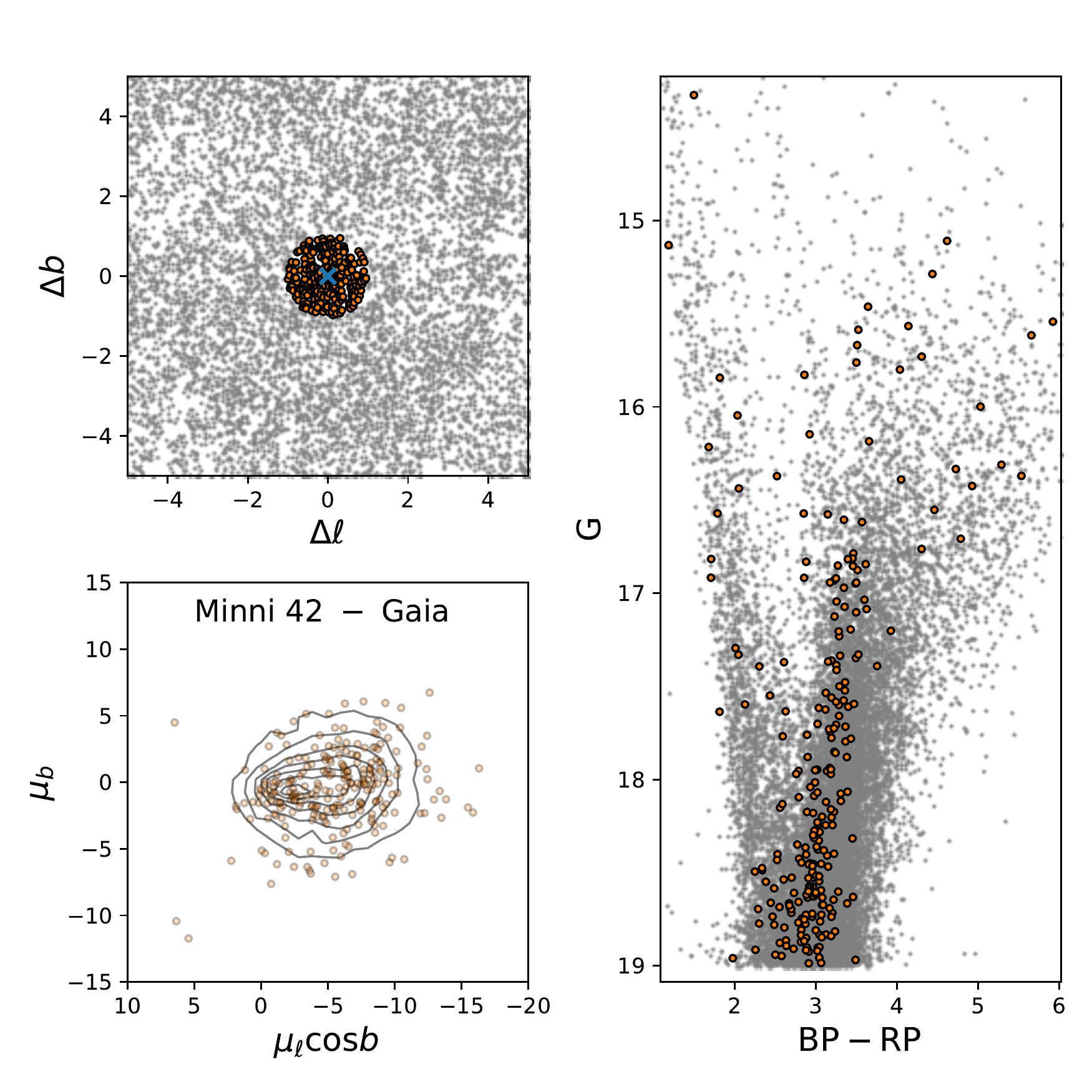} &  
\includegraphics[width=8cm]{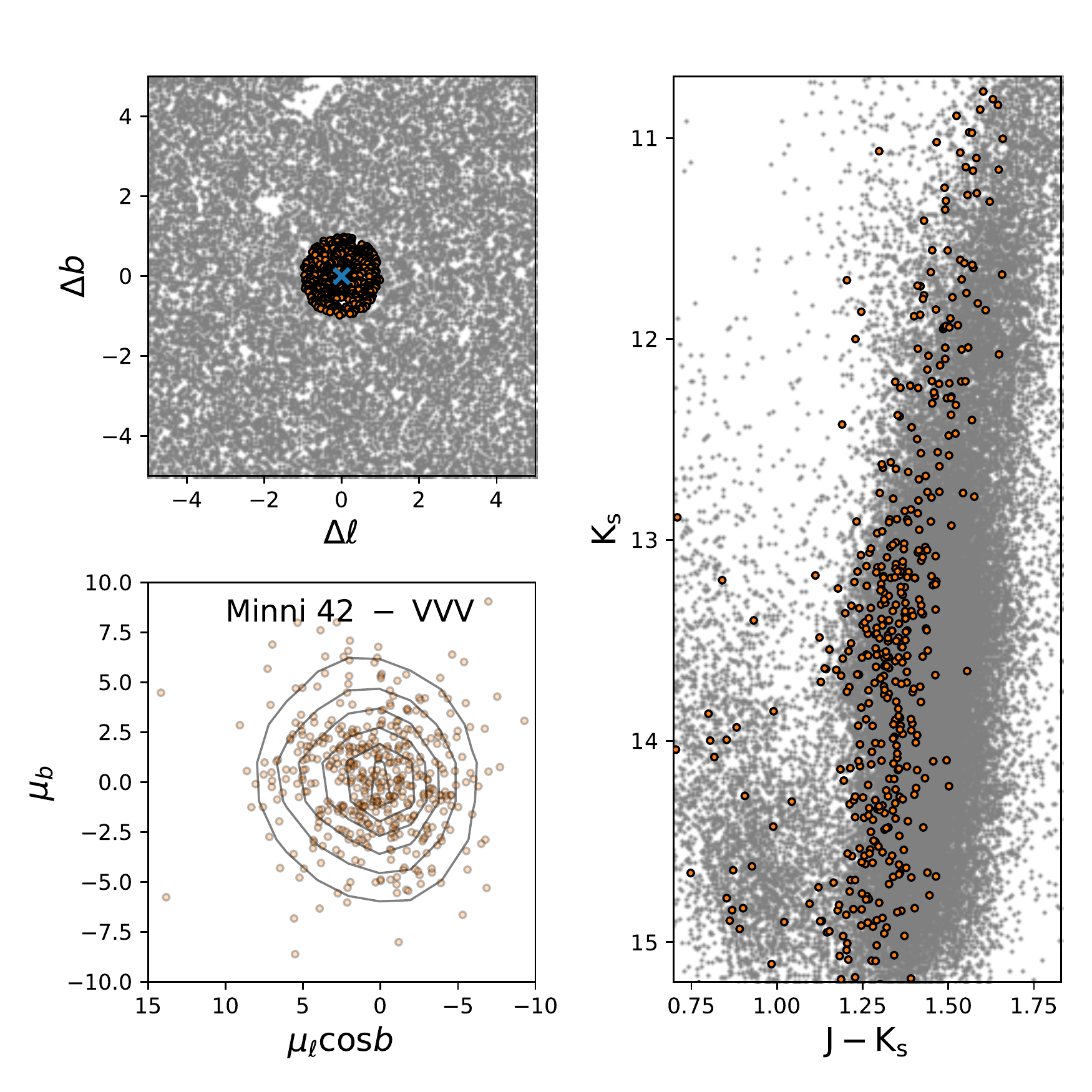} \\ 
\includegraphics[width=8cm]{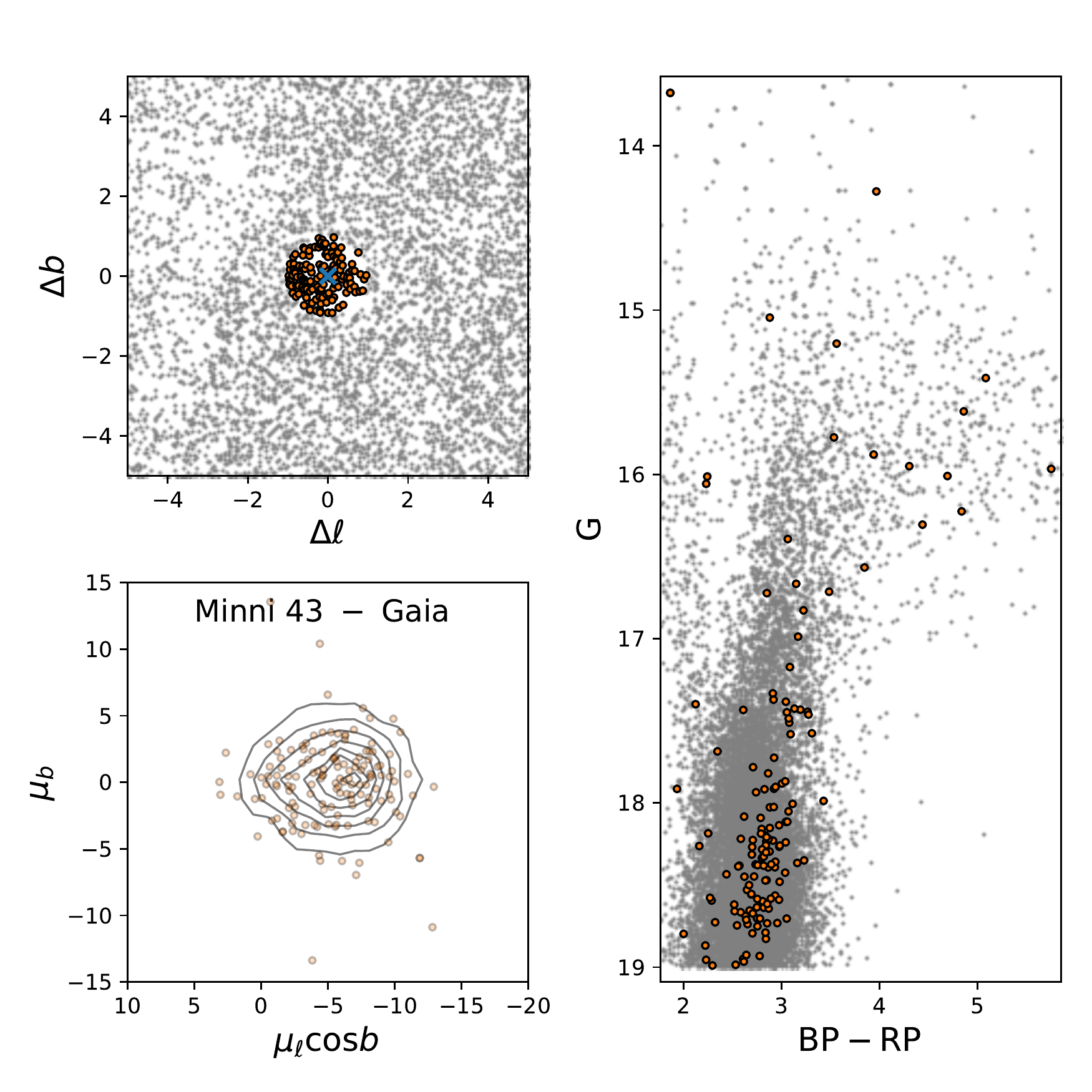} & \\
\end{tabular}
\end{table*}
\newpage
\begin{table*}
\begin{tabular}{cc}
\includegraphics[width=8cm]{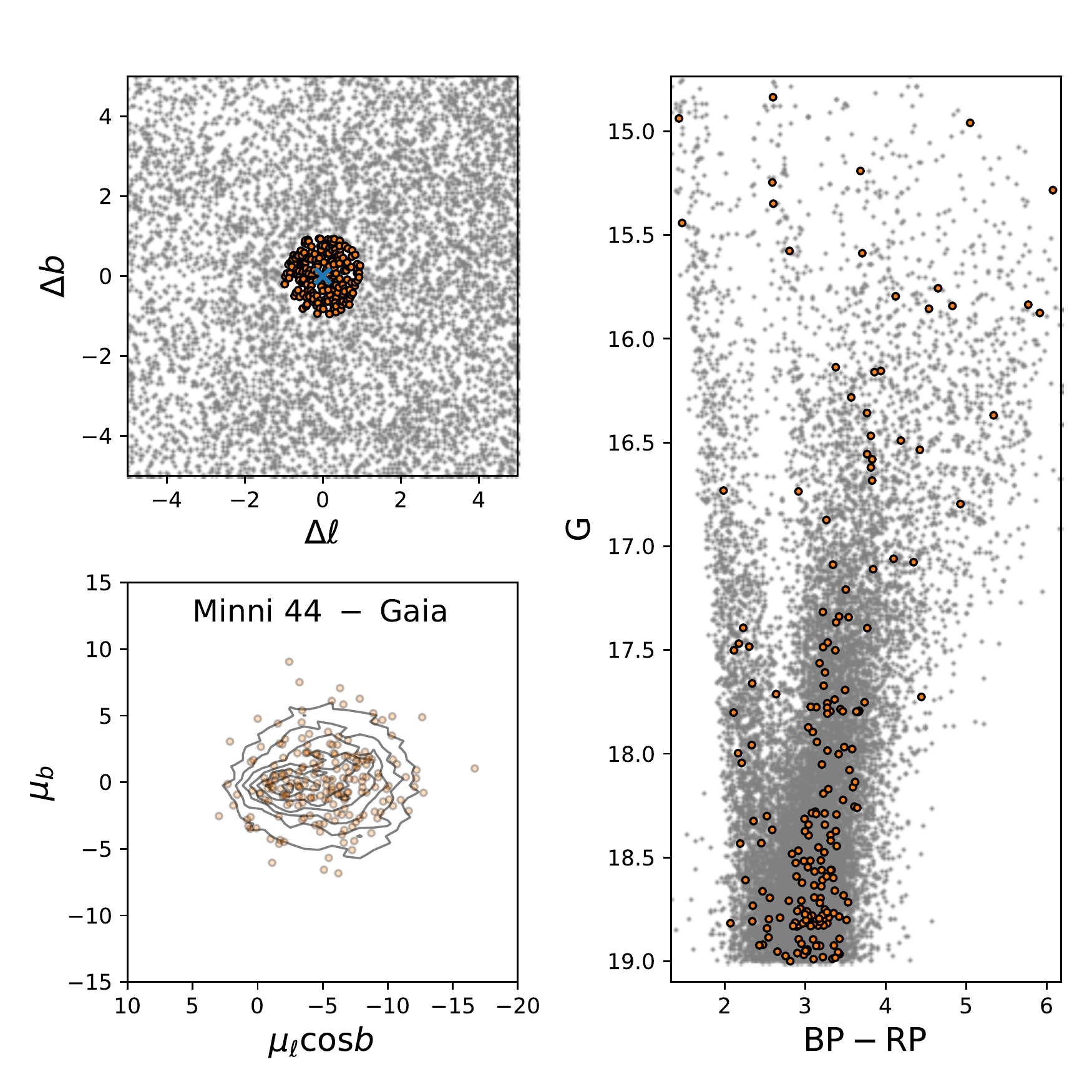} & \\ 
\includegraphics[width=8cm]{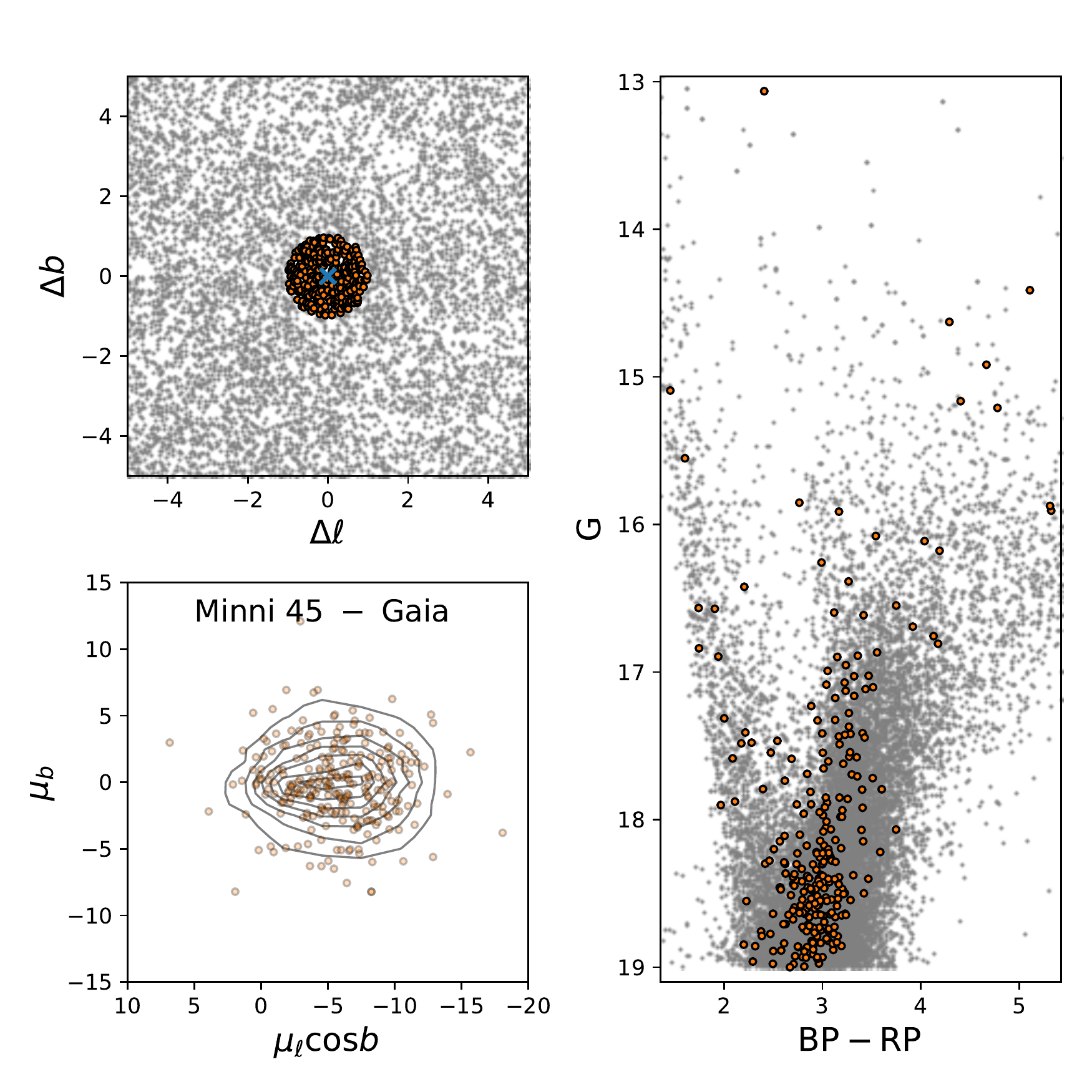} & \\
\includegraphics[width=8cm]{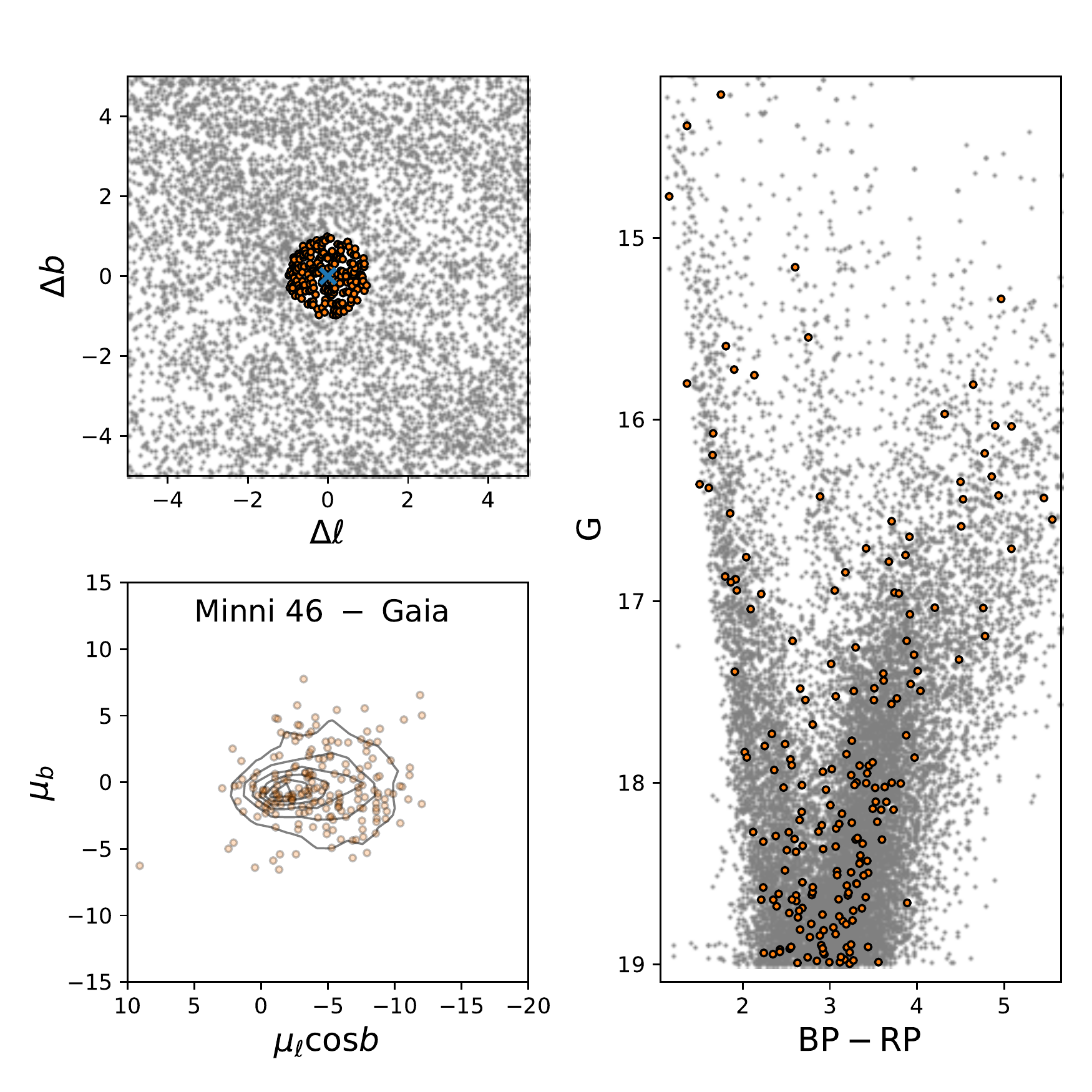} &  
\includegraphics[width=8cm]{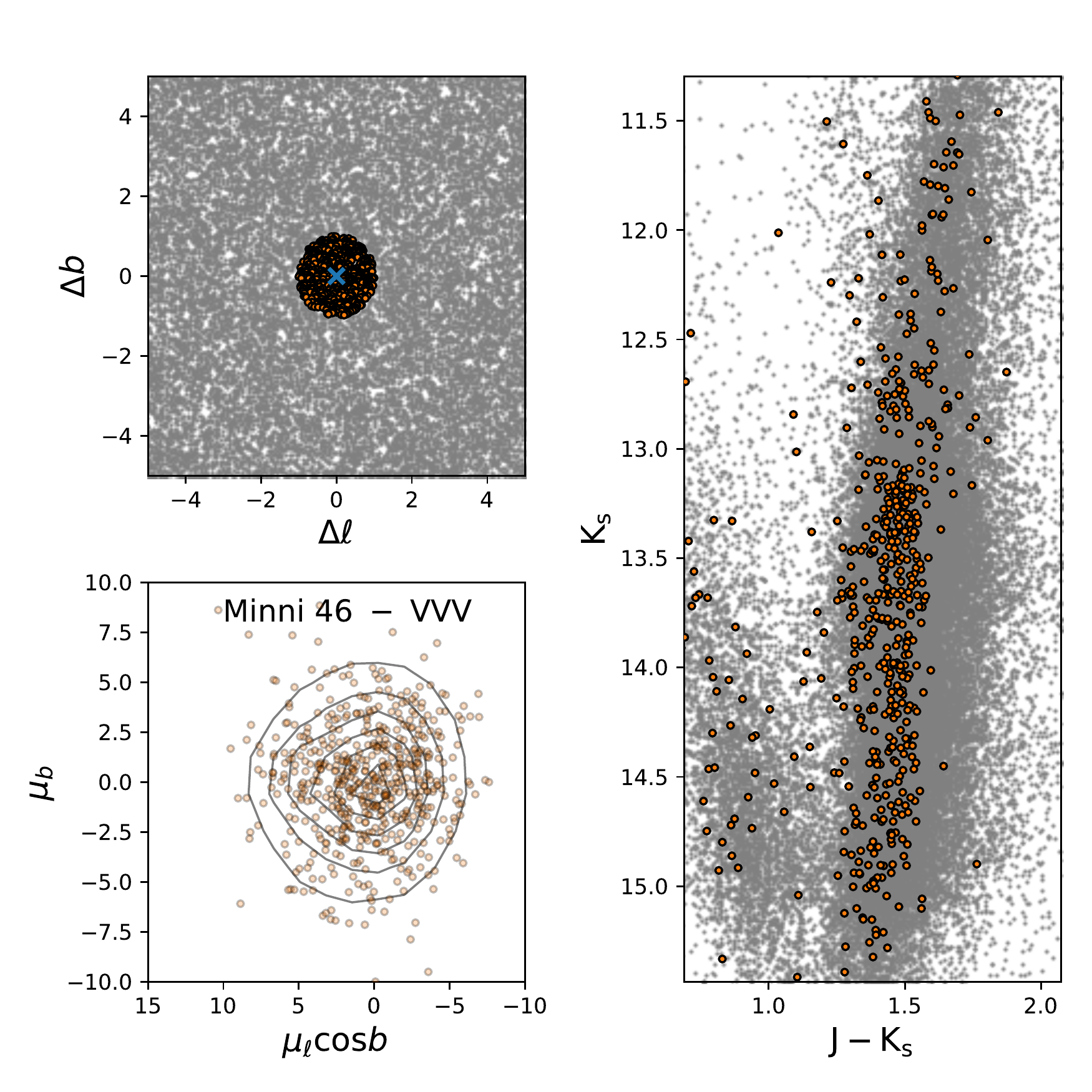} \\
\end{tabular}
\end{table*}
\newpage
\begin{table*}
\begin{tabular}{cc}
\includegraphics[width=8cm]{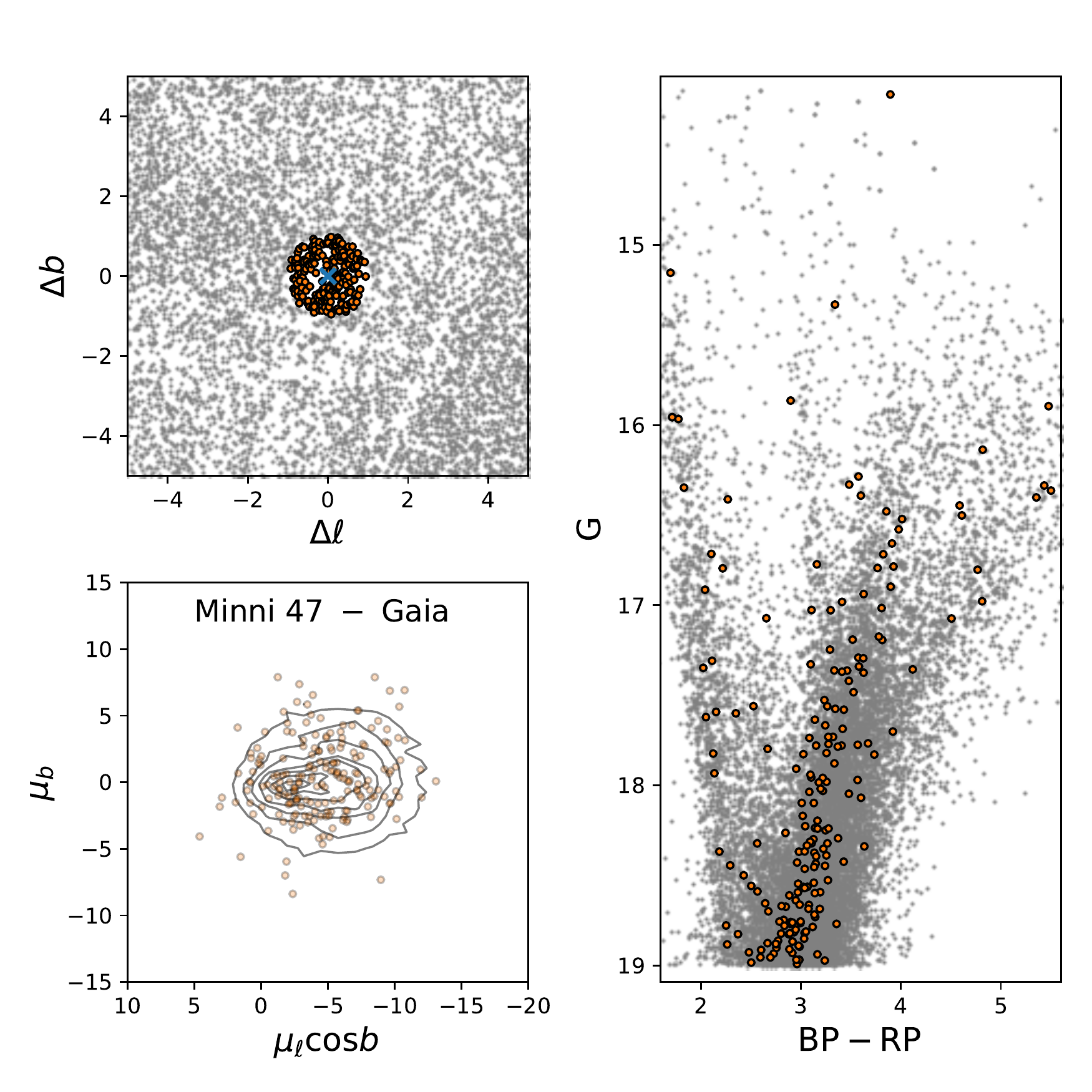} &  
\includegraphics[width=8cm]{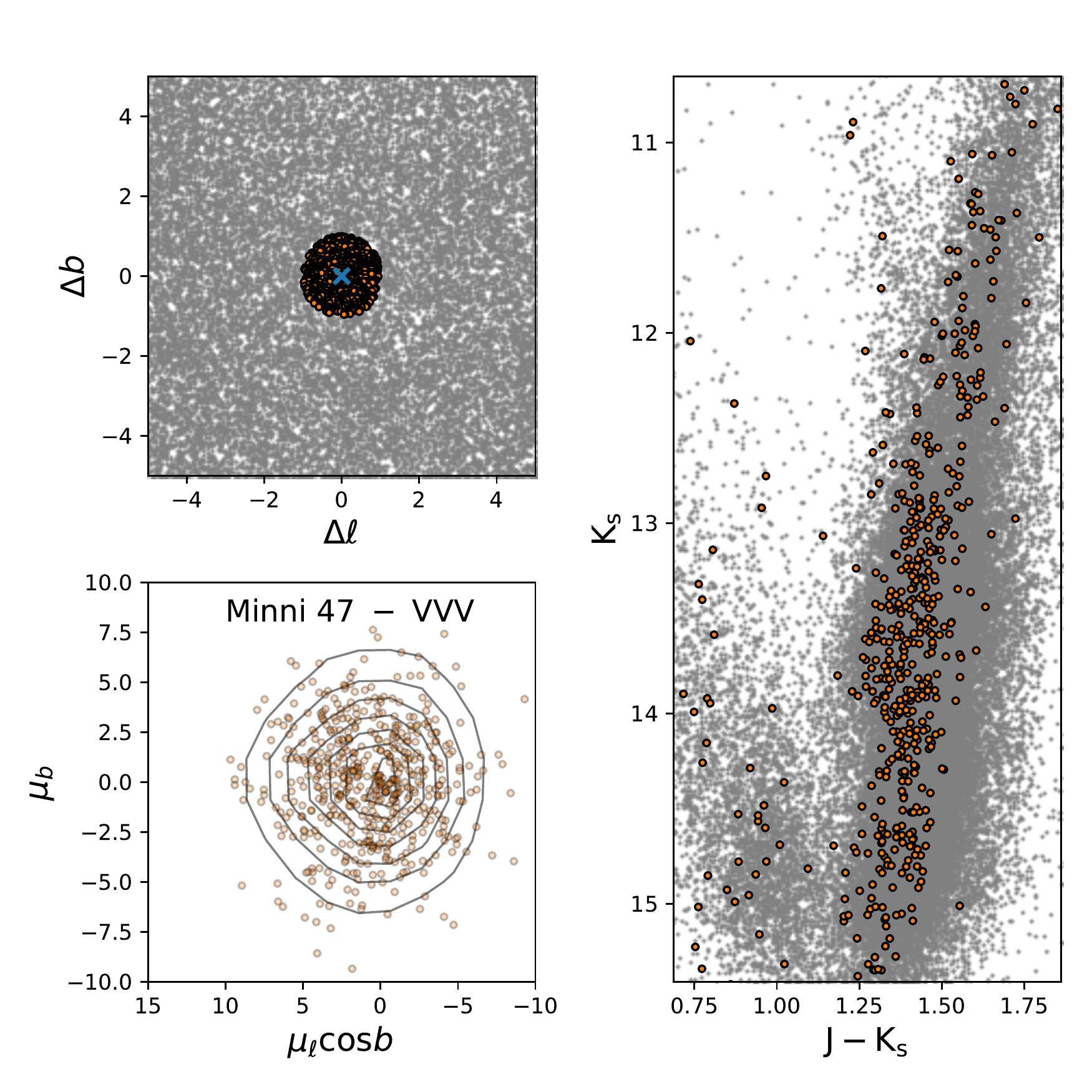} \\
\includegraphics[width=8cm]{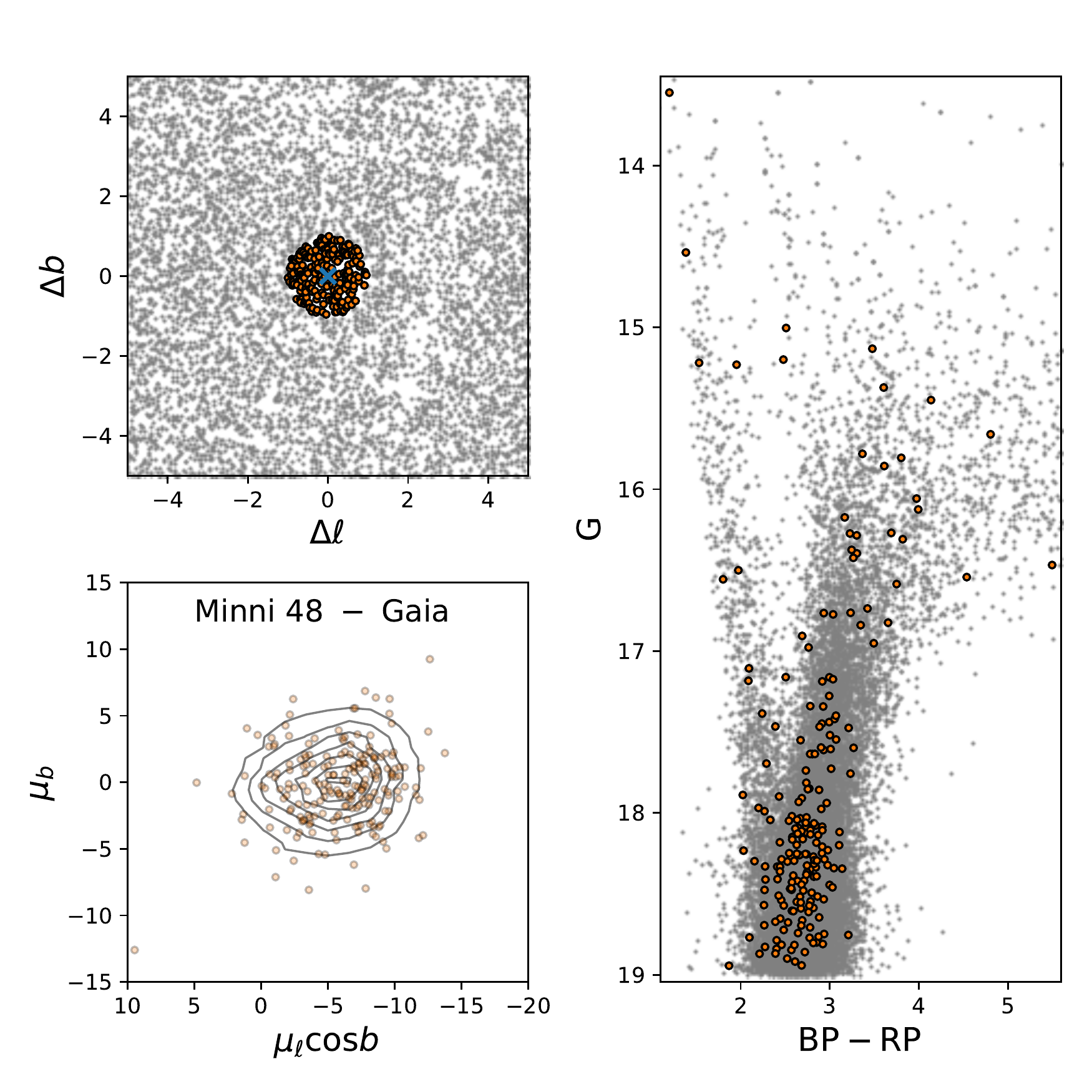} & 
\includegraphics[width=8cm]{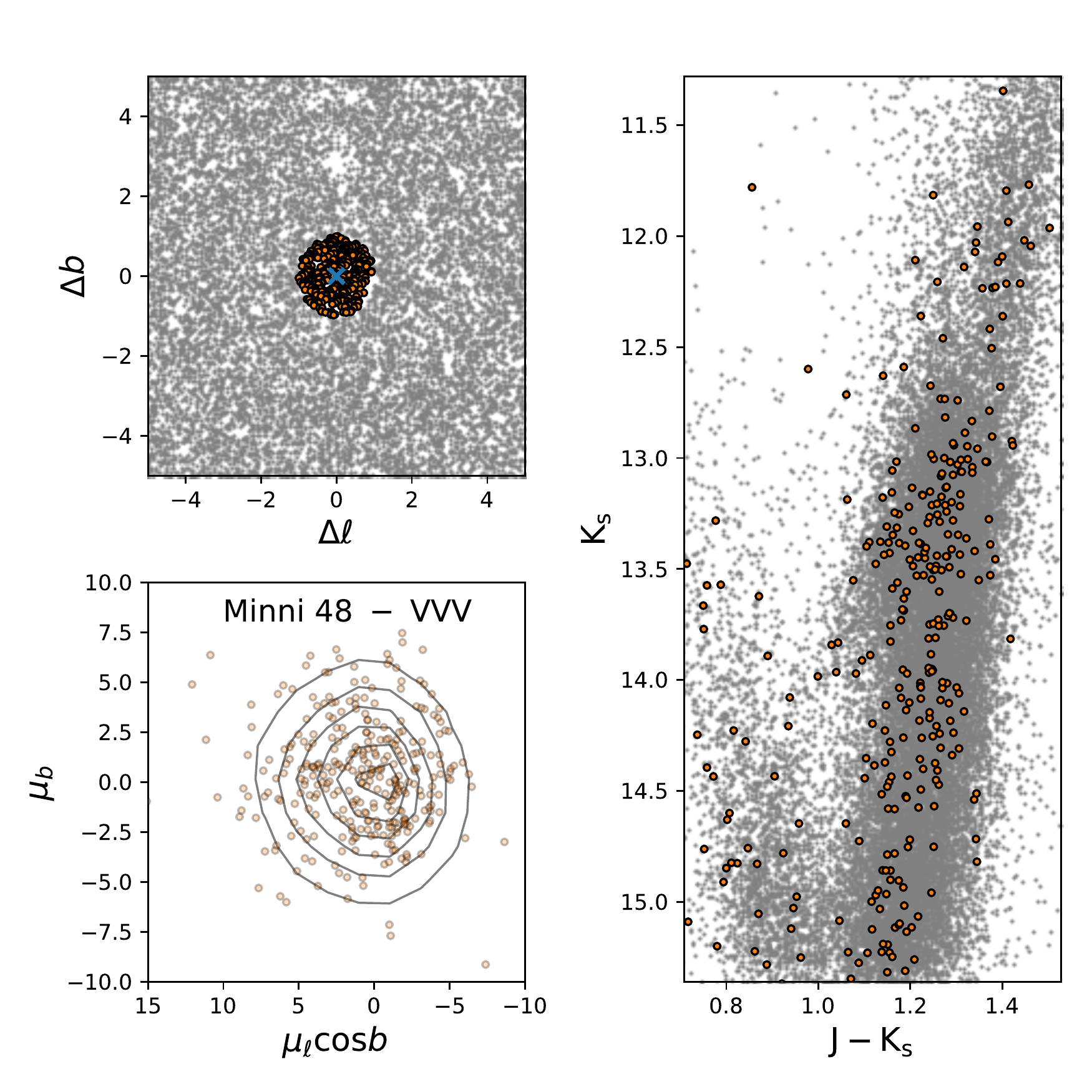} \\
\includegraphics[width=8cm]{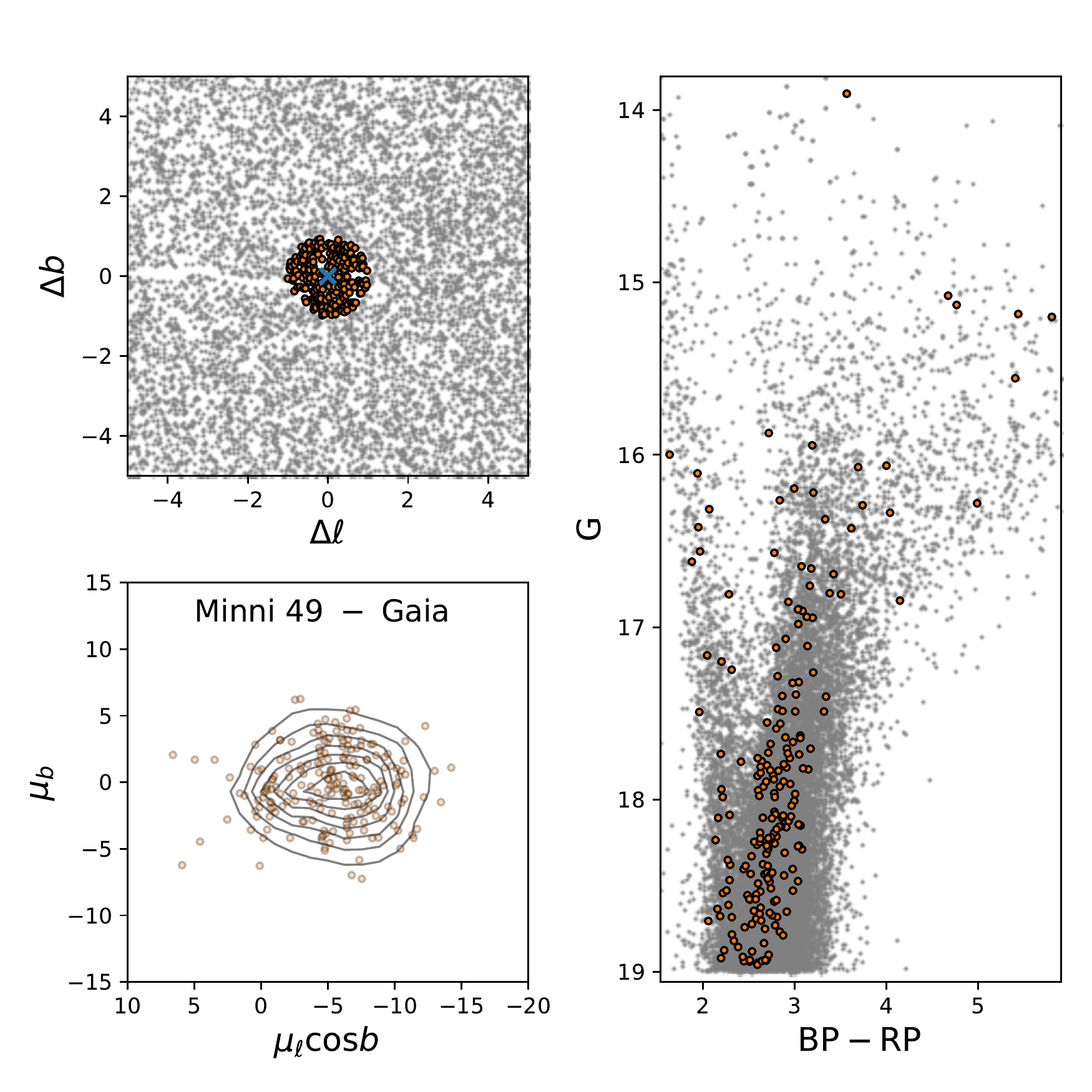} &
\includegraphics[width=8cm]{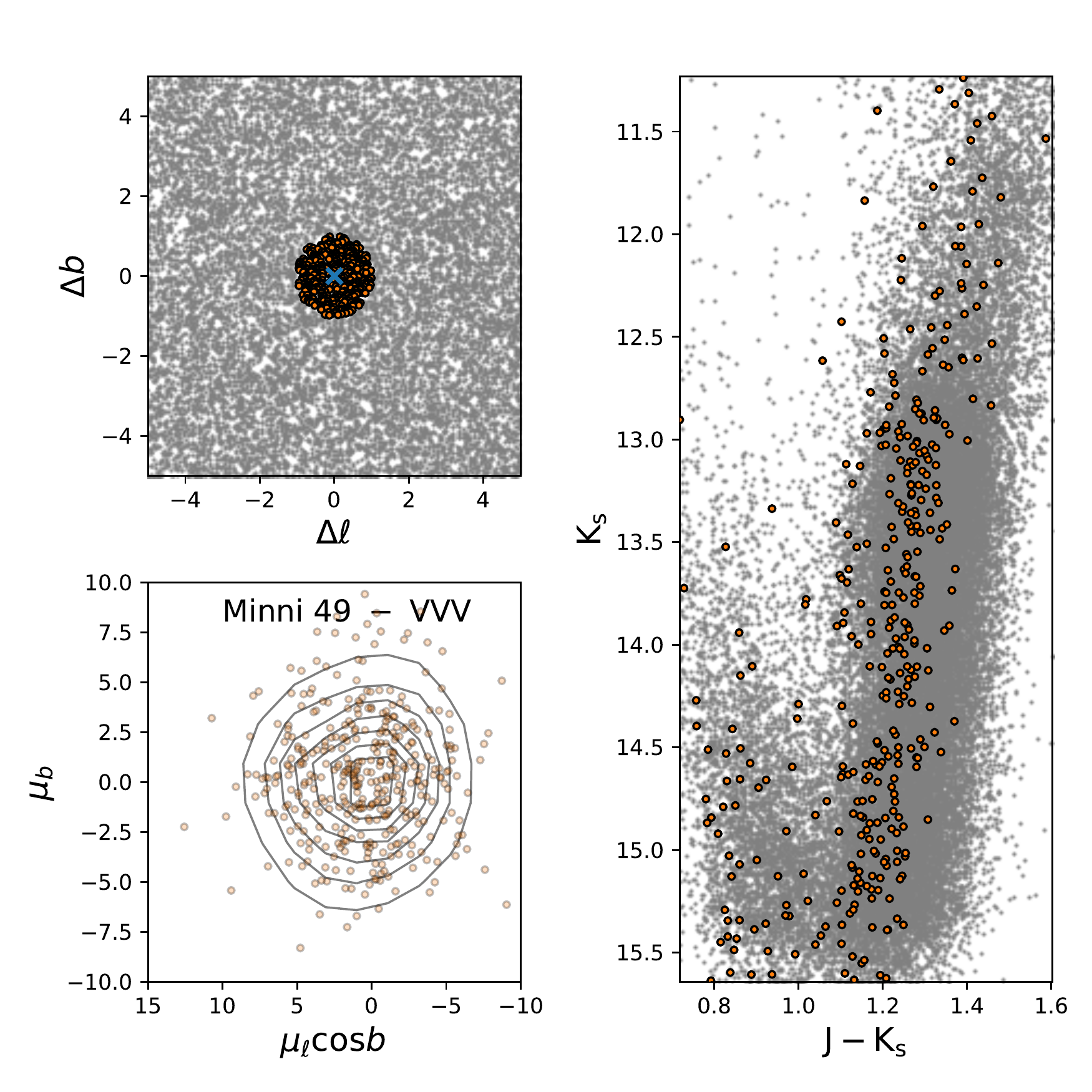} \\
\end{tabular}
\end{table*}
\newpage
\begin{table*}
\begin{tabular}{cc}
\includegraphics[width=8cm]{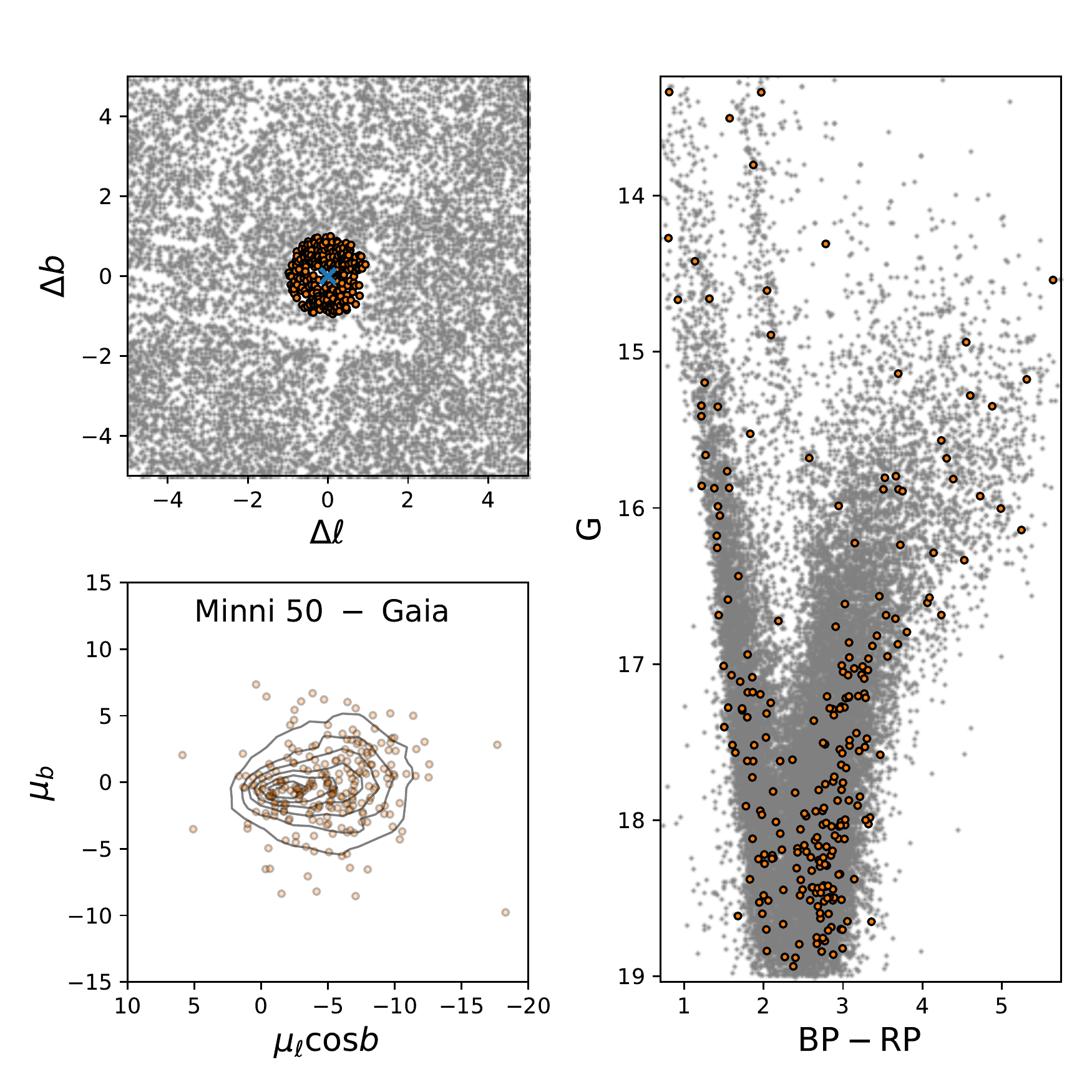} &  
\includegraphics[width=8cm]{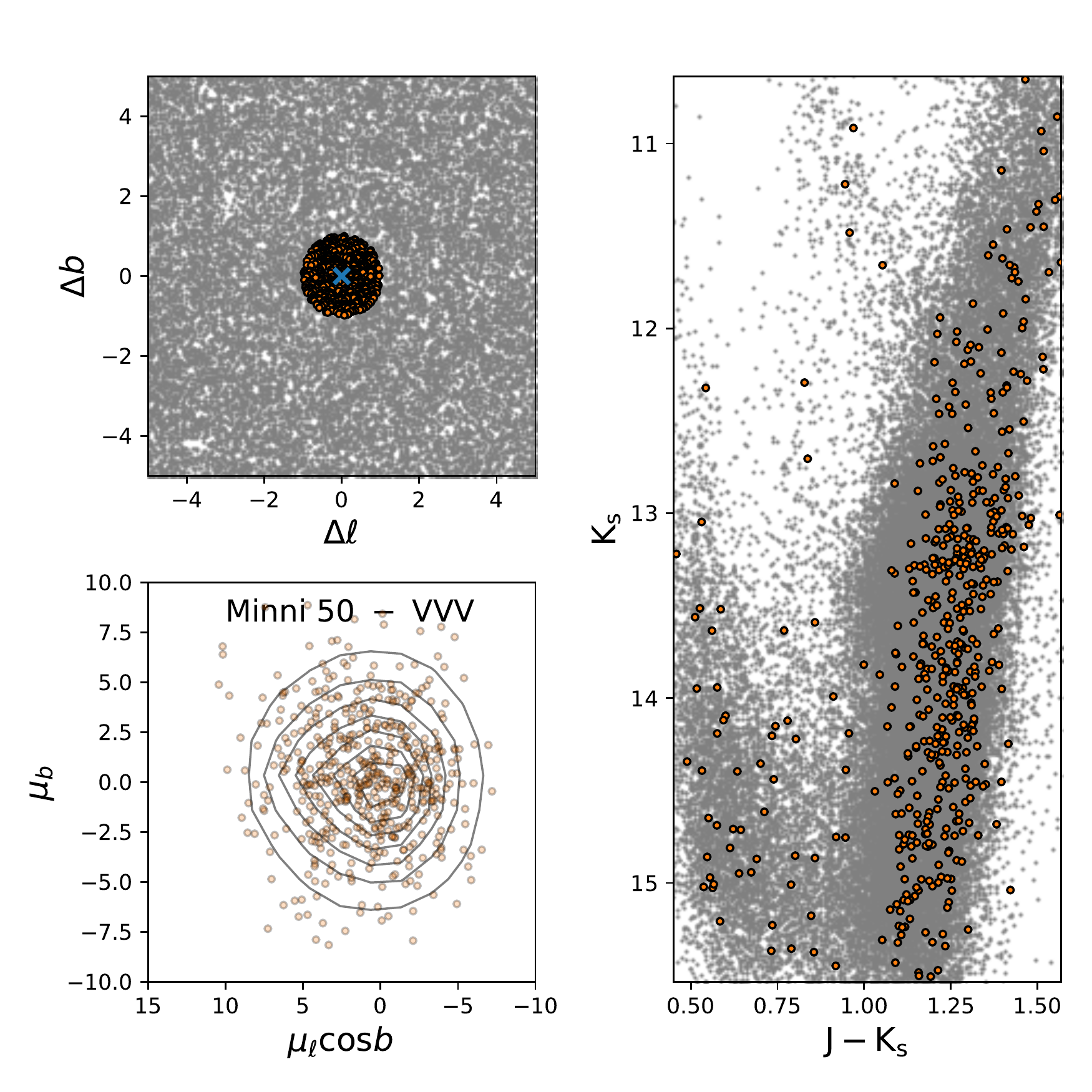} \\
\includegraphics[width=8cm]{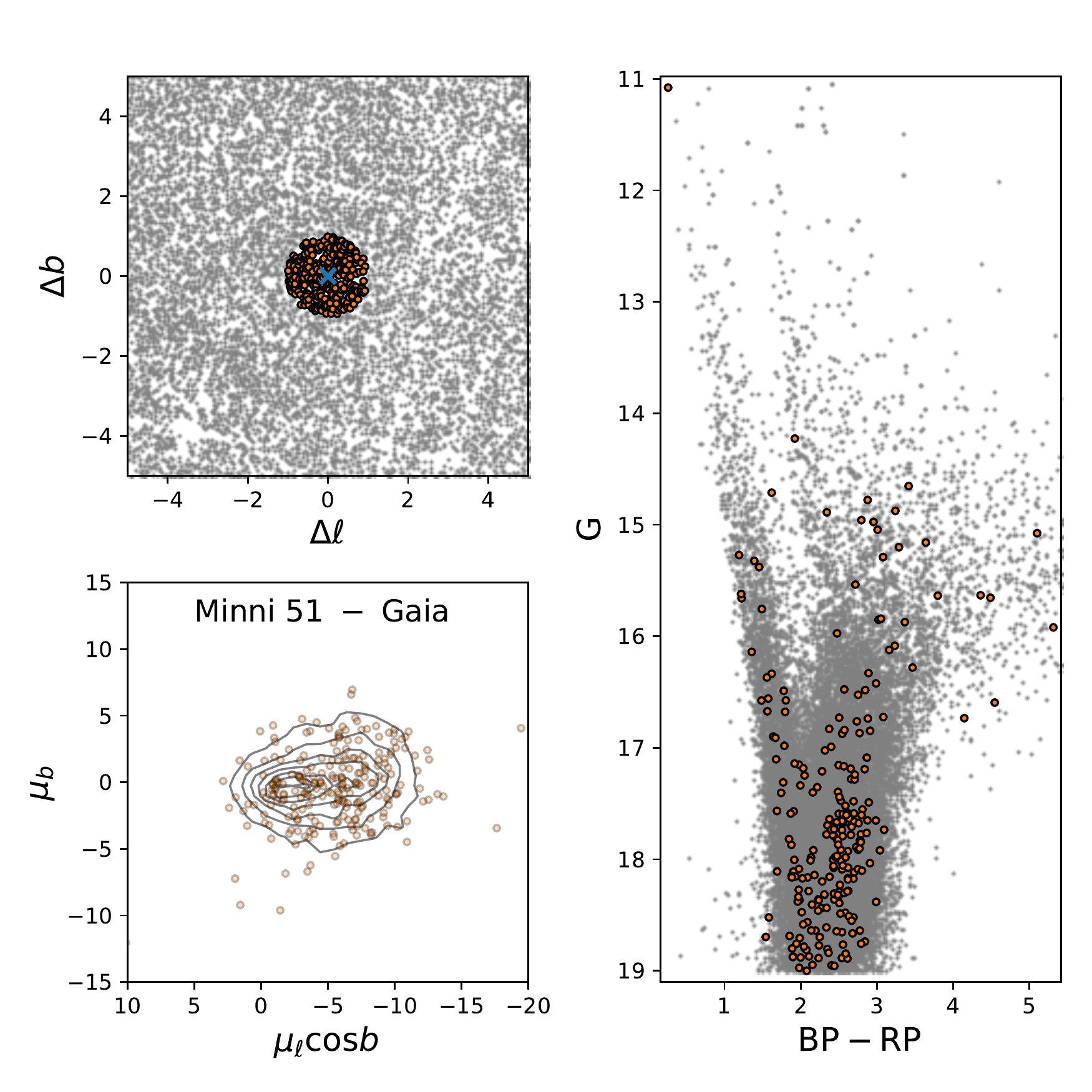} &  
\includegraphics[width=8cm]{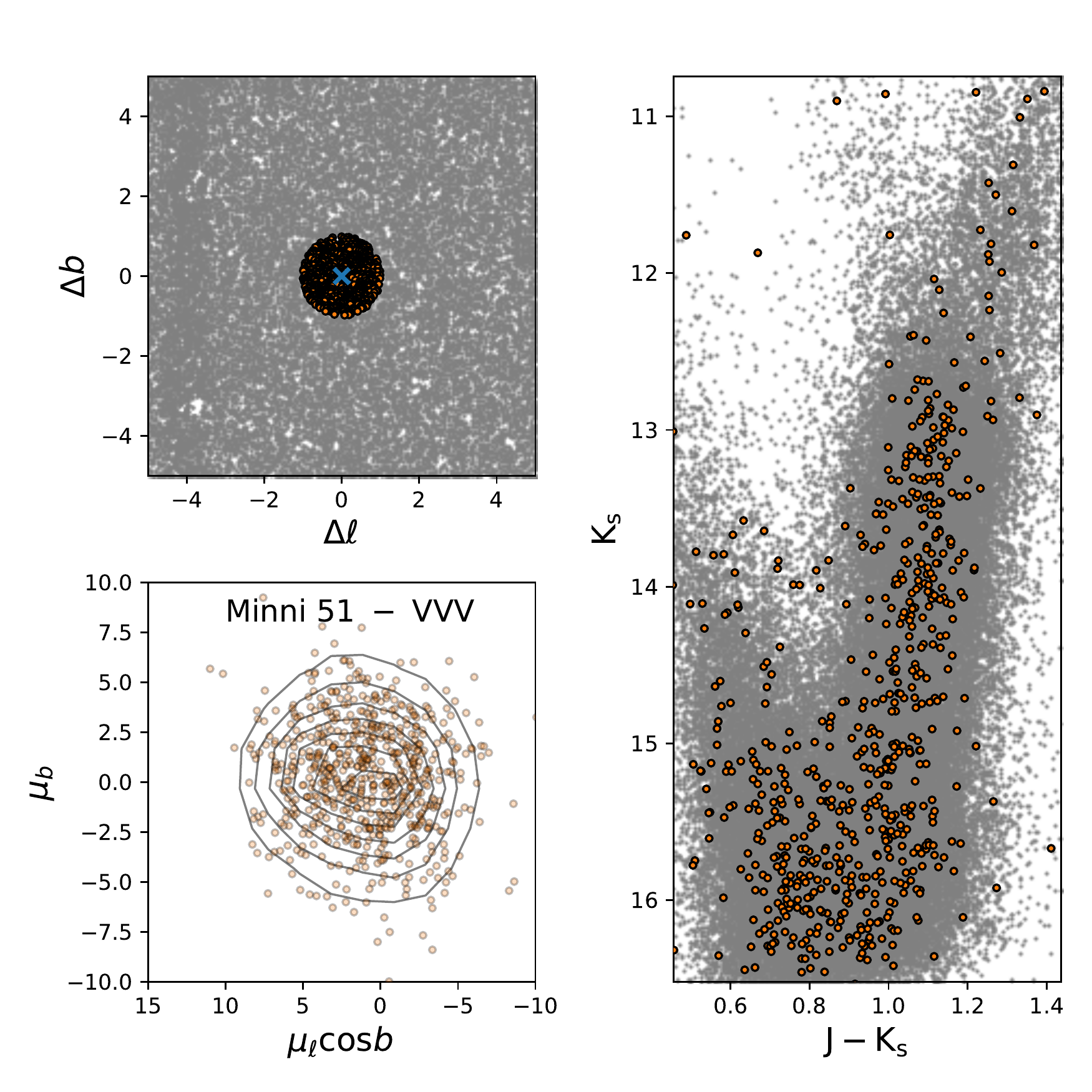} \\
\includegraphics[width=8cm]{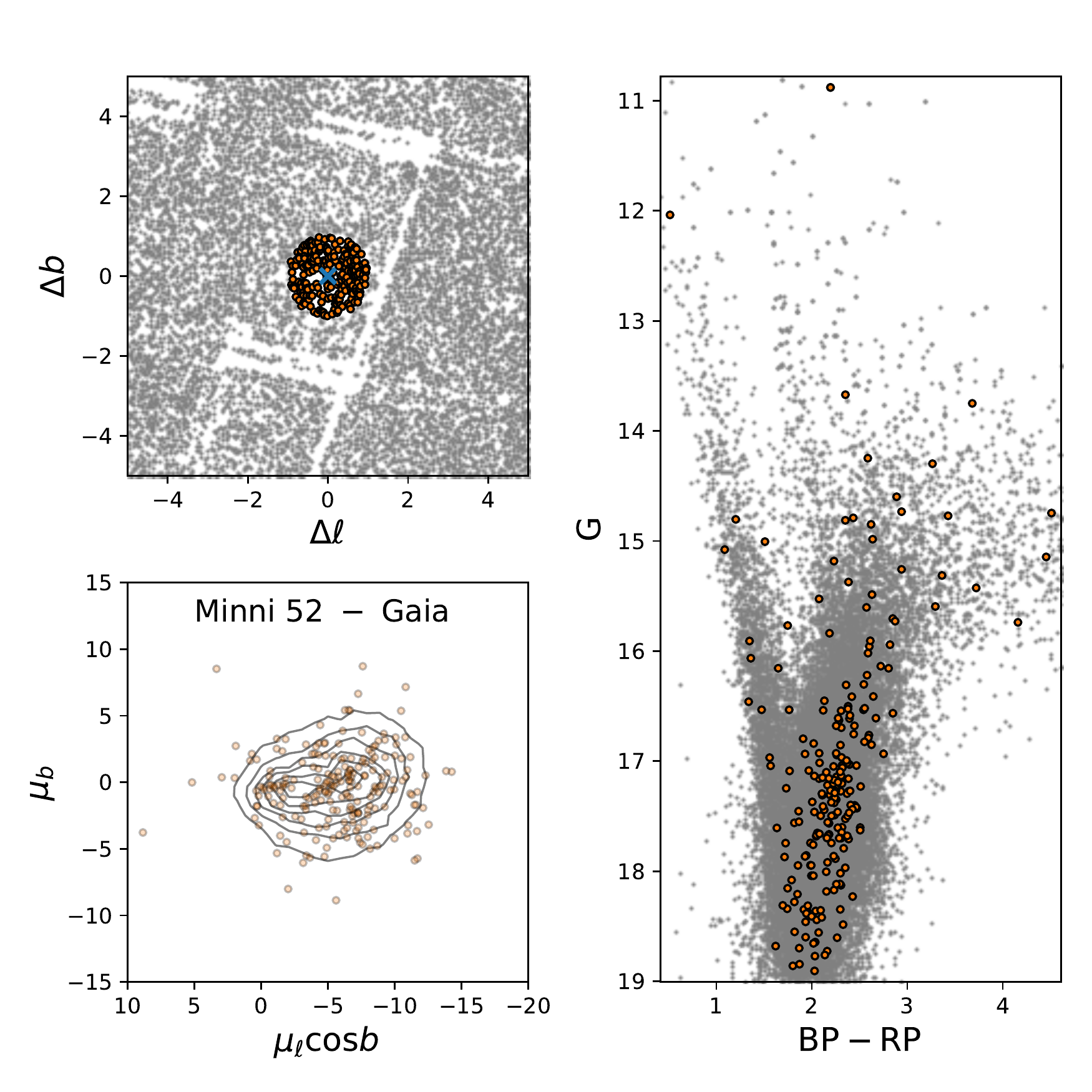} &  
\includegraphics[width=8cm]{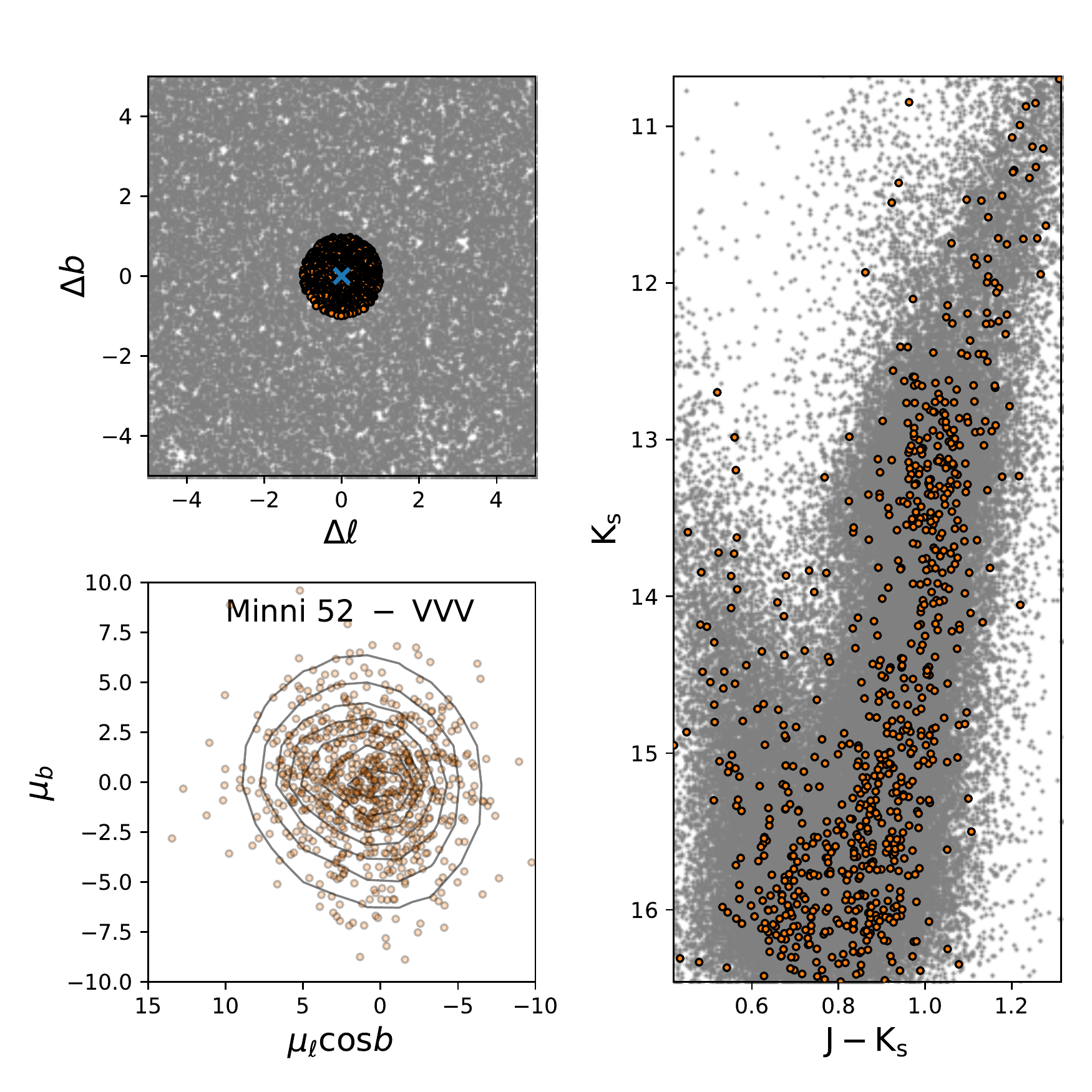} \\
\end{tabular}
\end{table*}
\newpage
\begin{table*}
\begin{tabular}{cc}
\includegraphics[width=8cm]{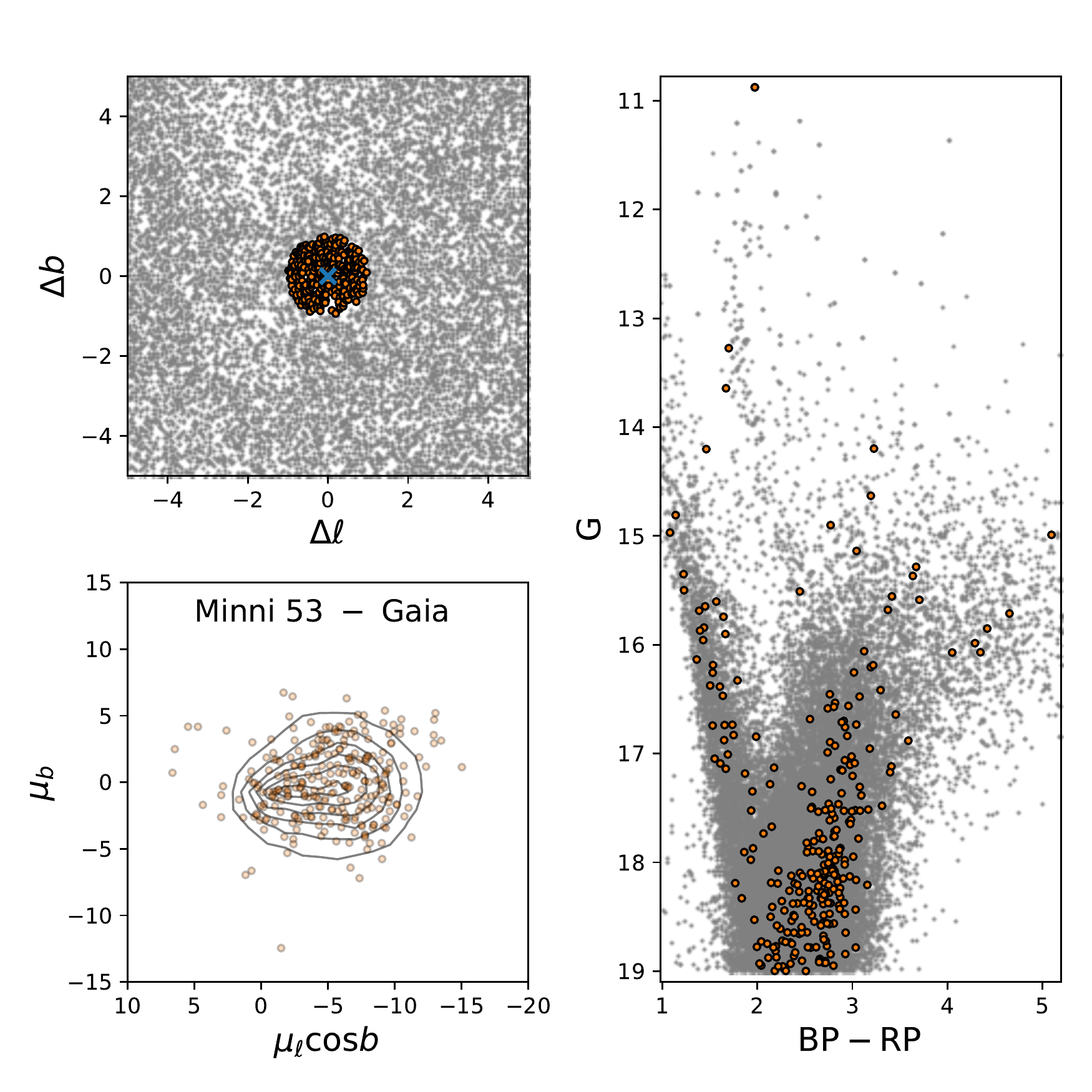} &
\includegraphics[width=8cm]{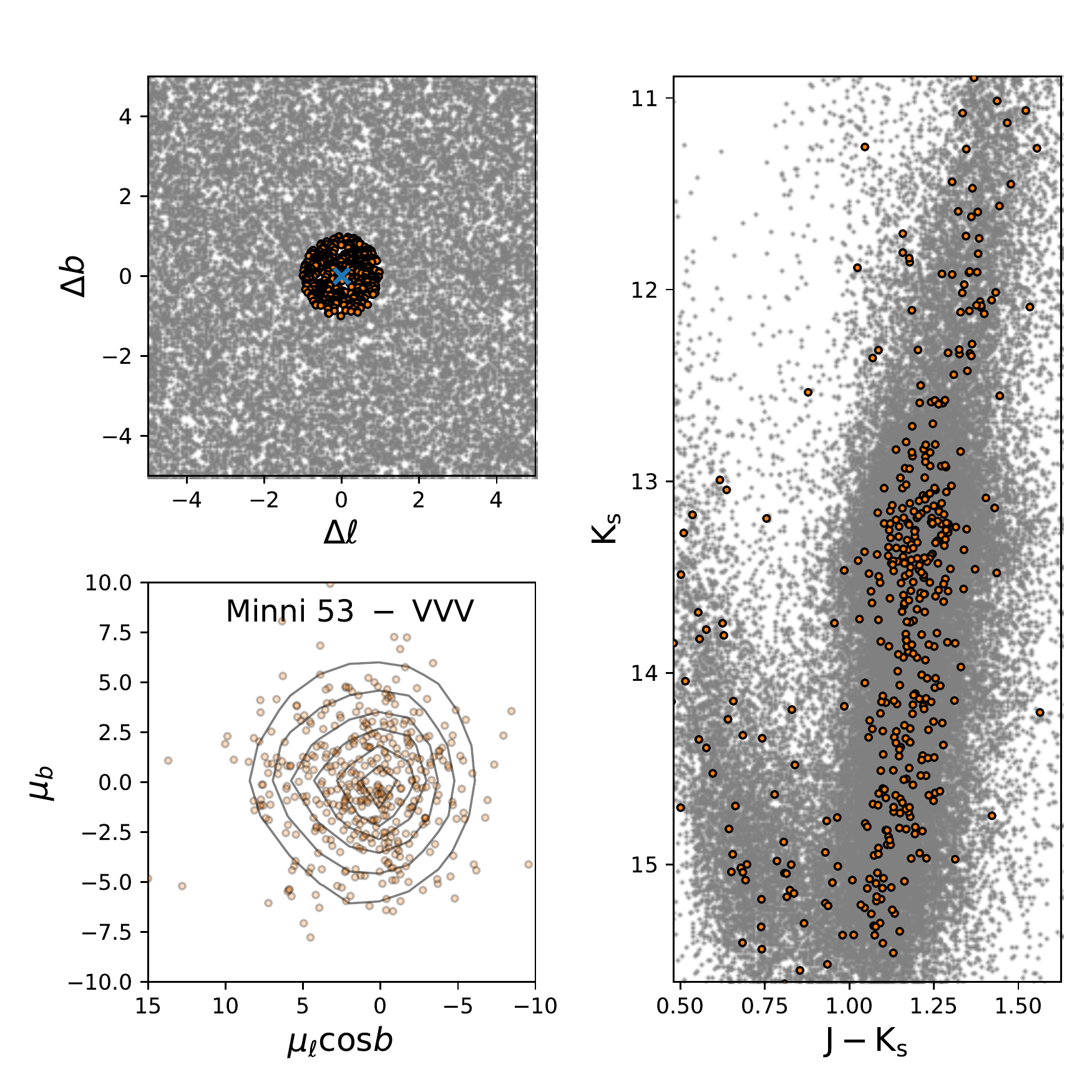} \\  
\includegraphics[width=8cm]{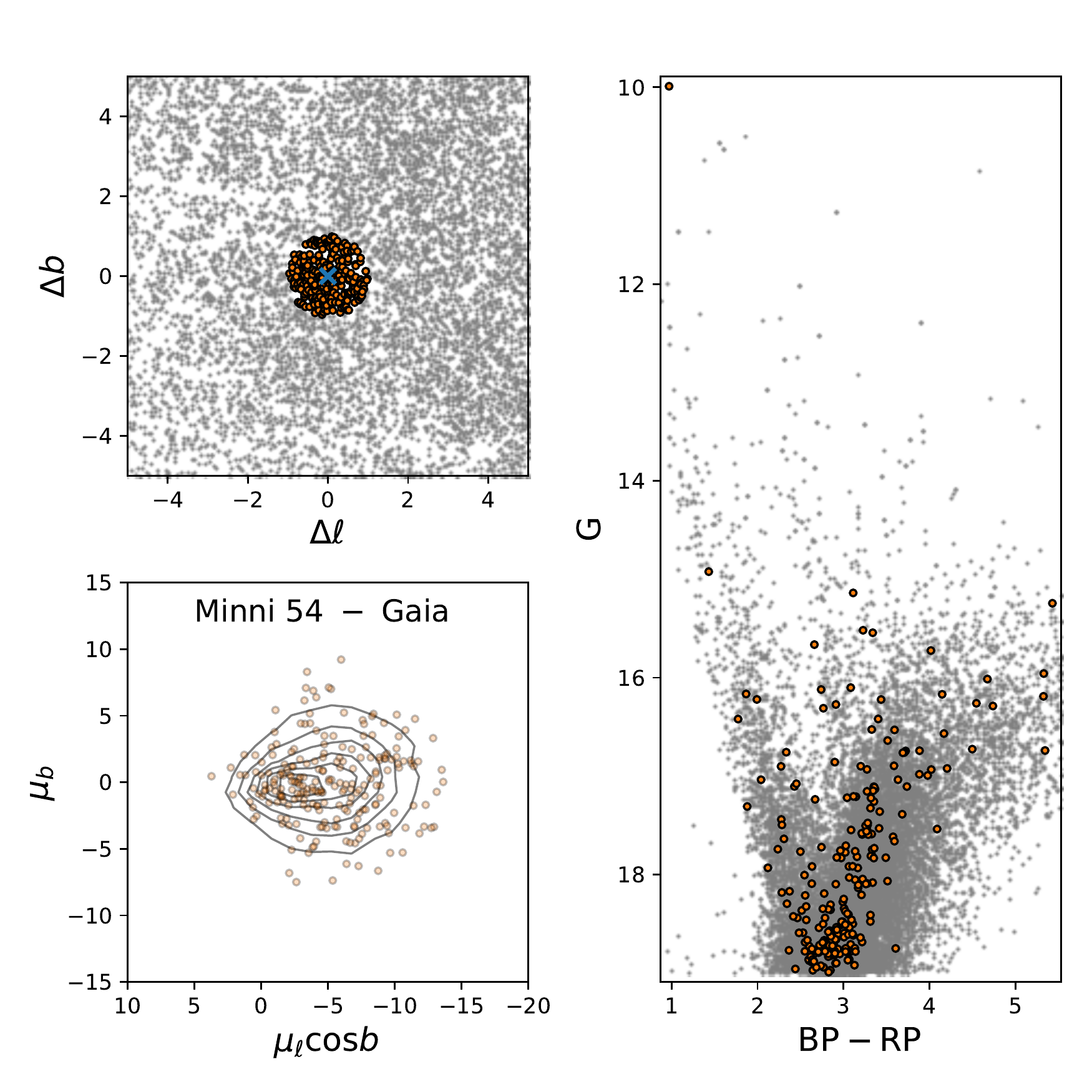} &  
\includegraphics[width=8cm]{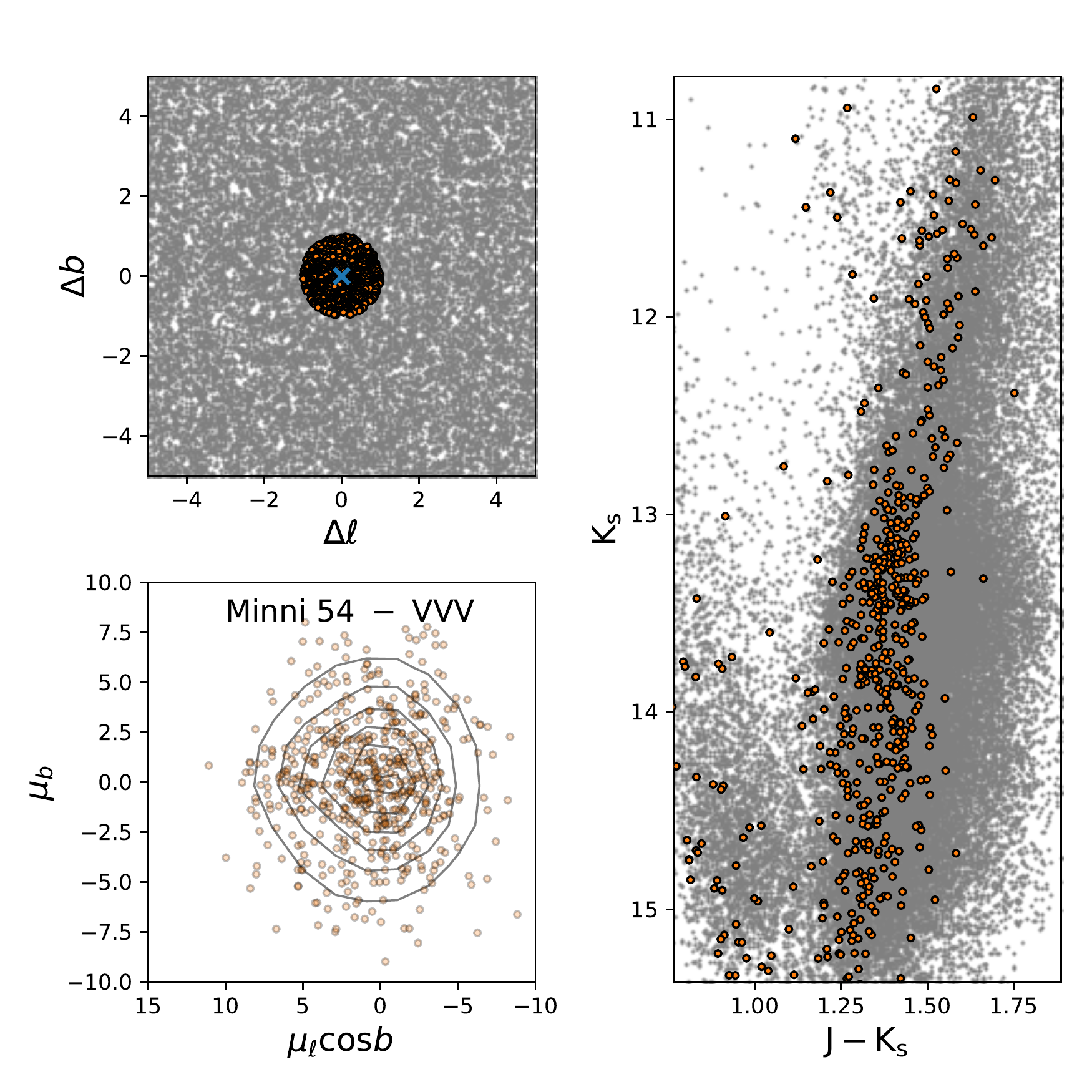} \\
\includegraphics[width=8cm]{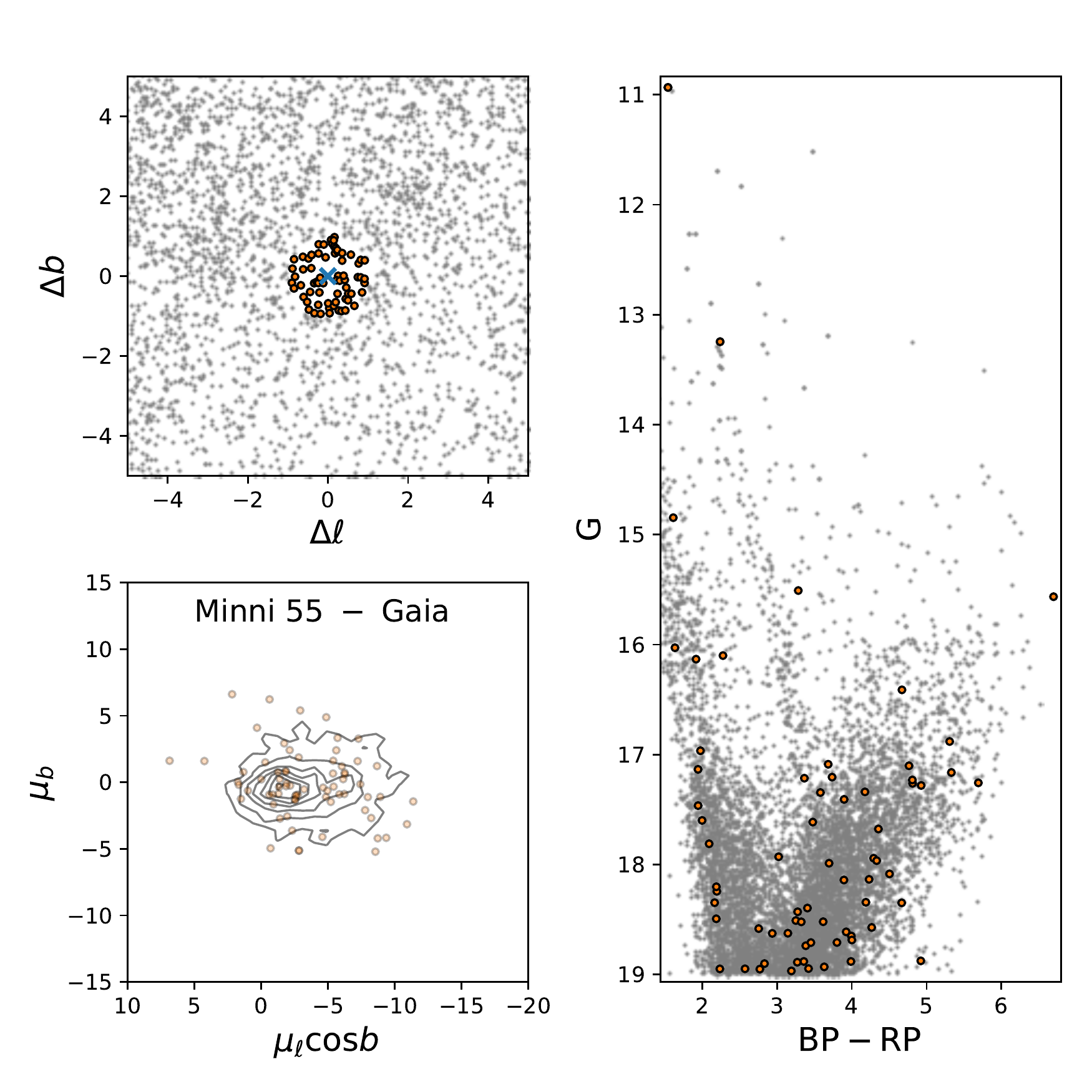} &  
\includegraphics[width=8cm]{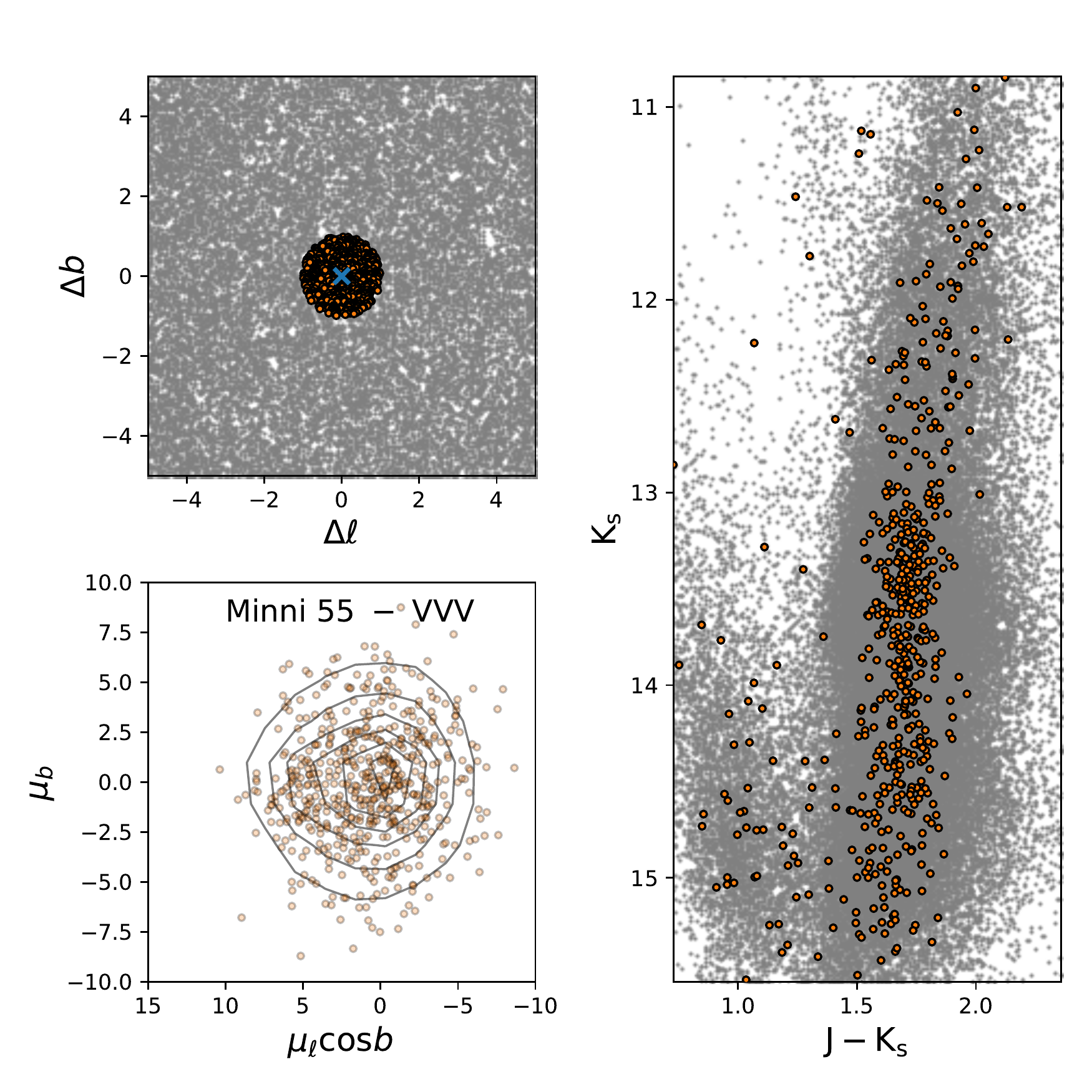} \\ 
\end{tabular}
\end{table*}
\newpage
\begin{table*}
\begin{tabular}{cc}
\includegraphics[width=8cm]{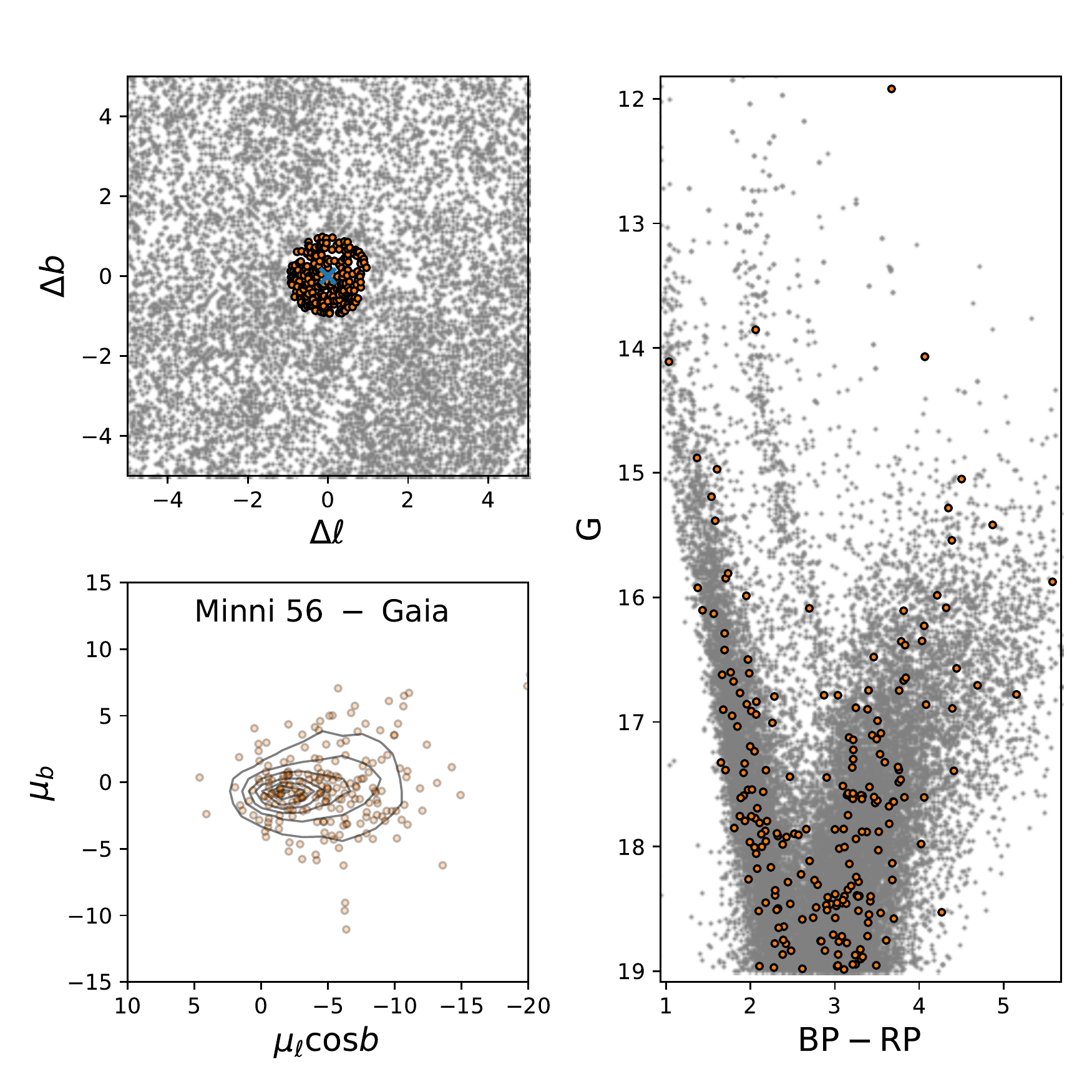} & 
\includegraphics[width=8cm]{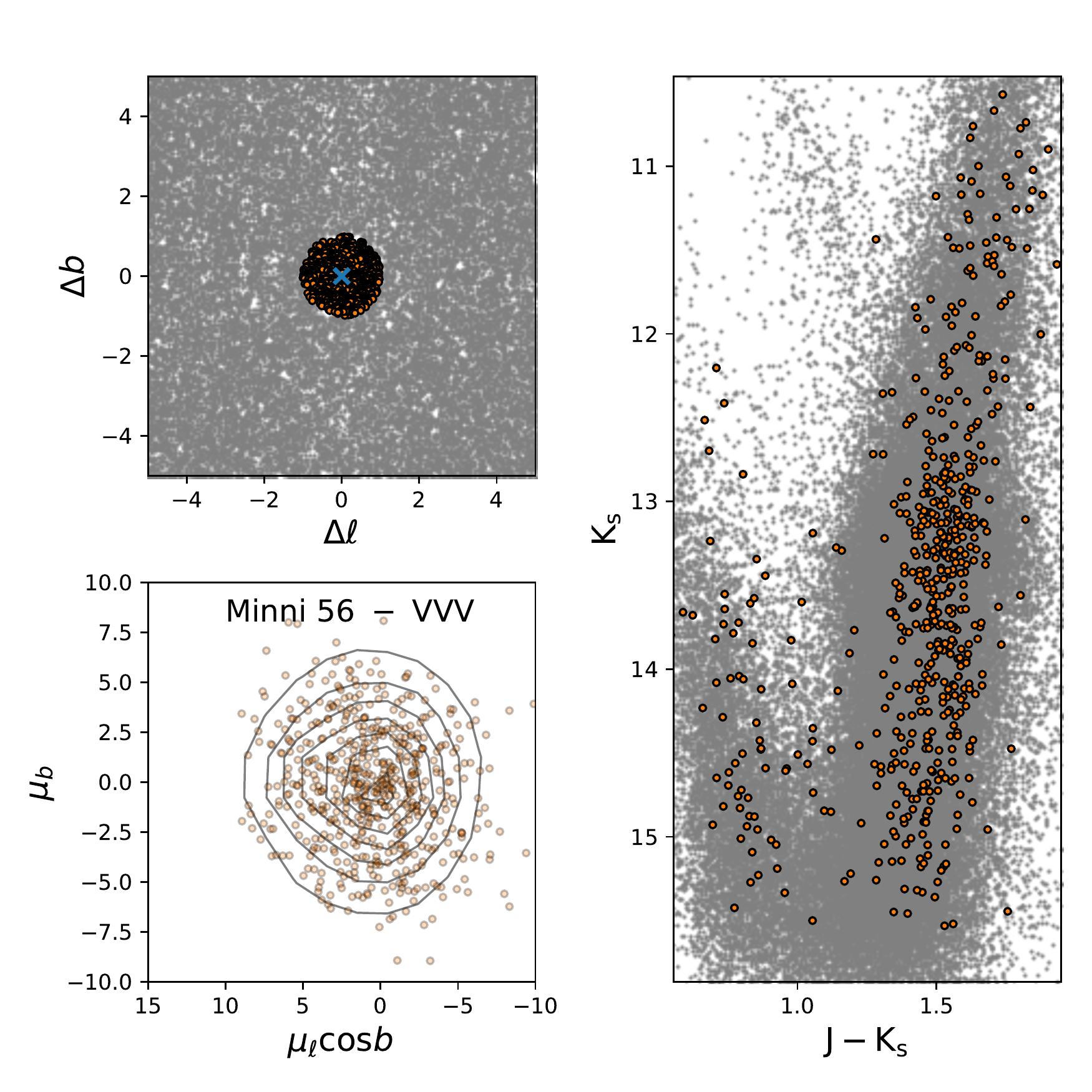} \\
\includegraphics[width=8cm]{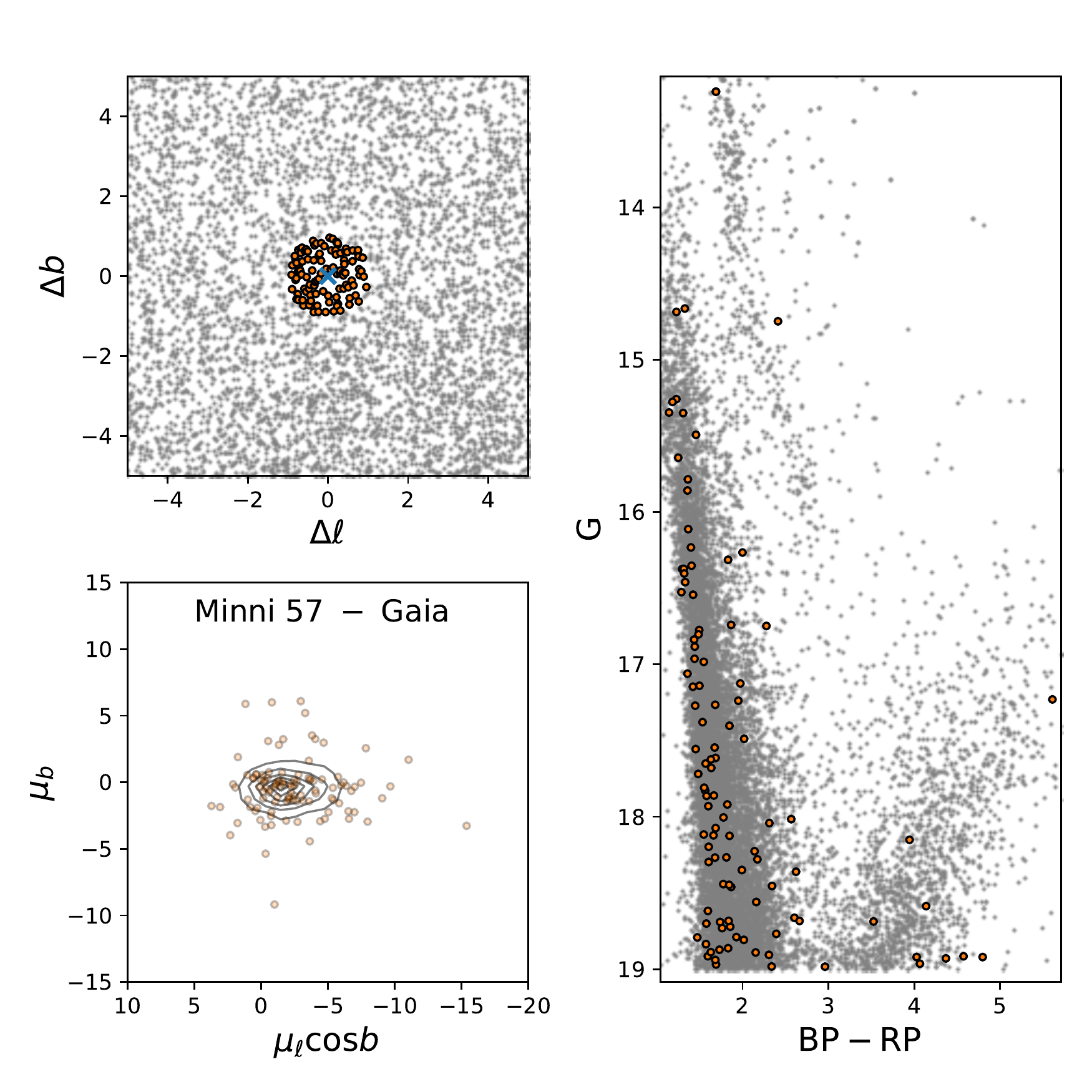} &
\includegraphics[width=8cm]{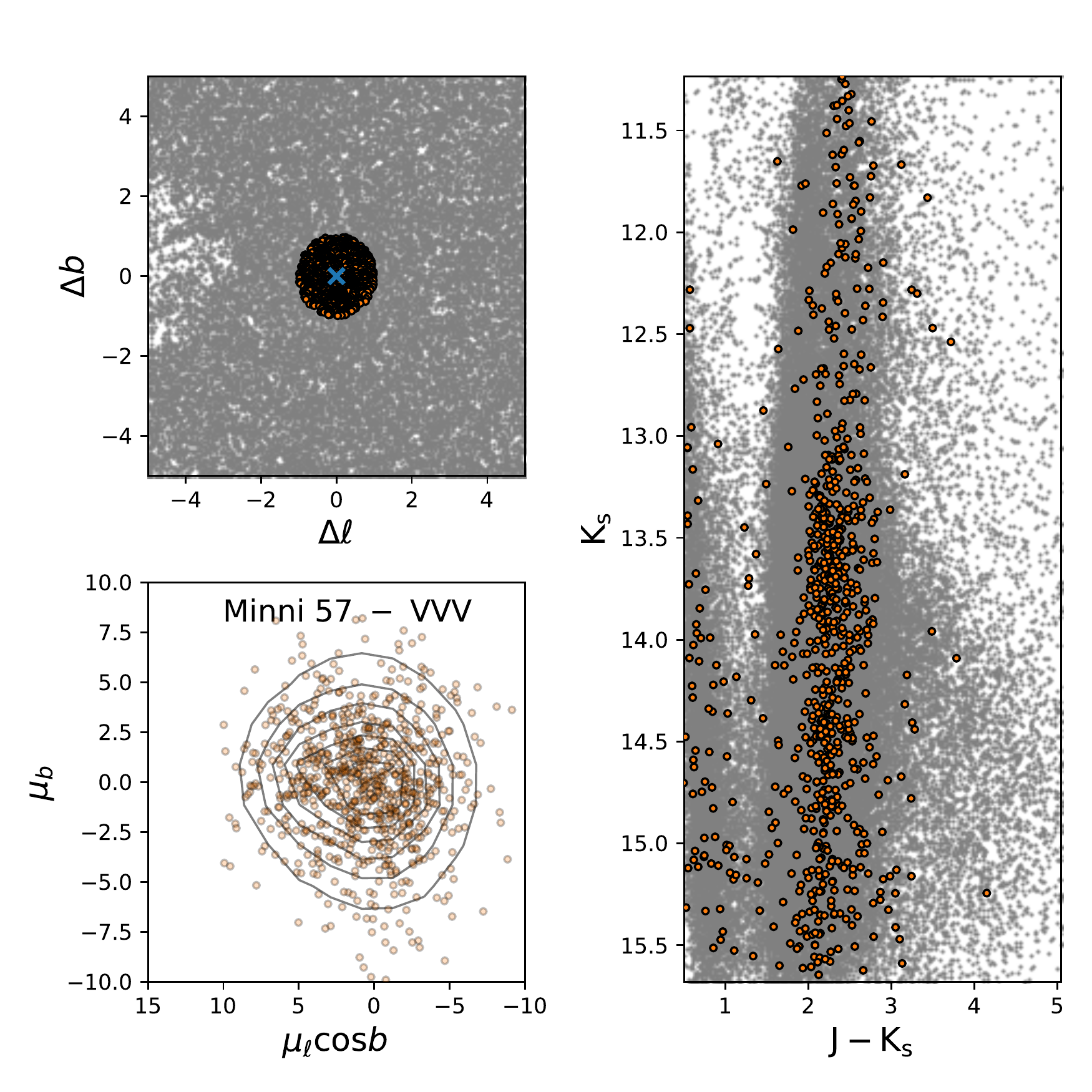} \\
\includegraphics[width=8cm]{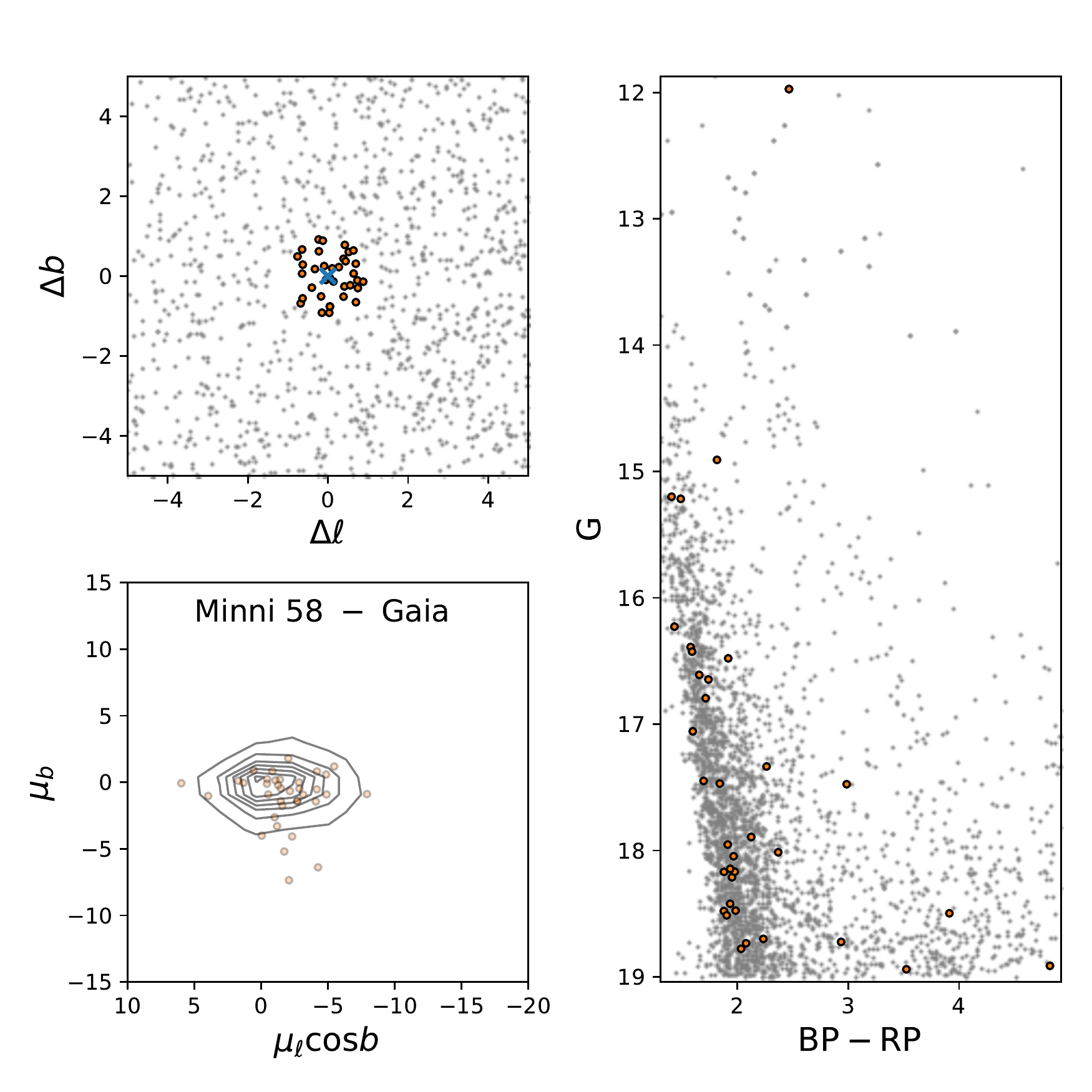} &  
\includegraphics[width=8cm]{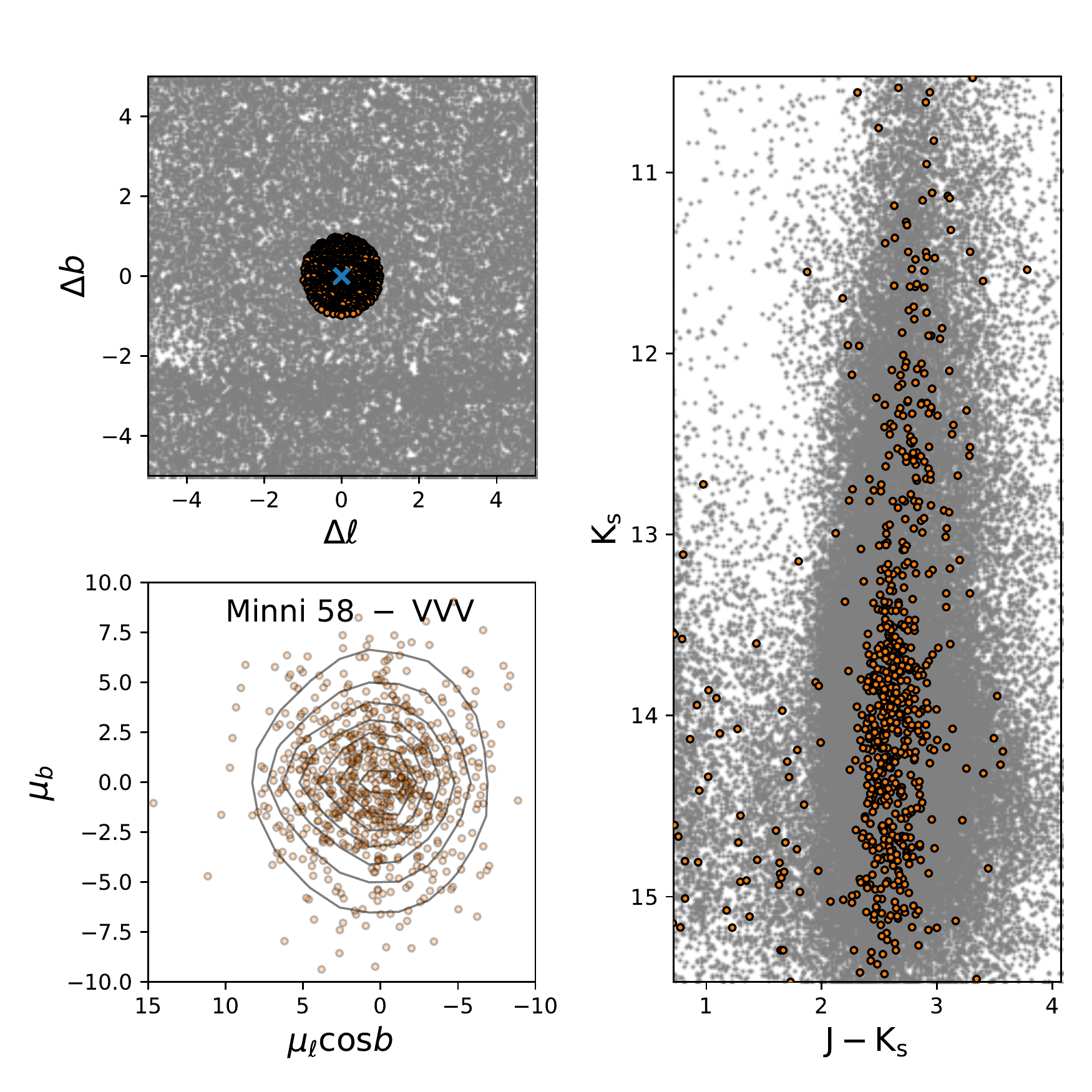} \\
\end{tabular}
\end{table*}
\newpage
\begin{table*}
\begin{tabular}{cc}
\includegraphics[width=8cm]{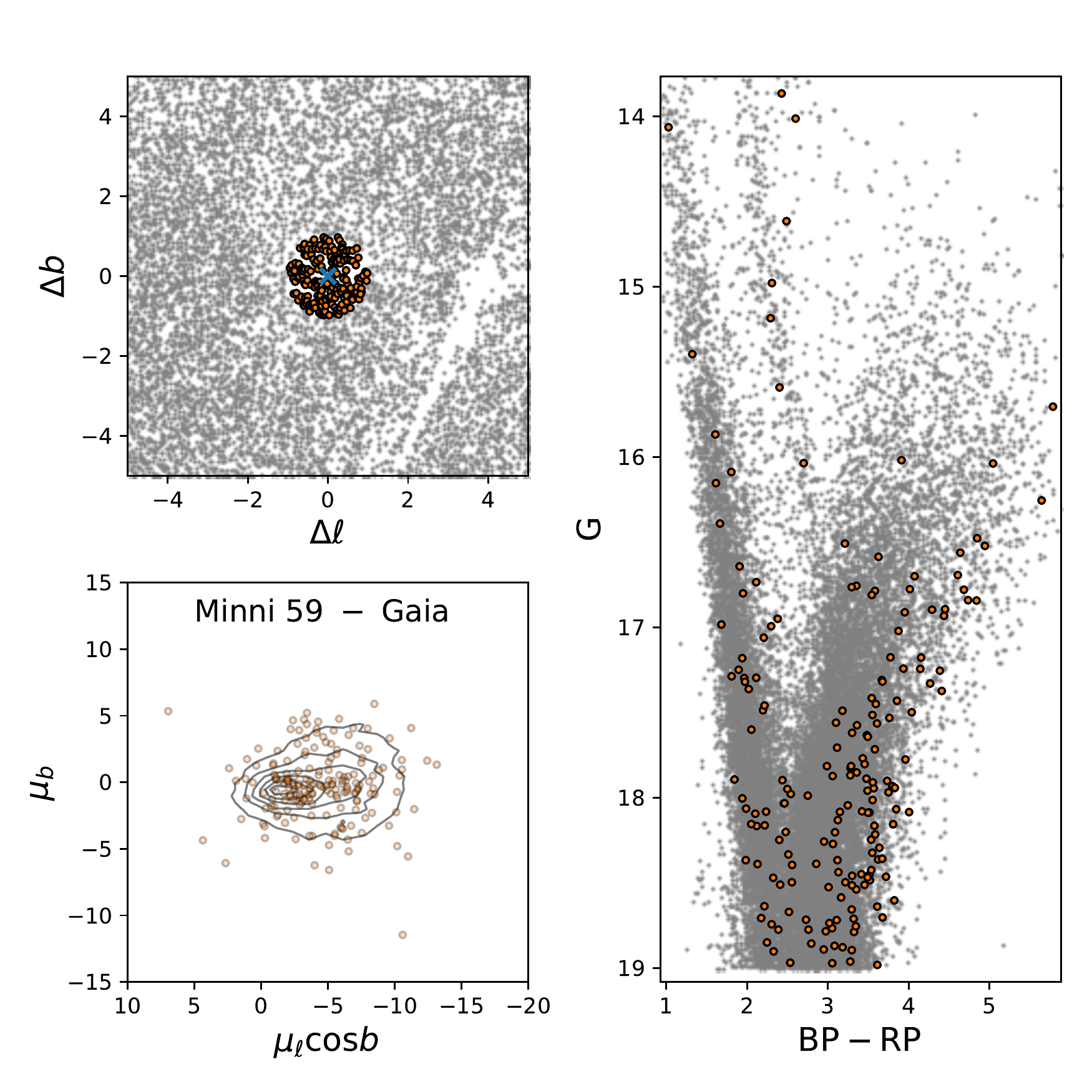} &  
\includegraphics[width=8cm]{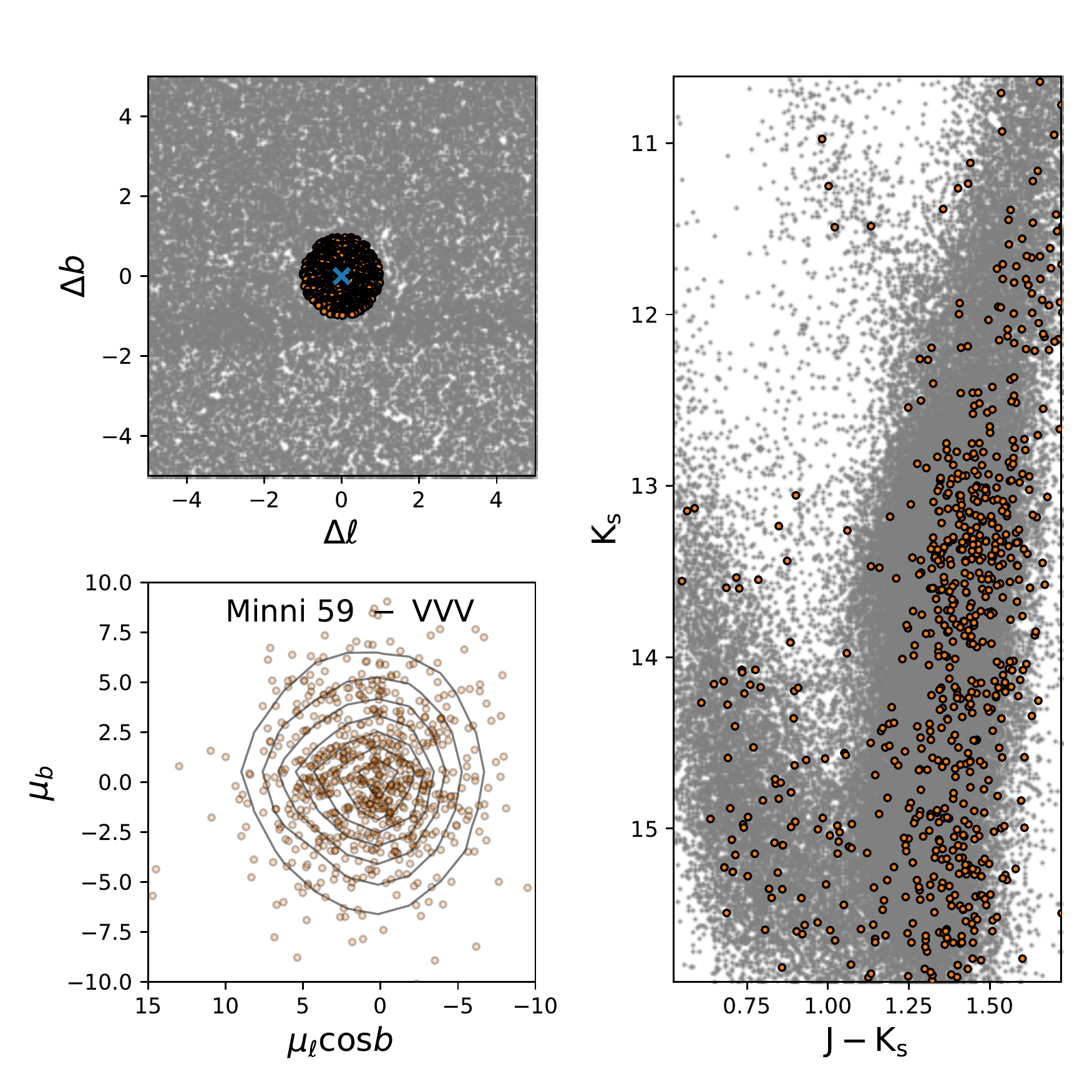} \\ 
\includegraphics[width=8cm]{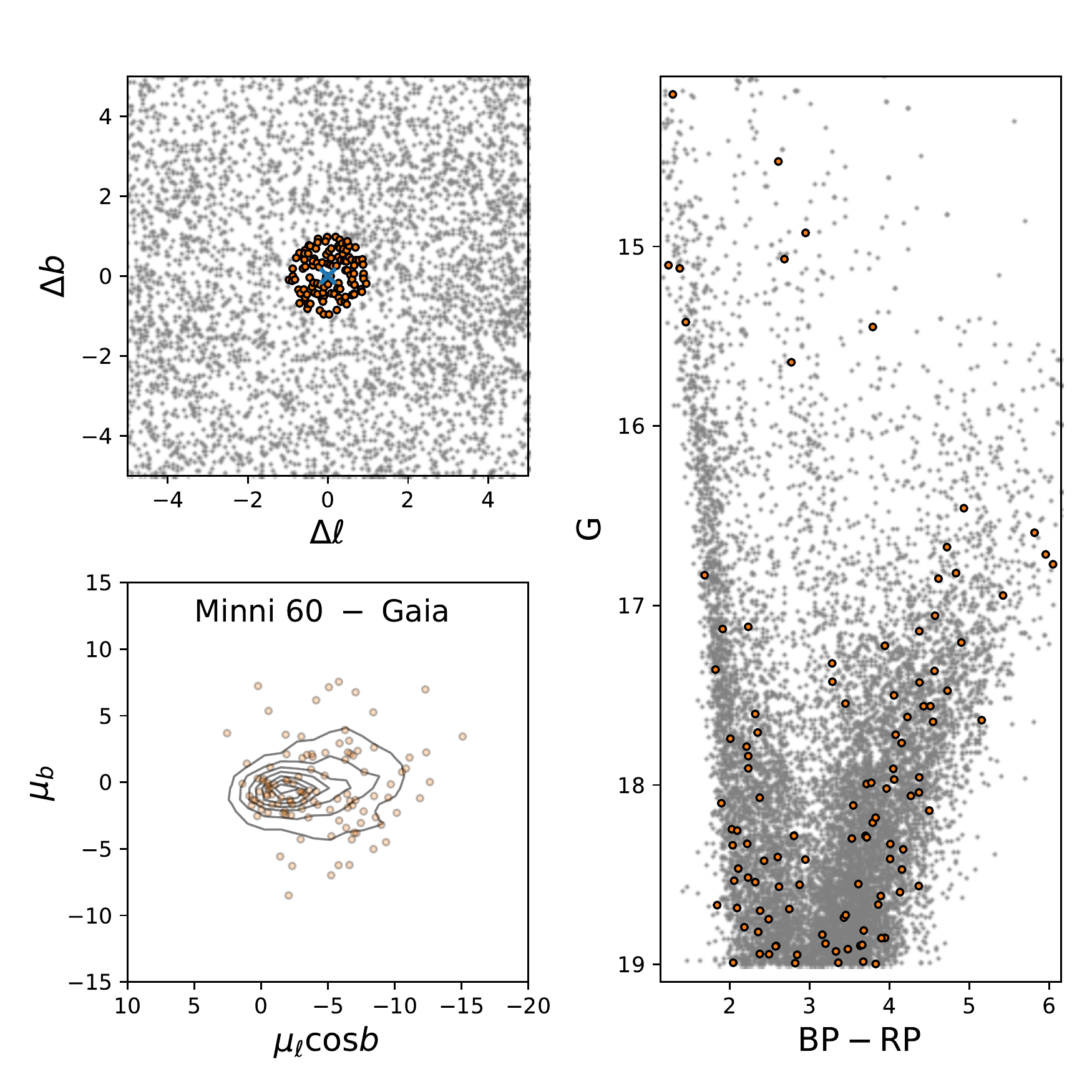} & 
\includegraphics[width=8cm]{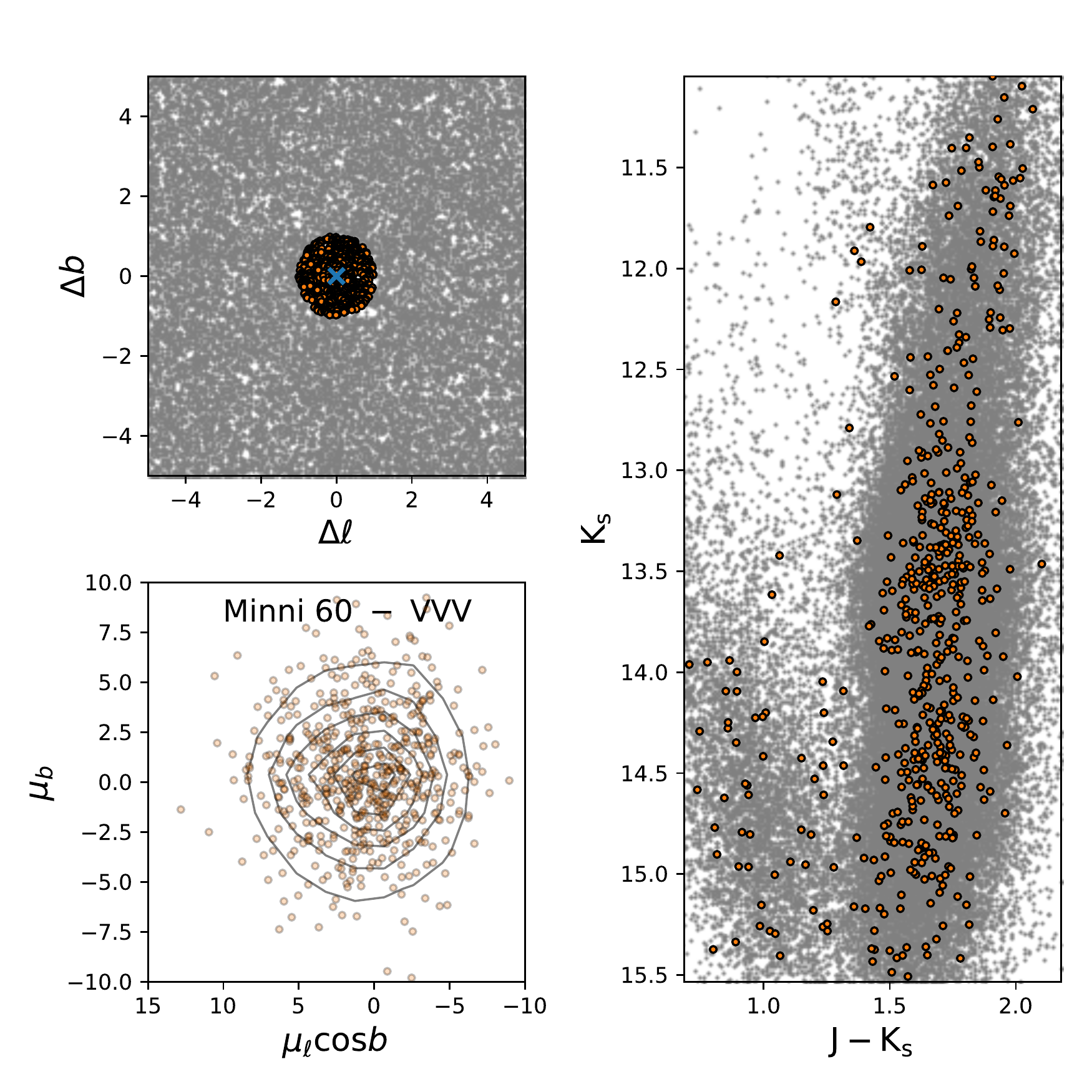} \\  
\includegraphics[width=8cm]{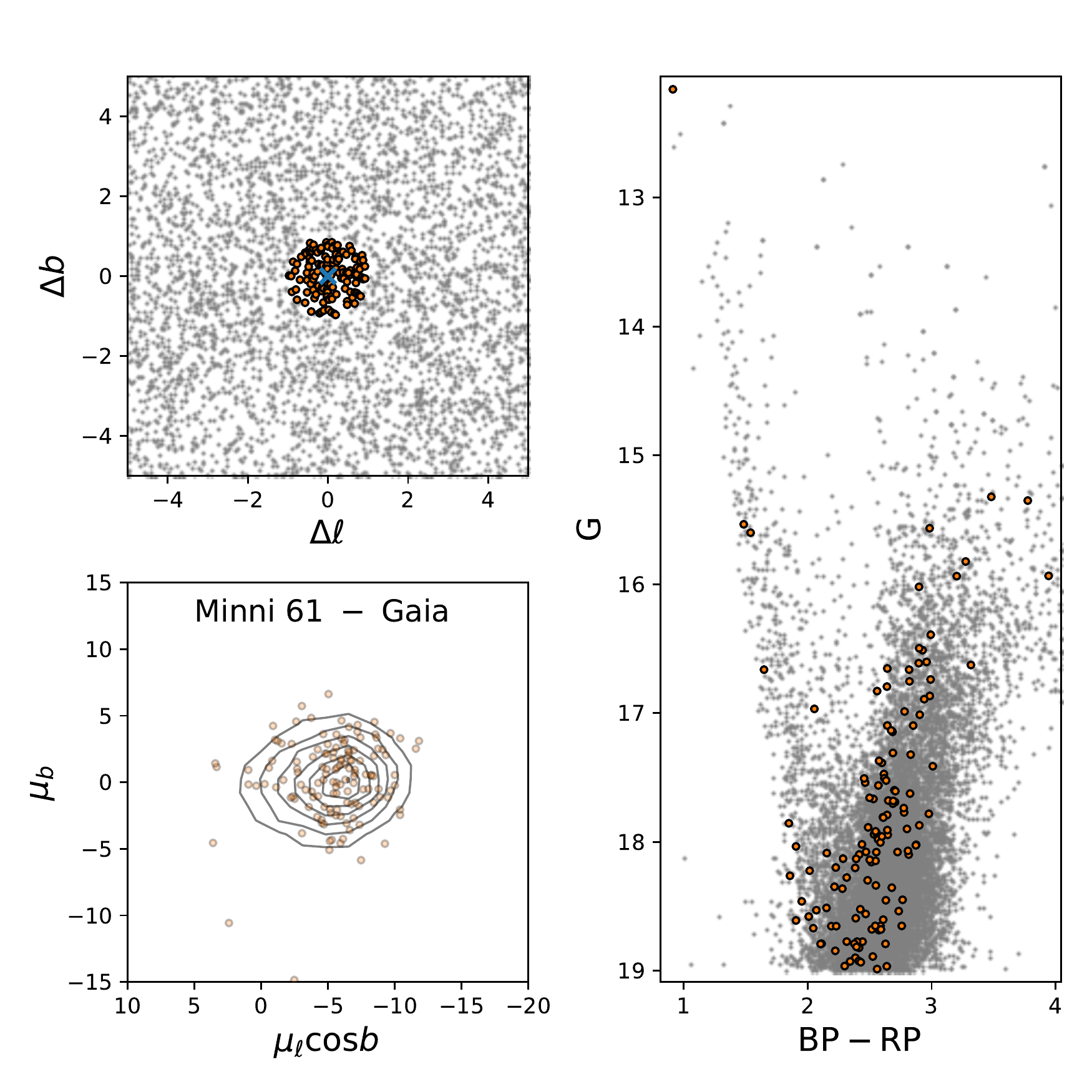} & \\
\end{tabular}
\end{table*}
\newpage
\begin{table*}
\begin{tabular}{cc}
\includegraphics[width=8cm]{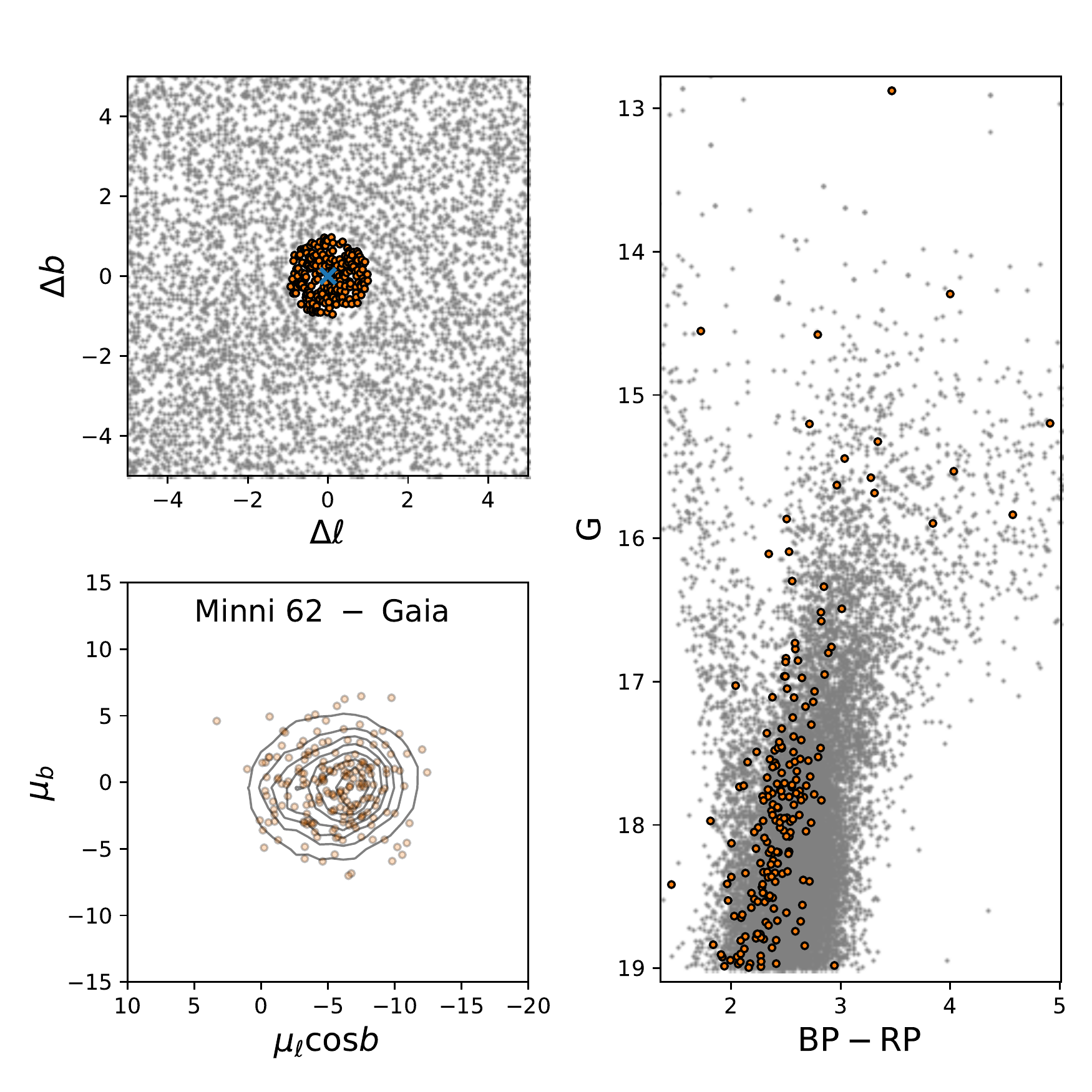} & \\
\includegraphics[width=8cm]{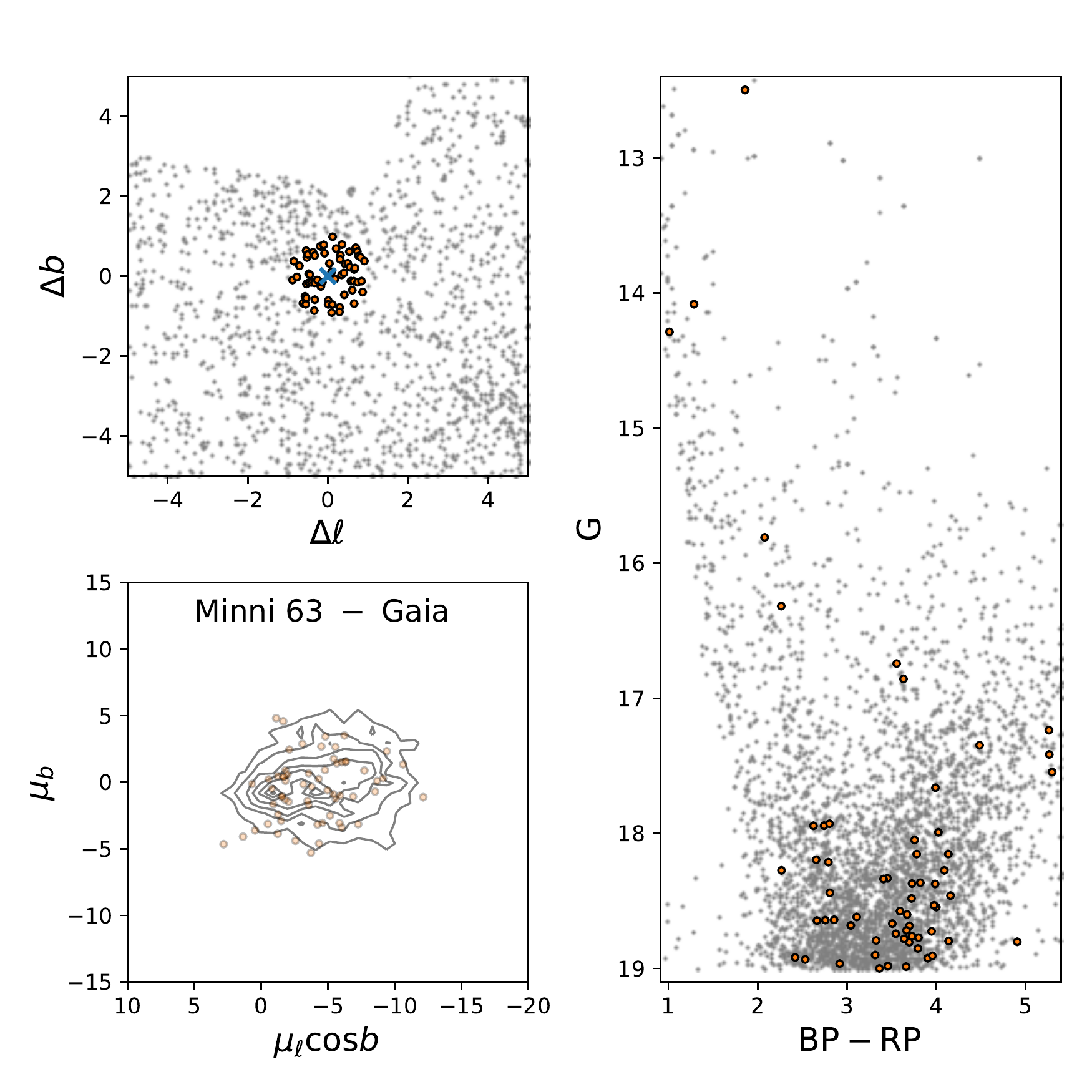} &  
\includegraphics[width=8cm]{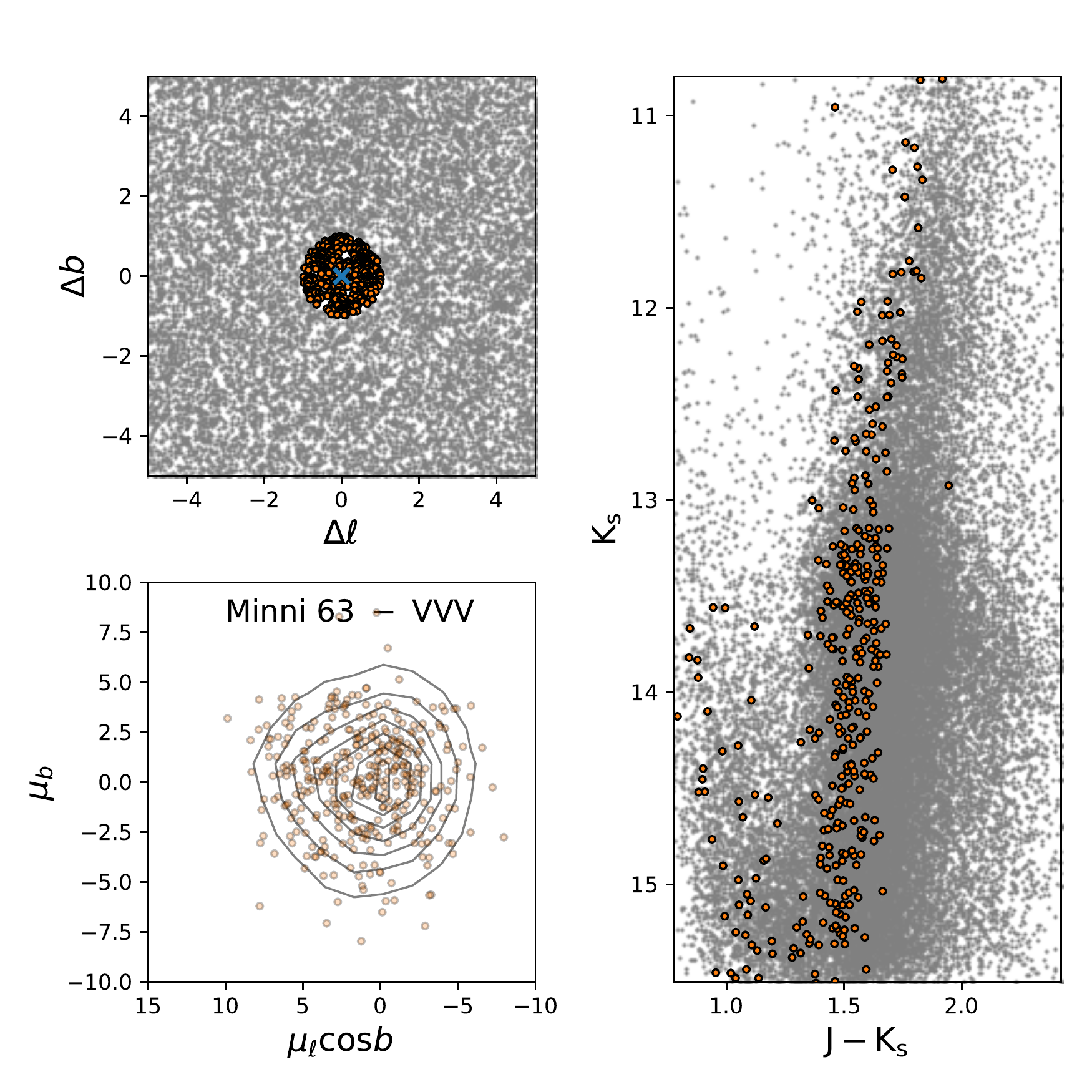} \\ 
\includegraphics[width=8cm]{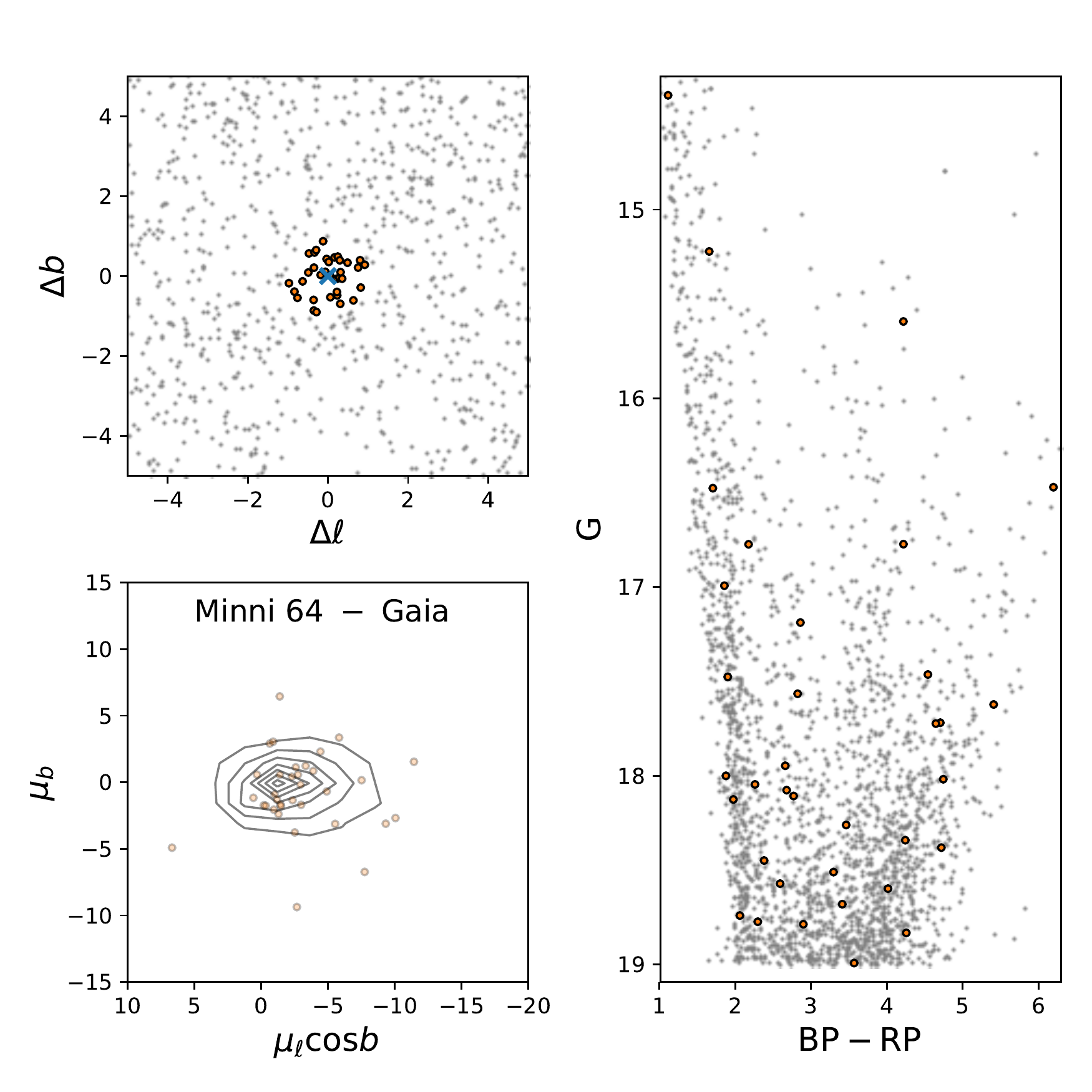} & 
\includegraphics[width=8cm]{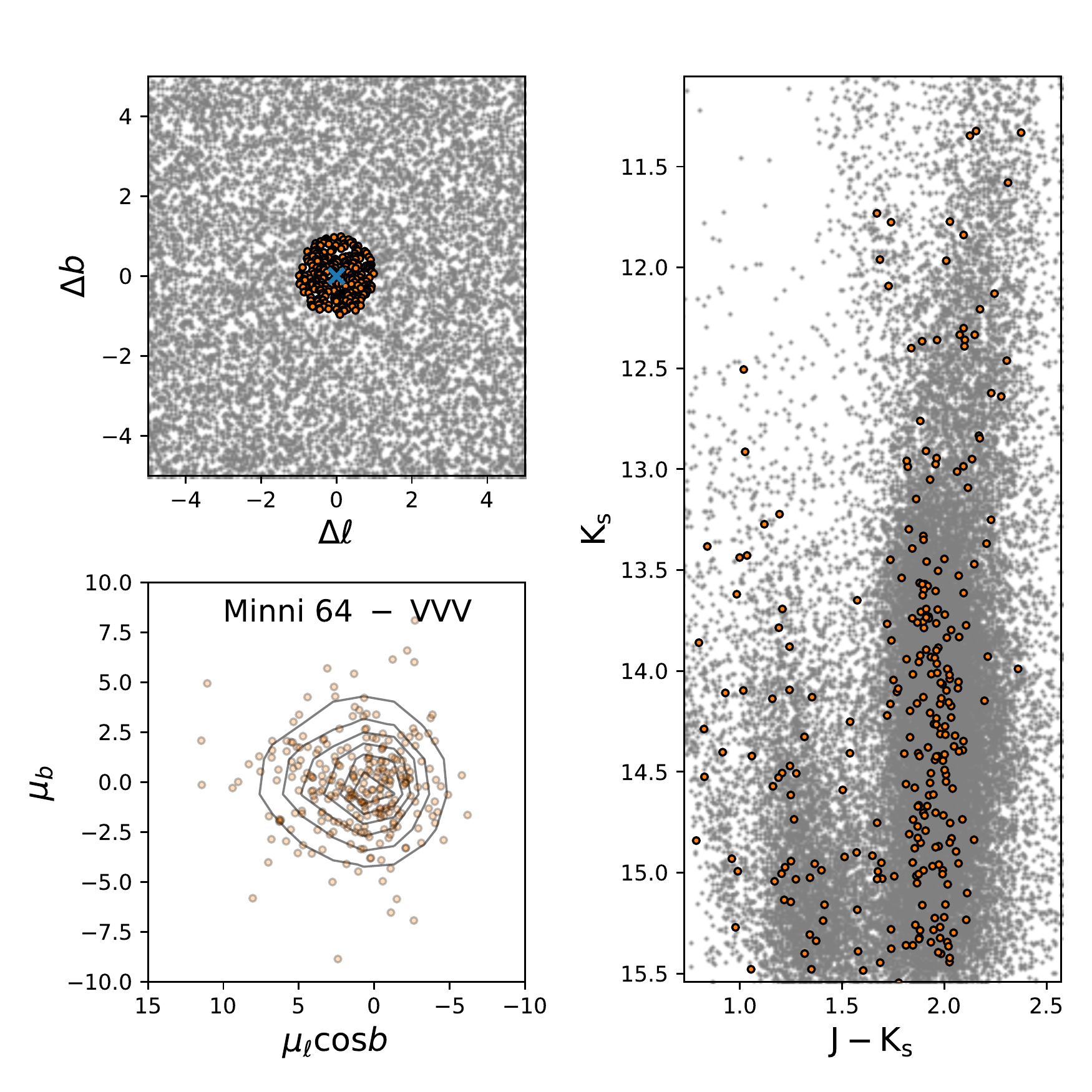} \\  
\end{tabular}
\end{table*}
\newpage
\begin{table*}
\begin{tabular}{cc}
\includegraphics[width=8cm]{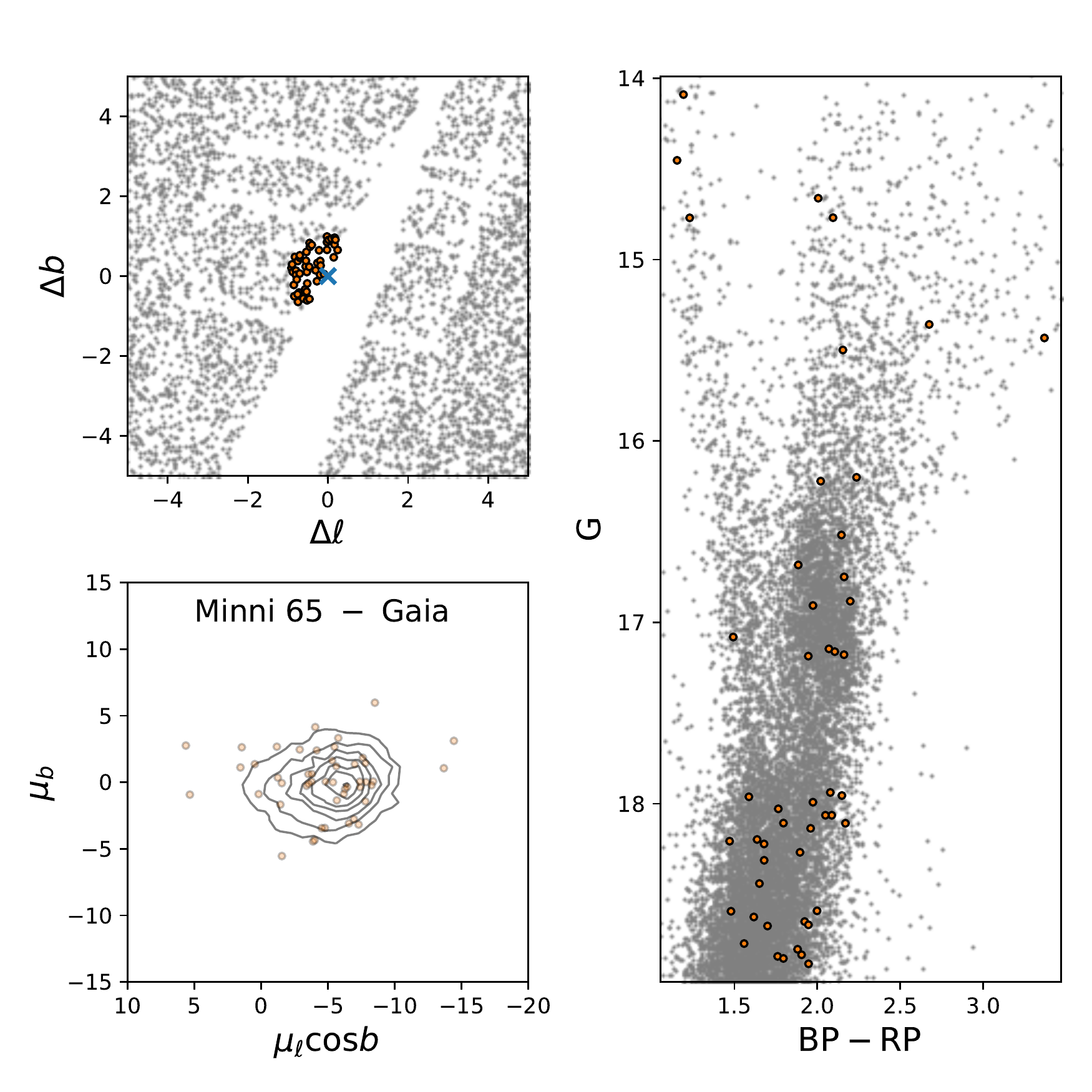} & \\
\includegraphics[width=8cm]{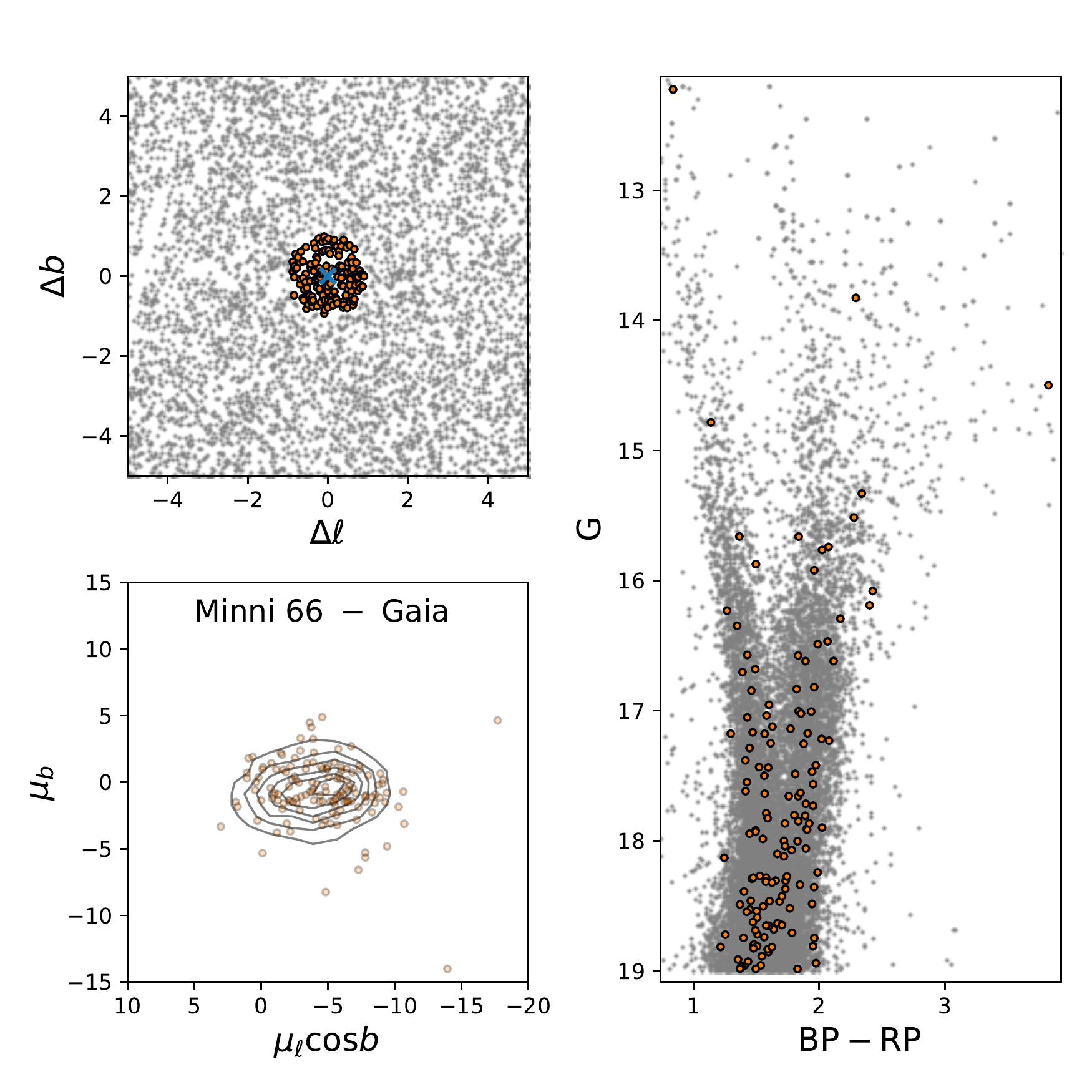} & \\
\includegraphics[width=8cm]{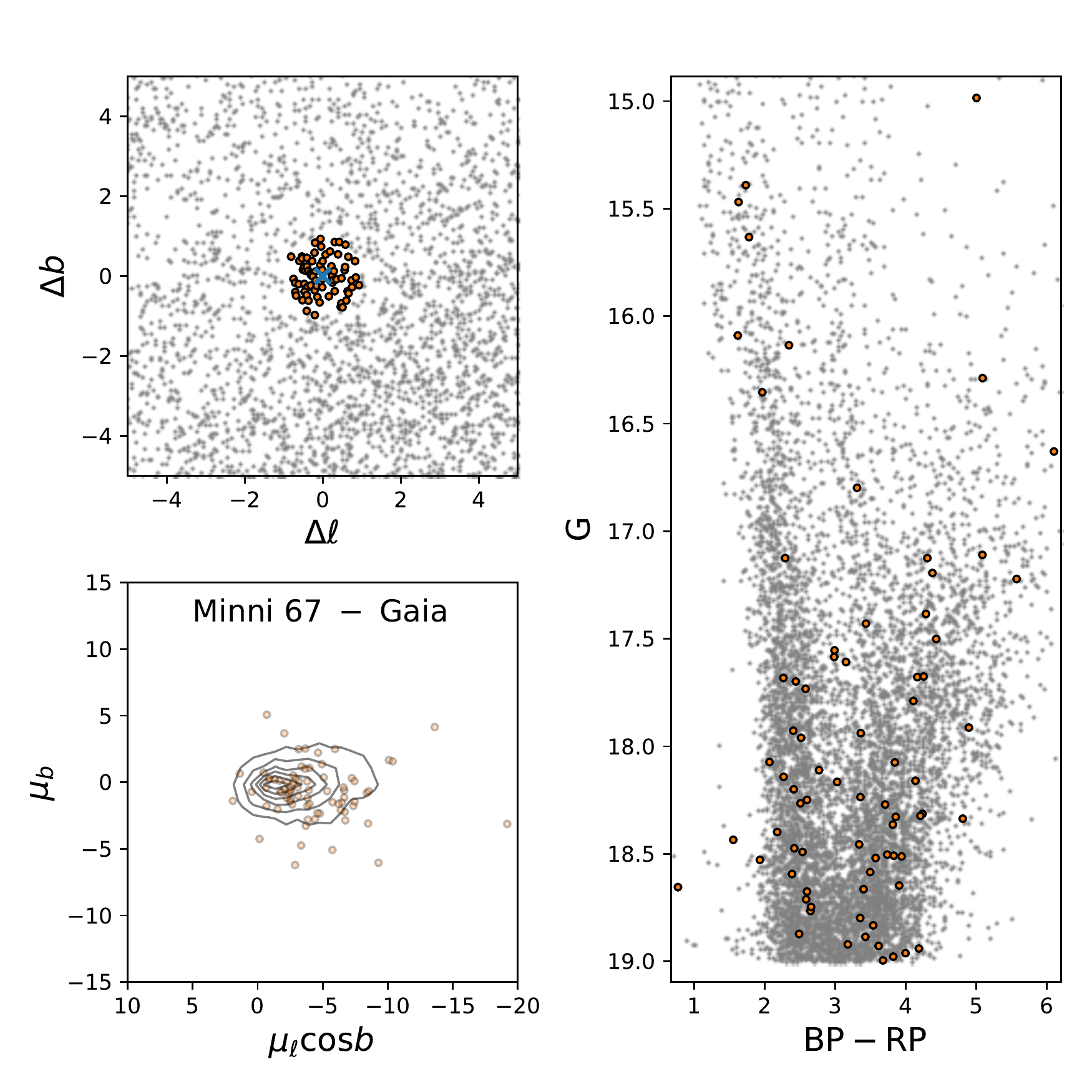} & 
\includegraphics[width=8cm]{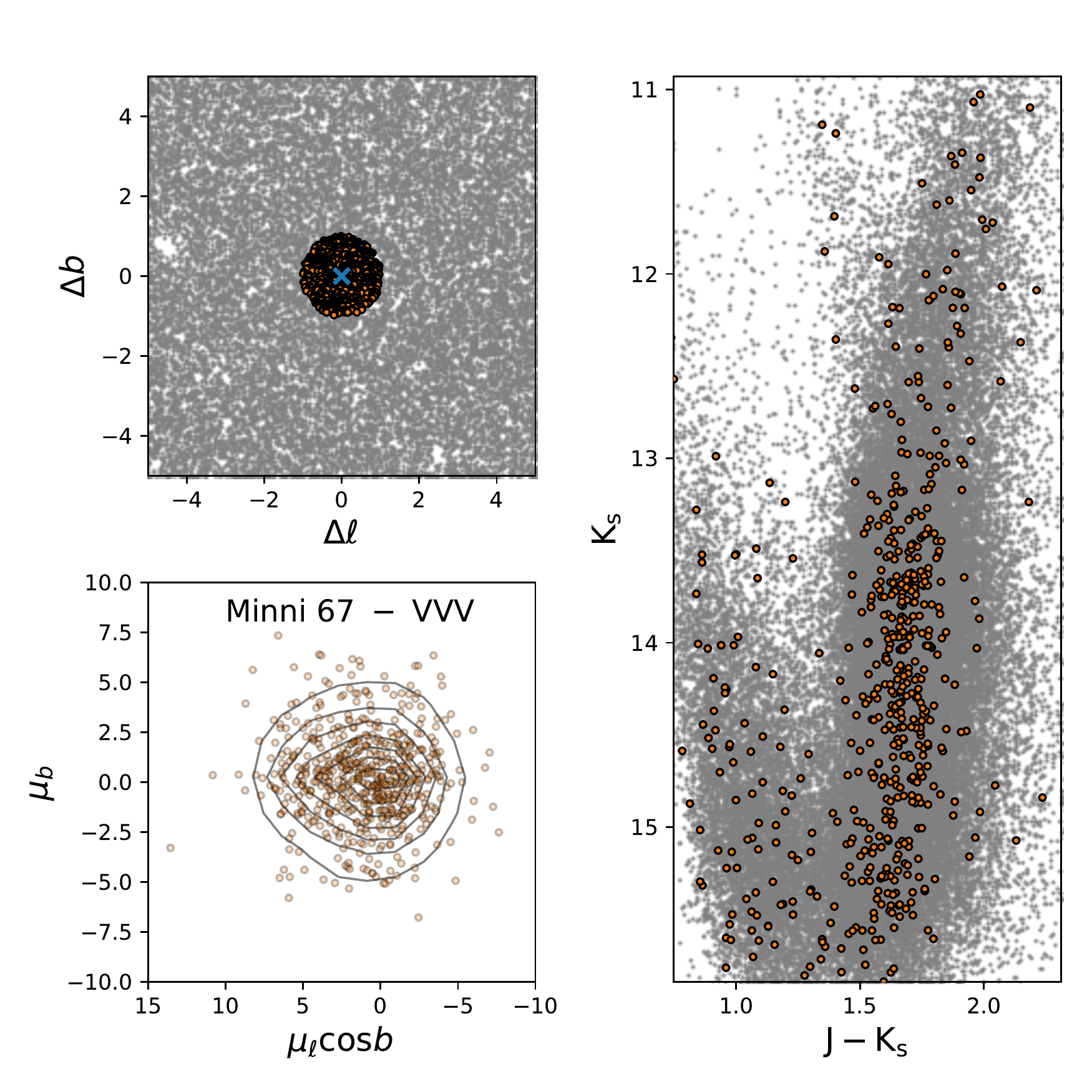} \\
\end{tabular}
\end{table*}
\newpage
\begin{table*}
\begin{tabular}{cc}
\includegraphics[width=8cm]{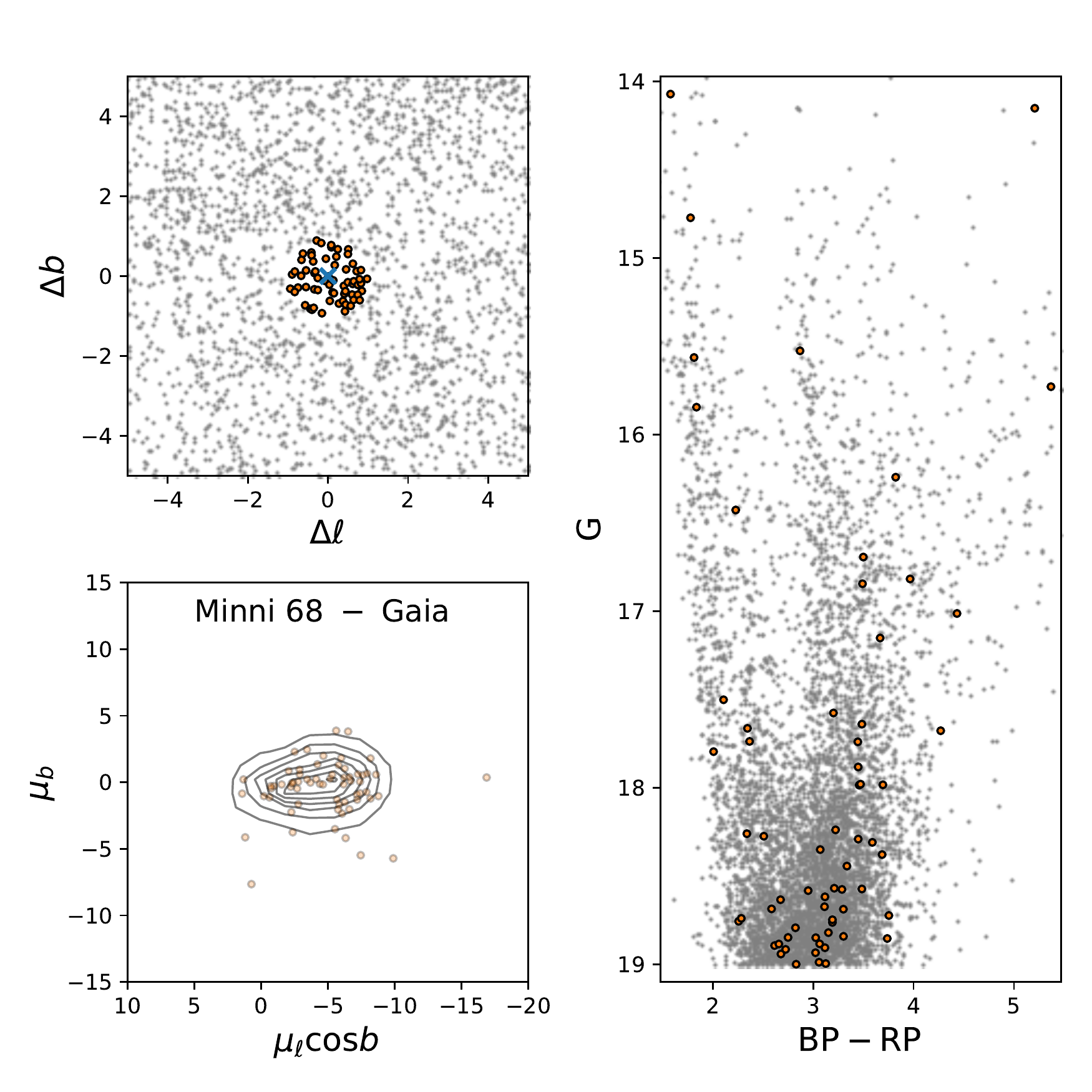} & 
\includegraphics[width=8cm]{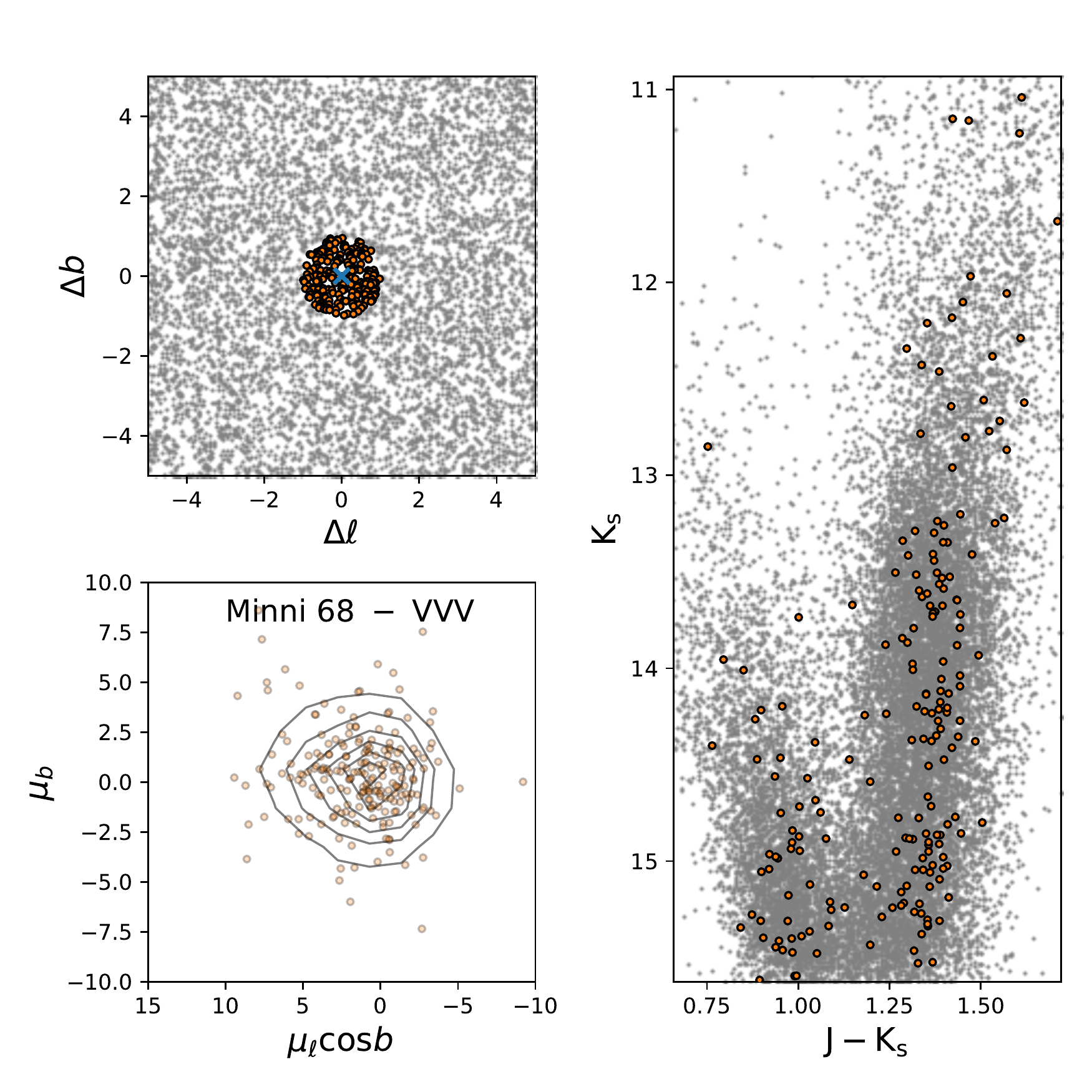} \\  
\includegraphics[width=8cm]{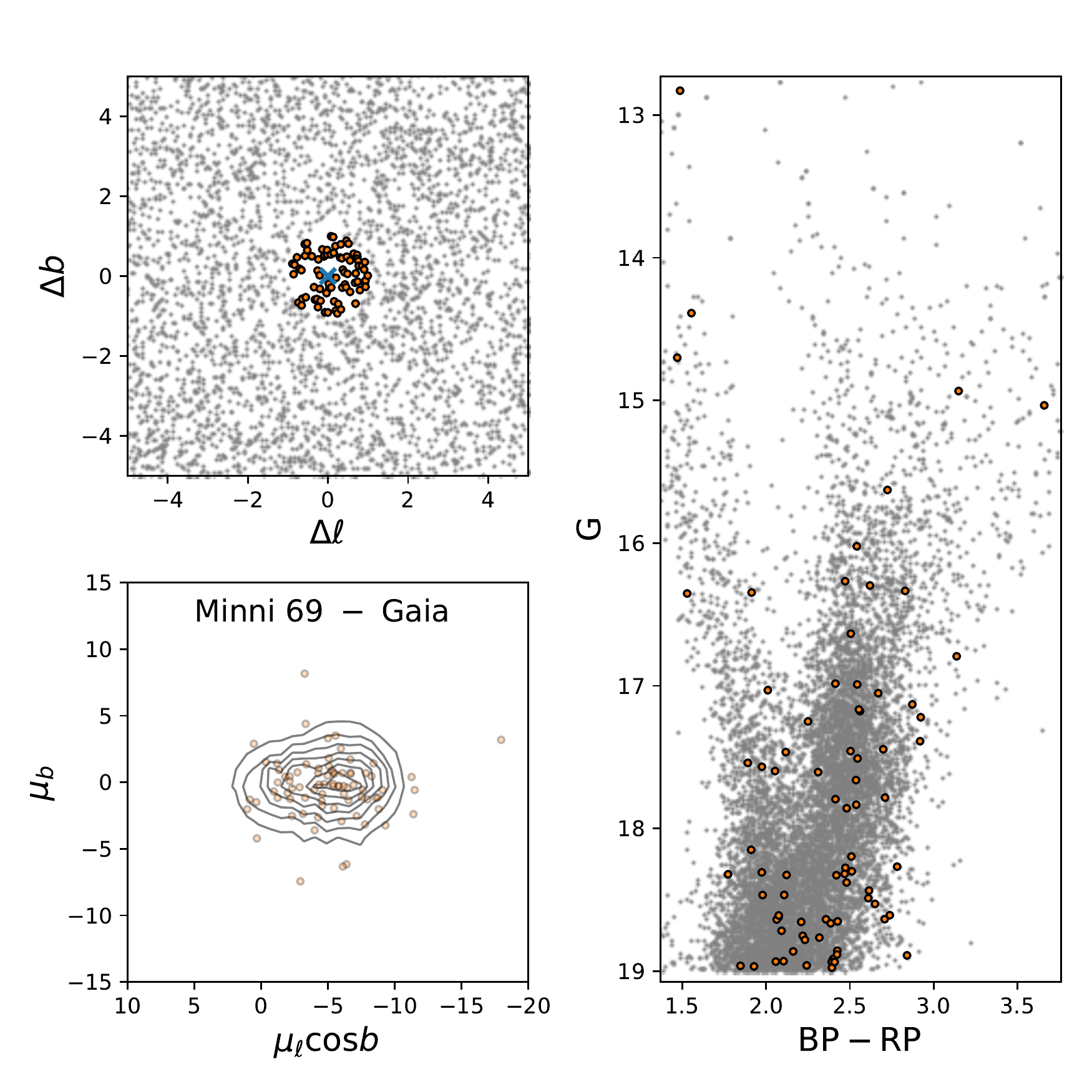} & \\
\includegraphics[width=8cm]{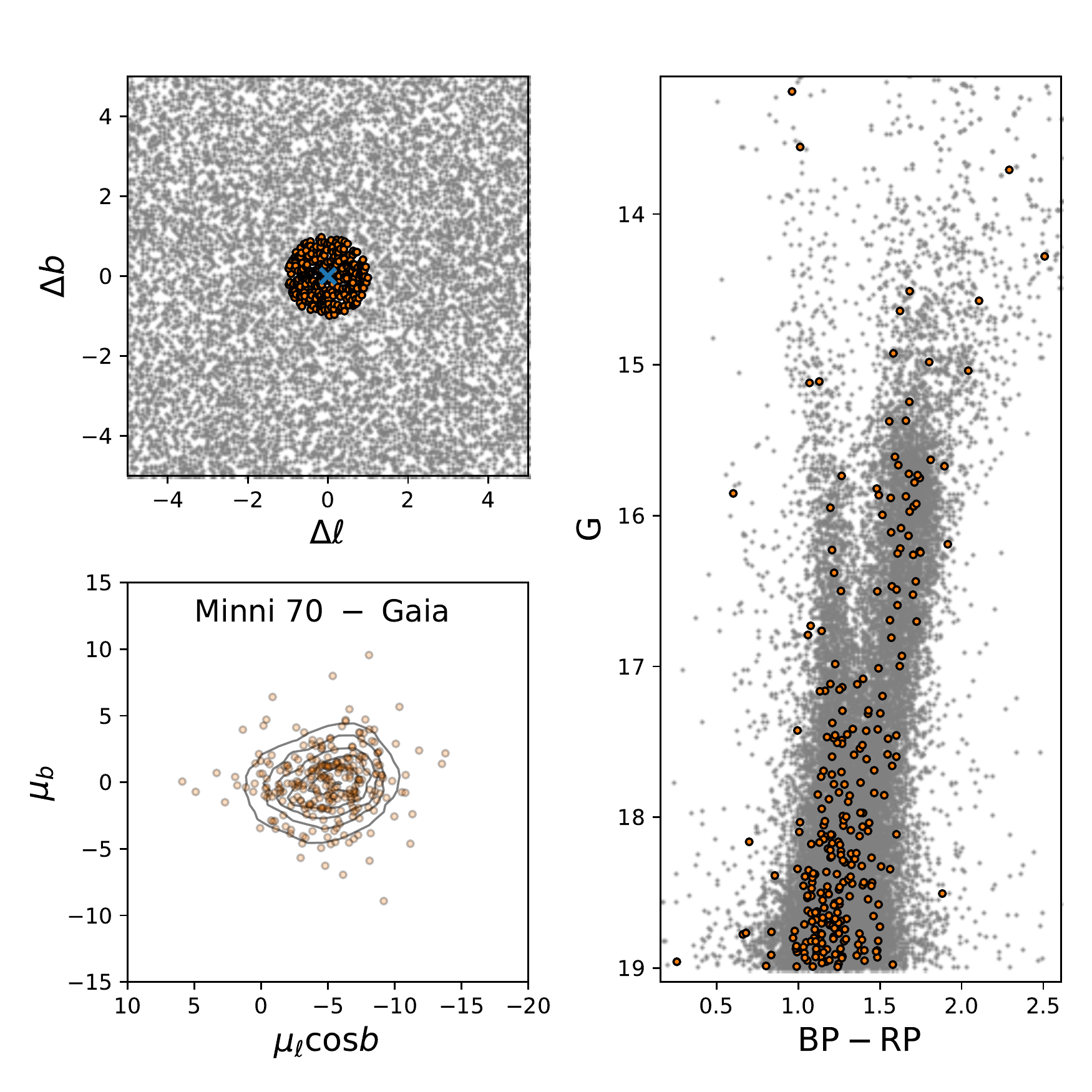} &  \\
\end{tabular}
\end{table*}
\newpage
\begin{table*}
\begin{tabular}{cc}
\includegraphics[width=8cm]{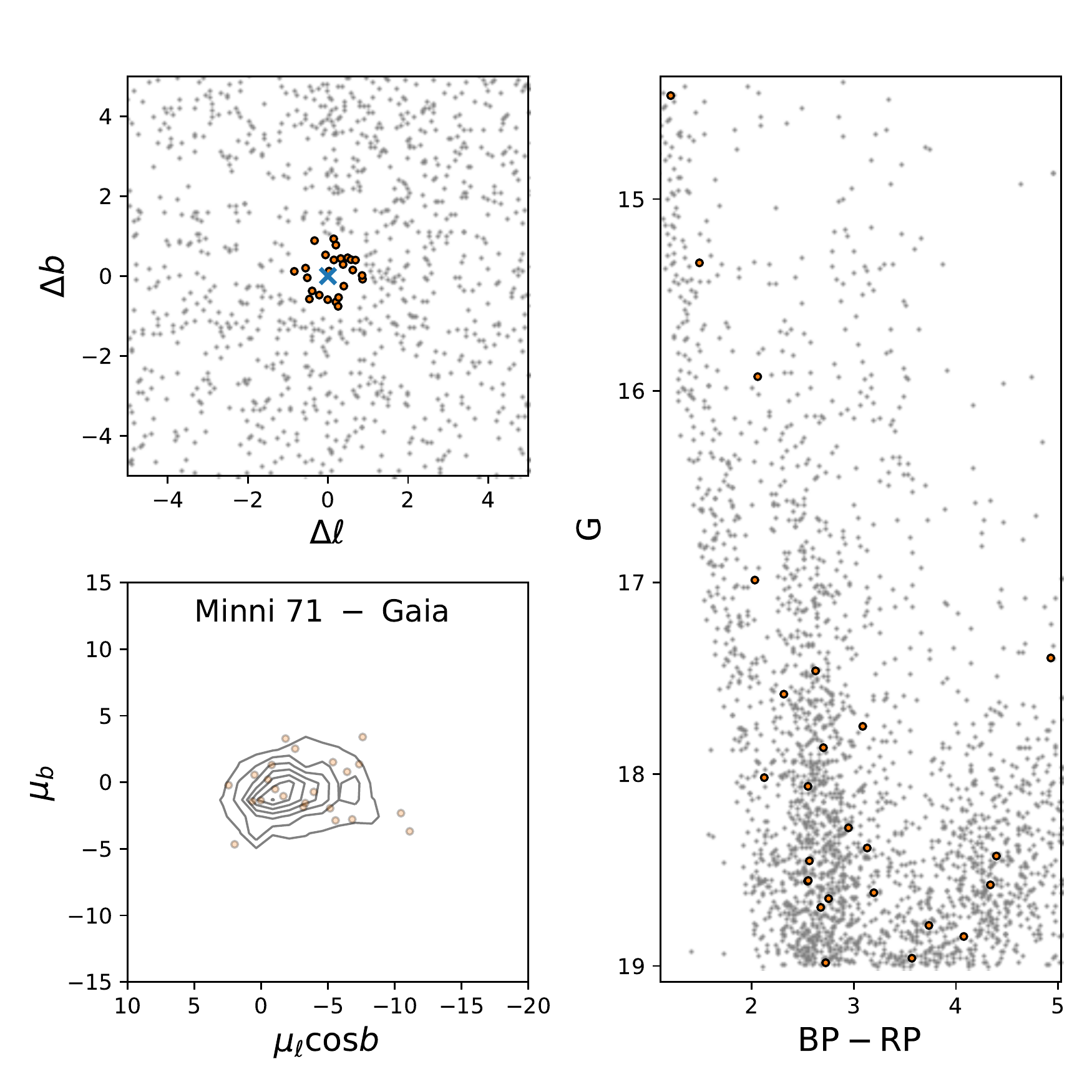} &  
\includegraphics[width=8cm]{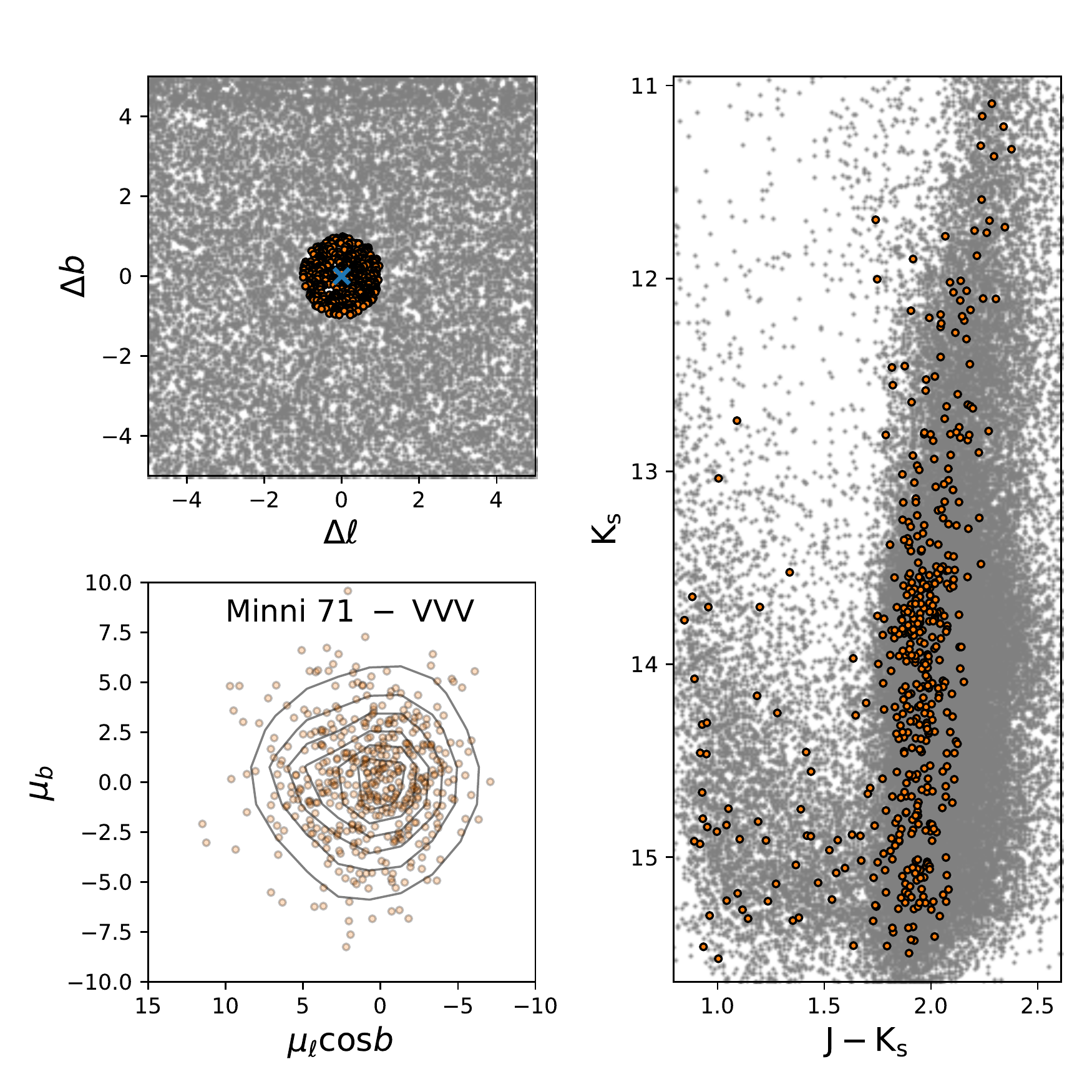} \\ 
\includegraphics[width=8cm]{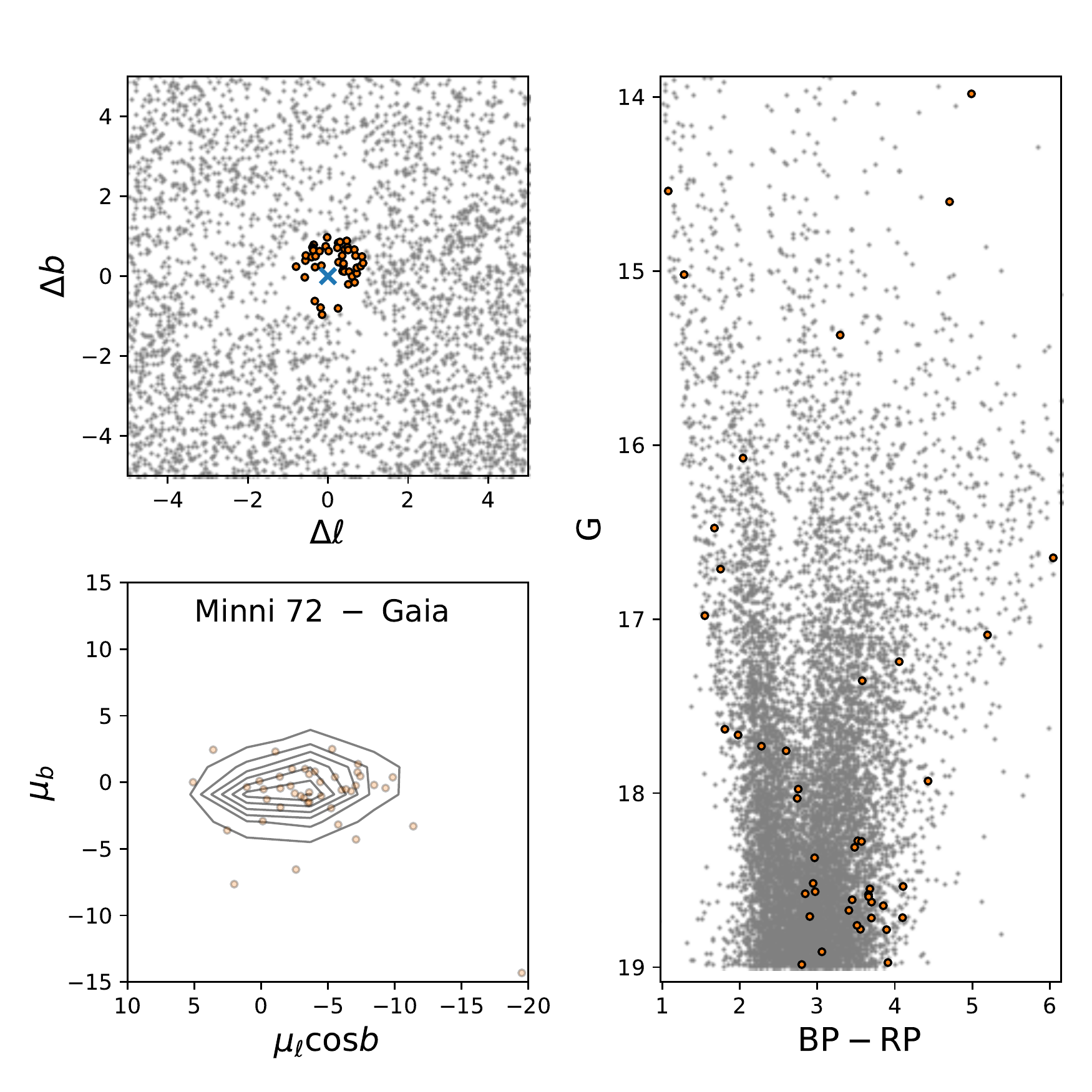} & 
\includegraphics[width=8cm]{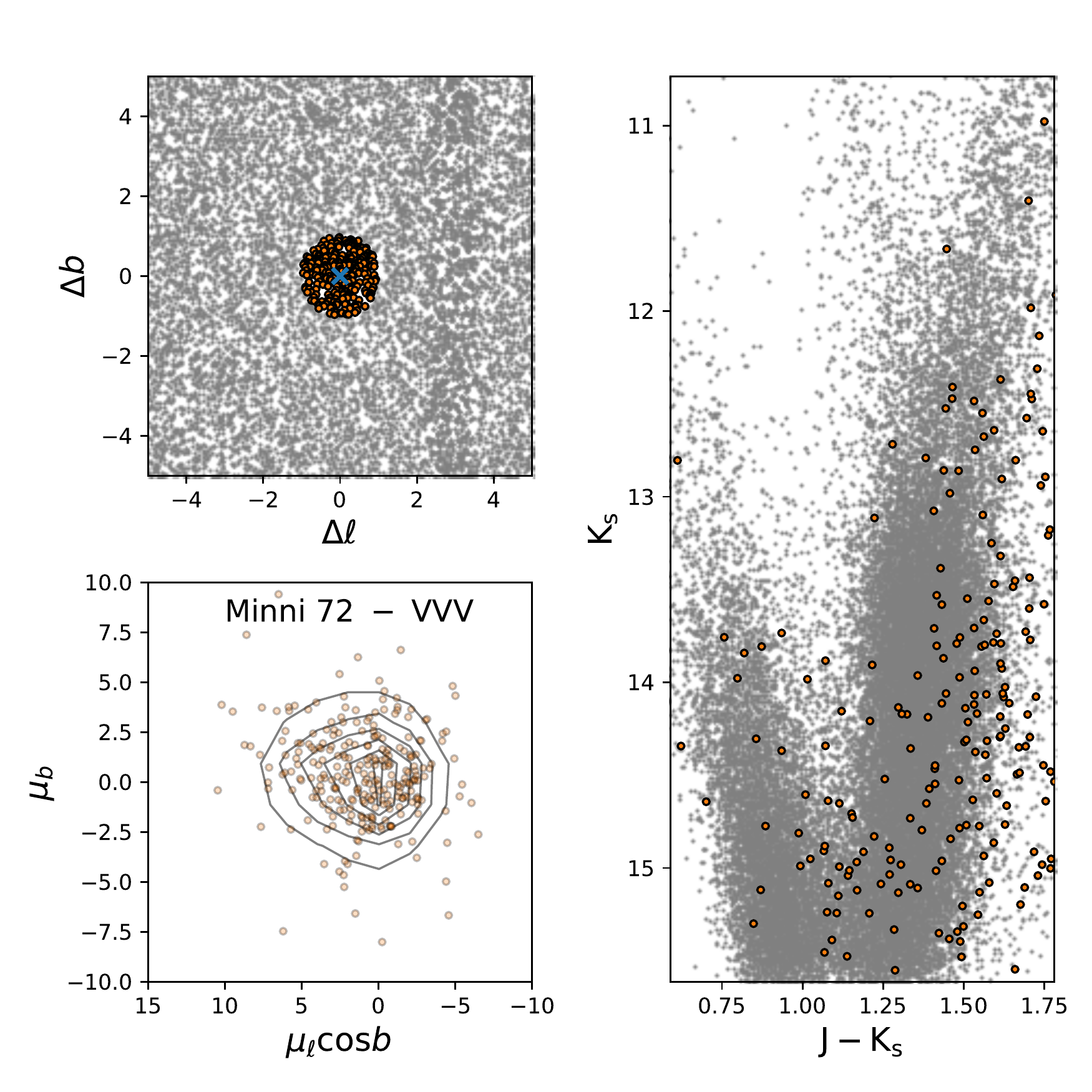} \\  
\includegraphics[width=8cm]{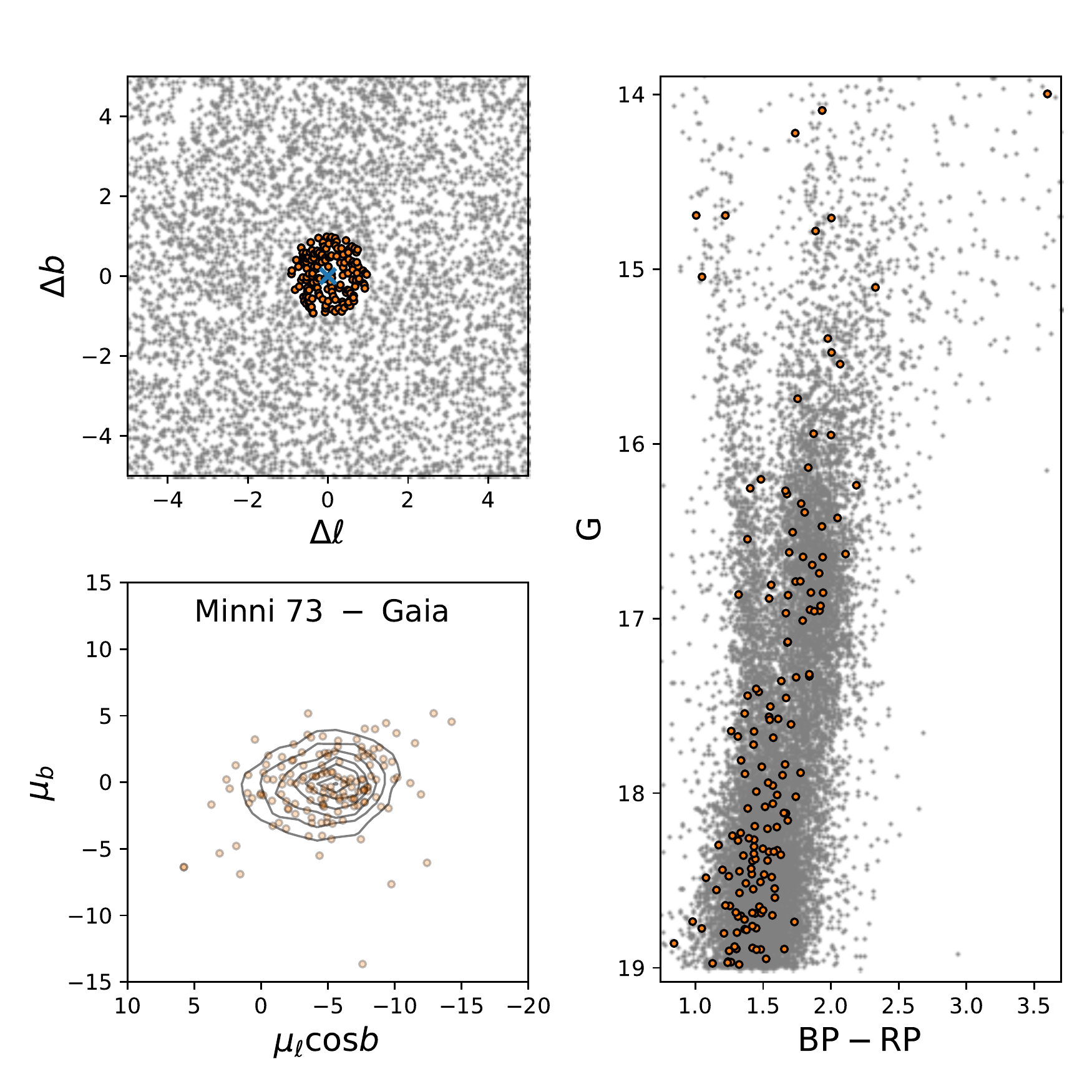} & \\
\end{tabular}
\end{table*}
\newpage
\begin{table*}
\begin{tabular}{cc}
\includegraphics[width=8cm]{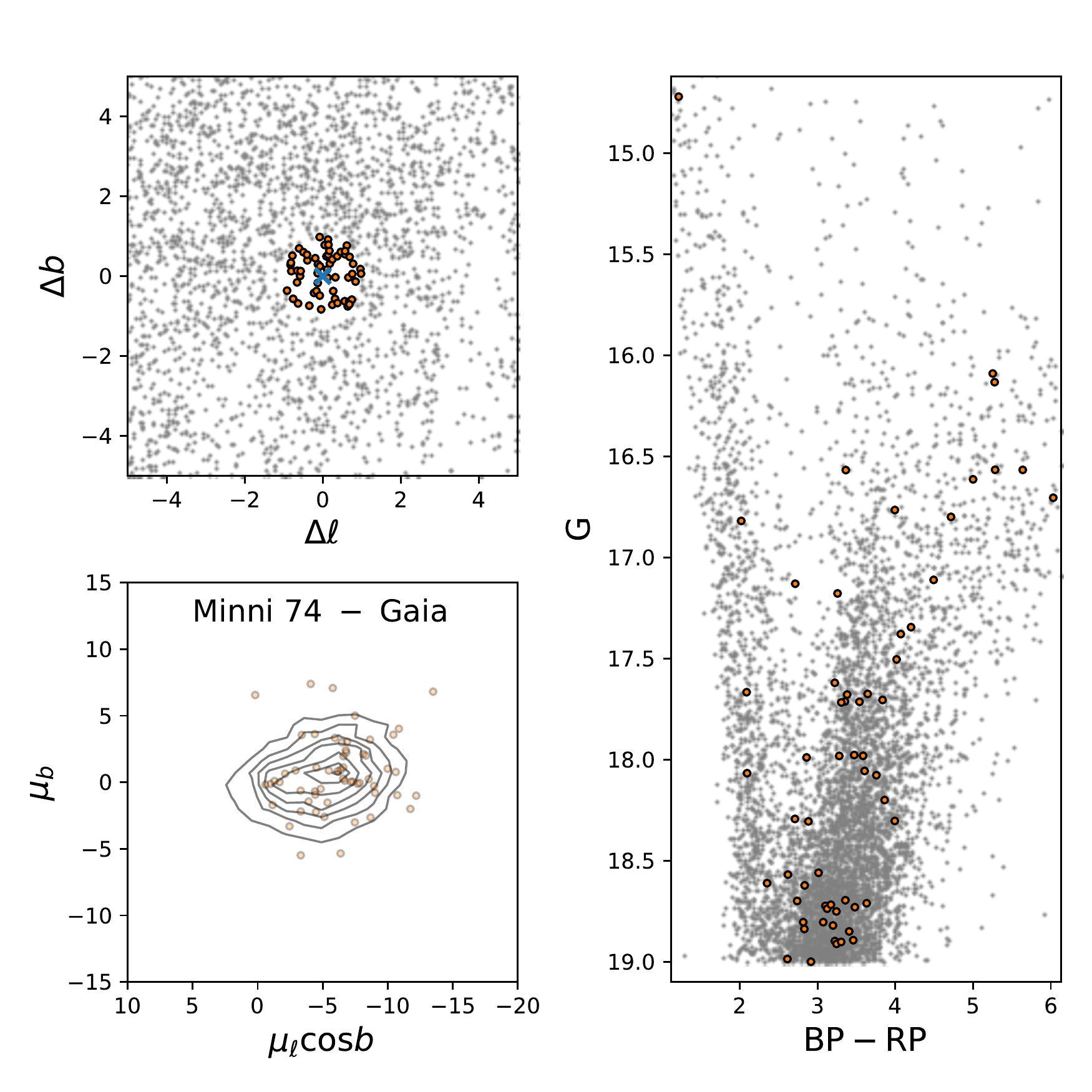} &  
\includegraphics[width=8cm]{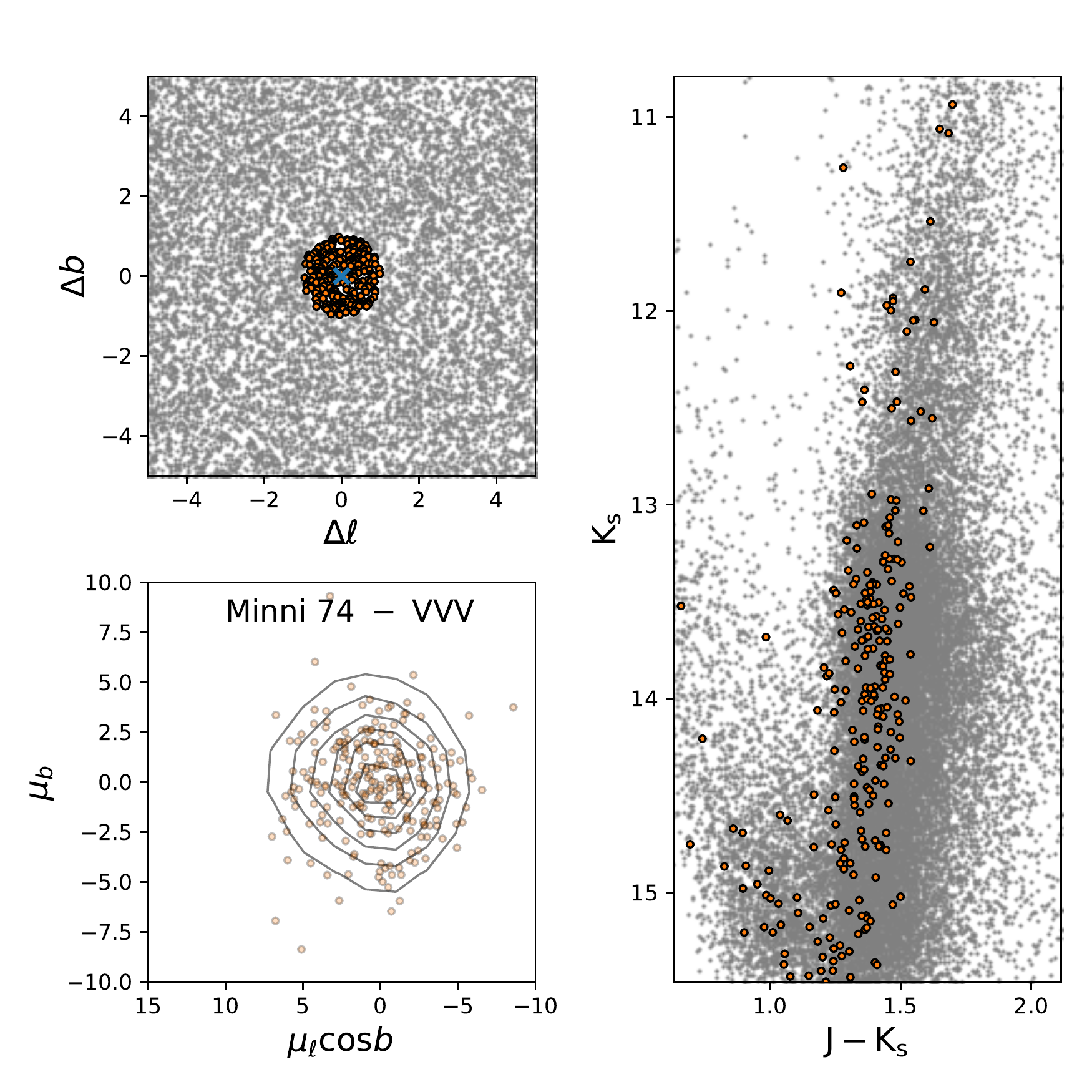} \\
\includegraphics[width=8cm]{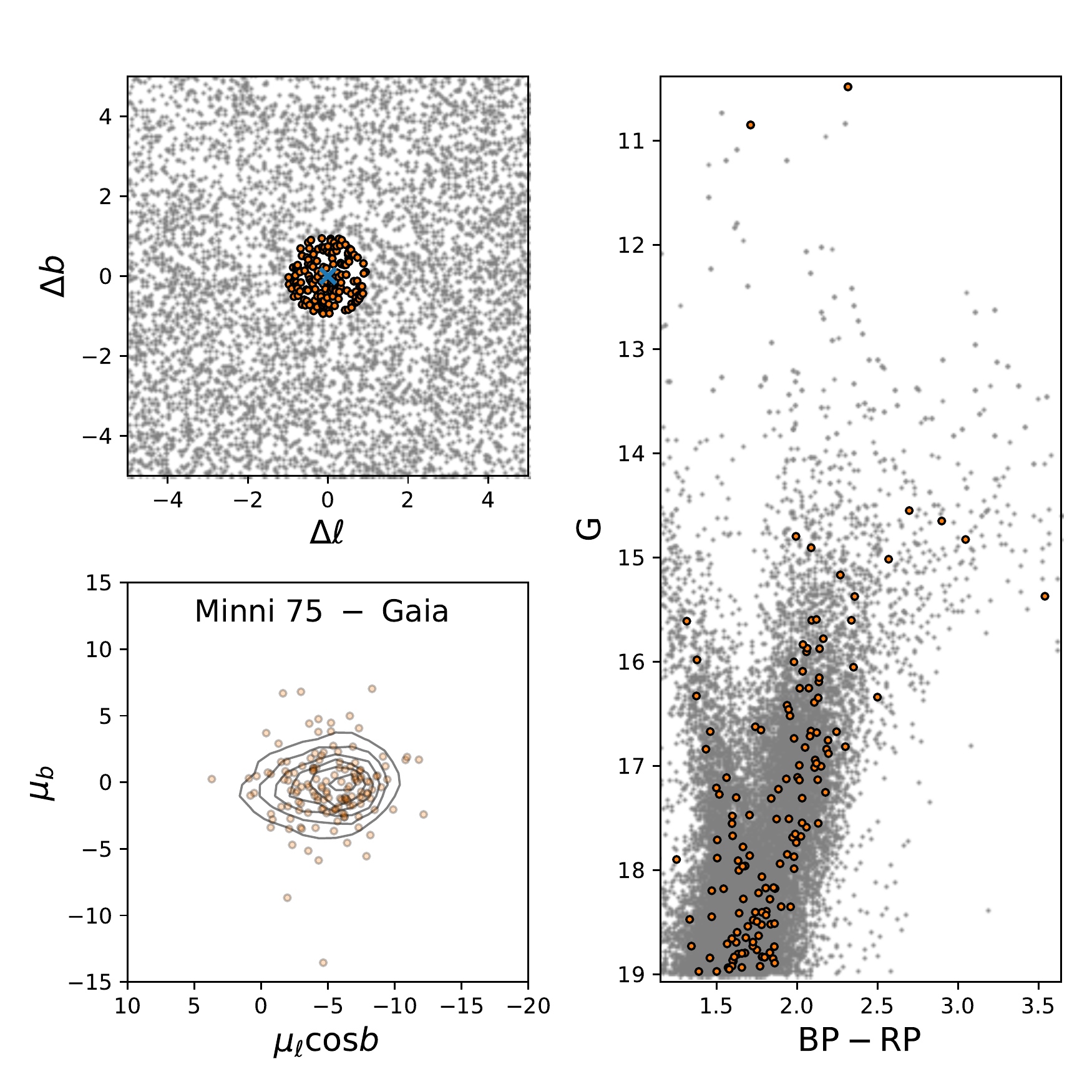} &  \\
\includegraphics[width=8cm]{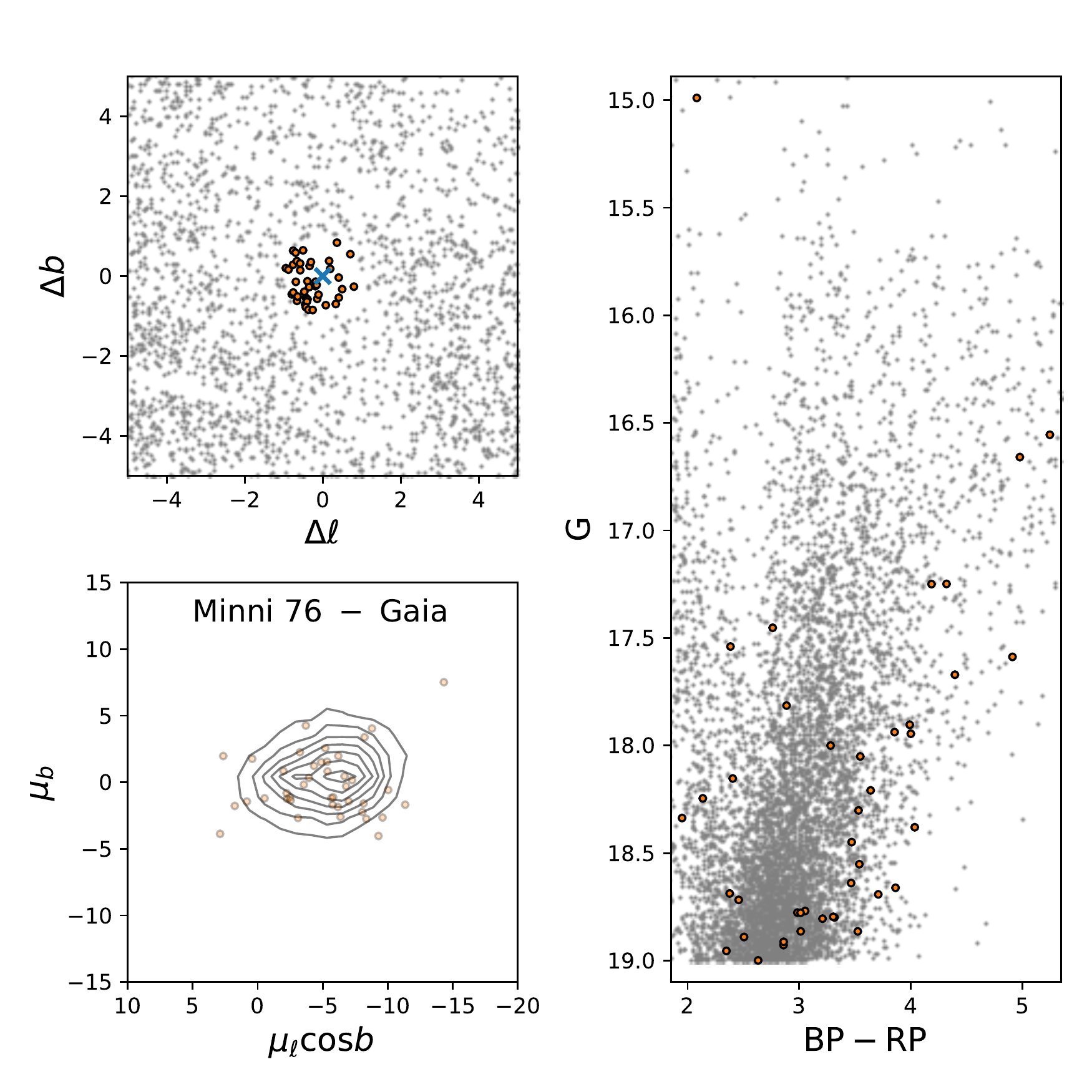} &
\includegraphics[width=8cm]{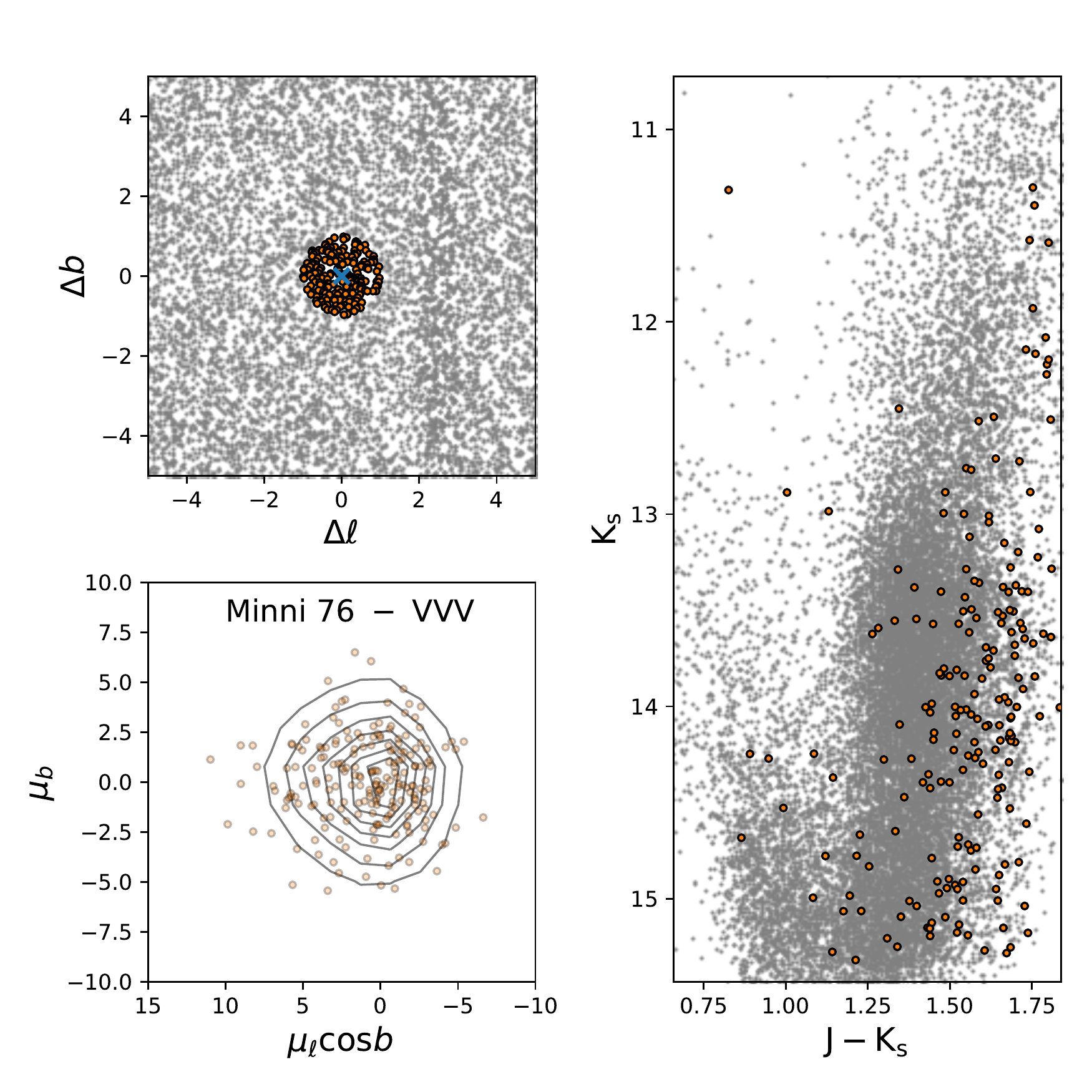} \\  
\end{tabular}
\end{table*}
\newpage
\begin{table*}
\begin{tabular}{cc}
\includegraphics[width=8cm]{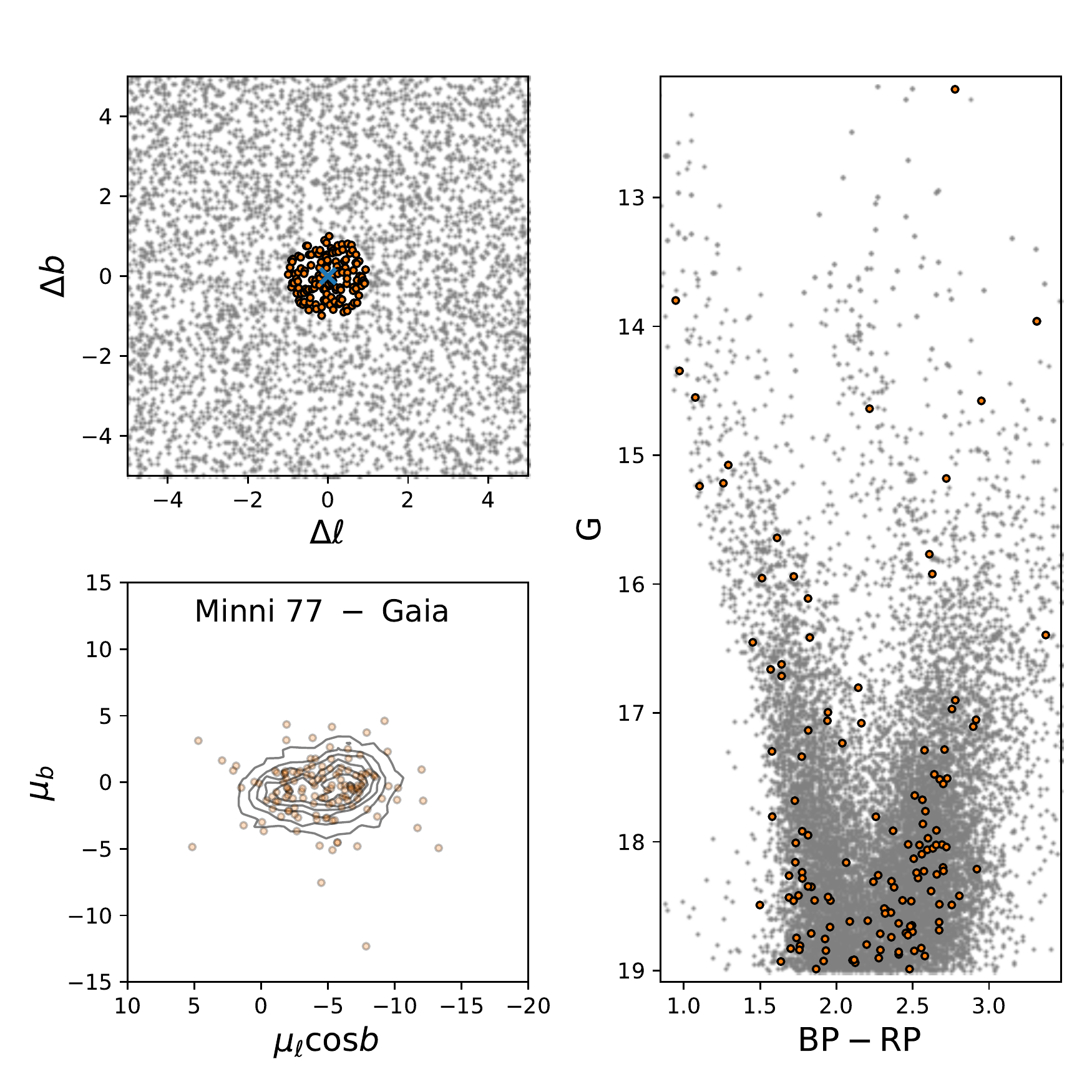} & \\
\includegraphics[width=8cm]{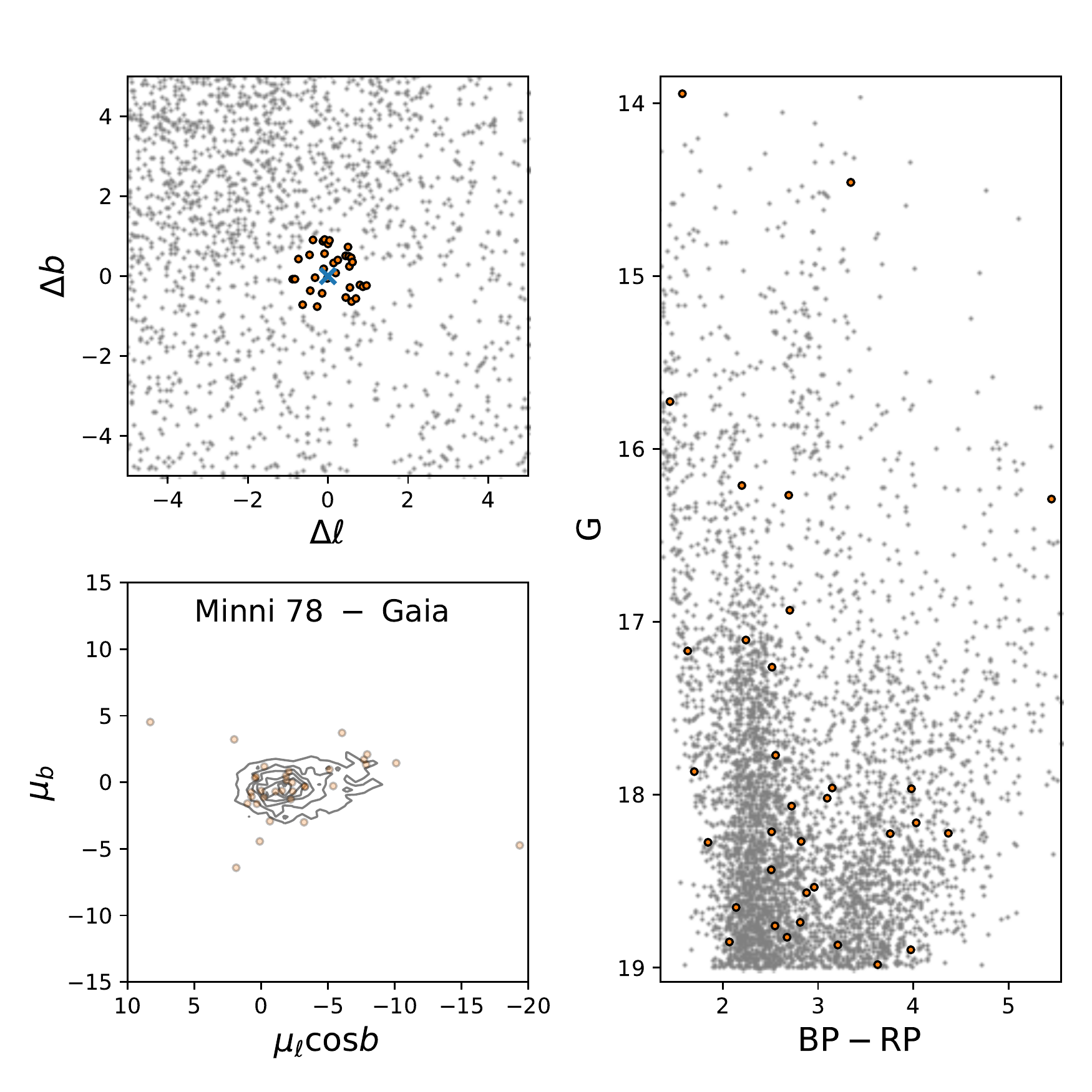} &  
\includegraphics[width=8cm]{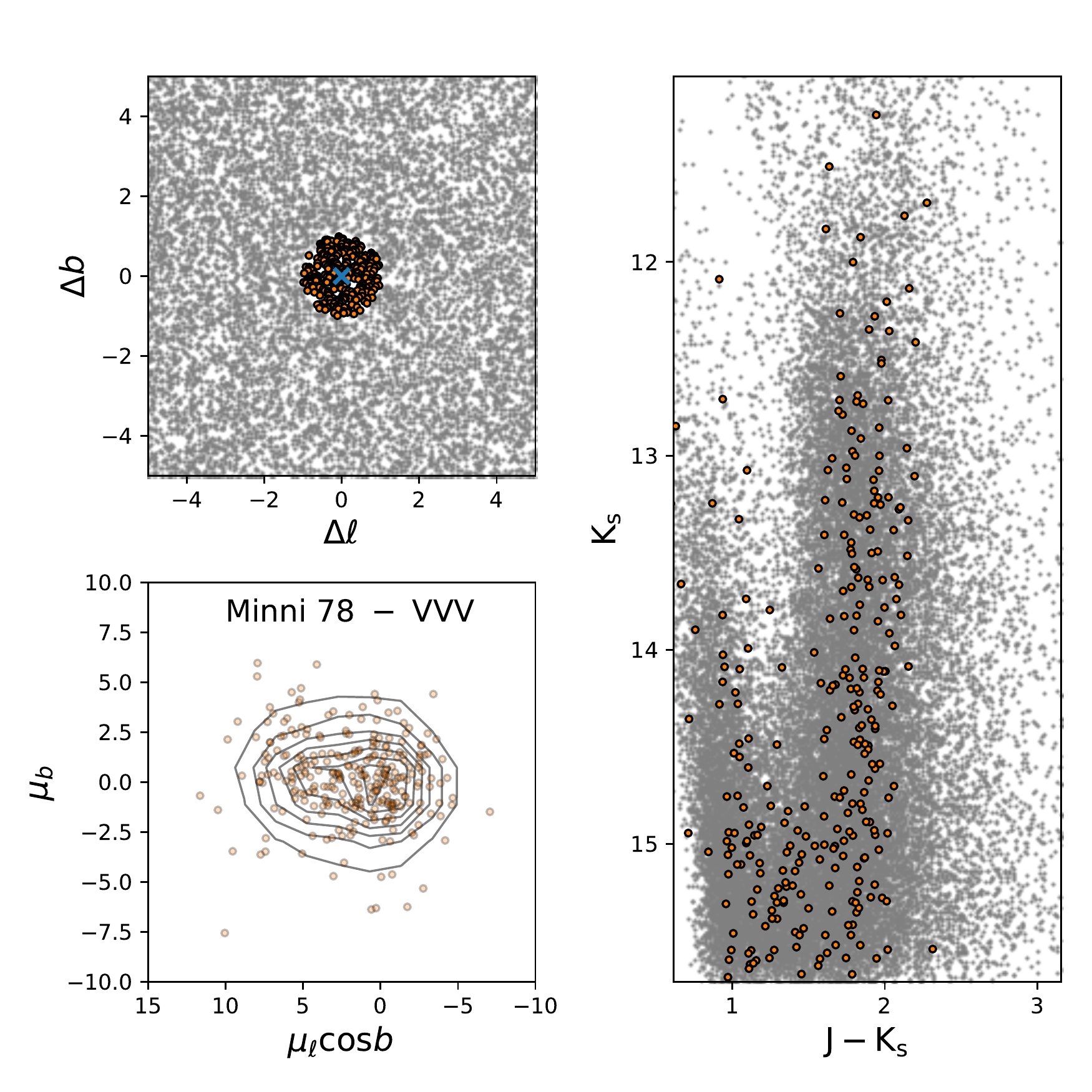} \\ 
\includegraphics[width=8cm]{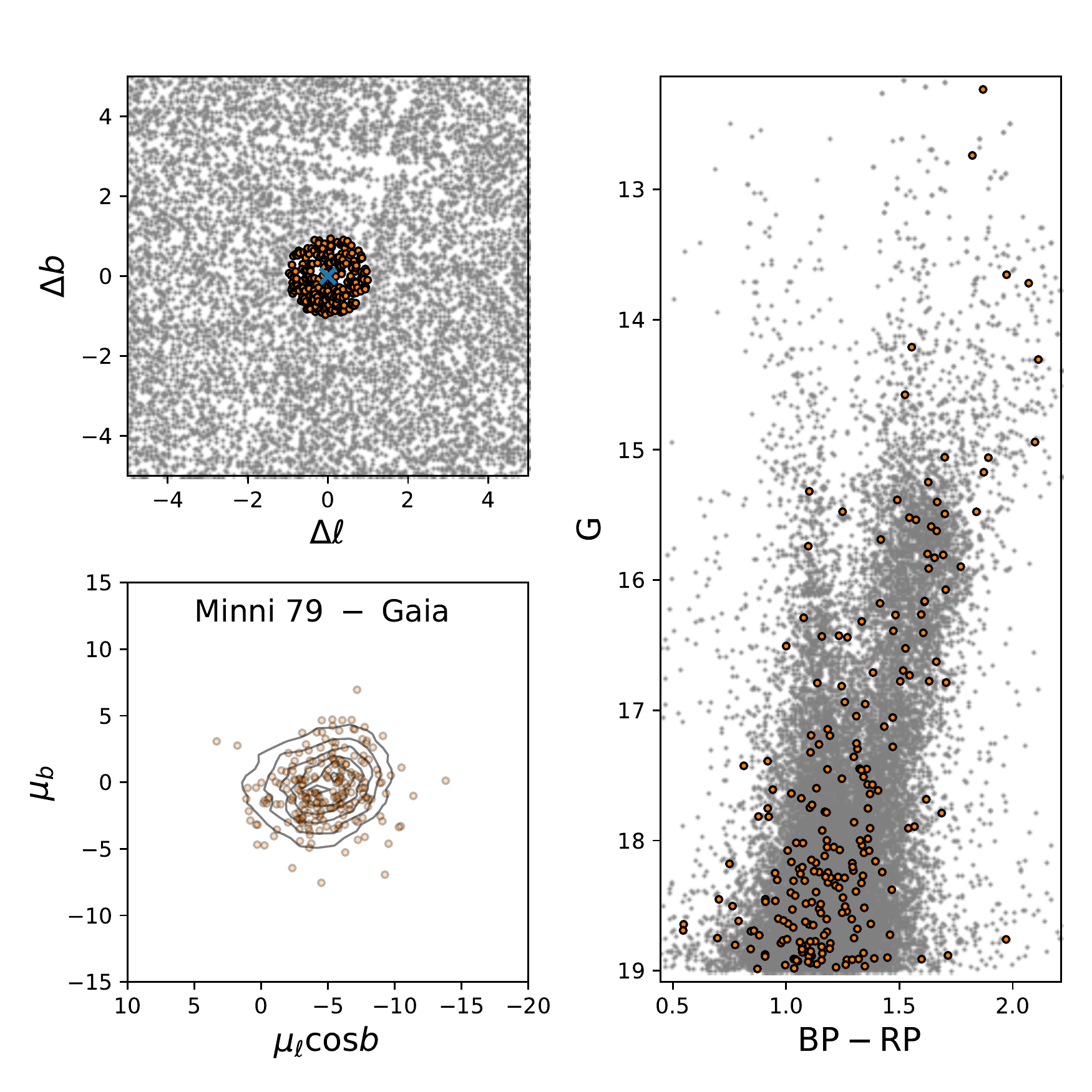} & \\
\end{tabular}
\end{table*}
\newpage
\begin{table*}
\begin{tabular}{cc}
\includegraphics[width=8cm]{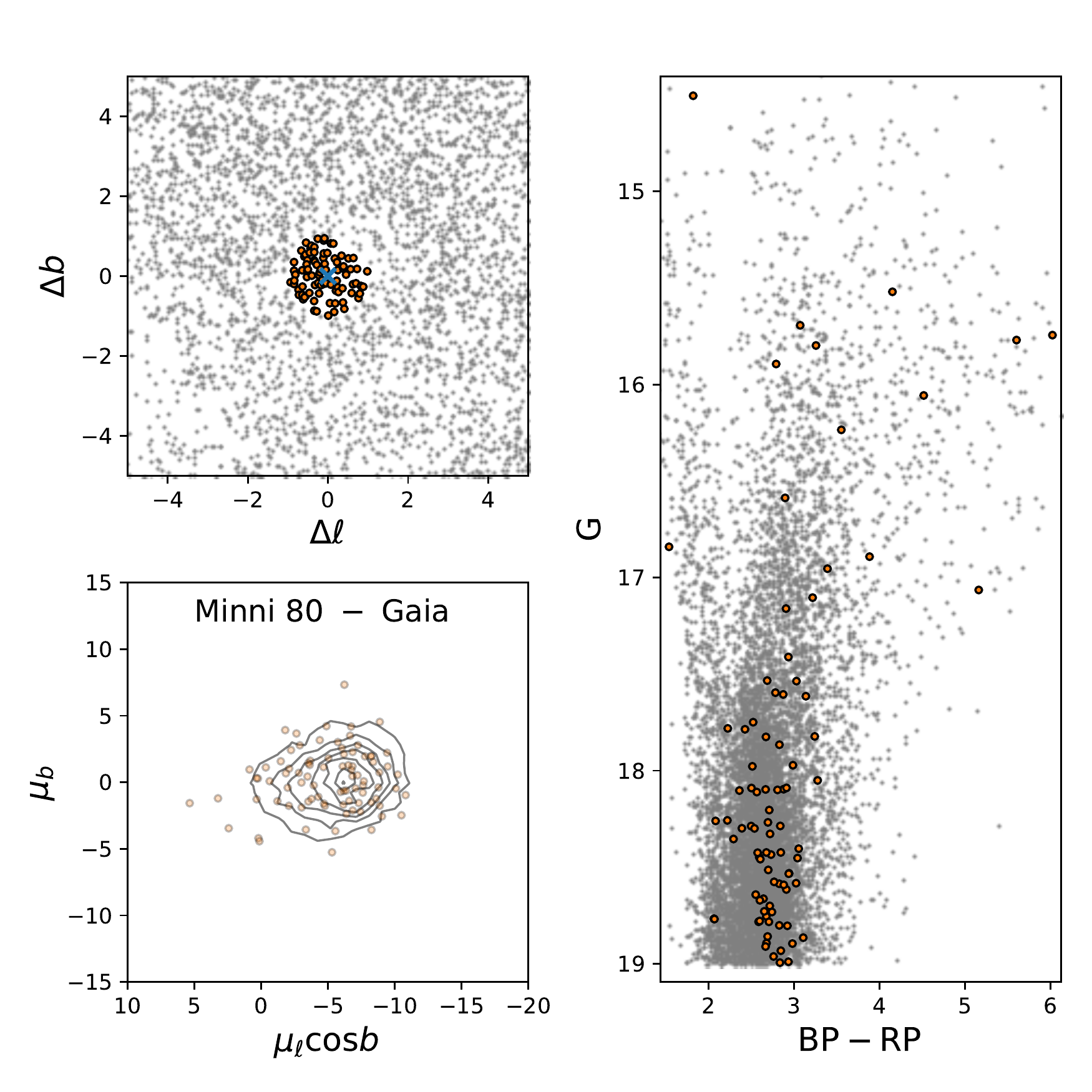} & \\ 
\includegraphics[width=8cm]{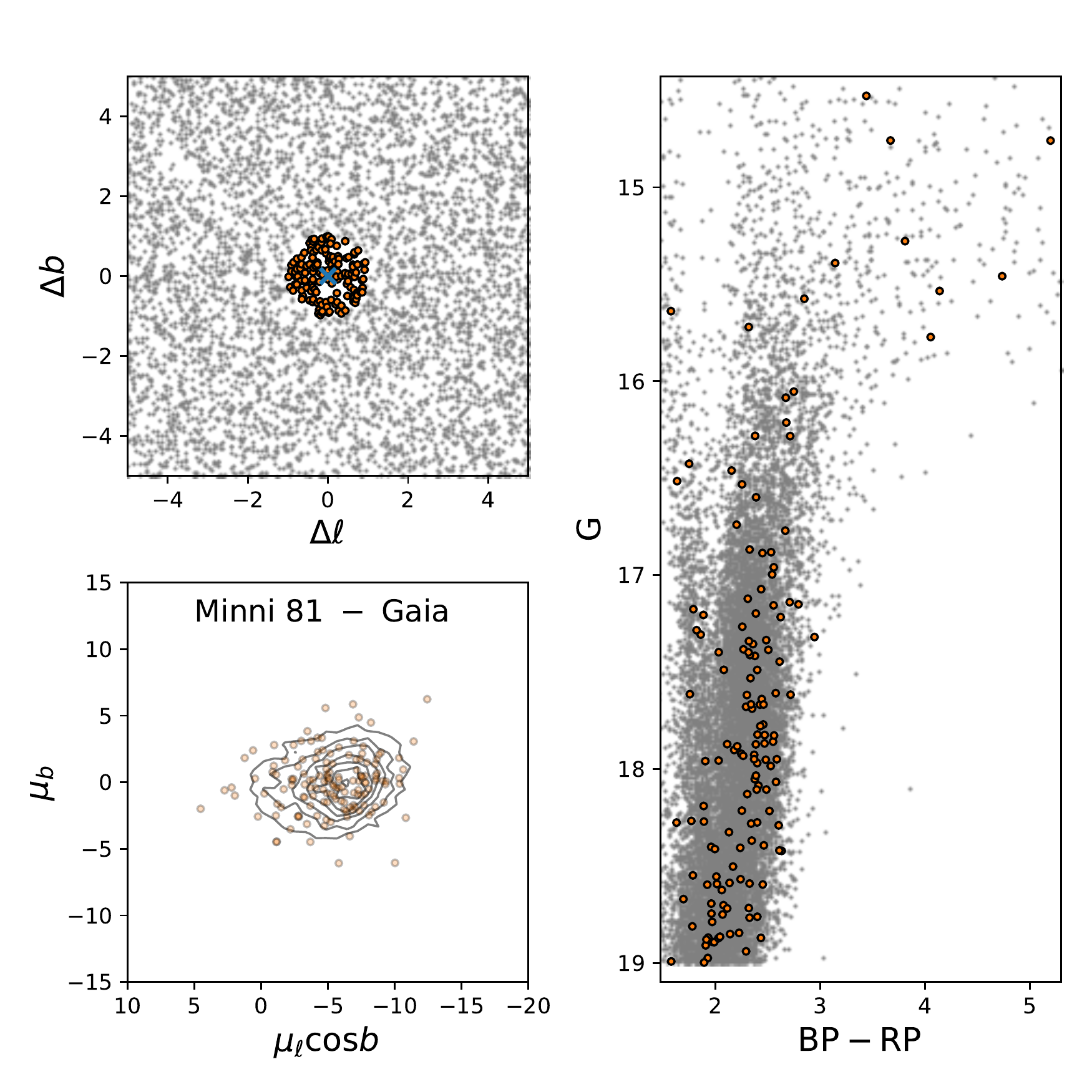} & \\
\includegraphics[width=8cm]{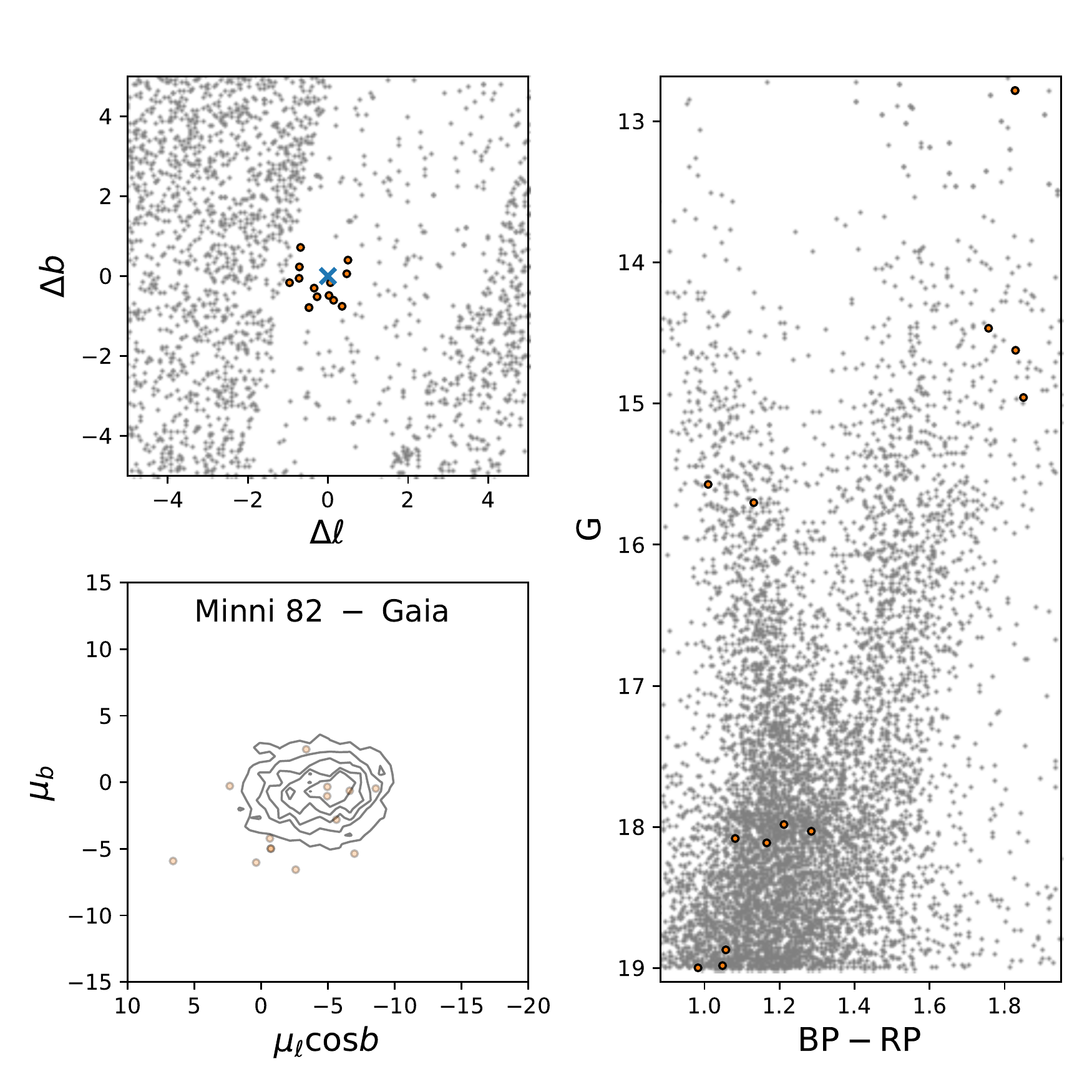} & \\
\end{tabular}
\end{table*}
\newpage
\begin{table*}
\begin{tabular}{cc}
\includegraphics[width=8cm]{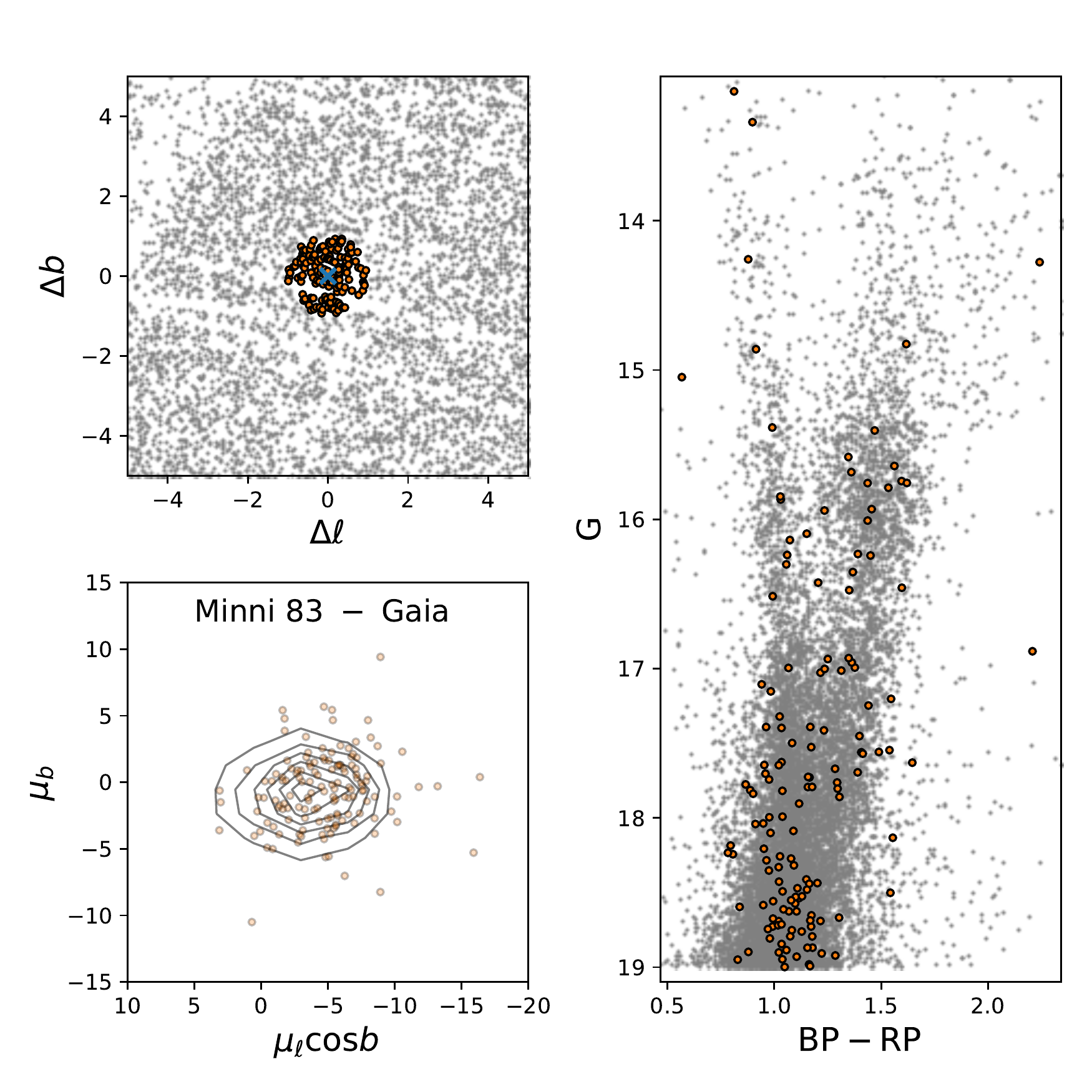} & \\ 
\includegraphics[width=8cm]{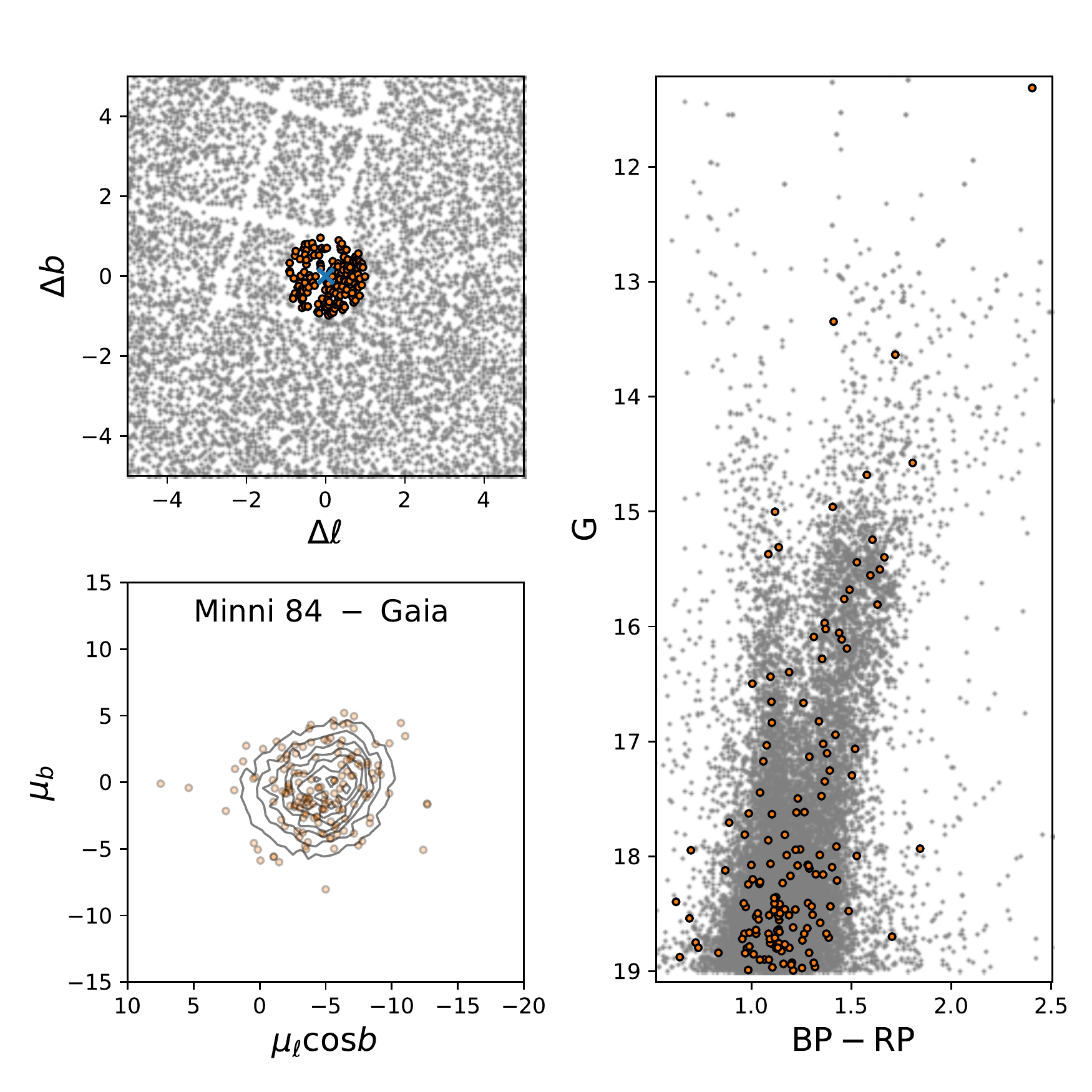} & \\ 
\includegraphics[width=8cm]{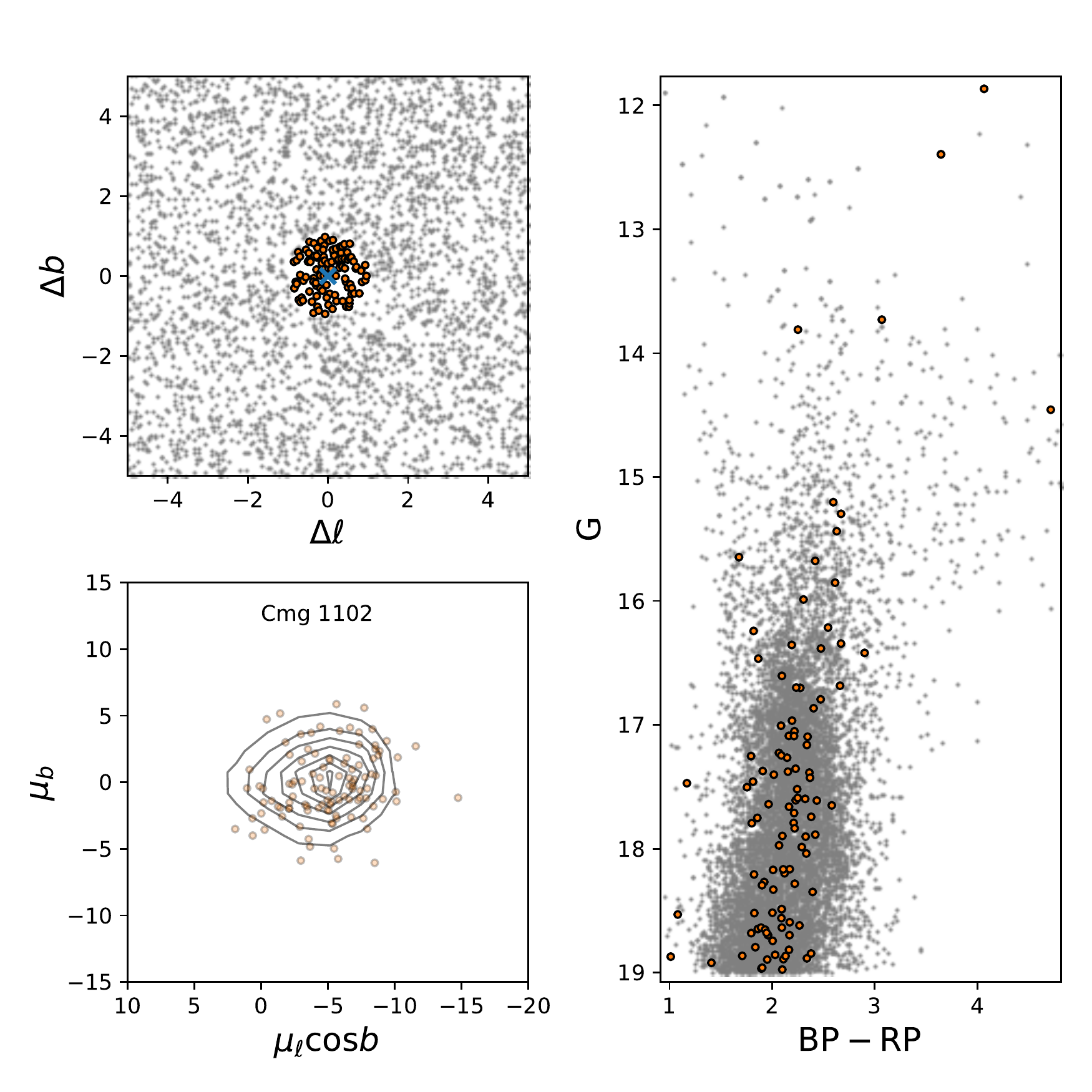} & \\
\end{tabular}
\end{table*}
\newpage
\begin{table*}
\begin{tabular}{cc}
\includegraphics[width=8cm]{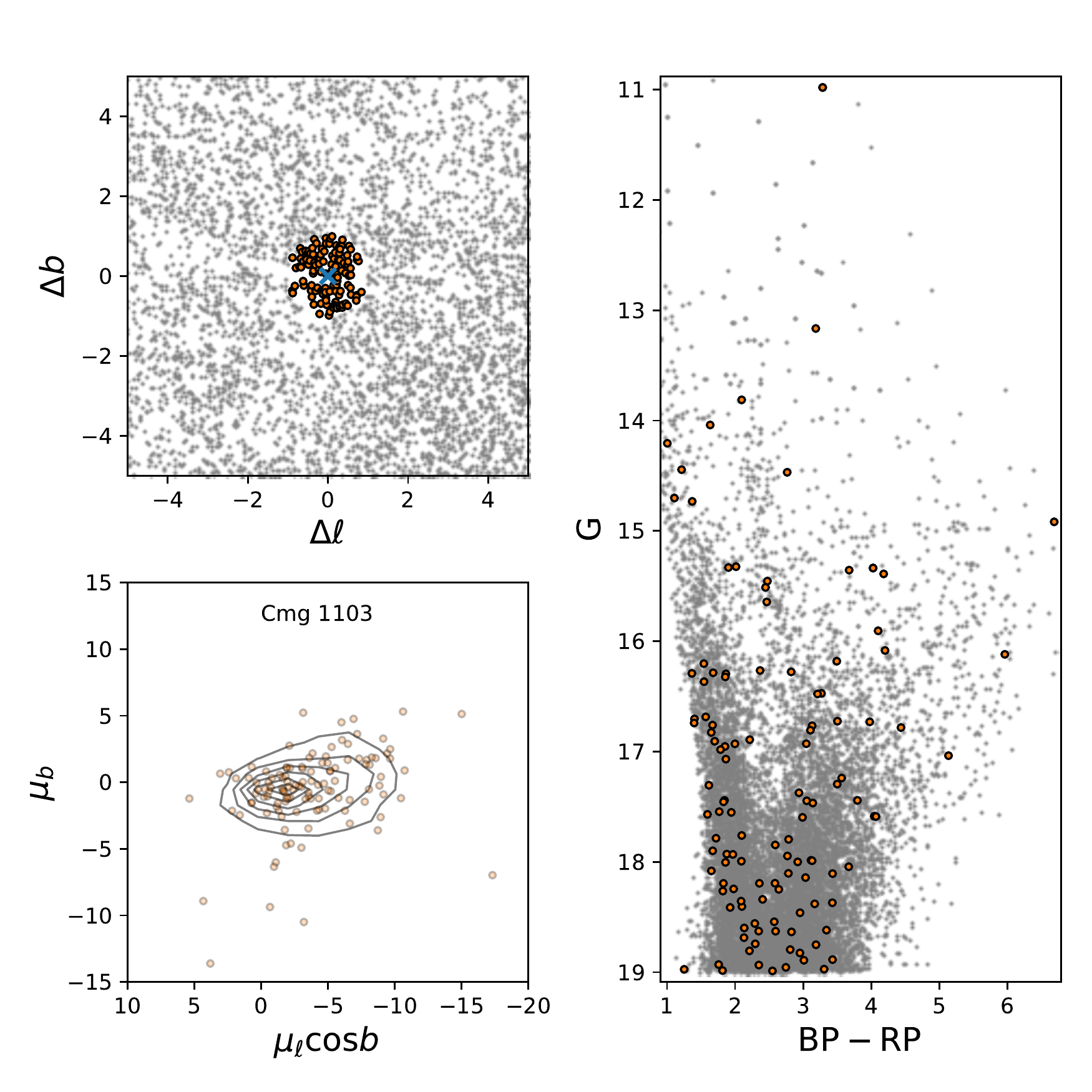} &
\includegraphics[width=8cm]{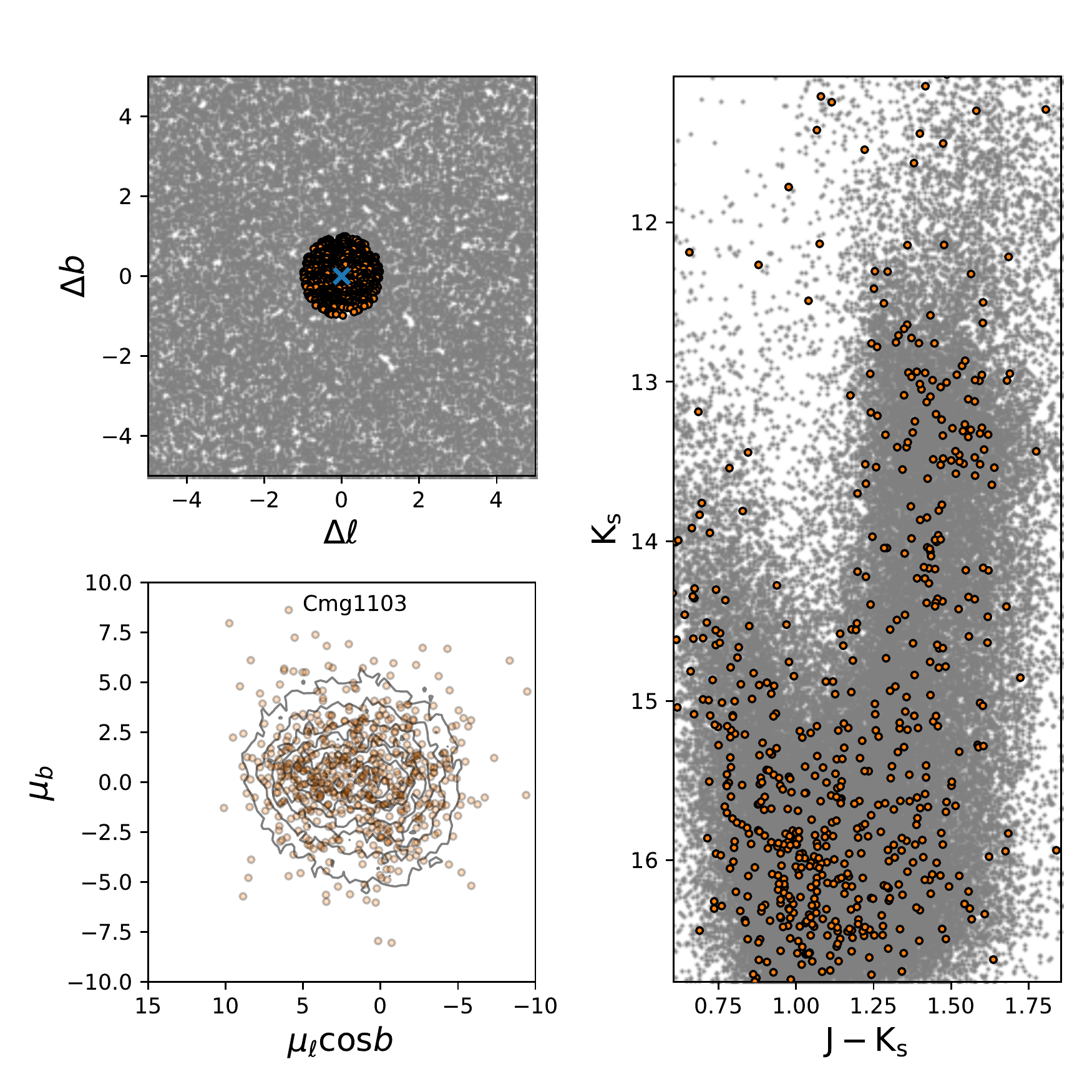}  \\
\includegraphics[width=8cm]{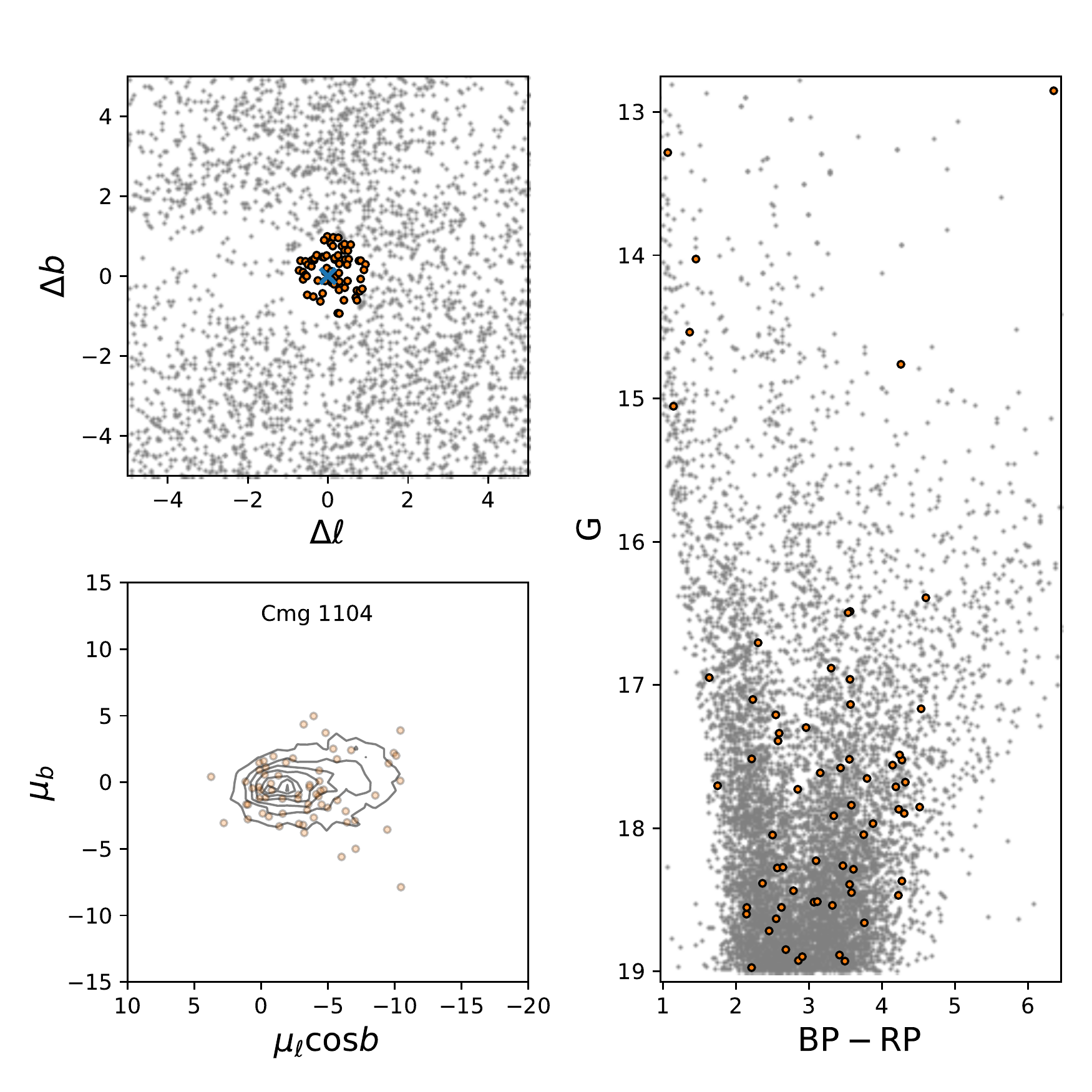} &
\includegraphics[width=8cm]{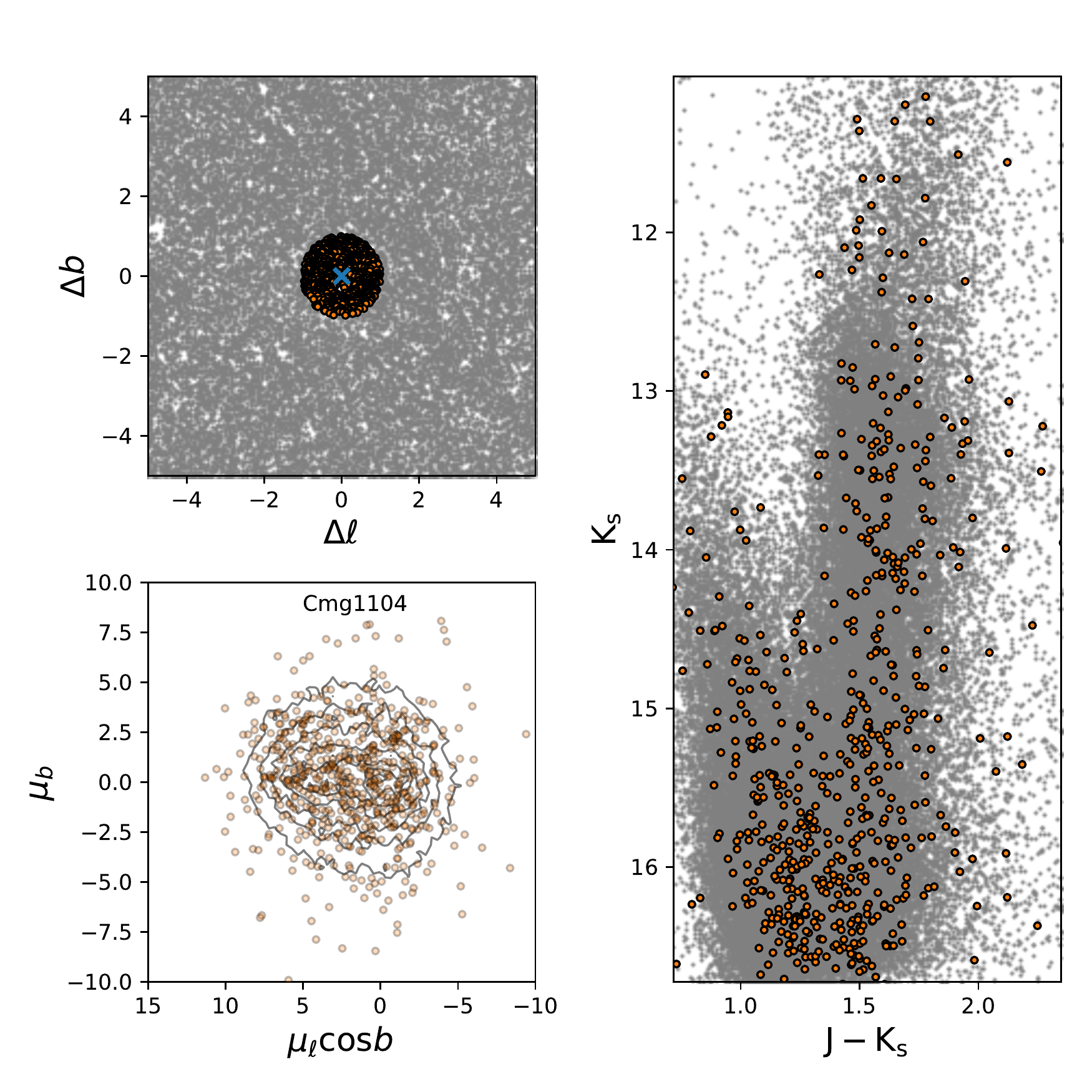}  \\
\includegraphics[width=8cm]{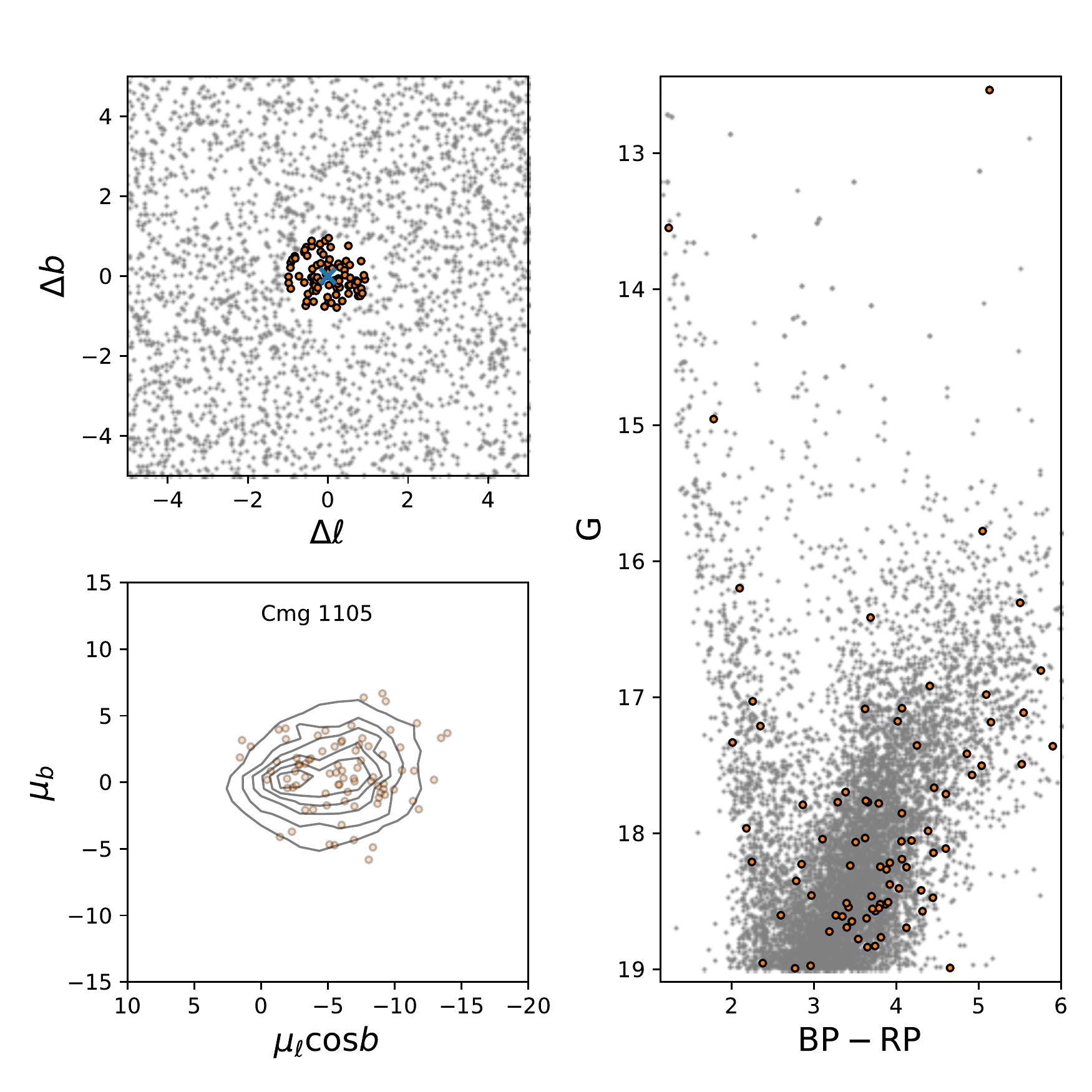} &
\includegraphics[width=8cm]{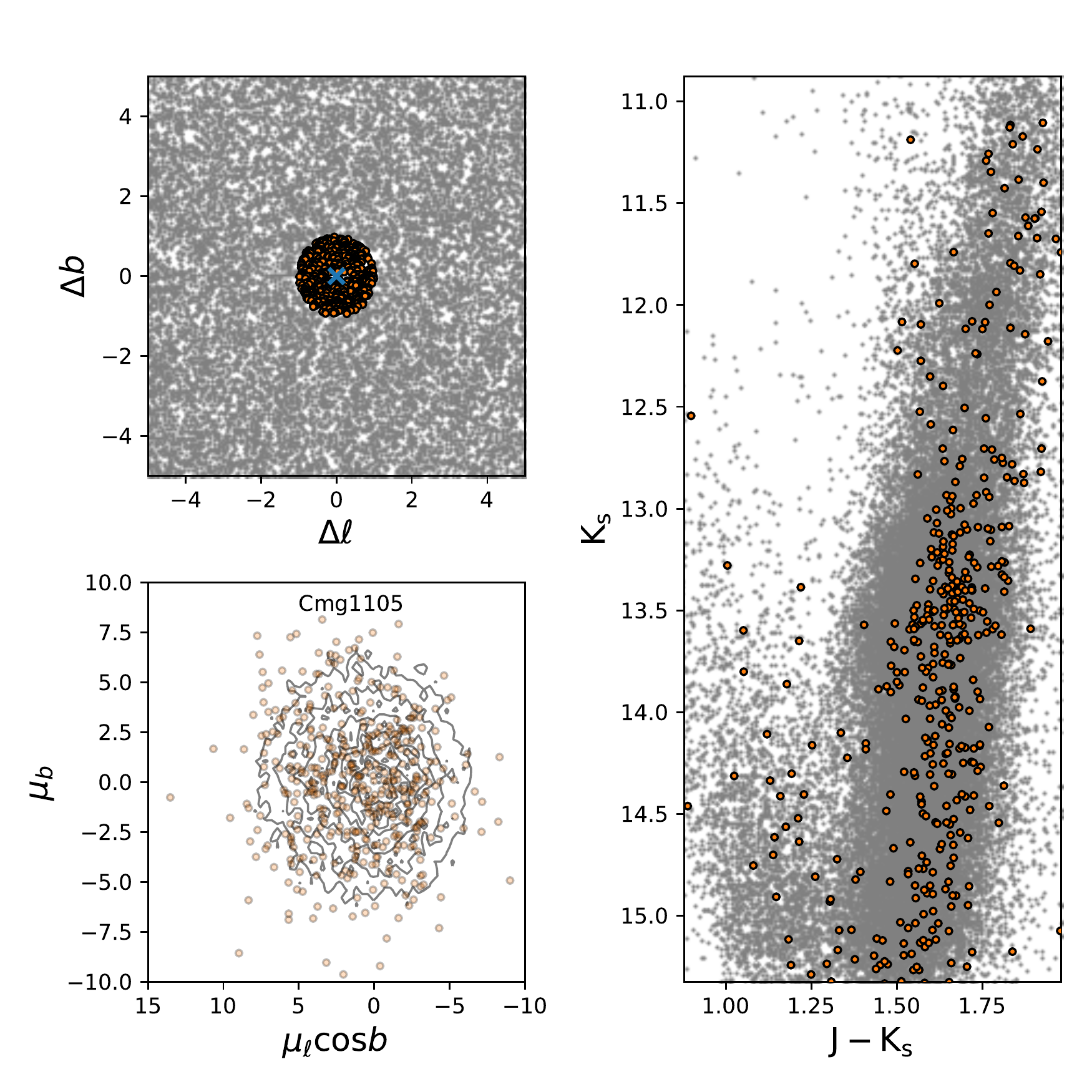}  \\
\end{tabular}
\end{table*}
\newpage
\begin{table*}
\begin{tabular}{cc}
\includegraphics[width=8cm]{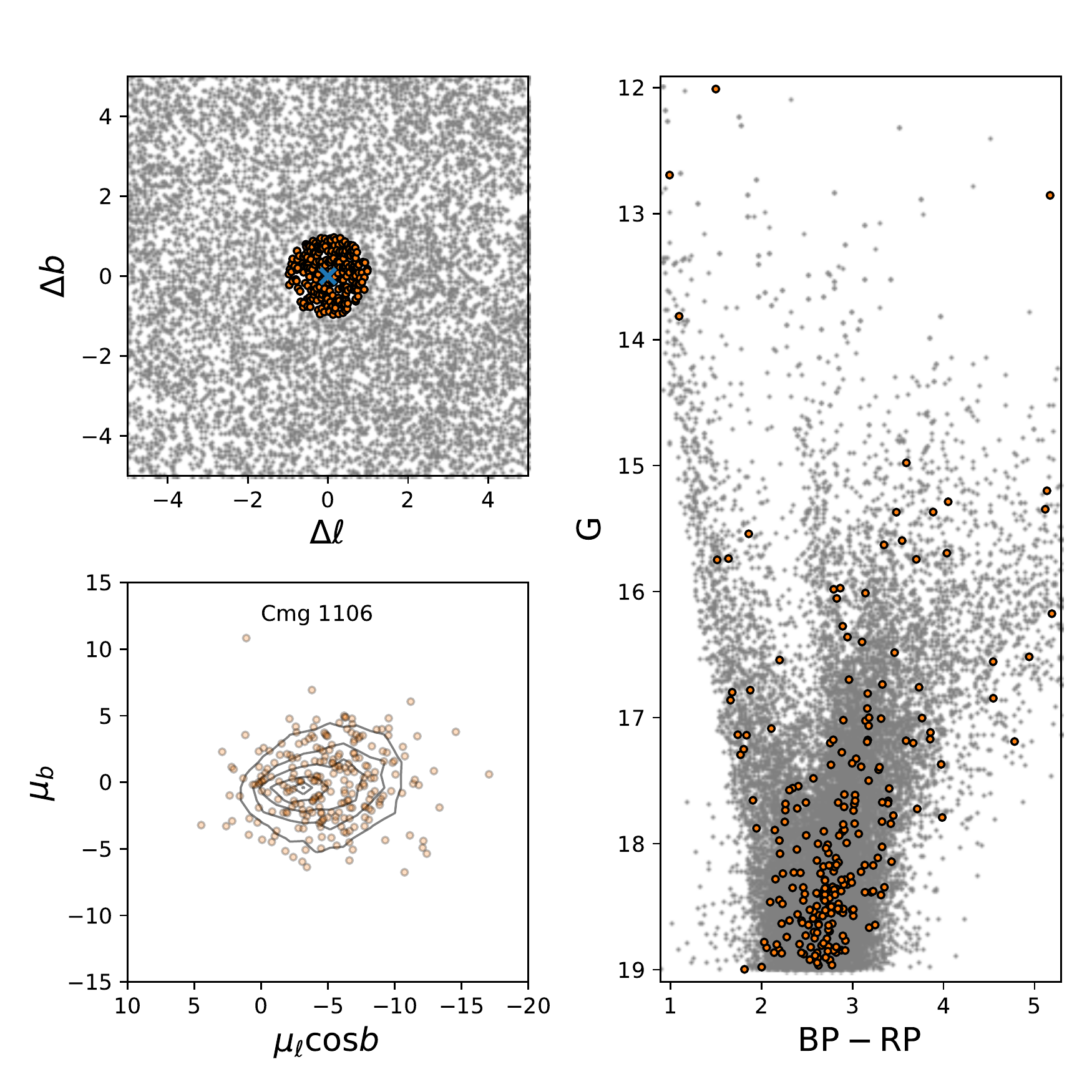} &
\includegraphics[width=8cm]{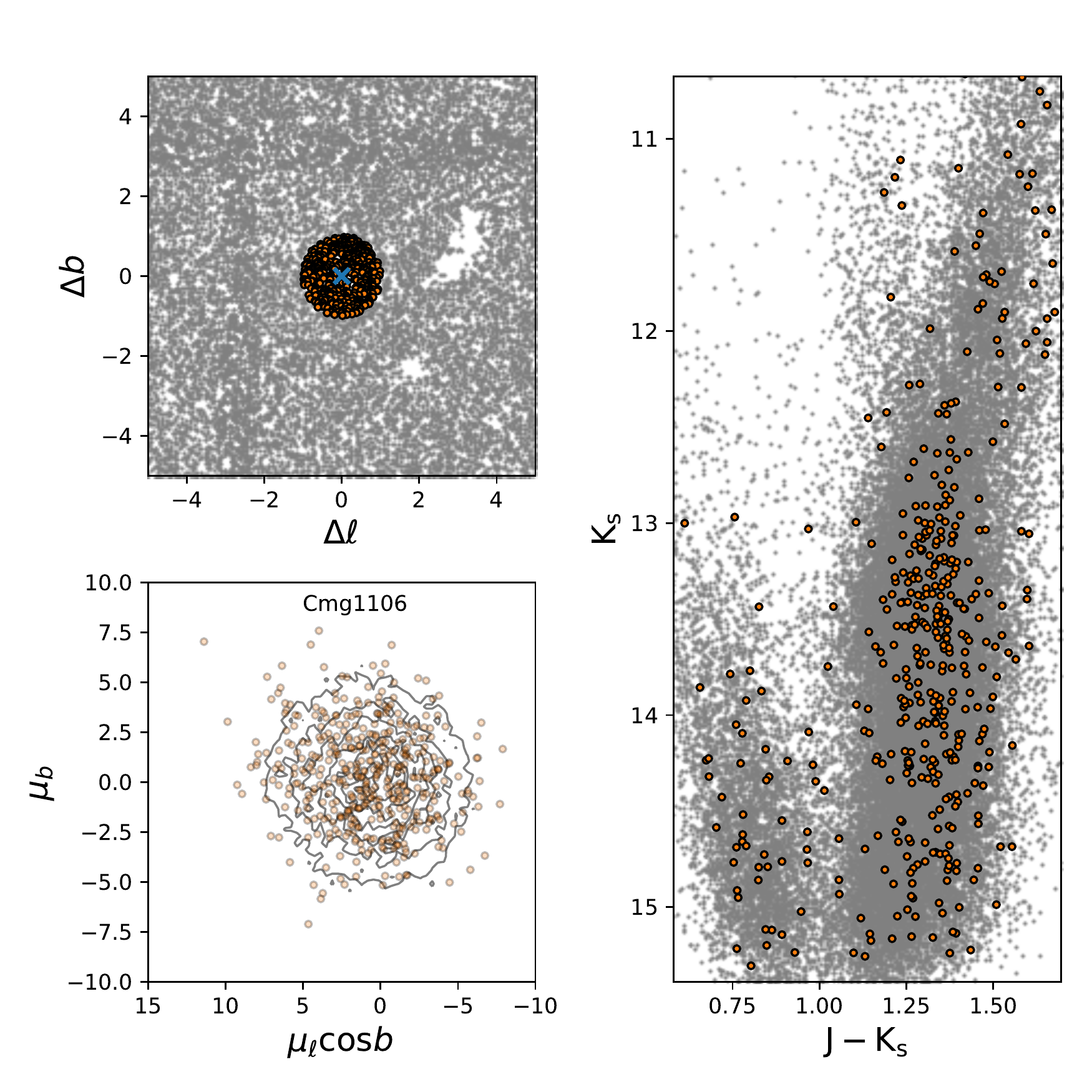} \\
\end{tabular}
\end{table*}

\end{appendix}

\end{document}